\documentclass[11pt,notitlepage]{report}
\usepackage[affil-it]{authblk}

\usepackage{theorem}
\usepackage{amssymb}
\usepackage{amsfonts}     

\usepackage{makeidx}         
\usepackage{graphicx}        
\usepackage{enumerate}

\makeindex             

\newcommand{\updow}{\hspace{0.5mm}{
{\ooalign{{\rotatebox{30}{\hspace{0.5mm}\raisebox{1mm}[0ex][0ex]{{\small$\wedge$}}}}\crcr{$)$}}}
{\ooalign{{\rotatebox{-30}{\hspace{-1.5mm}\raisebox{1.7mm}[0ex][0ex]{{\small$\wedge$}}}}\crcr{$($}}}}\hspace{0.5mm}}
\newcommand{\lefri}{\hspace{3mm}\rotatebox{90}{\hspace{-1.7mm}{
{\ooalign{{\rotatebox{30}{\hspace{0.5mm}\raisebox{1mm}[0ex][0ex]{{\small$\wedge$}}}}\crcr{$)$}}}\hspace{-0.5mm}
{\ooalign{{\rotatebox{-30}{\hspace{-1.5mm}\raisebox{1.7mm}[0ex][0ex]{{\small$\wedge$}}}}\crcr{$($}}}}} \hspace{0.3mm}}

\newcommand{\crosA}{{\hspace{0.5mm} \ooalign{{\large$\nearrow$}\crcr\hspace{-1.5mm}\raisebox{1.8mm}[0ex][0ex]{\footnotesize $\nwarrow$}\crcr \hspace{2mm}\raisebox{-1.6mm}[0ex][0ex]{\footnotesize $\diagdown$}}\hspace{1mm}}}

\newcommand{\crosB}{{\hspace{1mm}\ooalign{{\large$\nwarrow$}\crcr\hspace{2.2mm}\raisebox{1.8mm}[0ex][0ex]{\footnotesize $\nearrow$}\crcr \hspace{-0.9mm}\raisebox{-1.4mm}[0ex][0ex]{\footnotesize $\diagup$}}}}

\theorembodyfont{\normalfont}
\newtheorem{theorem}{Theorem}

\begin{document}
\setlength{\baselineskip}{15pt}
\author[1,2,3]{Keisuke Fujii}
\affil[1]{The Hakubi Center for Advanced Research, Kyoto University}
\affil[ ]{Yoshida-Ushinomiya-cho, Sakyo-ku, Kyoto 606-8501, Japan}
\affil[2]{Graduate School of Informatics, Kyoto University}
\affil[ ]{Yoshida Honmachi, Sakyo-ku, Kyoto 606-8501, Japan}
\affil[3]{Department of Physics, Graduate School of Science, Kyoto University}
\affil[ ]{Kitashirakawa Oiwake-cho, Sakyo-ku, Kyoto 606-8502, Japan}

\title{Quantum Computation with Topological Codes\\
--- from qubit to topological fault-tolerance ---}

\maketitle
\begin{abstract}
This is a comprehensive review on fault-tolerant topological quantum computation 
with the surface codes. The basic concepts and useful tools
underlying fault-tolerant quantum computation,
such as universal quantum computation,
stabilizer formalism, and measurement-based quantum computation, 
are also provided in a pedagogical way.
Topological quantum computation 
by brading the defects on the surface code
is explained in both circuit-based and measurement-based models
in such a way that their relation is clear.
The interdisciplinary connections between quantum error correction codes
and subjects in other fields such as 
topological order in condensed matter physics and 
spin glass models in statistical physics
are also discussed.
This manuscript will be appeared in SpringerBriefs.
\end{abstract}
\tableofcontents

%
%

\chapter*{Preface}
In 1982, Richard Feynman pointed out that
a simulation of quantum systems on classical computers
is generally inefficient because the dimension of the state space
increases exponentially with the number of particles~\cite{Feynman82}.
Instead, quantum systems could be simulated efficiently by other quantum systems.
David Deutsch put this idea forward
by formulating a quantum version of 
a Turing machine~\cite{Deutsch85}.
Quantum computation enables us 
to solve certain kinds of problems that 
are thought to be intractable with classical computers
such as the prime factoring problem
and an approximation of the Jones polynomial.
It has a great possibility to disprove the extended (strong) Church-Turing thesis,
i.e., that any computational process on realistic devices can be simulated 
efficiently on a probabilistic Turing machine.

However, for this statement to make sense,
we need to determine whether or not quantum computation
is a realistic model of computation.
Rolf Landauer criticized it (encouragingly)
by suggesting to put a footnote:
``{\it This proposal, like all proposals for quantum computation, 
relies on speculative technology, 
does not in its current form take into account 
all possible sources of noise, unreliability 
and manufacturing error, and
probably will not work.}"~\cite{LandauerLloyd}.
Actually, quantum coherence,
which is essential for quantum computation
is quite fragile against noise.
If we cannot handle the effect of noise,
quantum computation is of a limiting interest,
like classical analog computers,
as a realistic model of computation. 
To solve this, many researchers have investigated 
the fault-tolerance of quantum computation
with developing quantum error correction techniques.
One of the greatest achievements of this approach 
is topological fault-tolerant quantum computation
using the surface code
proposed by R. Raussendorf {\it et al.}~\cite{RaussendorfAnn,RaussendorfNJP,RaussendorfPRL}.
It says that 
nearest-neighbor two-qubit gates and single-qubit operations
on a two-dimensional array of qubits can perform 
universal quantum computation fault-tolerantly
as long as the error rate per operation is less than $\sim 1\%$.

In this book, 
I present 
a self-consistent review of 
topological fault-tolerant quantum computation
using the surface code.
The book covers 
everything required to 
understand topological 
fault-tolerant quantum computation,
ranging from the definition of the surface code
to topological quantum error correction and 
topological operations on the surface code.
The basic concepts and powerful tools for understanding 
topological fault-tolerant quantum computation, such as 
universal quantum computation, quantum algorithms, stabilizer formalism, 
and measurement-based quantum computation,
are also introduced in the first part (Chapter 1 and Chapter 2) of the book.
In particular, in Chapter 1, I also mention
a quantum algorithm for approximating 
the Jones polynomials,
which is also related to topological quantum computation
with braiding non-Abelian anyons.
In Chapter. 3, the definition of the surface code and 
topological quantum error correction on it is explained. 
In Chapter 4, topological quantum computation on the surface code
is described in the circuit-based model,
where topological diagrams are introduced
to understand the logical operations on the surface code diagrammatically.
In Chapter. 5, 
I explain the same thing in the measurement-based model,
as done in the original proposal~\cite{RaussendorfAnn}.
Hopefully, it would be easy to see how these two viewpoints are related.

Throughout the book, 
I have tried to explain the quantum operations 
using circuit and topological diagrams 
so that the readers can get a graphical understanding of the operations.
The graphical understanding should be helpful to study
the subjects more efficiently.
Topological quantum error correction codes 
are a nice play ground for studying
the interdisciplinary connections between 
quantum information and other fields of physics,
such as condensed matter physics and statistical physics.
Actually, there is a nice correspondence between 
topological quantum error correction codes and topologically ordered
systems in condensed matter physics.
Furthermore, if we consider a decoding problem of 
a quantum error correction code,
a partition function of a random statistical mechanical model
is naturally appeared as a posterior probability for the decoding.
These interdisciplinary topics are also included in Chapter 3.

Almost all topics, except for the basic concepts 
in the first part, are based on the results
achieved after the appearance of the standard textbook
of quantum information science
entitled {\it``Quantum Computation and Quantum Information"} (Cambridge University Press 2000) by M. A. Nielsen and I. L. Chuang.
In this sense, the present comprehensive review 
on these topics would be helpful to learn and update
the recent progress efficiently.
In this book, 
I concentrated on the quantum information aspect 
of topological quantum computation.
Unfortunately, I cannot cover 
the more physical and condensed matter aspects of 
topological quantum computation, such as non-Abelian anyons
and topological quantum field theory.
In this sense, this book is complemented by the book 
{\it``Introduction to topological quantum computation"}
(Cambridge University Press 2012)
written by J. K. Pachos.
Readers who are interested in the more physical aspects
of topological quantum computation are recommended to read it.

Hopefully, this review will encourage 
both theoretical and experimental
researchers to find a more feasible way of quantum computation.
It will also bring me great pleasure if this review provides 
an opportunity 
to reunify and refine
various subdivided fields of modern physics 
in terms of quantum information.

\vspace{\baselineskip}
\begin{flushright}\noindent
Kyoto, Japan
\hfill {\it Keisuke  Fujii}, \\
April, 2015
\\
\end{flushright}




\chapter{Introduction to quantum computation}
In this chapter, we introduce the basic concepts of quantum computation.
We first describe the minimum unit of quantum information, the qubit, and define several gate operations for it.
Then, we explain the Solovay-Kitaev algorithm, which provides a way to decompose an arbitrary single-qubit unitary operation into an elementary set of single-qubit gates.
Using multi-qubit gates, we construct an arbitrary $n$-qubit unitary operation from an elementary universal set of gates, which we call universal quantum computation.
Quantum algorithms, which run on a universal quantum computer, are also presented. 
One example of this is Shor's prime factoring algorithm based on the phase estimation algorithm.
Another is an approximation of the Jones polynomial.
Finally, we will introduce quantum noise and see how we can describe a quantum system coupled with an environment.

\section{Quantum bit and elementary operations}
\begin{figure}[t]
\centering
\includegraphics[width=100mm]{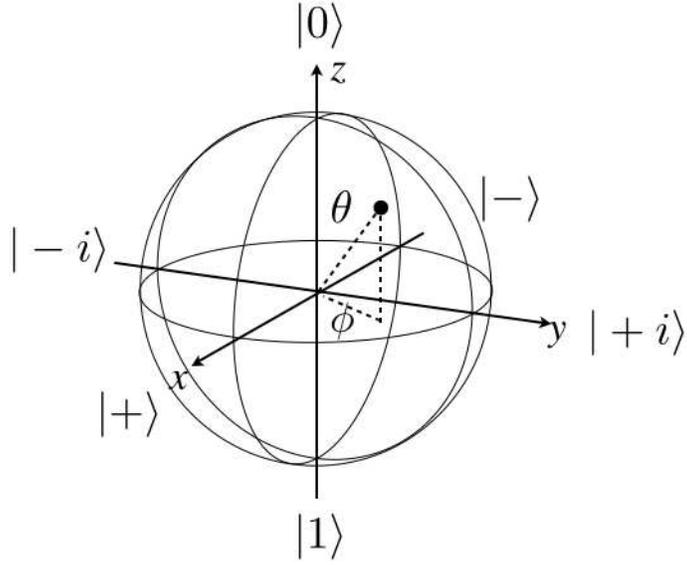}
%
%
\caption{The Bloch sphere.}
\label{fig01}       
\end{figure}

In classical information science, the minimum unit of information is described by a binary digit or {\it bit}, which takes the value 0 or 1.
Its quantum counterpart is a quantum bit, the so-called {\it qubit}\index{qubit}.
The qubit is defined as a linear superposition of two orthogonal quantum states $|0\rangle = \left( \begin{array}{c}1\\0 \end{array} \right)$ and 
$|1\rangle = \left( \begin{array}{c}0\\1 \end{array} \right)$,
\begin{eqnarray}
|\psi\rangle = \alpha |0\rangle + \beta |1\rangle,
\end{eqnarray}
where $\alpha$ and $\beta$ are arbitrary complex values satisfying $|\alpha|^2 + |\beta|^2=1$.
The complex amplitudes can be expressed as
\begin{eqnarray}
\alpha = \cos \frac{\theta}{2} , \;\;\; \beta = e^{i \phi} \sin \frac{\theta}{2},
\end{eqnarray}
up to an unimportant global phase.
By using the angles $\theta$ and $\phi$, the qubit can be mapped onto a point on a three-dimensional (3D) sphere, the so-called {\it Bloch sphere}\index{Bloch sphere}, as shown in Fig.~\ref{fig01}.

The time evolution, which maps a quantum state into another, is given as a unitary operator in quantum mechanics.
The most important operators are the Pauli operators\index{Pauli operator}
\begin{eqnarray}
I
=
\left(
\begin{array}{cc}
1 & 0
\\
0 & 1
\end{array}
\right),\ 
X
=
\left(
\begin{array}{cc}
0 & 1
\\
1 & 0
\end{array}
\right),\ 
Y
=
\left(
\begin{array}{cc}
0 & -i
\\
i & 0
\end{array}
\right),\ 
Z
=
\left(
\begin{array}{cc}
1 & 0
\\
0 & -1
\end{array}
\right).
\end{eqnarray}
The computational basis states $\{ |0\rangle , |1\rangle\}$ are eigenstates of the Pauli $Z$ operator.
The Pauli $X$ operator flips the computational basis state:
\begin{eqnarray}
|1\rangle = X|0\rangle, \;\;\; |0\rangle = X|1\rangle.
\end{eqnarray}
We define the eigenstates of the Pauli $X$ operator as
\begin{eqnarray}
|+\rangle = \frac{|0\rangle + |1\rangle}{\sqrt{2}},
\;\;\;
|-\rangle = \frac{|0\rangle - |1\rangle}{\sqrt{2}},
\end{eqnarray}
which we call the $X$-basis states.
Similarly, the eigenstates of the Pauli $Y$ operator are defined as 
\begin{eqnarray}
|+i\rangle = \frac{|0\rangle + i|1\rangle}{\sqrt{2}},
\;\;\;
|-i\rangle = \frac{|0\rangle - i|1\rangle}{\sqrt{2}},
\end{eqnarray}
which we call the $Y$-basis state.

The second most important operators are the Hadamard $H$ and phase $S$ operators
\begin{eqnarray*}
H=
\frac{1}{\sqrt{2}}
\left(
\begin{array}{cc}
1 & 1
\\
1 & -1
\end{array}
\right)
\textrm{ and }
S=
\left(
\begin{array}{cc}
1 & 0
\\
0 & i
\end{array}
\right).
\end{eqnarray*}
These gates transform between the different Pauli-basis states, i.e., $H: \{ |0\rangle , |1\rangle\} \leftrightarrow \{ |+\rangle, |-\rangle \}$ and
$S: \{ |+ \rangle , |-\rangle\} \leftrightarrow \{ |+i \rangle, |-i\rangle\}$.
Equivalently, we may say that these operators 
transform a Pauli operator into another Pauli operator under their conjugations:
\begin{eqnarray}
X=HZH, \;\;\; Y=SXS^{\dag}.
\end{eqnarray}
From this property, the $H$ and $S$ gates are called Clifford gates\index{Clifford gate}.
The Clifford gates are depicted as circuit diagrams as follows:
\begin{center}
\includegraphics[width=100mm]{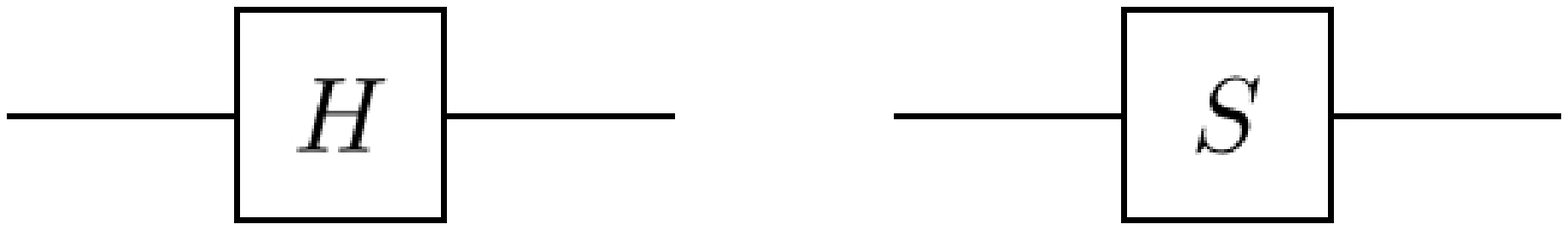}
\end{center}

If we measure the qubit $|\psi \rangle$ in the $Z$-basis $\{ |0\rangle, |1\rangle \}$, we obtain the measurement outcomes 0 and 1 with probabilities 
\begin{eqnarray}
p_0 &=& |\langle 0 | \psi \rangle |^2 = {\rm Tr}\left[
|0\rangle\langle 0| \psi \rangle \langle \psi | \right] ,
\\
p_1 &=& |\langle 1 | \psi \rangle |^2 
= {\rm Tr}\left[|1\rangle\langle 1| \psi \rangle \langle \psi | \right],
\end{eqnarray}
respectively.
More generally, we may use the measurement operators $\{ M_i \}$, satisfying that $E_i \equiv M_i^{\dag}M_i$ is positive semidefinite and that $\sum _i E_i =I $.
(If an operator $A$ satisfies ${}^\forall |u\rangle$, $\langle u|A|u\rangle \geq 0$, it is said to be positive semidefinite.)
The probability of obtaining the measurement outcome $i$ is given by
\begin{eqnarray}
p_i = \langle \psi | M^{\dag}_i M_i | \psi \rangle 
= {\rm Tr} \left[ M^{\dag}_i M_i | \psi \rangle \langle \psi | \right]
= {\rm Tr} \left[ E_i | \psi \rangle \langle \psi | \right].
\end{eqnarray}
Such a measurement and the set of positive operators $\{ E_i = M_i ^{\dag} M_i\}$, are called a positive-operator-valued measure (POVM) measurement 
\index{positive-operator-valued measure measurement, POVM measurement} and POVM elements, respectively.
The post-measurement state conditioned on the measurement outcome $i$ is 
\begin{eqnarray}
\frac{M_i | \psi \rangle}{\sqrt{{\rm Tr} [ M^{\dag}_i M_i | \psi \rangle \langle \psi | ]}}.
\end{eqnarray}

Suppose we have quantum states $|\psi\rangle$ and $|\phi\rangle$ with probability $p_\psi$ and $p_{\phi}$, respectively.
We perform a measurement with the measurement operator $\{ M_i \}$.
If we assume that the measurement outcome $i$ is obtained with probability 
\begin{eqnarray}
p_i = p_{\psi} {\rm Tr} \left[ M^{\dag}_i M_i | \psi \rangle \langle \psi | \right]
+ p_{\phi} {\rm Tr} \left[ M^{\dag}_i M_i | \phi \rangle \langle \phi | \right],
\end{eqnarray}
we can express the classical mixture of $|\psi\rangle$ and $|\phi\rangle$ as an operator
\begin{eqnarray}
\rho =  p_\psi | \psi \rangle \langle \psi | + p_{\phi} | \phi \rangle \langle \phi |,
\end{eqnarray}
which is called the {\it density matrix} or {\it density operator}.
More generally, if we have pure quantum states $\{ |\psi _ k\rangle \}$ with probability $\{ p_{k}\}$, the density matrix is given by 
$\rho = \sum _k p_k | \psi \rangle \langle \psi |$.
A density matrix, which has information of a quantum state including its statistical property, has to be a positive hermitian (self-adjoint) operator $\rho$ satisfying 
${\rm Tr}\left[ \rho \right]=1$. 
Note that for a given density matrix $\rho$,
the pure state decomposition of it is not uniquely determined,
and any decomposition has the same statistical property.

A mixed state of a qubit $\rho$ can be represented as a point inside the Bloch sphere through the three coordinates calculated from the density matrix
\begin{eqnarray}
(r_x, r_y, r_z) = ( {\rm Tr}[X \rho],{\rm Tr}[Y \rho], {\rm Tr}[Z \rho] ).
\end{eqnarray}
If the state is a pure state $|\psi \rangle = \cos \frac{\theta }{2}|0\rangle+ e^{i \phi} \sin \frac{\theta}{2} |1\rangle$, the coordinates can be calculated to be
\begin{eqnarray}
(r_x, r_y, r_z) =(\sin \theta  \cos \phi,\sin \theta \sin \phi,\cos \theta ),
\end{eqnarray}
which is consistent with the previous definition.
The completely mixed state $I/2$ corresponds to the origin of the coordinate system.

\section{The Solovay-Kitaev algorithm}
The Pauli operators $\{I, X,Y,Z\}$ and single-qubit Clifford operators $\{H,S\}$ form a finite group, and hence cannot cover all unitary operations for a qubit.
If a non-Clifford operation exists, e.g., $e^{-i (\pi/8)Z}$, we can generate an arbitrary single-qubit unitary operation using the Solovay-Kitaev algorithm~\cite{DawsonNielsen05}.
The underlying Solovay-Kitaev theorem states that if a set of single-qubit operations generates a dense subset of $SU(2)$, then that set is guaranteed to fill $SU(2)$ quickly.

Suppose we have a basic ($0$th order) approximation $U_0$ of an arbitrary unitary operator $U$, and that it approximates $U$ with a certain constant error $\epsilon _0$:
\begin{eqnarray}
\Vert  U_0 - U \Vert  \leq \epsilon _0,
\end{eqnarray}
where $\Vert  ...\Vert $ indicates an operator norm.
The Solovay-Kitaev algorithm takes the $(n-1)$th order approximation $U_{n-1}$ with an error $\epsilon _{n-1}$ and returns the $n$th order approximation $U_n$ with an error $\epsilon _n$ as follows.

First, $U U^{\dag}_{n-1}$ is decomposed in terms of the unitary operators $V$ and $W$ as a group commutator:
\begin{eqnarray}
U U_{n-1}^{\dag}= VWV^{\dag}W^{\dag} ,
\label{eq:BGC}
\end{eqnarray}
where $V$ and $W$ are chosen such that 
\begin{eqnarray}
\Vert  V - I \Vert  < c \sqrt{\epsilon _{n-1}}, \;\;\; \Vert  W - I \Vert  < c \sqrt{\epsilon _{n-1}},
\end{eqnarray}
with a constant $c$.
The above decomposition always exists from the following argument.
Let $V$ and $W$ be rotations with an angle $\phi$ about the $x$- and $y$-axes, respectively.
Then $VWV^{\dag}W^{\dag}$ is a rotation with an angle $\theta$ about some axis, where
\begin{eqnarray}
\sin (\theta /2) = 2 \sin ^2 (\phi /2) \sqrt{1-\sin ^4 (\phi /2)}.
\end{eqnarray}
We assume that $\phi , \theta \ll 1$, and hence $\theta \simeq \phi^2$.
By inverting the above argument, if
\begin{eqnarray}
\Vert  I- VWV^{\dag}W^{\dag} \Vert  \simeq \theta /2 + O(\theta ^3)
\end{eqnarray}
we can find $V$ and $W$ such that
\begin{eqnarray}
\Vert  I - V\Vert  ,\;\; \Vert  I-W \Vert  \simeq \phi/2 + O(\phi^3).
\end{eqnarray}
Because
\begin{eqnarray}
\Vert I -U U_{n-1}^{\dag} \Vert  = \Vert  I- VWV^{\dag}W^{\dag} \Vert < \epsilon _{n-1},
\end{eqnarray}
we can find $V$ and $W$ such that $\Vert  I -V\Vert , \Vert I-W\Vert  <c \sqrt{\epsilon _{n-1}}$.

Second, we calculate the $(n-1)$th order approximations $V_{n-1}$ and $W_{n-1}$ of $V$ and $W$, respectively. 
Then the Solovay-Kitaev algorithm returns the $n$th order approximation 
\begin{eqnarray}
U_n = V_{n-1} W_{n-1} V^{\dag}_{n-1} W^{\dag}_{n-1} U_{n-1},
\end{eqnarray}
which satisfies 
\begin{eqnarray}
\Vert  U_n - U \Vert  \leq \epsilon _{n}.
\end{eqnarray}
Next, we will calculate $\epsilon _{n}$ as a function of $\epsilon _{n-1}$.
By using the property of the operator norm, we obtain 
\begin{eqnarray}
\Vert  U_n - U \Vert  &\leq &
\Vert V_{n-1} W_{n-1} V^{\dag}_{n-1} W^{\dag}_{n-1} - U U_{n-1}^{\dag} \Vert  \Vert  U_{n-1} \Vert 
\\
&\leq& 
\Vert V_{n-1} W_{n-1} V^{\dag}_{n-1} W^{\dag}_{n-1} - VWV^{\dag}W^{\dag} \Vert 
 \Vert  U_{n-1} \Vert 
\end{eqnarray}
By denoting
$V_{n-1} = V+ \Delta _V, \;\;\;  W_{n-1} = W+ \Delta _W$,
we obtain
\begin{eqnarray}
\Vert  U_n - U \Vert  &\leq & \Vert V_{n-1} W_{n-1} V^{\dag}_{n-1} W^{\dag}_{n-1} - VWV^{\dag}W^{\dag} \Vert 
 \\ &\leq &
\Vert   \Delta _V W V^{\dag} W^{\dag} 
+ VW  \Delta_V^{\dag} W^{\dag}   \Vert 
+
\Vert   V \Delta _W V^{\dag} W^{\dag} 
+  VW V^{\dag} \Delta_W^{\dag} \Vert 
+O(\Delta ^2).
\nonumber \\
\end{eqnarray}
Moreover, in terms of $\delta _V$ and $\delta _W$
defined by $V= I+ \delta _V$ and $W = W+ \delta _W$
respectively,
we obtain
\begin{eqnarray}
 \Vert  U_n - U \Vert  &\leq &
\Vert   \Delta _V  V^{\dag} 
+ V  \Delta_V^{\dag}    \Vert  
+
\Vert   \Delta _W  W^{\dag} 
+  W \Delta_W^{\dag} \Vert  
\\
&&
+ \Vert \Delta _V \delta _{W}\Vert 
+\Vert \Delta _V^{\dag} \delta _{W}^{\dag}\Vert 
+\Vert  \delta _V \Delta _W \Vert  + \Vert  \delta _V^{\dag} \Delta_W^{\dag}\Vert + O(\Delta ^2). 
\\ &\leq &
4 c\epsilon _{n-1}^{3/2}+O(\Delta ^2, \delta^2 \Delta) .
\end{eqnarray}
Here we have used that 
\begin{eqnarray}
\Delta _V  V^{\dag} 
+ V  \Delta_V^{\dag}   = \Delta _V \Delta _{V}^{\dag},
\end{eqnarray}
which can be derived from the unitarity of $V$ and $V_{n-1}=V+\Delta _V$.
To leading order, the error is given by $\epsilon _n \leq c' \epsilon _{n-1}^{3/2}$ with a certain constant $c'$ and is calculated to be 
\begin{eqnarray}
\epsilon _n \leq (2c' \epsilon _0)^{(3/2)^n}/(2c').
\label{eq:KSerror}
\end{eqnarray}
If $\epsilon _0 < 1/(2c')$, the error decreases super-exponentially in $n$.
On the other hand, the $n$th order approximation calls the $(n-1)$th order approximation three times, i.e., through $U_{n-1}$, $V_{n-1}$, and $W_{n-1}$.
Including the resource $R_D$ required for the decomposition (\ref{eq:BGC}), the overhead $R_n$ for the $n$th order approximation is given by
\begin{eqnarray}
&&R_n = 3R_{n-1} + R_D
\\
&\Leftrightarrow& R_n = O(3^n).
\label{eq:KSresource}
\end{eqnarray} 
Similarly, the number of unitary operations employed in the $n$th order approximation is calculated to be $M_n = O(5^n)$.
By using (\ref{eq:KSerror}) and (\ref{eq:KSresource}), the overhead $R_{\epsilon}$ and the number $M_{\epsilon}$ of gates required to obtain an approximation of $U$ with an error $\epsilon$ can be estimated: 
\begin{eqnarray}
R_{\epsilon} &=& O(\ln ^{\ln 3/\ln (3/2)} (1/\epsilon)),
\\
M_{\epsilon} &=& O(\ln ^{\ln 5/\ln (3/2)} (1/\epsilon)).
\end{eqnarray}
Thus, both $R_{\epsilon}$ and $M_{\epsilon}$ scale as polylogarithmic functions of $1/\epsilon$.

\section{Multi-qubit gates}
An $n$-qubit state is given by a superposition of tensor product states
\begin{eqnarray*}
| \Psi \rangle = \sum _{i_1 , i_2 ,... , i_n} C_{i_1 i_2 ... i_n} | i_1 i_2 ... i_n\rangle,
\label{eq01}
\end{eqnarray*}
where $i_k =0,1$ and $| i_1 i_2 ... i_n\rangle \equiv |i_1 \rangle \otimes |i_2 \rangle \otimes ... \otimes |i_n\rangle$.
A single qubit gate $A$ acting on the $k$th qubit is denoted by
\begin{eqnarray}
A_k = \overbrace{I \otimes \cdots \otimes I}^{k-1} \otimes A \otimes \overbrace{I \otimes \cdots I}^{n-k-1}.
\end{eqnarray}
An important two-qubit gate is the controlled-NOT (CNOT) gate,
\begin{eqnarray*}
\Lambda_{c,t} (X) = |0\rangle \langle 0| _{c} 
I_{t} + |1\rangle \langle 1|_{c}  X_{t}.
\end{eqnarray*}
For a computational basis input state $|i\rangle _{c} | j \rangle _{t}$, the CNOT gate acts as $\Lambda_{c,t}(X) |i\rangle _c |j \rangle _t = |i \oplus j \rangle$.
In this sense, the CNOT gate is a quantum generalization of the XOR operation in classical computation.
If the input state is $|+\rangle _{c} |0\rangle_{t}$, the output of the CNOT gate is a maximally entangled state:
\begin{eqnarray}
\Lambda _{c,t} (X) |+\rangle _{c}|0\rangle _{t} 
= (|00\rangle + |11\rangle )/\sqrt{2}.
\end{eqnarray}
The CNOT gate is depicted by a circuit diagram as follows:
\begin{center}
\includegraphics[width=50mm]{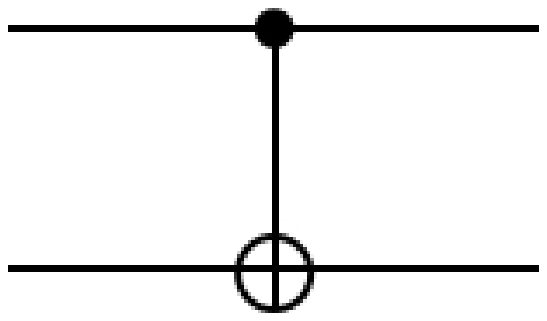}
\end{center}

We may also define the controlled-$Z$ (CZ) gate,
\begin{eqnarray*}
\Lambda_{c,t} (Z) = |0\rangle \langle 0| _{c} 
I_{t} + |1\rangle \langle 1|_{c}  Z_{t}.
\end{eqnarray*}
The CZ operation is symmetric; $\Lambda _{c,t}(Z) = \Lambda _{t,c}(Z)$, and is represented by the circuit diagram
\begin{center}
\includegraphics[width=50mm]{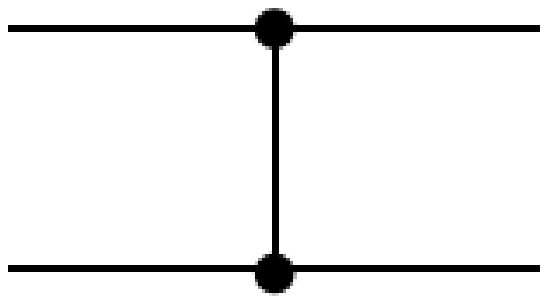}
\end{center}

Because the Pauli $Z$ operator is transformed into the Pauli $X$ operator by the Hadamard gate, we have the following relation between the CZ and CNOT gates:
\begin{center}
\includegraphics[width=100mm]{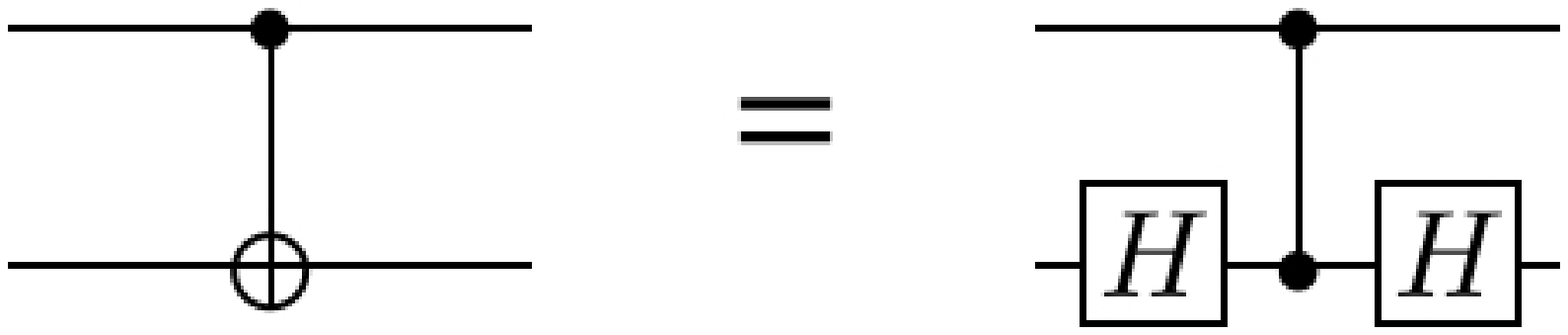}
\end{center}

These CNOT and CZ gates are both Clifford gates, i.e., $\Lambda (A)$ ($A=X,Z$) transforms the two-qubit Pauli group onto itself under the conjugation 
$\Lambda (A) [ \cdots ] \Lambda (A)^{\dag}$.
For example,
$\Lambda (X) _{c,t} X_c \otimes I_t \Lambda (X) _{c,t}^{\dag} = X_c \otimes X_t$,
$\Lambda (X) _{c,t} I_c \otimes Z_t \Lambda (X) _{c,t}^{\dag} = Z_c \otimes Z_t$,
$\Lambda (Z) _{c,t} X_c \otimes I_t \Lambda (Z) _{c,t}^{\dag} = X_c \otimes Z_t$,
etc.

For an arbitrary unitary operator $U$, the controlled-$U$ gate is denoted by
\begin{eqnarray}
\Lambda _{c,t} (U) = |0\rangle \langle 0|_c I_t
+ |1\rangle \langle 1|_c  U_t,
\end{eqnarray}
where the qubits $c$ and $t$ are called the control and target qubits, respectively.
By decomposing a single-qubit unitary gate into $U=e^{ i \alpha }A X BXC$ with the unitary operators $A,B,C$ satisfying $ABC=I$, the controlled-$U$ operation $\Lambda(U)$ can be implemented as follows:
\begin{center}
\includegraphics[width=120mm]{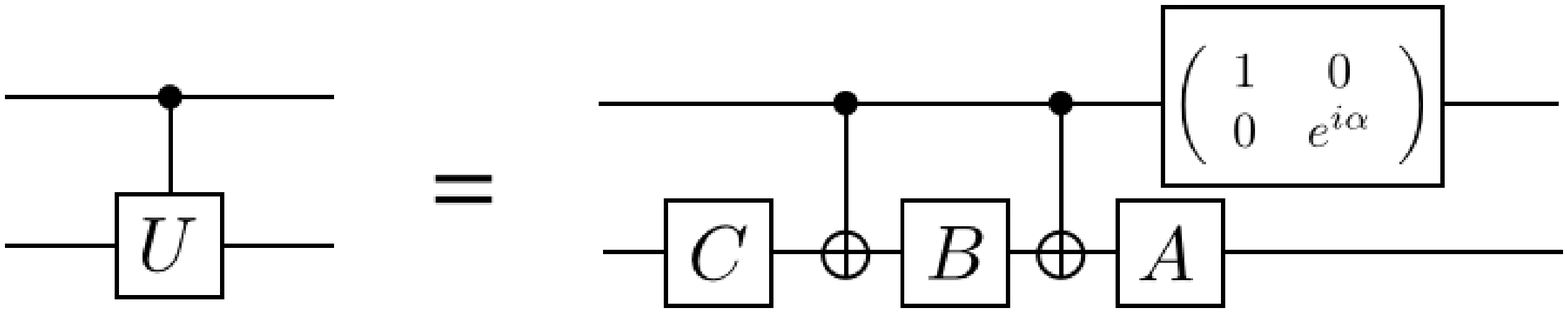}
\end{center}
(Note that we always have such a decomposition for an arbitrary single-qubit 
gate~\cite{Barenco95,NielsenChuang}.)

For example, the controlled-Hadamard gate $\Lambda(H)$ can be represented by
\begin{center}
\includegraphics[width=100mm]{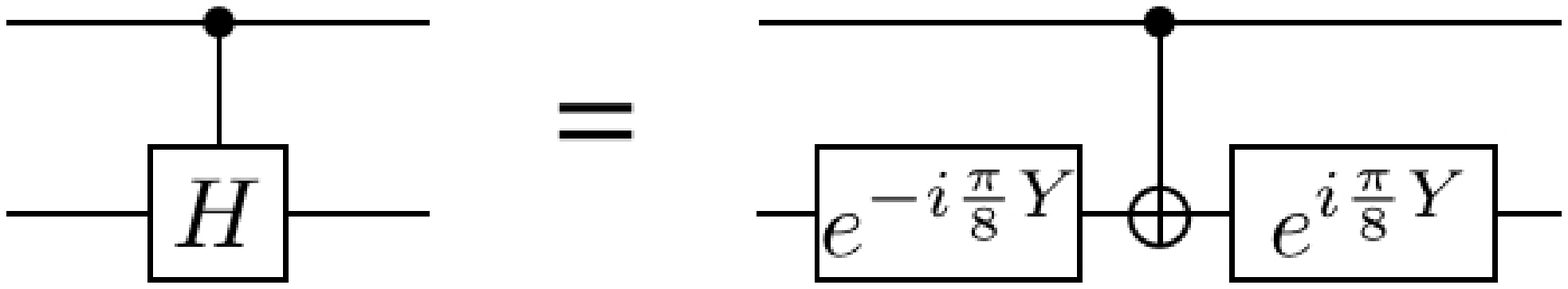}
\end{center}

Next, we will discuss one of the most important multi-qubit gates, the Toffoli gate:
\begin{eqnarray}
\Lambda^2 _{c_1, c_2,t} (X) = (I_{c_1}I_{c_2} - |1\rangle \langle 1|_{c_1}|1\rangle \langle 1|_{c_2} ) I_{t} + |1\rangle \langle 1|_{c_1}|1\rangle \langle 1|_{c_2}   X_t,
\end{eqnarray}
which is depicted as a circuit diagram:
\begin{center}
\includegraphics[width=30mm]{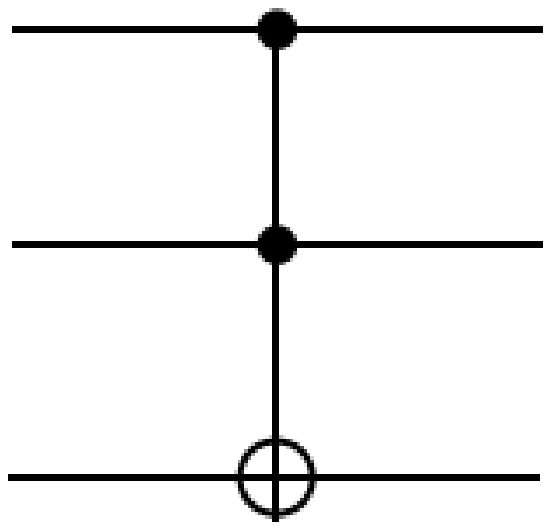}
\end{center}
For a computational basis input state $|i_1\rangle _{c_1} |i_2 \rangle_{c_2} |j\rangle _{t}$, the Toffoli gate acts as
\begin{eqnarray}
\Lambda^2 _{c_1, c_2,t} (X)|i_1\rangle _{c_1} |i_2 \rangle_{c_2} |j\rangle _{t} = |i_1\rangle _{c_1} |i_2 \rangle_{c_2} |j\oplus (i_1 \cdot i_2) \rangle _{t}.
\end{eqnarray}
The state of the third qubit is equivalent to the output of the NAND operation in classical computation.
In this sense, the Toffoli operation can be regarded as a quantum extension of the NAND operation.
The NAND operations are known to be universal in classical computation in the sense that any logic gate (boolean function) can be constructed from them.
This implies that quantum computation trivially includes classical computation.
More importantly, because unitary operations are reversible, quantum computation can simulate classical computation in a reversible way.
Suppose we want to calculate a boolean function $f(x)$ for an input state $x$.
Then, we can construct a quantum circuit $U_f$ consisting of Toffoli and Pauli $X$ gates:
\begin{eqnarray}
U_f|x\rangle _{\rm input}|0...0\rangle _{\rm ancilla} |0\rangle_{\rm answer} = |x\rangle _{\rm input} |g(x) \rangle _{\rm ancilla} |f(x) \rangle_{\rm answer},
\end{eqnarray}
where the qubits $|\cdot\rangle_{\rm input}$, $|\cdot\rangle_{\rm ancilla}$, and $|\cdot\rangle_{\rm answer}$ indicate the registers for the input state, the ancillae for the Toffoli operations, and the answer of the calculation, respectively.
The output state of the ancilla register $|g(x)\rangle _{\rm ancilla}$ is the garbage of the computation.
However, the garbage can be uncomputed as follows (see also the circuit diagram below):
\begin{eqnarray}
&& U_f^{\dag} \Lambda _{\rm answer,out}(X) U_f|x\rangle _{\rm input}|0...0\rangle _{\rm ancilla} |0\rangle_{\rm answer} |0\rangle_{\rm out} 
\nonumber \\
&=& U_f^{\dag}|x\rangle _{\rm input} |g(x) \rangle _{\rm ancilla} |f(x) \rangle_{\rm answer} |f(x) \rangle_{\rm out}
\nonumber \\
&=&
|x\rangle _{\rm input} |0...0 \rangle _{\rm ancilla} |0 \rangle_{\rm answer} |f(x) \rangle_{\rm out}.
\end{eqnarray}
\begin{center}
\includegraphics[width=110mm]{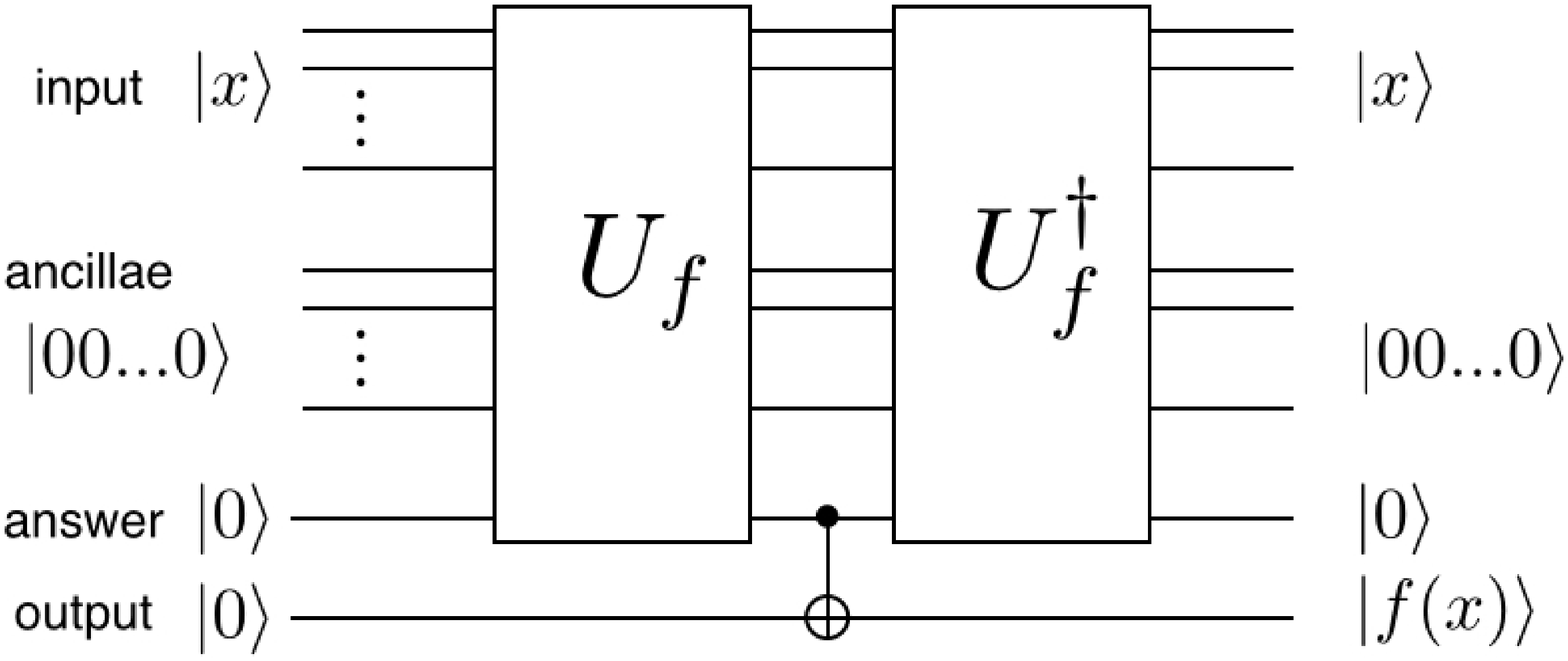}
\end{center}
In this way, we can calculate an arbitrary boolean function in a reversible way.

Finally, we introduce the multi-controlled unitary gate
\begin{eqnarray}
\Lambda ^{k} (U) = (I^{\otimes k} - |1\rangle \langle 1|^{\otimes k}) \otimes I + |1\rangle \langle 1|^{\otimes k} \otimes U,
\end{eqnarray}
where $U$ is applied to the target qubit if all $k$ control qubits are $|1\rangle$.
The multi-controlled gate can be implemented by using the Toffoli gates and $k$ ancilla qubits as follow:
\begin{center}
\includegraphics[width=110mm]{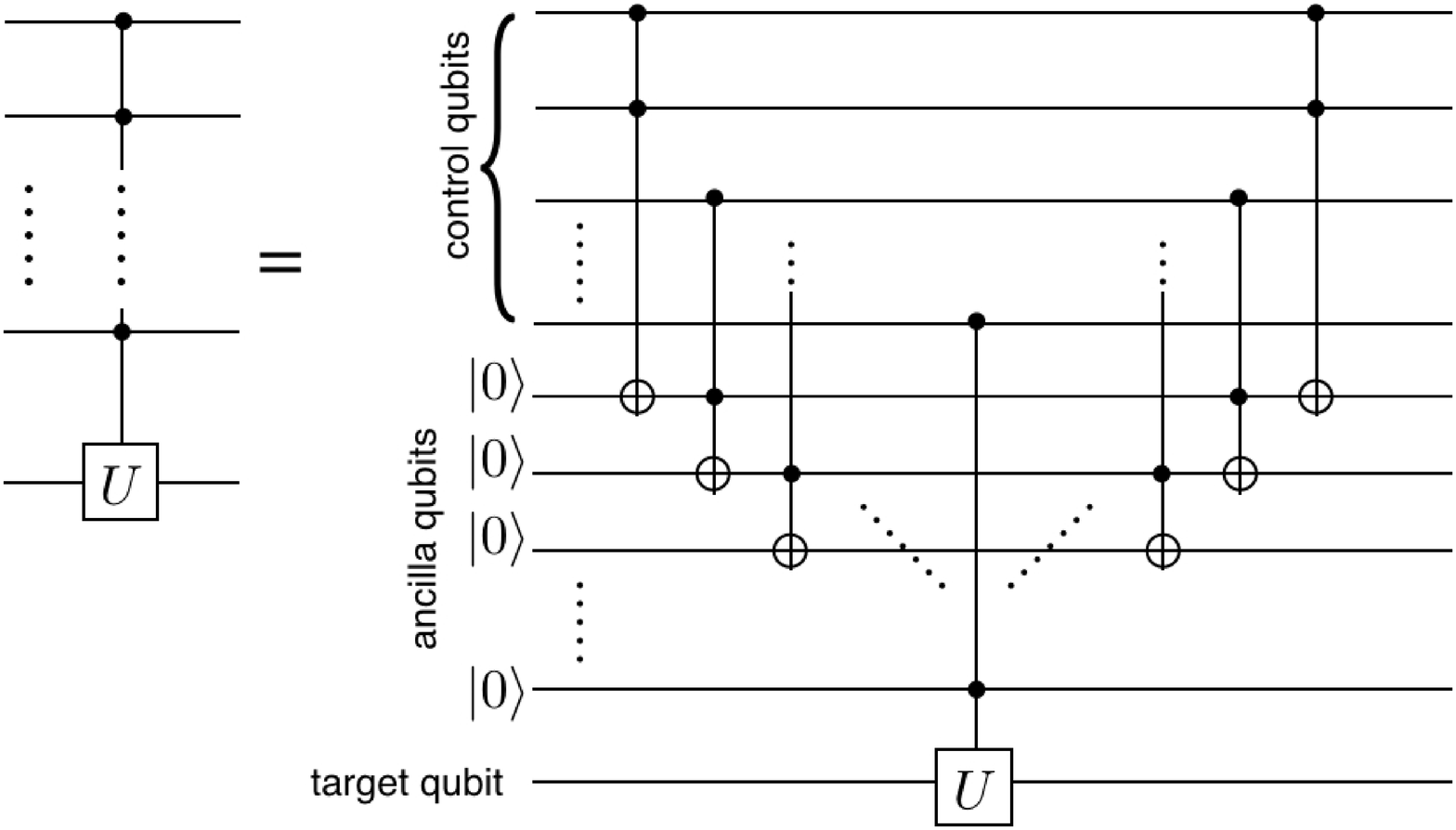}
\end{center}
The $\Lambda ^2(U)$ gate can be decomposed into CNOT and single-qubit gates by using an idea similar to the decomposition of the $\Lambda (U)$ gate:
\begin{center}
\includegraphics[width=100mm]{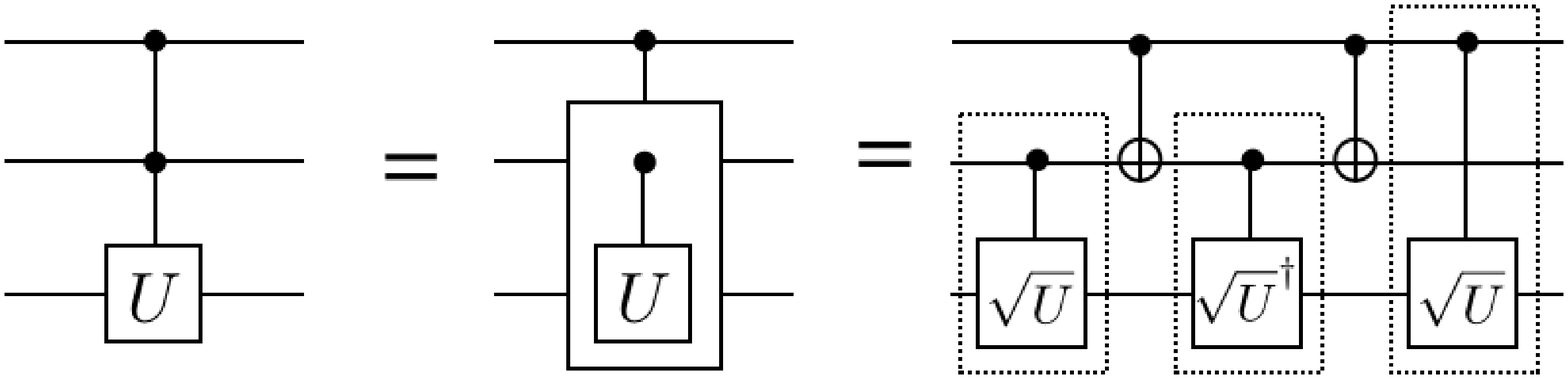}
\end{center}
For example, the Toffoli gate can be constructed from the CNOT, Hadamard, and $\pi/8$ operations as follows: 
\begin{center}
\includegraphics[width=100mm]{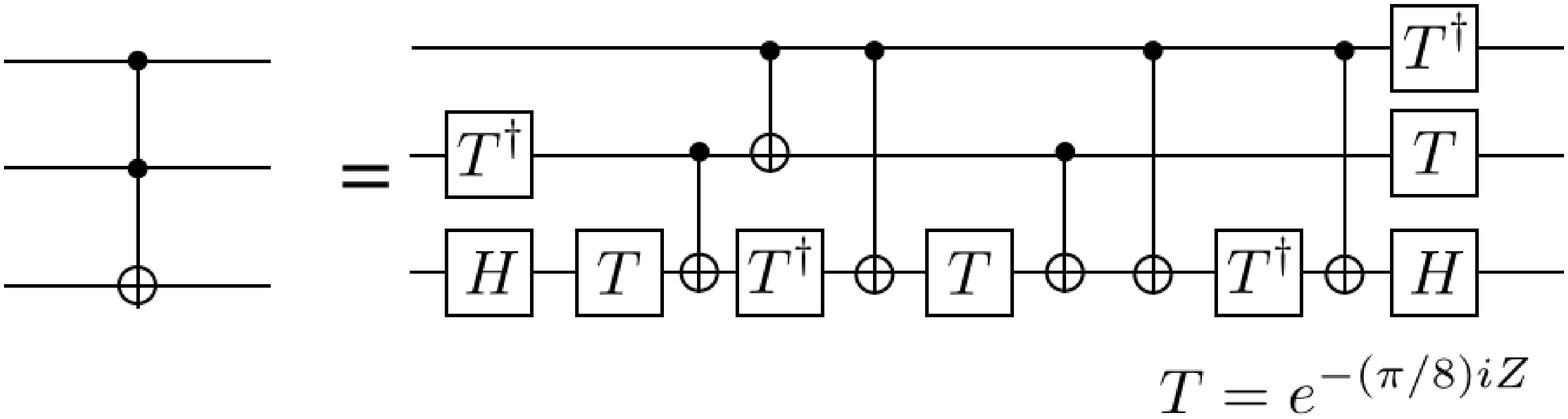}
\end{center}
where we have used the fact that that the controlled-$\sqrt{X}$ gate is decomposed into the CNOT, Hadamard, and $\pi/8$ ($T= e^{-i (\pi/8)Z}$) gates as follows:
\begin{center}
\includegraphics[width=80mm]{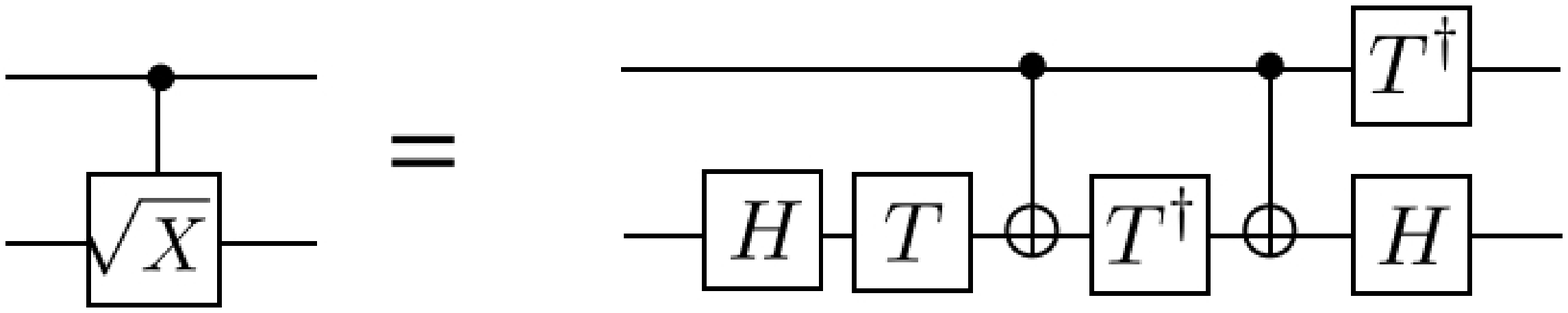}
\end{center}
It should be noted that
a multi-controlled gate $\Lambda ^k (U)$
can also be constructed
without any ancilla qubit 
from multi-controlled gates $\Lambda ^{(k-1)} (U)$ 
with lower controlled qubits
by using the trick employed for decomposing 
$\Lambda ^2(U)$ into $\Lambda (U)$~\cite{Barenco95}.

\section{Universal quantum computation}
Here we will show that an arbitrary unitary operation 
can be decomposed into single-qubit and CNOT gates.
To this end, we will first show how to decompose an arbitrary unitary operation into a product of two-level unitary gates.
Second, the two-level gates are decomposed into 
the multi-controlled gates,
which can be constructed from single-qubit and CNOT gates
as seen in the previous section.

Let $U$ be an arbitrary $n$-qubit unitary operator, represented by an $m\times m$ unitary matrix with $m\equiv 2^n$.
Let $T_{ij}$ be a unitary operator such that $(T_{ij})_{kl} = \delta _{kl}$ if $k,l \neq i,j$, which we call a two-level unitary gate.
(The $kl$ element of a matrix $A$ is denoted by $(A)_{kl}$.)
The $ii$, $ij$, $ji$, and $jj$ elements define a unitary operation on the two-level subsystem spanned by $|j\rangle$ and $|i\rangle$.
By choosing $T_{m\; m-1}$ appropriately, we have
\begin{eqnarray}
U T_{m \;m-1} = \left( 
\begin{array}{cccc}
	u_{11} & \cdots & u_{1\; m-1} & u_{1 \; m}
\\
	\vdots & \ddots & \vdots & \vdots
\\
	u_{m-1 \; 1} & \cdots & u'_{m-1 \; m-1} & u'_{m-1 \; m}
\\
	u_{m \; 1} & \cdots & 0  & u'_{m \; m}
\end{array}
\right),
\end{eqnarray}
where $(U)_{kl} = u_{kl}$.
By repeating this procedure, we obtain 
\begin{eqnarray}
U T_{m \; m-1} T_{m \; m-2} \cdots T_{m \; 1} 
= 
\left(
\begin{array}{cccc}
	u''_{11} & \cdots & u''_{1\; m-1} & u''_{1 \; m}
\\
	\vdots & \ddots & \vdots & \vdots
\\
	u''_{m-1 \; 1} & \cdots & u''_{m-1 \; m-1} & u''_{m-1 \; m}
\\
	0 & \cdots & 0  & u''_{m \; m}
\end{array}
\right).
\end{eqnarray}
Due to unitarity, $u'' _{1\;m}=\cdots = u''_{m-1 \; m}=0$ and $|u''_{mm}|=1$.
Defining $R_{m} \equiv T_{m \; m-1} T_{m \; m-2} \cdots T_{m \; 1} $, we can decompose $U$ into a product of $R_k$ and a diagonal unitary operator $D$:
\begin{eqnarray}
U= D (R_{m} \cdots R_1)^{\dag}.
\end{eqnarray}
It is obvious that $D$ can be decomposed into two-level unitary gates.
Thus, an arbitrary unitary operator $U$ can be decomposed into two-level unitary gates.

Next, we show that any two-level unitary operator $T_{ij}$ can be implemented by using CNOT and single-qubit gates.
Let us rewritte $i$ and $j$ ($i,j= 0,..., m-1$) by using the $n$-bit strings $\mathbf{s}=s_1 s_2 ... s_n$ and $\mathbf{t}=t_1 t_2 ... t_n$, respectively.
It is easy to find a sequence of $n$-bit strings $\{ \mathbf{g}_{k} \}_{k=1}^{d}$ such that $\mathbf{s}=\mathbf{g}_1$, $\mathbf{t}=\mathbf{g}_{d}$, and $\mathbf{g}_k$ and $\mathbf{g}_{k+1}$ differ by only one bit.
By using the Pauli $X$ gate and the multi-controlled-NOT gate
controlled by the same $n-1$ bits and targeting the one different bit,
we can transform the basis $|\mathbf{g}_k\rangle$ to $|\mathbf{g}_{k+1}\rangle$.
In this way, the basis $| i \rangle = |\mathbf{s}\rangle = |\mathbf{g}_1\rangle$ is transformed into $|\mathbf{g}_{d-1}\rangle$.
After this basis transformation, we now want to apply a two-level unitary gate between $|\mathbf{g}_{d-1}\rangle$ and $|j\rangle =|\mathbf{t}\rangle = |\mathbf{g}_{d}\rangle$.
Because $\mathbf{g}_{d-1}$ and $\mathbf{g}_{d}$ differ by only one bit, we can perform such a two-level unitary gate by the multi-conditional gate, using the Pauli $X$ gate for bit flips.
Finally, the basis $|\mathbf{g}_{d-1}\rangle$ is returned to $|\mathbf{s}\rangle$ by applying the inverse of the basis transformation.

For example, a two-level unitary operator acting on a subspace spanned by $\{ |000\rangle , |111\rangle\}$ can be implemented as follows:
\begin{center}
\includegraphics[width=120mm]{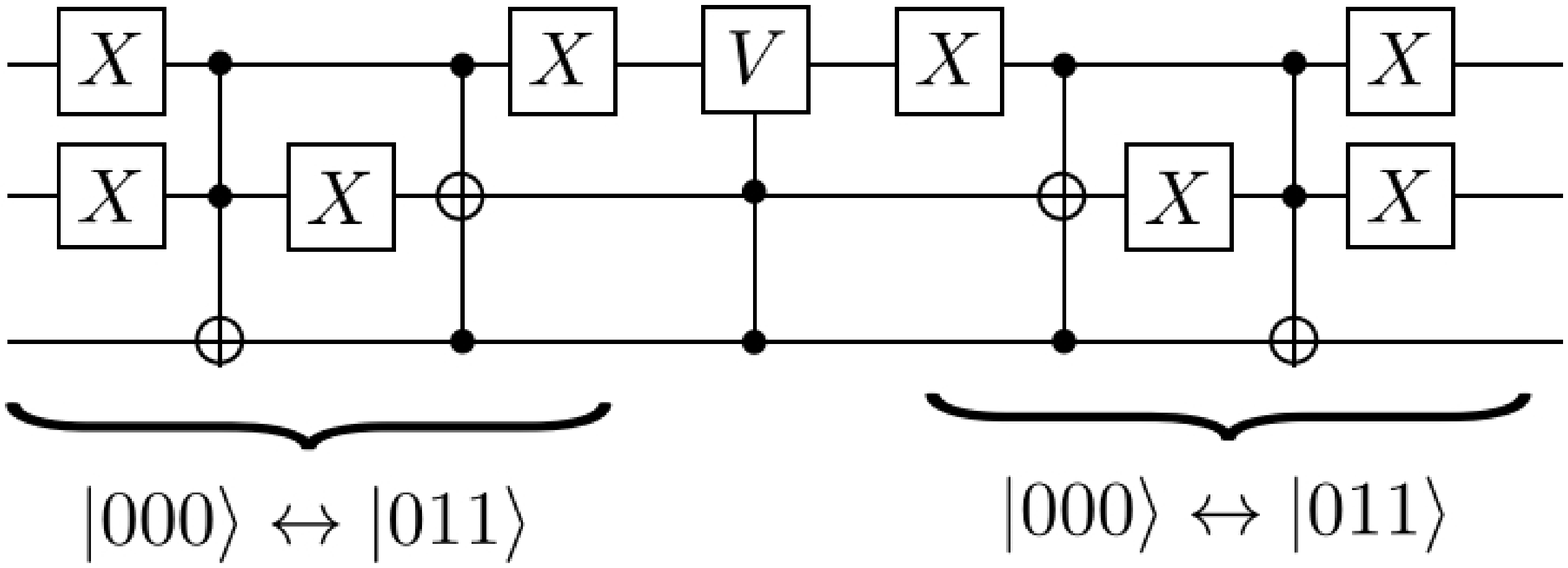}
\end{center}

As seen previously, the multi-conditional gate can be decomposed into CNOT and single-qubit gates.
Moreover, an arbitrary single-qubit unitary operation can be approximated by using the Hadamard and $\pi/8$ operations by virtue of the Solovay-Kitaev algorithm.
Thus, the CNOT, Hadamard, and $\pi/8$ operations form a universal set of operations for quantum computations.
The Toffoli and Hadamard operations also form a universal set as shown in~\cite{Reck94,Barenco95,DiVincenzo95,BernsteinVazirani,NielsenChuang}. 

\section{Quantum algorithms}
\label{sec:Qgadget}
In this section, we explain two representative quantum algorithms; Shor's prime factorization algorithm~\cite{Shor,Shor97} and the Aharonov-Jones-Landau algorithm for an additive approximation of the Jones polynomial~\cite{AharonovJones,AharonovJonesHard}.
We do not go deep into the mathematically rigorous details, but we aim to understand how they work.
The readers who are interested in more details should read Refs.~\cite{Shor,Shor97,KitaevAbelian,NielsenChuang,AharonovJones,AharonovJonesHard}.

\subsection{Indirect measurement and the Hadamard test}
\label{sec:HadamardTest}
An indirect measurement of an observable (hermitian) $A$ with eigenvalues $\pm 1$ can be performed by using $\Lambda(A)$:
\begin{center}
\includegraphics[width=50mm]{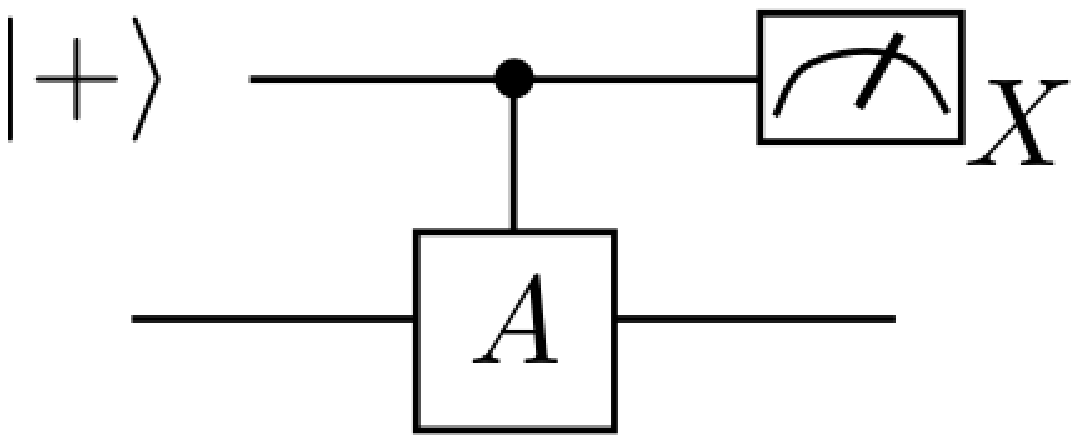}
\end{center}
By denoting the input state $|\psi \rangle$, the post-measurement state is given by
\begin{eqnarray}
\frac{I+(-1)^{s}A}{2}|\psi \rangle /\sqrt{{\rm Tr}[(I+(-1)^s A)/2|\psi \rangle \langle \psi | ]},
\end{eqnarray}
where $s=0,1$ is the measurement outcome.
For example, a circuit measuring the eigenvalue of the operator $X_1X_2X_3$ is given by
\begin{center}
\includegraphics[width=50mm]{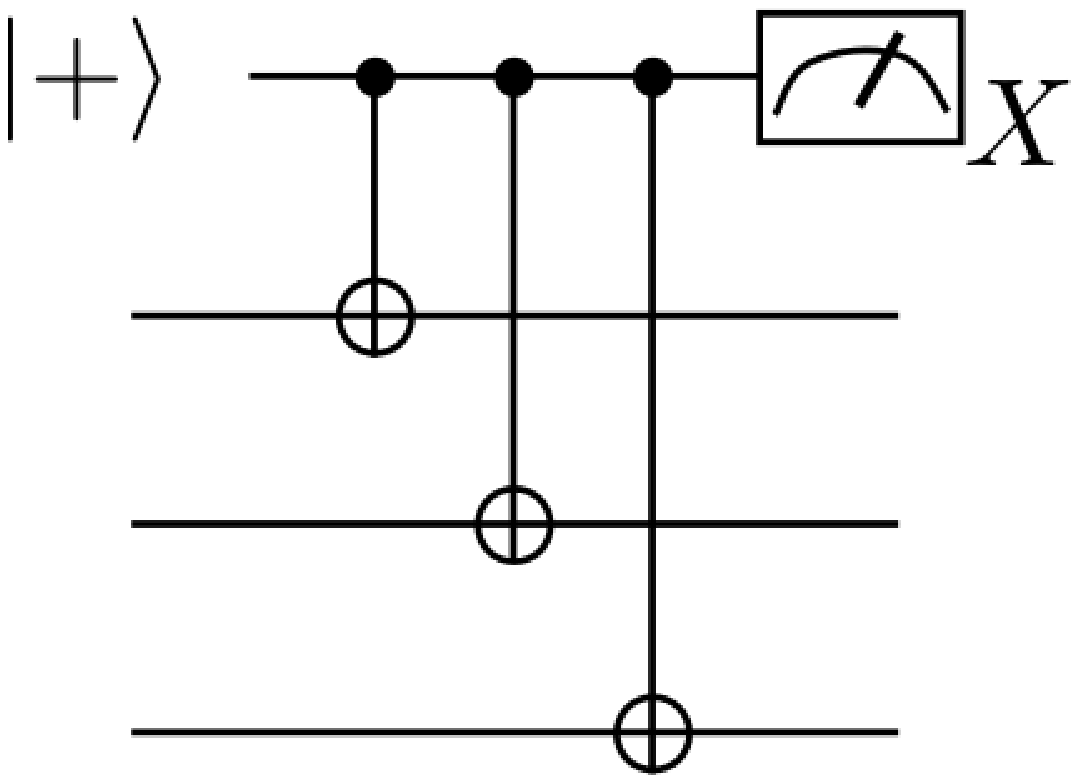}
\end{center}
According to the measurement outcome $s=0,1$, the post-measurement state is projected by $\frac{I +(-1)^{s}X_1 X_2 X_3}{2}$.
The indirect measurement will be employed frequently in quantum error correction to measure the eigenvalues of the stabilizer operators.

The Hadamard test\index{Hadamard test} of an arbitrary unitary operator $U$ is defined by the following circuit:
\begin{center}
\includegraphics[width=50mm]{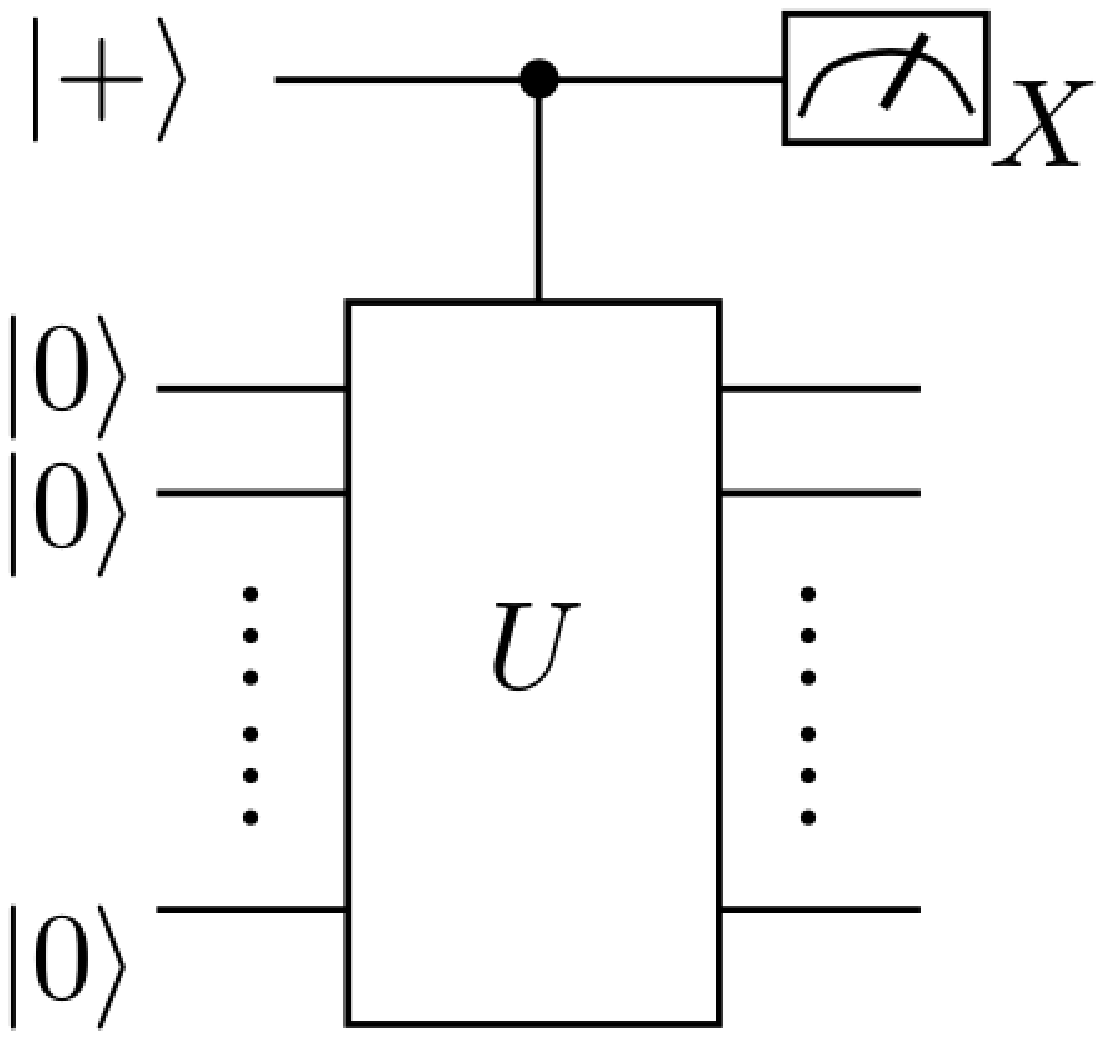}
\end{center}
The probabilities of the measurement outcomes $0,1$ of the $X$-basis measurement are calculated to be
\begin{eqnarray}
p_0 &=& \frac{1}{2}\left( 1+ {\rm Re} \langle 0 | ^{\otimes n} U |0 \rangle ^{ \otimes n}\right),
\\
p_1 &=& \frac{1}{2}\left( 1- {\rm Re} \langle 0 | ^{\otimes n} U |0 \rangle ^{ \otimes n}\right).
\end{eqnarray}
Similarly, the Hadamard test for the imaginary part is defined:
\begin{center}
\includegraphics[width=55mm]{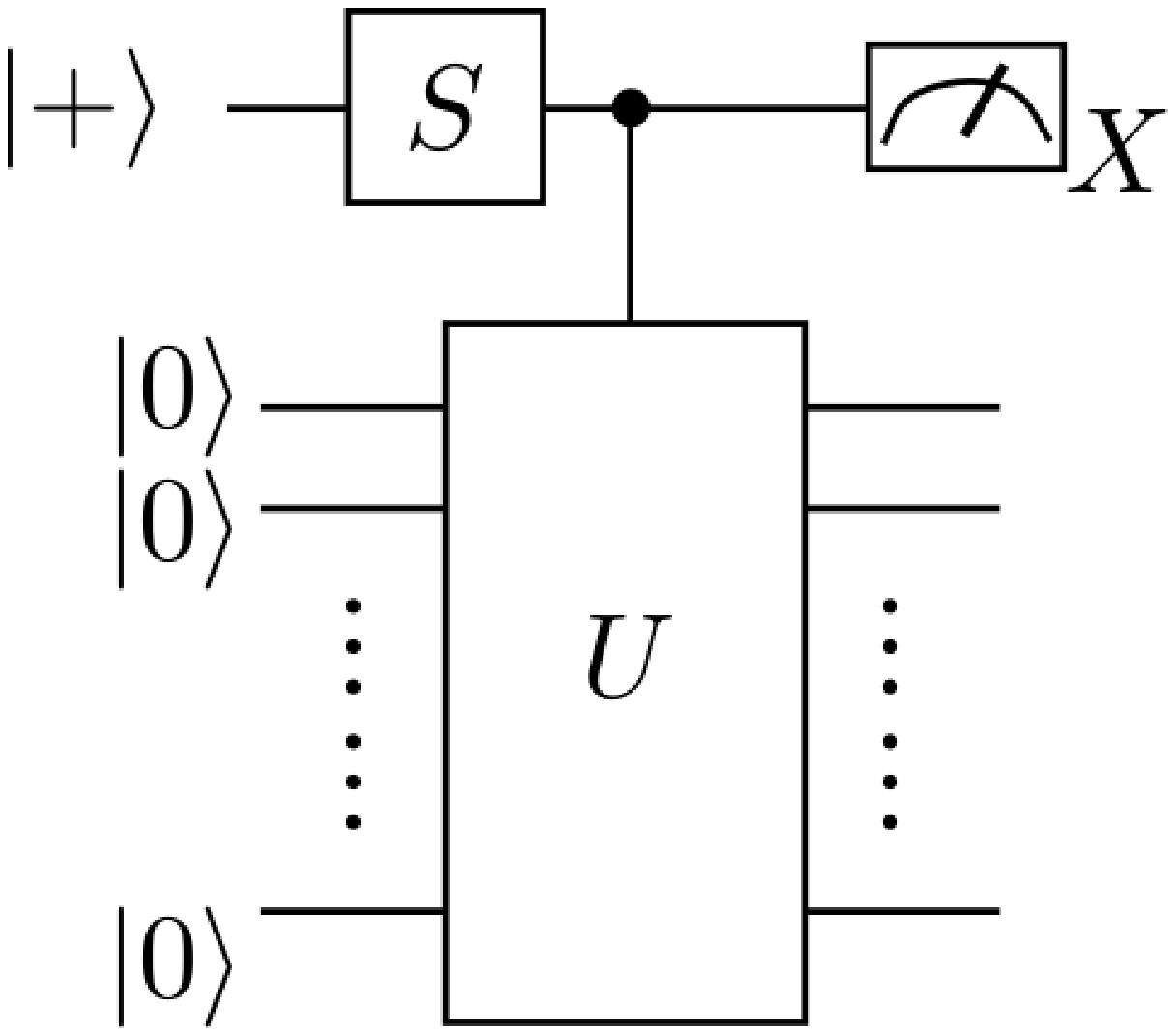}
\end{center}
Suppose we perform the Hadamard test $N$ times and obtain the measurement outcome 0, $N_0$ times.
By using the Chernoff-Hoeffding bound\index{Chernoff-Hoeffding bound},
\begin{eqnarray}
{\rm Prob} \left( \left| \frac{N_0}{N} - p_0\right| > \epsilon \right) < 2 e^{-2 \epsilon ^2 N},
\end{eqnarray}
we can estimate the matrix element $\langle 0 | ^{\otimes n} U |0 \rangle ^{ \otimes n}$ with an error $\epsilon$ by repeating the Hadamard test $N={\rm poly}(1/\epsilon)$ times.
The Hadamard test is employed in various quantum algorithms such as approximations of the Jones and Tutte polynomials~\cite{AharonovJones,AharonovJonesHard,AharonovTutte} and the partition functions of statistical mechanical models~\cite{Cuevas11,Iblisdir,MatsuoFujii}.

Specifically, if we choose the input state to be a completely mixed $n$-qubit state,
\begin{center}
\includegraphics[width=55mm]{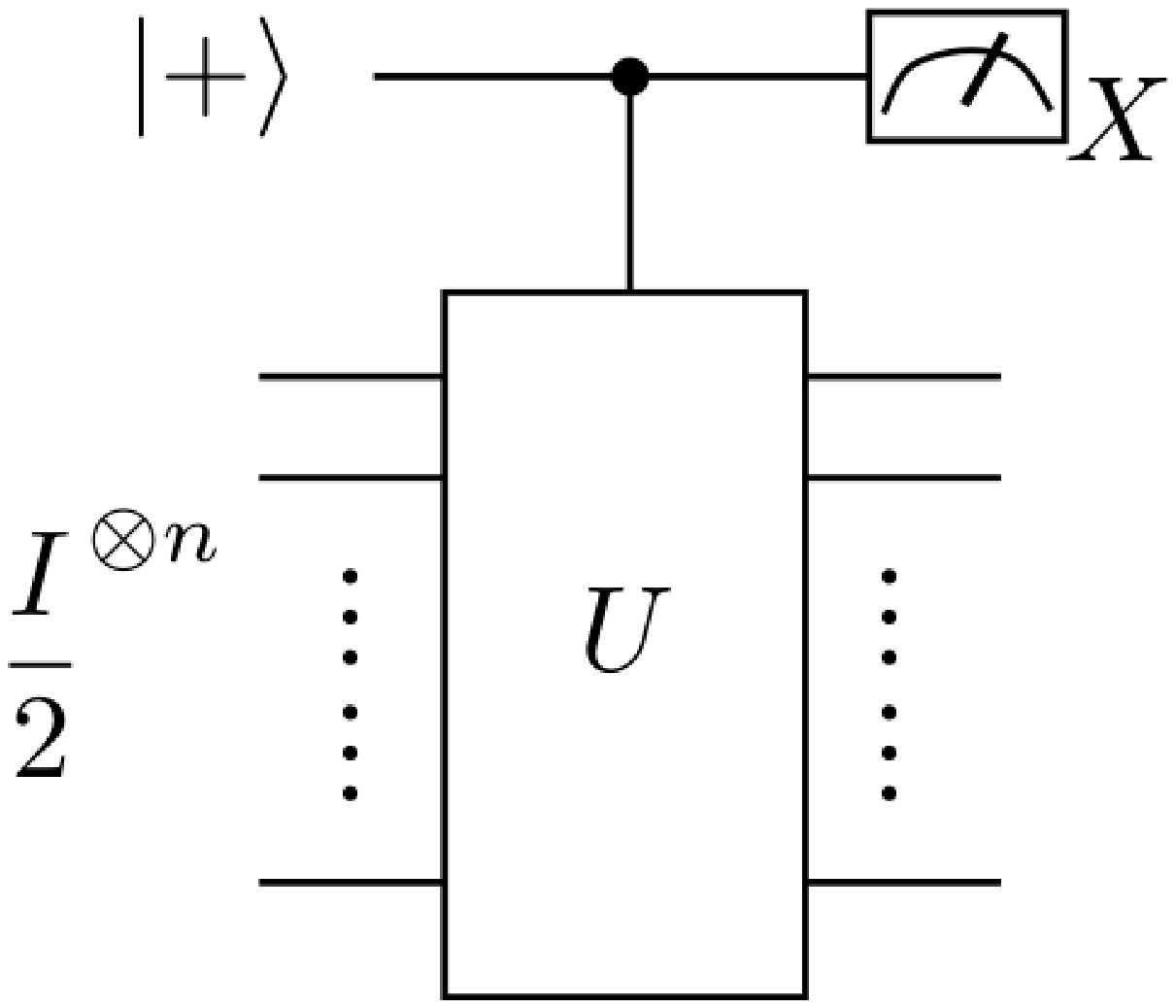}
\end{center}
then, the Hadamard test provides the trace ${\rm Tr}[U]/2^n$ of the unitary operator $U$.
Such a restricted type of quantum computation is called a deterministic quantum computation with one clean qubit (DQC1) \index{deterministic quantum computation with one clean qubit, DQC1}~\cite{DQC1}.
DQC1 seems to be less powerful than universal quantum computation, because only one qubit is a pure state.
However, DQC1 can evaluate functions, which would be intractable on a classical computer, such as the Jones and Homefly polynomials~\cite{ShorJordan,JordanWacjan}, spectral density function~\cite{DQC1}, and fidelity decay~\cite{FidelityDecay}.
Recently, classical sampling of the output of DQC1 with a few qubits measurements has been shown to be intractable unless the polynomial hierarchy collapses to the third level~\cite{DQC1hardness,DQC12}.

\subsection{Phase estimation, quantum Fourier transformation, and factorization}
\index{Shor's factorization algorithm}
If we have an eigenstate $|E_i\rangle$ of a unitary operator $U$, for which $\Lambda (U)$ is described by a polynomial number of gates, we can estimate the eigenvalue $\lambda _i$ of $U$ with a polynomial accuracy using the Hadamard test as follows:
\begin{center}
\includegraphics[width=60mm]{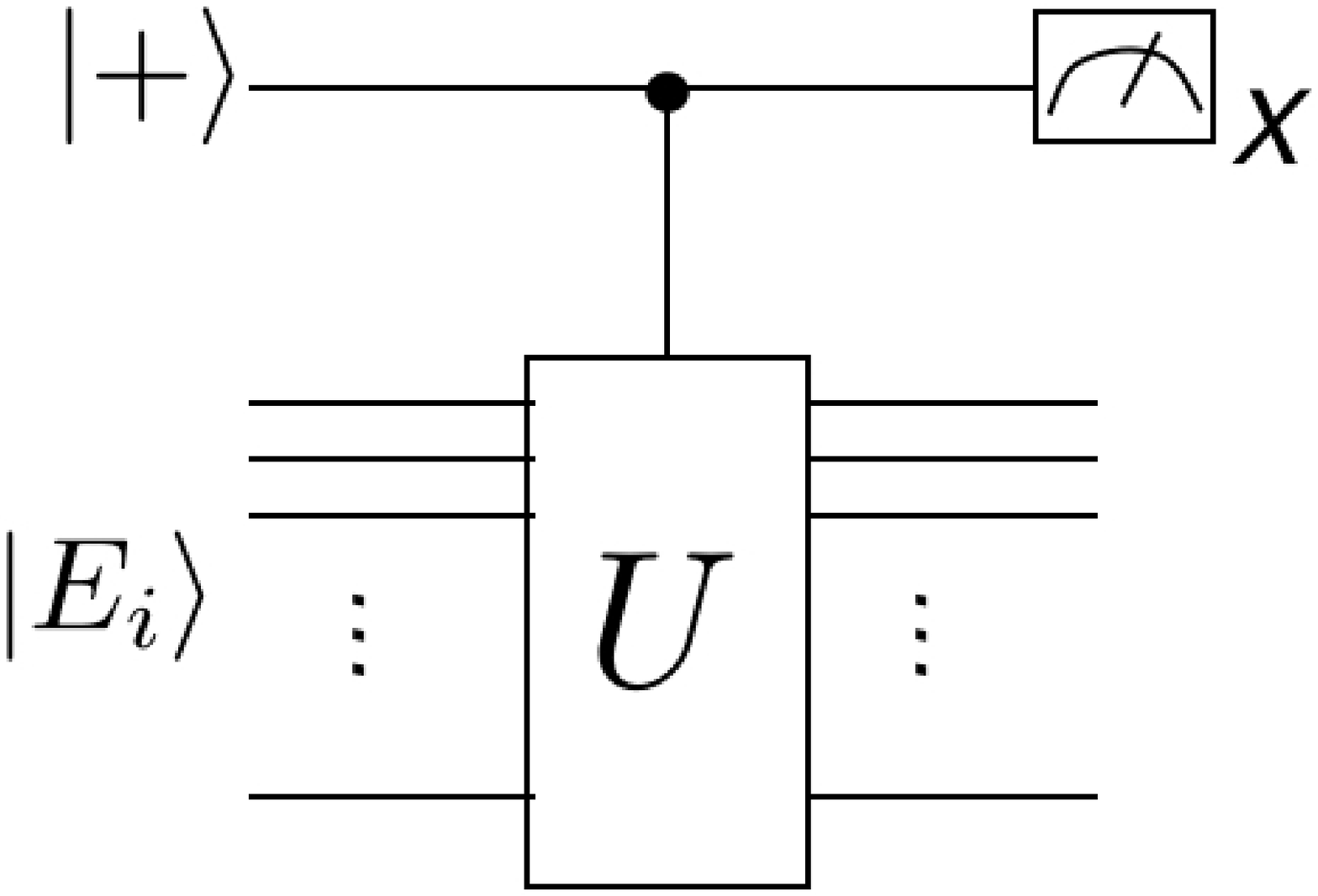}
\end{center}
When we do not have the eigenstate, the input state is projected onto one of the eigenstates by the Hadamard test.
Thus, by repeating the Hadamard test, we can obtain one of the eigenvalues with polynomial accuracy.

Moreover, if a controlled-$U^{2^k}$ gate $\Lambda (U^{2^k})$ can be described by a polynomial number of gates, we can estimate the eigenvalue with exponential accuracy~\cite{KitaevPhase}.
Suppose the eigenvalue is given by $\lambda _i =e^{i \phi}=e^{2 \pi i 0.j_1 j_2 ... j_n} $, where we employ a decimal of a binary number,
\begin{eqnarray}
0.j_1 j_2 ... j_n = \sum _{k=1}^{n} j_k (1/2)^k.
\end{eqnarray}
Then, the Kitaev's phase estimation algorithm\index{Kitaev's phase estimation algorithm} is given by 
\begin{center}
\includegraphics[width=110mm]{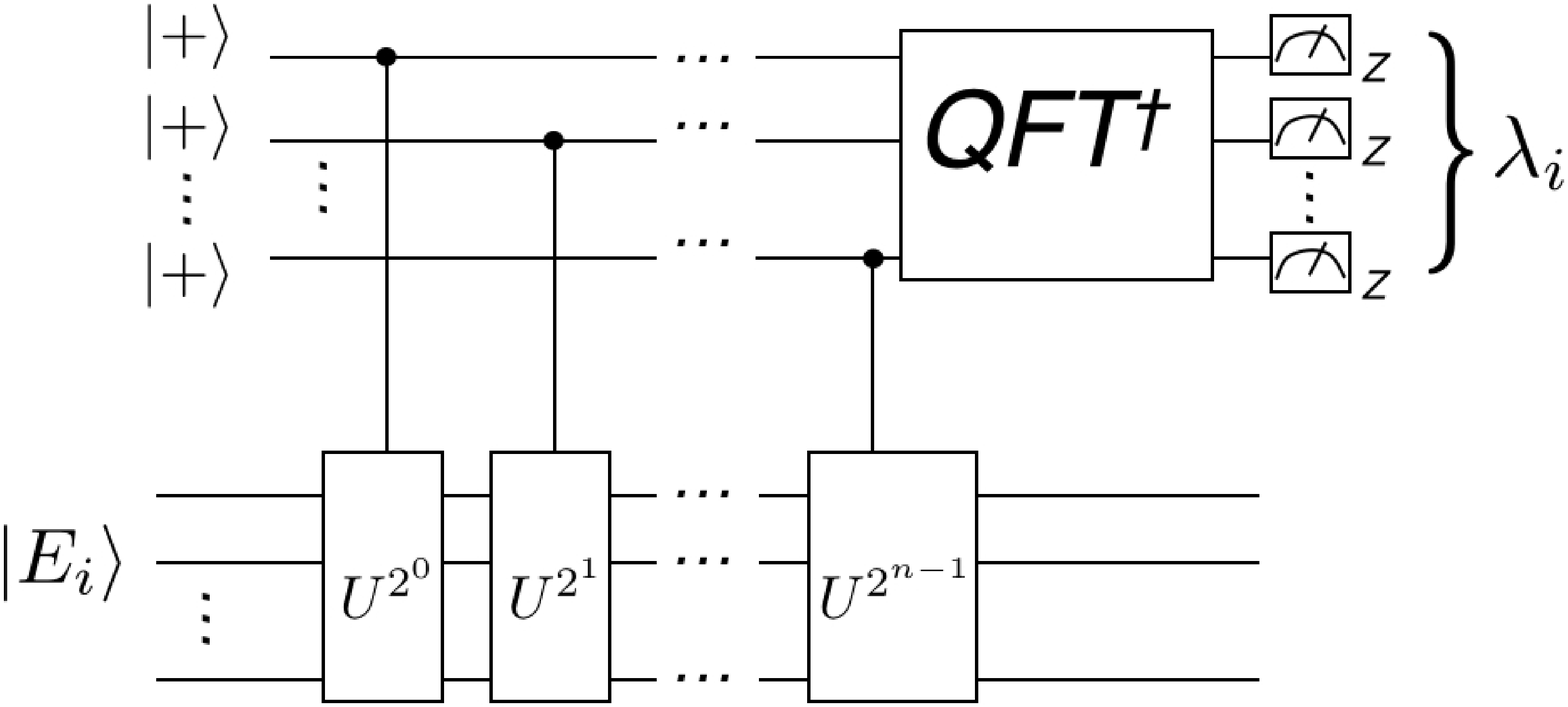}
\end{center}
where QFT stands for the quantum Fourier transform\index{quantum Fourier transform}.
The controlled gate $\Lambda (U^{2^k})$ kicks back the phase $e^{2^k \phi}=e^{2 \pi i j_k.j_{k+1}...  j_n}$ to the ancilla state $|+\rangle$.
The phase information is transformed into a computational basis state by using the inverse quantum Fourier transformation:
\begin{eqnarray}
QFT^{\dag} (|0\rangle + e^{2\pi i 0.j_1 j_2 ... j_n }|1\rangle)
(|0\rangle + e^{2\pi i 0.j_2 ... j_n }|1\rangle)
...
(|0\rangle + e^{2\pi i 0.j_n }|1\rangle)
=
|j_1 j_2 ... j_n\rangle.
\nonumber \\
\end{eqnarray}
Here QFT is given by
\begin{center}
\includegraphics[width=150mm]{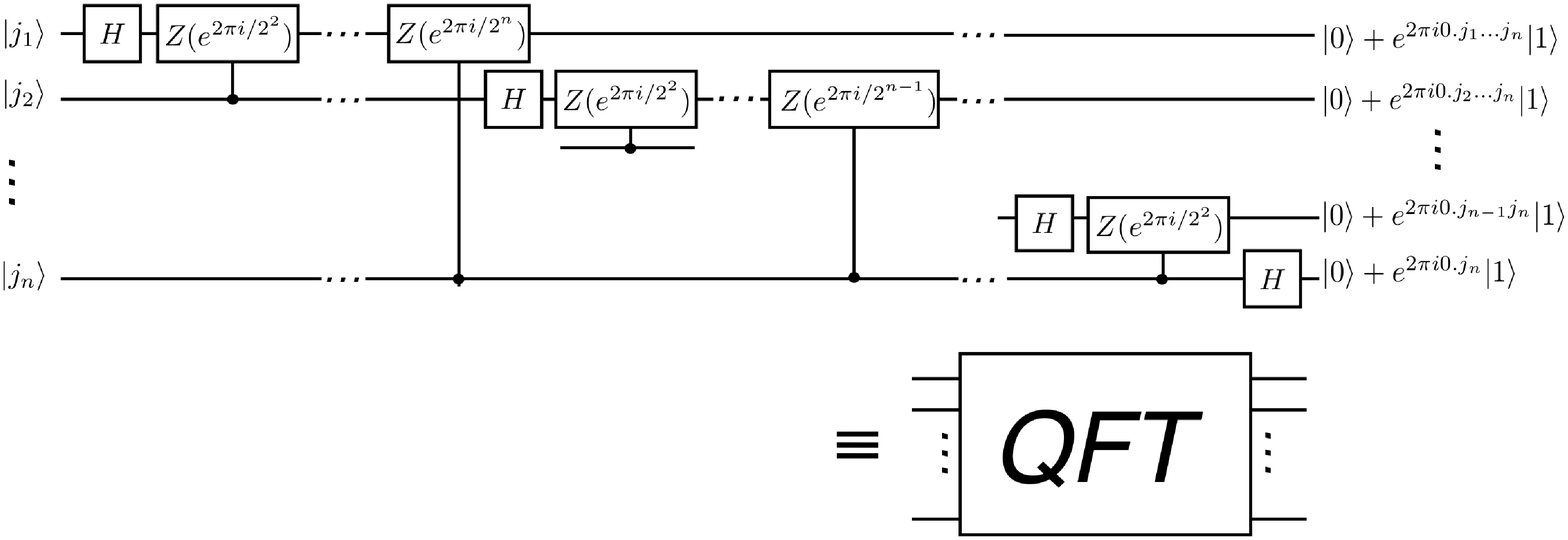}
\end{center}
Then, we can obtain the phase $\phi = (2\pi i)0.j_1 j_2 ...j_n$ through the $Z$-basis measurements.
Note that the accuracy of the estimate is improved exponentially compared to the Hadamard test (provided that we have a polynomial size description of $\Lambda (U^{2^k})$).

Let $N$ and $x$ be co-prime integers.
The order-finding problem is about finding an order $r$ such that $x^r =1 \textrm{ mod } N$.
This problem can be solved by using the phase estimation against a unitary operator
\begin{eqnarray}
U_x = \sum _{y} |x y \textrm{ mod } N\rangle \langle y| ,
\end{eqnarray}
because we have
\begin{eqnarray}
U_x |u_s \rangle = e^{2 \pi i (s/r))} |u_s\rangle,
\end{eqnarray}
for a state
\begin{eqnarray}
|u_s\rangle = \frac{1}{\sqrt{r}}
\sum _{k=0}^{r-1} e^ {-2 \pi i (s/r) k }|x^k {\rm mod}\; N\rangle.
\end{eqnarray}
Note that we can prepare the initial state $|u_s\rangle$ randomly by using the fact that $|1\rangle= \sum _{s=0}^{r-1}|u_s\rangle$.
After estimating the phase $2 \pi i (s/r)$, the continued fraction provides $r$ with high probability.
The modular exponentiation $x^{2^k}\; \textrm{mod} \;N$ can be calculated by using the square $k$ times.
Thus we can implement controlled-$U_x^{2^k}$, $\Lambda (U_x^{2^k})$, by a polynomial number of gates.

If we randomly choose $s$, we obtain an even order $r$ with high probability.
Thus, we have $(x^{r/2}-1)(x^{r/2}+1)=0 \; {\rm mod} \; N$.
Finally, the euclidean algorithm gives us the greatest common divisor of $x^{r/2}-1$ and $N$ or $x^{r/2}+1$ and $N$, and it is a factor of $N$.
This is the so-called Shor's prime factoring algorithm.
Kitaev, who formulated the phase estimation algorithm, generalized this idea for the more general Abelian stabilizer problem~\cite{KitaevAbelian}.

\subsection{A quantum algorithm to approximate Jones polynomial}
\label{sec:JonesPoly}
\index{Aharonov-Jones-Landau algorithm}
Next we will explain Aharonov-Jones-Landau algorithm 
for approximating the Jones polynomial\index{Jones polynomial}
~\cite{AharonovJones,AharonovJonesHard},
which is an algorithmic version of 
the original proposal based on topological quantum field theory~\cite{TQFT,Witten89}.
We will show that the approximation of the Jones polynomial 
is BQP-complete;
the approximation of the Jones polynomial
can be done efficiently by using a universal quantum computer,
and inversely it is as hard as any problems (BQP) solvable by 
a universal quantum computation.
Below we will introduce the braid diagram and the braid group,
which are related to link diagrams with an appropriate closure.
A unitary representation of the braid group 
is constructed by embedding the braid group to 
the Temperley-Lieb (TL) algebra.
Using the constructed representation,
the Jones polynomial and quantum algorithm are connected.

\begin{figure}[t]
\centering
\includegraphics[width=110mm]{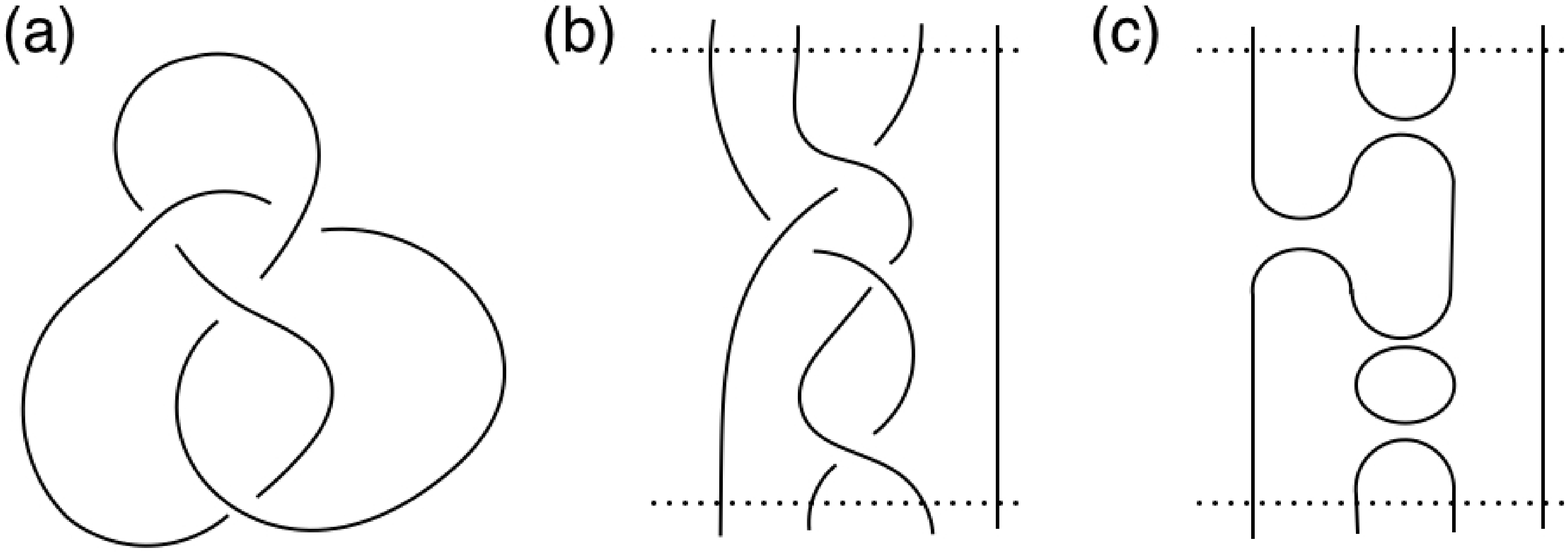}
%
%
\caption{(a) A link diagram. (b) A braid diagram. (c) A tangle diagram.}
\label{fig122}       
\end{figure}
\index{Jones polynomial}
Let us first define the Jones polynomial.
The Jones polynomial is an invariant of 
a link, that is, closed loops 
in a 3D space, which are tangled in general~\cite{Lickorish97}.
It is convenient to describe the link
on a 2D projected space as shown in Fig.~\ref{fig122} (a),
which we call a link diagram.
Specifically two links are equivalent 
if two link diagrams can be transformed into each others 
under the Reidemeister moves:
\begin{center}
\includegraphics[width=120mm]{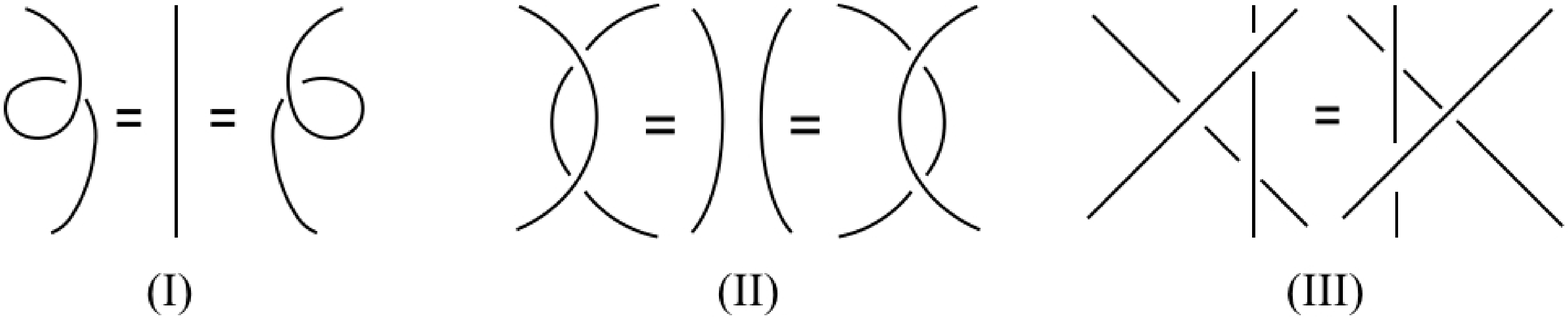}
\end{center}
which appropriately reflect the continuous deformations in the 3D space.
The Jones polynomial is calculated from 
a directed link diagram $L$ as follows:
(i) Smooth each crossing \crosA in two ways 
$\{$\updow,\lefri $\}$.
Let $s$ be the resultant diagram consisting of 
closed loops with no crossing, which we call a state.
(ii) For each state $s$, we assign a weight 
\begin{eqnarray}
W(s) = A^{s^{+}- s^{-}} d^{|s|-1},
\end{eqnarray}
where $d=-(A^2+A^{-2})$ is a complex constant called a loop value,
$|s|$ is the number of the crossing,
and $s^{+}$ and $s^{-}$ are the number of smoothings by \updow and \lefri,
respectively.
By taking the summation over all states,
the Kauffman bracket $\langle L \rangle$ of the link $L$ is defined:
\begin{eqnarray}
\langle L \rangle = \sum _{s} W(s).
\end{eqnarray}
\index{Kauffman bracket}
The Jones polynomial is defined 
as a function of $t=A^{-4}$ by 
multiplying a factor to the Kauffman bracket: 
\begin{eqnarray}
V_L(t) = (-A)^{3 \omega (L)} \langle L \rangle,
\end{eqnarray}
where the writhe $\omega (L)$ of the link $L$ is
defined as the number of \crosA type crossings  minus 
that of \crosB type. 
It is easy and a good exercise to confirm that the Jones polynomial is invariant 
under the Reidemeister moves based on the above definition~\cite{Lickorish97}.

\begin{figure}[t]
\centering
\includegraphics[width=120mm]{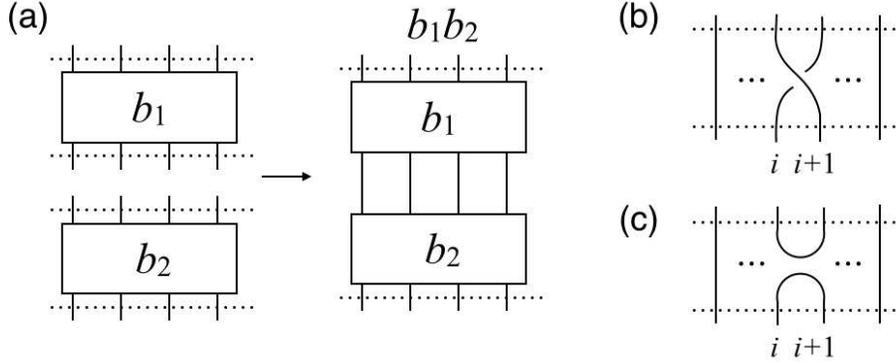}
%
%
\caption{(a) A multiplication for two braid diagrams $b_1$ and $b_2$. (b) The generator $\sigma _i$ of the braid group. (c) The generator $E_i$ of 
the TL algebra.}
\label{fig124}       
\end{figure}

\index{braid group}
Let $B_n$ be a braid group 
consisting of the braid diagrams
of $n$ strands, where two endpoints of each strand
are tied at the top and bottom, respectively
as shown in Fig.~\ref{fig122} (b).
The multiplication of two braids $b_1$ and $b_2$
are defined by connecting the bottom and top endpoints 
of two braid diagrams as shown in Fig.~\ref{fig124} (a).
The braid group $B_n$ is generated 
by $n-1$ generators $\{ \sigma _i \}$
subject to
\begin{eqnarray}
&&\sigma _i \sigma _j  = \sigma _j \sigma _i \textrm{ for } |i-j|\geq 2
\\
&&\sigma _i \sigma _{i+1} \sigma _i = \sigma _{i+1}\sigma _i \sigma _{i+1}.
\end{eqnarray}
The generator $\sigma _i$ corresponds to 
the braid diagram with only one crossing of $i$th and $i+1$th strands
as shown in Fig.~\ref{fig124} (b).
Then, the latter equality is nothing but the 
Reidemeister move III.
By introducing an appropriate closure,
which closes the endpoints of the strands,
a braid diagram is related to a link diagram
as we will see later.

In order to construct a unitary representation of $B_n$,
we embed it into the Temperley-Lieb (TL) algebra $TL_n(d)$~\cite{TLalgebra}
on tangle diagrams with no crossing as shown in Fig.~\ref{fig122} (c).
\index{Temperley-Lieb (TL) algebra}
Similarly to the previous case,
$TL_n(d)$ is generated by a set of generators $\{E_1,...,E_{n-1}\}$
subject to
\begin{eqnarray}
&&E_i E_j = E_j E_i \textrm{ for } |i-j | \geq 2,
\\
&&E_i E_{i\pm1} E_i =  E_i,
\\
&&E_i^2=d E_i.
\end{eqnarray}
The generator $E_i$
corresponds to $\cup$ and $\cap$ for the $i$th and $(i+1)$th strands
as shown in Fig.~\ref{fig124} (c).
The braid group is embedded into the TL algebra by
\begin{eqnarray}
\sigma _i = A E_i + A^{-1} \mathbf{1}.
\label{eq:BraidTL}
\end{eqnarray}

\begin{figure}[t]
\centering
\includegraphics[width=120mm]{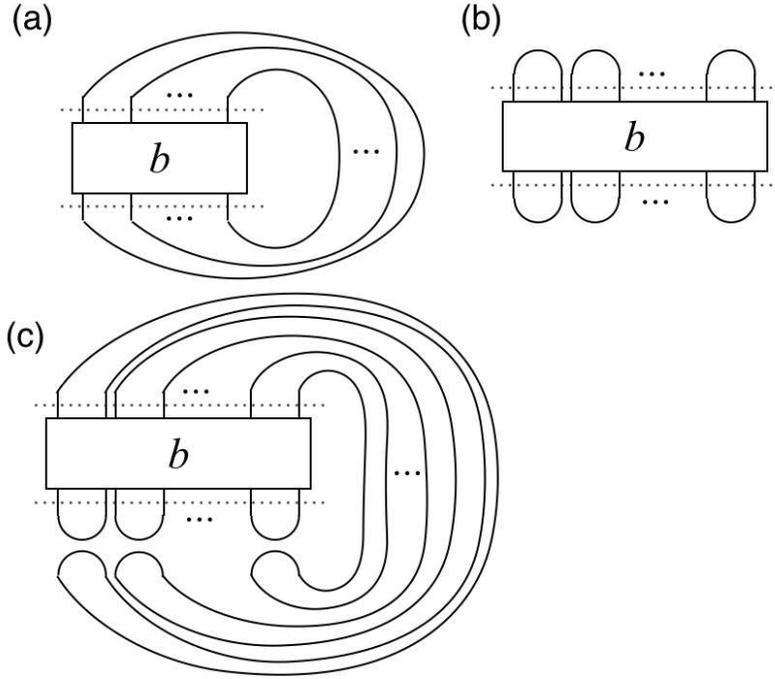}
%
%
\caption{(a) The Markov trace closure for the braid $b$. (b) The plat closure 
for the braid $b$. (c) The plat closure 
is deformed into the Markov trace closure.}
\label{fig123}       
\end{figure}
In order to relate the braid diagram 
and the Jones polynomial,
we employ a Markov trace closure on the tangle diagram as shown in Fig.~\ref{fig123} (a).
The Markov trance closure denoted by mtr has the following property: 
\begin{eqnarray}
&&{\rm mtr} (\mathbf{1})=1,
\\
&&{\rm mtr} (AB)={\rm mtr} (BA) \textrm{ for } A,B \in TL_n(d),
\\
&&{\rm mtr} (A) = d{\rm mtr} (A E_{n-1}) \textrm{ for }
A \in TL_{n-1}(d).
\end{eqnarray}
Now the representation of 
the braid group and the Jones polynomial are related.
Eq. (\ref{eq:BraidTL}) corresponds to 
superposition of two smoothings $\{$\updow,\lefri$\}$ 
of a crossing $\crosA$.
More precisely,
let $\rho$ and $\tilde \rho$ be a representation of the TL algebra 
and an induced representation 
of the braid group, respectively.
Moreover, the matrix trace 
has the same property as the Markov trace closure.
Thus,
for a given braid diagram $b$, we have
\begin{eqnarray} 
V_{b^{{\rm mtr}}}(A^{-4}) =  \Delta{\rm Tr}[\tilde \rho (b)] ,
\end{eqnarray}
where $b^{\rm mtr}$ is a link diagram 
generated from the braid diagram $b$ with the Markov trace closure,
and $\Delta _{mtr}= (-A)^{3 \omega (b^{\rm mtr})}d^{n-1}$.

Let us construct 
a representation of the TL algebra,
the so-called path-model representation.
\index{path-model representation}
Suppose $G$ is a one-dimensional graph with $k$ vertices
labeled by $1,...,k$ from left to right and $k-1$ edges.
We consider an $n$ step walk on $G$ starting from the left endpoint.
Let $p=(v_1,...,v_n)$ be a path of such an $n$ step walk,
where $v_1=1$ and $v_j =0$ and $=1$ mean moving to the left and right neighboring vertices,
respectively.
Then we define a Hilbert space $\mathcal{H}_{n,k}$ 
spanned by all possible paths as a basis $\{ |p\rangle\}$.
For each generator $E_i$, we define 
its representation $\Phi_i =\rho(E_i)$ on $\mathcal{H}_{n,k}$
as follows:
\begin{eqnarray}
\Phi _i | ... v_{i-1} 00 v_{i+1}...\rangle &=& 
0,
\\
\Phi _i | ... v_{i-1} 01 v_{i+1}...\rangle &=& 
\frac{\lambda_{z_i -1}}{\lambda_{z_i}}| ... v_{i-1} 01 v_{i+1}...\rangle+
\frac{\sqrt{\lambda_{z_i +1} \lambda_{z_i -1}}}{\lambda_{z_i}}
| ... v_{i-1} 10 v_{i+1}...\rangle,
\nonumber \\
\\
\Phi _i | ... v_{i-1} 10 v_{i+1}...\rangle &=&
\frac{\lambda_{z_i +1}}{\lambda_{z_i}}| ... v_{i-1} 10 v_{i+1}...\rangle+
\frac{\sqrt{\lambda_{z_i +1} \lambda_{z_i -1}}}{\lambda_{z_i}}
|... v_{i-1} 01 v_{i+1}...\rangle,
\nonumber \\
\\
\Phi _i |... v_{i-1} 11 v_{i+1}...\rangle &=& 0,
\end{eqnarray}
where $z_i \in \{1,...,k\}$ is the label of the vertex at the $i$th step
and $\lambda _j = \sin( j\theta)$ with $\theta =\pi  / k$. 
The representation $\tilde \rho (\sigma _i)=A\Phi _i +A^{-1} \mathbf{1}$ 
for the braid group $B_n$ is induced
by $\rho$.
It is easy to confirm that,
with $A=ie^{-i \theta /2}$ and $d=2\cos\theta$,
the representation $\rho$ is hermitian for all $E_i$, and hence
the induced representation $\tilde \rho$ is unitary.
Since $\rho (\sigma _i)$ is a unitary operator
acting on the neighboring two-qubit,
a multiplication of such unitary operators
can be implemented by using universal quantum computer.
Since, by using the Hadamard test in Sec.~\ref{sec:HadamardTest},
we can evaluate the Jones polynomial $V_{b^{\rm mtr}}(A^{-4})$
with an additive error $\Delta 2^{n} \epsilon$
taking ${\rm poly}(n,m,k,1/\epsilon)$ overhead,
where $m$ is the number of the crossing.

The approximation scale is further improved 
by considering another closure, the so-called plat closure
as shown in Fig.~\ref{fig123} (b).
We deform the link diagram obtained by the plat closure
to another link diagram obtained 
by the Markov trace closure as shown in Fig.~\ref{fig123} (c),
Then, we have $\cup$ and $\cap$ for all strands.
In this case, the support of $\rho (b)$ 
is only on $|10101...\rangle$,
which allows us to replace the matrix trace 
as follows:
\begin{eqnarray}
V_{b^{\rm plt}}(A^{-4}) = \Delta_{\rm plt} 
\langle 10101...| \rho (b) | 10101...\rangle,
\end{eqnarray}
where $b^{\rm plt}$ is a link generated from 
the braiding $b$ with the plat closure
and $\Delta _{\rm plt} = (-A)^{3 \omega (b^{\rm plt})}d^{n/2-1}$.
(Note that for each $\rho(E_i)$ of $n/2$ $\cup$s and $\cap$s, 
a factor ${\lambda_{2}/\lambda_{1}} = d$ is substituted.)
Again using the Hadamard test,
we can estimate $V_{b^{\rm plt}}(A^{-4})$ 
with an additive error $\Delta _{\rm plt} \epsilon$
taking ${\rm poly}(n,m,k,1/\epsilon)$ overhead.

Finally, we show that the approximation of the Jones polynomial 
with an additive error $\Delta _{\rm plt} \epsilon$ is BQP-hard.
Unfortunately, the computational basis defined by the path $p=(v_1,...,v_n)$
is inappropriate for this purpose,
since the $n$-step walks cannot span the whole $2^n$-dimensional Hilbert space.
Instead, we employ the four-step encoding~\cite{wocjan06}, 
where the computational basis states are defined by the following 
two four-steps:  
\begin{eqnarray}
|\bar 0 \rangle \equiv |1100\rangle, \;\;\; |\bar 1 \rangle 
\equiv |1010\rangle.
\end{eqnarray}
The space $\mathcal{H}_{4n,k}$ spanned by the $4n$-step walks
can support the whole $2^n$-dimensional space for $\{ |\bar 0 \rangle , |\bar 1\rangle \}^{\otimes n}$.
The dimension of the space spanned by the 8-step walks
starting from vertex 1 ending at vertex 1
is 14. Thus the unitary representation $\rho$ 
of the braid group $B_8$
results in unitary operators in $SU(14)$.
If they are dense in $SU(14)$,
we can approximate an arbitrary unitary operator in $SU(4)$ spanned by two four-step encoded qubits
by using the Solovay-Kitaev algorithm~\cite{AharonovJonesHard},
which are enough to implement universal quantum computation.
The density is achieved by $k>4$ and $k\neq 6$.
For such a parameter, 
the approximation of the Jones polynomial with an additive 
error $\Delta _{\rm plt} \epsilon$ is enough to solve 
a BQP-complete problem.

The approximation of the Jones polynomial 
can be done by universal quantum computer
and also is enough to solve the problems solvable by 
universal quantum computer. 
Thus the approximation of the Jones polynomial is a 
BQP-complete problem~\cite{AharonovJones,AharonovJonesHard}.
The AJL algorithm for approximation of the Jone polynomial 
was extended for the Tutte polynomial in Ref.~\cite{AharonovTutte},
where the Solovay-Kitaev algorithm for non-unitary linear operators 
was developed.

\section{Quantum noise}
Quantum coherence, one of the essential properties of quantum systems, is quite fragile against noise, due to interactions between the system and the environment.
Suppose that the system $S$ of interest interacts via a unitary operation $U$ with the environment $E$, where the system and environment are initially uncorrelated. 
The reduced density matrix $\rho' _S$ of the system after the interaction is calculated to be
\begin{eqnarray}
\rho' _S   = {\rm Tr}_{E} 
\left[U (\rho_{S} \otimes \rho _{E})U^{\dag} \right].
\end{eqnarray}
Using a spectral decomposition of the initial state in the environment, $\rho _{E} = \sum _{k} p_k |e_k \rangle \langle e_k |$, we obtain a map of the system $S$:
\begin{eqnarray}
\rho'_S = \sum _{(k',k)} K_{(k',k)} \rho _S K_{(k',k)}^{\dag},
\end{eqnarray}
where 
\begin{eqnarray}
K_{(k,k')} = \sqrt{p_k}\langle e_k' | U |e_k \rangle.
\end{eqnarray}
This map satisfies 
\begin{eqnarray}
\sum _{k}K_{(k',k)}^{\dag} K_{(k',k)} &=& 
p_k \langle e_k | U^{\dag} |e_k' \rangle 
\langle e_k' | U^{\dag} |e_k \rangle
\nonumber \\
&=& I_S,
\end{eqnarray}
where $I_S$ is the identity operator in the system $S$.

In general, a map of a quantum state is given as a completely-positive-trace-preserving (CPTP) \index{completely-positive-trace-preserving (CPTP)} map $\mathcal{E}$, subject to
\begin{itemize}
	\item ${\rm Tr} [\mathcal{E}  \rho ] =1$ for any density matrix $\rho$ (preservation of probability)
	\item $\mathcal{E}(\sum _{i} q_i \rho _i ) = \sum _i q_i \mathcal{E}\rho_i$ (convex linear map)
	\item $\mathcal{I}_{A} \otimes \mathcal{E}_S (\rho_{AS}) \geq 0$ (complete positivity) for an arbitrary ancilla system $A$ and density matrix $\rho _{AS}$ on the composite system $AS$.
\end{itemize}
Such a CPTP map can always be written by Kraus operators\index{Kraus operators} $\{ K_k\}$~\cite{NielsenChuang}:
\begin{eqnarray}
\mathcal{E} \rho = \sum _{k=1}^{M} K_{k} \rho K_{k}^{\dag}.
\end{eqnarray}

Under the Born-Markov (with rotating-wave) approximation, the time evolution of a two-level system coupled with an environment is given by a master equation of the Lindblad form~\cite{Breuer}:
\begin{eqnarray}
\dot \rho  (t)
&=& -  \frac{\gamma _{+}}{2} \bigl[ \sigma _{-}\sigma _{+} \rho(t) + \rho (t) \sigma _{-} \sigma _{+} - 2 \sigma _{+} \rho (t) \sigma _{-} \bigr]
 \nonumber \\
 &&  -  \frac{\gamma _{-}}{2} \bigl[ \sigma _{+}\sigma _{-} \rho(t) + \rho (t) \sigma _{+} \sigma _{-} - 2 \sigma _{-} \rho (t) \sigma _{+} \bigr]
 \nonumber \\
 &&  -  \frac{\gamma _{0}}{2} \bigl[ \sigma _{z} \rho(t) + \rho (t) \sigma _{z} - 2 \sigma _{z} \rho (t) \sigma _{z} \bigr]
 \equiv \mathcal{L} \rho (t),
\end{eqnarray}
where $\sigma _{+} = |0\rangle \langle 1|$, $\sigma _{-} = |1\rangle \langle 0|$ and $\gamma _{\alpha}$ ($\alpha = 0, +,-$) are the decay rates of the decay channels.
One can easily find the eigenoperators of the Lindblad super-operator. 
These eigenoperators form the damping basis \cite{Briegel93}:
\begin{eqnarray}
\mathcal{L} \sigma _{1} &=& 
\frac{ \gamma _{+} + \gamma _{-} + 2 \gamma _{0}}{2} \sigma _{1}
\equiv \lambda _{1} \sigma _{1},
\\
\mathcal{L} \sigma _{2} &=& 
\frac{ \gamma _{+} + \gamma _{-} + 2 \gamma _{0}}{2} \sigma _{2}
\equiv \lambda _{2} \sigma _{2},
\\
\mathcal{L} \sigma _{3} &=& 
( \gamma _{+} + \gamma _{-} ) \sigma _{3}
\equiv \lambda _{3} \sigma _{3},
\\
\mathcal{L} \rho _{eq} &=& 0,
\end{eqnarray}
where $\rho _{eq} = (\gamma _{+} |0\rangle \langle 0| + \gamma _{-} |1\rangle \langle 1|)/ (\gamma _{+} +\gamma _{-}) \equiv (\sigma _{0} + a \sigma _{3})/2$
with $a= (\gamma_{+}-\gamma _{-})/(2\gamma _{+} +2\gamma _{-})$.
The solution of this master equation is given by the CPTP map $\mathcal{E} (t)$:
\begin{eqnarray}
\mathcal{E} (t) \rho = 
p_{0} (t) \rho  + \sum _{i=1,2,3}
p_{i} (t) \sigma _{i} \rho \sigma _{i}
+ f(t) ( \sigma _{3} \rho + \rho \sigma _{3} - i \sigma _{1} \rho \sigma _{2} + i \sigma _{2} \rho \sigma _{1}),
\nonumber \\
\label{eq:noisemap}
\end{eqnarray}
where
\begin{eqnarray}
p_{0}(t) &=& \frac{1}{4} ( 1+ e ^{-\lambda _{1} t} + e^{-\lambda _{2} t} + e^{-\lambda _{3} t}),
\\
p_{1}(t) &=& \frac{1}{4} ( 1+ e ^{-\lambda _{1} t} - e^{-\lambda _{2} t} - e^{-\lambda _{3} t}),
\\
p_{2}(t) &=& \frac{1}{4} ( 1- e ^{-\lambda _{1} t} + e^{-\lambda _{2} t} - e^{-\lambda _{3} t}),
\\
p_{3}(t) &=& \frac{1}{4} ( 1- e ^{-\lambda _{1} t} - e^{-\lambda _{2} t} + e^{-\lambda _{3} t}),
\\
f(t) &=& \frac{a}{4}(1- e^{- \lambda _{3} t}).
\end{eqnarray}
If we consider a high temperature case (i.e., $a \rightarrow 0$), Eq. (\ref{eq:noisemap}) can be rewritten as
\begin{eqnarray}
\mathcal{E} (t) \rho = \left[1-   p_{1}(t) - p_{2}(t) - p_{3}(t) \right]   \rho  + \sum _{i=1}^{3}  p_{i}(t) \sigma _{i} \rho \sigma _{i}.
\label{StochasticPauli}
 \end{eqnarray}
Hence, the CPTP map can be viewed as a stochastic Pauli error with probabilities $p_{i}(t)$.
In general, noise cannot be written by a stochastic Pauli error.
However, by performing an appropriate operation, one can depolarize the CPTP map into a stochastic Pauli error as a standard form, where the noiseless part of the evolution is not altered~\cite{Dur05}.
Otherwise, the Pauli basis measurements can collapse the CPTP map into a stochastic Pauli error, as we will see later.
In the rest of this book, therefore, we will consider Markovian stochastic Pauli errors only.

The fault-tolerance against more general noise has been discussed in Refs.~\cite{Aharonov97,Aharonov08,Aliferis06,Aliferis08,Aliferis09}.
If the decoherence is non-Markovian, the dynamical decoupling or quantum Zeno effect can be used to suppress the decoherence \cite{Zurek84,Vaidman96,Duan98,Viola98,Viola99,Facchi05}.
Besides, if the decoherence is spatially correlated, one can utilize a passive error-prevention scheme, the so-called {\it decoherence free subspace} (DFS), which is immune to collective noise \cite{Zanardi97,Duan97,Lidar98}.
\chapter{Stabilizer formalism and its applications}
In general, the description of quantum states is a difficult task because it requires exponentially many parameters in the number of qubits as shown in Eq.\ (\ref{eq01}).
To understand these complex quantum systems, it is essential to have efficient tools.
The stabilizer formalism is one such powerful tool to describe an important class of entangled states.
It also provides a diagrammatic understanding of quantum states and operations.
The stabilizer states, described by the stabilizer formalism, play important roles in quantum computation, such as for quantum error correction codes and resource states in MBQC.
In this chapter, we introduce the stabilizer formalism\index{stabilizer formalism}, especially focusing on its diagrammatic understanding.
Based on the stabilizer formalism, we explain quantum error correction, magic state distillation, and MBQC.

\section{Stabilizer formalism}
We first define an $n$-qubit Pauli group\index{Pauli group} $\mathcal{P}_n$:
\begin{eqnarray}
\mathcal{P}_n := \{ \pm 1, \pm i \} \times \{ I, X, Y, Z\}^{\otimes n} .
\end{eqnarray}
An element of the Pauli group is called a Pauli product\index{Pauli product}.
For example, the two-qubit Pauli group is given by
\begin{eqnarray}
\mathcal{P}_2 &:= & \{ \pm 1, \pm i \} 
\nonumber \\
&&\times \{II, IX,IY,IZ,XI,XX,XY,XZ,YI,YX,YY,YZ,ZI,ZX,ZY,ZZ\} ,
\end{eqnarray}
where $A \otimes B$ is denoted by $AB$ for simplicity.
(We will frequently use this notation when there is no possibility for confusion.) 
Next, we define an $n$-qubit stabilizer group\index{stabilizer group} $\mathcal{S}$ as an Abelian (commutative) subgroup of the $n$-qubit Pauli group:
\begin{eqnarray}
\mathcal{S}:= \{S_i \} 
\textrm{ s.t.  } 
-I \notin \mathcal{S}
\textrm{ and }
{}^\forall S_i, S_j \in \mathcal{S},
[S_i,S_j]=0.
\end{eqnarray}
Because $-I$ is not included in the stabilizer group, all elements are hermitian $S_i = S_i ^{\dag}$, which guarantees that the eigenvalues $=\pm 1$.
An element of the stabilizer group is called a stabilizer operator.
The maximum independent subset $\mathcal{S}_g$ of 
the stabilizer group is called stabilizer generators\index{stabilizer generator}.
Here, independence means that any element of $\mathcal{S}_g$ cannot be expressed as a product of other elements in $\mathcal{S}_g$.
Any element of the stabilizer group can be generated as a product of the stabilizer generators.
The stabilizer group $\mathcal{S}$ 
generated by the generators $\mathcal{S}_g$ is denoted by $\mathcal{S} = \langle \mathcal{S}_g \rangle$.

Let us, for example, consider a two-qubit stabilizer group:
\begin{eqnarray}
\mathcal{S}_{\rm Bell}=\{ II, XX, ZZ, -YY\}.
\end{eqnarray} 
Because they contain two anticommuting Pauli operators, $XX$ and $ZZ$ commutes.
The stabilizer group $\mathcal{S}_{\rm Bell}$ is generated by $\{XX,ZZ\}$, because $-YY$ can be expressed as a product of $XX$ and $ZZ$.
Thus, we can write $\mathcal{S}_{\rm Bell} = \langle \{ XX,ZZ\} \rangle$.

For a given stabilizer group $\mathcal{S}$, the stabilizer state\index{stabilizer state} is defined as a simultaneous eigenstate of all stabilizer elements $S_i \in \mathcal{S}$ with the eigenvalue +1:
\begin{eqnarray}
{}^{\forall} S_i \in \mathcal{S}, \;\;\; S_i | \psi \rangle = |\psi \rangle   .
\end{eqnarray}
It is sufficient that the state is an eigenstate of all stabilizer generators:
\begin{eqnarray}
{}^{\forall} S_i \in \mathcal{S}_g, \;\;\; S_i | \psi \rangle = |\psi \rangle.
\end{eqnarray}
Let $k$ be the number of elements in the stabilizer generator $\mathcal{S}_g$.
Each stabilizer generator divides an $n$-qubit system (Hilbert space) into two orthogonal subspaces associated with the eigenvalues $\pm1$.
Because all stabilizer operators commute with each other, the $k$ stabilizer generators divide the $n$-qubit system into $2^{k}$ orthogonal subspaces.
Thus, the dimension of the space spanned by the stabilizer states, which we call a {\it stabilizer subspace}\index{stabilizer subspace}, is $2^d=2^{n-k}$.
When $n=k$, we can define the quantum state uniquely.
The number of stabilizer generators is at most $n$ for an $n$-qubit stabilizer group.
In the case of $k<n$, the degrees of freedom in the stabilizer subspace can be addressed by using {\it logical operators}\index{logical operator}, which commute with all stabilizer generators and also are independent of them.

Let us consider the stabilizer group $\mathcal{S}_{\rm Bell}$ again.
The stabilizer state is the eigenstate of $XX$ and $ZZ$ with eigenvalue $+1$, and hence given by the Bell state $(|00\rangle + |11\rangle )/\sqrt{2}$~\cite{EPR}.
If $XX$ is removed from the generators, the two-dimensional subspace spanned by $|00\rangle$ and $|11\rangle$ is stabilized.
By choosing logical operators $L_X=XX$ and $L_Z=ZI$, we can specify the state in the subspace.
For example, the eigenstate of $L_X$ with the eigenvalue $+1$ is the Bell state.
The eigenstate of $L_Z$ with the eigenvalue $+1$ is $|00\rangle$.
Another representative example of the stabilizer states is an $n$-qubit cat state,
\begin{eqnarray}
| {\rm cat } \rangle = \frac{1}{\sqrt{n}} ( | 00 ... 0 \rangle + |11 ... 1\rangle ),
\label{eq:cat}
\end{eqnarray}
whose stabilizer group is given by 
\begin{eqnarray}
\left\langle Z_1 Z_2, \; ... , \; Z_{n-1}Z_{n}, \;\prod _{i=1}^{n} X_i \right\rangle.
\end{eqnarray}
The cat state is a representative example of a macroscopically entangled state.
If it is determined whether a particle is $|0\rangle$ or $|1\rangle$, the superposition is completely destroyed.
If an element $\prod _{i=1}^{n} X_i$ is removed from the stabilizer generator, it defines a stabilizer subspace spanned by $|00..0\rangle$ and $|11..1\rangle$.
We can choose $L_X=\prod _{i=1}^{n} X_i$ and $L_Z=Z_i$ as logical operators, which anti-commute with each other and behave as logical Pauli operators.

\begin{figure}[t]
\centering
\includegraphics[width=120mm]{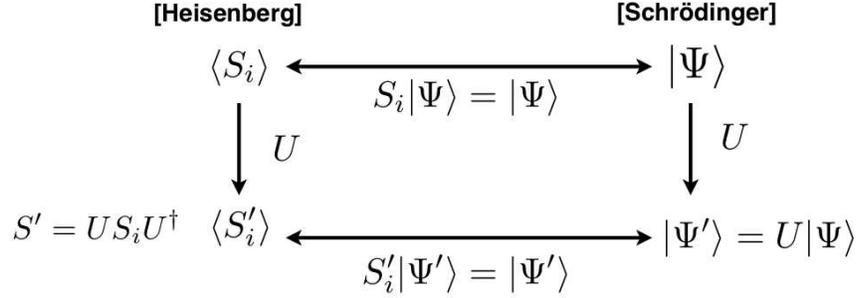}
%
%
\caption{The stabilizer formalism: a Heisenberg picture of quantum computation. A Clifford operation is represented as a transformation of the stabilizer group by the conjugation of $U$.}
\label{fig19}       
\end{figure}
\section{Clifford operations}
In the stabilizer formalism, 
we can describe a restricted class of unitary operations,
the so-called Clifford operations,\index{Clifford operation}
acting on the stabilizer states quite efficiently.
The Clifford operation is defined as an operation $U$ 
that transforms a Pauli product into another Pauli product under its conjugation, 
$[...] \rightarrow U[...]U^{\dag}$.
Let us consider the action of a Clifford operation\index{Clifford operation} $U$ on the stabilizer state $|\psi \rangle$ defined by a stabilizer group 
$\mathcal{S}=\langle \{S_i\}  \rangle$:
\begin{eqnarray}
U | \psi \rangle = US_i |\psi \rangle  = U S_i U^{\dag}U  | \psi \rangle  = S'_i  U| \psi \rangle,
\end{eqnarray}
where we define $S'_i \equiv U S_i U^{\dag} $.
The above equality indicates that the state $U | \psi \rangle $ is an eigenstate of the operator $S'_i$ with an eigenvalue +1 for all $S'_i$.
Because $U$ is a Clifford (unitary) operation, the group $\{ S'_i \}$ is also an Abelian subgroup of the Pauli group.
Accordingly, the state $U | \psi \rangle $ is a stabilizer state with respect to the stabilizer group $\{ S'_i\}$.
In this way, the action of $U$ on the stabilizer state can be represented as a transformation of the stabilizer groups under the conjugation of $U$ as shown in Fig.~\ref{fig19}.
For example, the stabilizer state stabilized by $\langle X_1I_2, I_1Z_2 \rangle$ is $|+\rangle_1 |0\rangle_2$.
The stabilizer group is transformed by $\Lambda (X)_{1,2}$ into $\langle X_1X_2, Z_1Z_2 \rangle$, whose stabilizer state is $(|00\rangle  + |11\rangle)/\sqrt{2}$.

The stabilizer formalism corresponds to the Heisenberg picture of quantum computation, where a minimum number of operators are employed to describe a restricted type of quantum states and operations~\cite{Gottesman,GottesmanHeisenberg}. 
This representation is powerful because it requires us to keep a time evolution of at most $n$ operators, while a straightforward state-based approach needs exponentially many states.
For example, let us consider the following quantum circuit:
\begin{center}
\includegraphics[width=70mm]{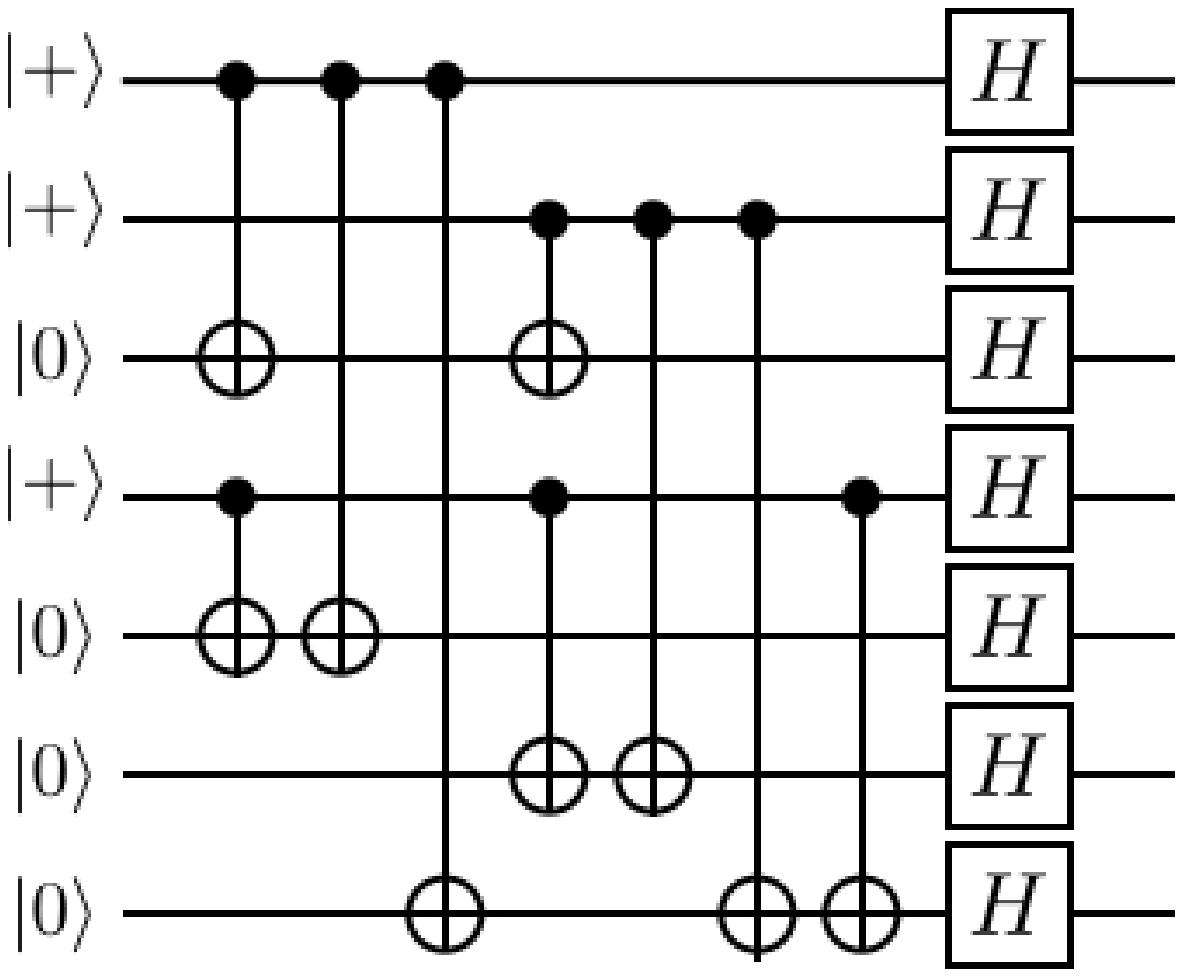}
\end{center}
A straightforward calculation yields the output state $|\psi \rangle$,
\begin{eqnarray}
|\psi \rangle &=& (|0000000\rangle + |1010101\rangle + |0110011\rangle+|1100110\rangle
\nonumber \\
&& + |0001111\rangle + |1011010\rangle+|0111100\rangle+|1101001\rangle 
\nonumber \\
&&+ |1111111\rangle + |0101010\rangle + |1001100\rangle+ |0011001\rangle 
\nonumber \\
&&+ |1110000\rangle + |0100101\rangle+|1000011\rangle+|0010110\rangle)/4.
\end{eqnarray}
It is rather cumbersome to write down the above state.
Instead, we can understand the output state as a stabilizer state whose stabilizer generators are 
\begin{eqnarray}
\{ ZIZIZIZ, IZZIIZZ, IIIZZZZ, XXXIIII, XXIXXII, IXIXIXI, XIIXIIX \}.
\label{eq:7-qubit}
\end{eqnarray}
Equivalently, we may also choose the following stabilizer generators because they generate the same stabilizer group:
\begin{eqnarray}
\{ ZIZIZIZ, IZZIIZZ, IIIZZZZ, XXXXXXX, IIIXXXX, XIXIXIX, IXXIXXI \}.
\end{eqnarray}
Actually, these stabilizer generators are enough to understand the properties of the quantum state $|\psi \rangle$.
If an explicit description of the state is required, we can systematically write it down as follows:
\begin{eqnarray}
|\psi \rangle = 4\frac{I+S_4}{2}\frac{I+S_3}{2}\frac{I+S_2}{2}\frac{I+S_1}{2}|0000000\rangle ,
\end{eqnarray}
where $S_1 = XIXIXIX$, $S_2 = IXXIIXX$, $S_3 = IIIXXXX$, and  $S_4 = XXXXXX$.
The above equation means that $|0000000\rangle$ is an eigenstate for all $Z$'s stabilizer operators.
By projecting it into the $+1$ eigenstate of the stabilizer generator $S_i$ by the projection $\frac{I+S_i}{2}$, we obtain the stabilizer state $|\psi \rangle$.

In order for the above calculation to work, 
we have to obtain the stabilizer generators of the output state.
This can easily be done graphically.
We introduce commutation rules between 
the Pauli operators and Clifford operations below.
In the case of the Hadamard operation, $HX=ZH$ and $ZH=HX$, and hence we have
\begin{center}
\includegraphics[width=60mm]{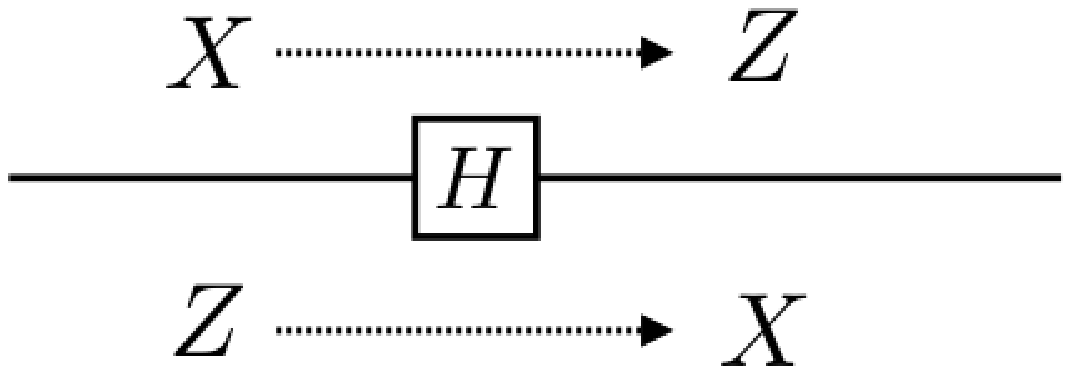}
\end{center}
meaning that the Pauli $X$ operator acting before the Hadamard operation is equivalent to the Pauli $Z$ operator acting after the Hadamard operation and so on.
Similarly, for the phase operation $X$, we have 
\begin{center}
\includegraphics[width=60mm]{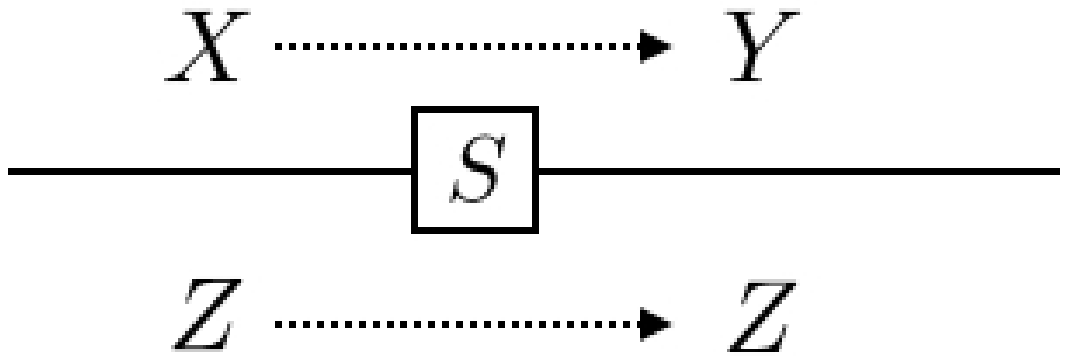}
\end{center}

The CNOT operation transforms the Pauli operators under its conjugation as follows:
\begin{eqnarray}
\Lambda _{c,t} (X) X_c  \Lambda_{c,t}(X) &=& X_c X_t,
\label{eq:CNOT1}
\\
\Lambda _{c,t} (X) X_t  \Lambda_{c,t}(X) &=&  X_t,
\\
\Lambda _{c,t} (X) Z_c  \Lambda_{c,t}(X) &=& Z_c, 
\\
\Lambda _{c,t} (X) Z_t  \Lambda_{c,t}(X) &=& Z_c Z_t.
\label{eq:CNOT4}
\end{eqnarray}
The commutation relation between the CNOT operation and the Pauli operators is understood as follows:
\begin{center}
\includegraphics[width=80mm]{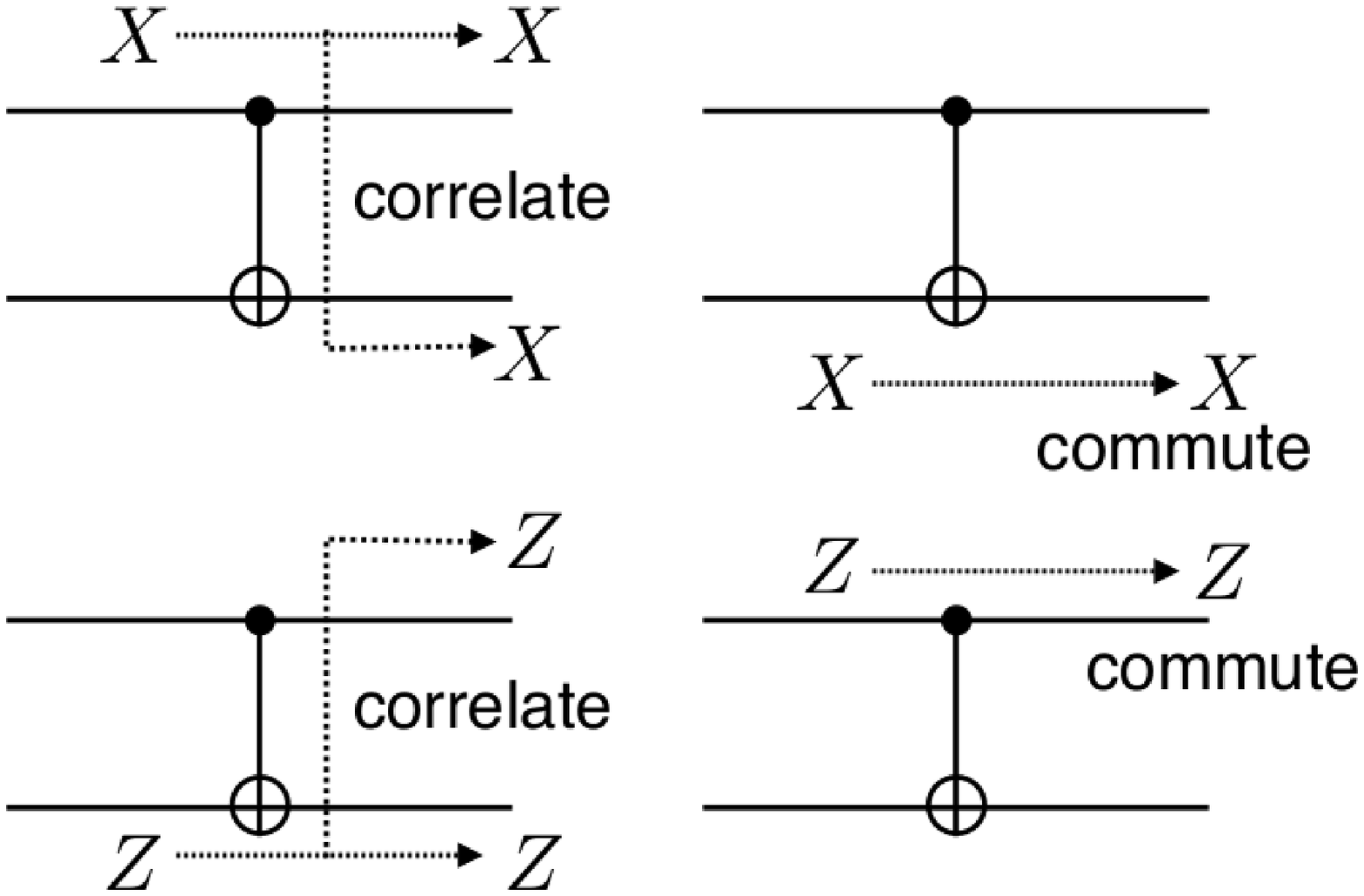}
\end{center}
In the above circuit diagram, the solid circle commutes with the Pauli $Z$ operator, while the Pauli $X$ operator is propagated as the Pauli $X$ operator on the target qubit, making a correlation.
Similarly, the open circle commutes with the Pauli $X$ operator, while the Pauli $Z$ operator is propagated as the Pauli $Z$ operator on the control qubit, making a correlation.
By recalling that the CNOT operation is transformed into the CZ operation by the Hadamard operations on the target qubit, the commutation relation between the CZ operation and the Pauli operators are obtained straightforwardly.
This is described graphically as follows:
\begin{center}
\includegraphics[width=80mm]{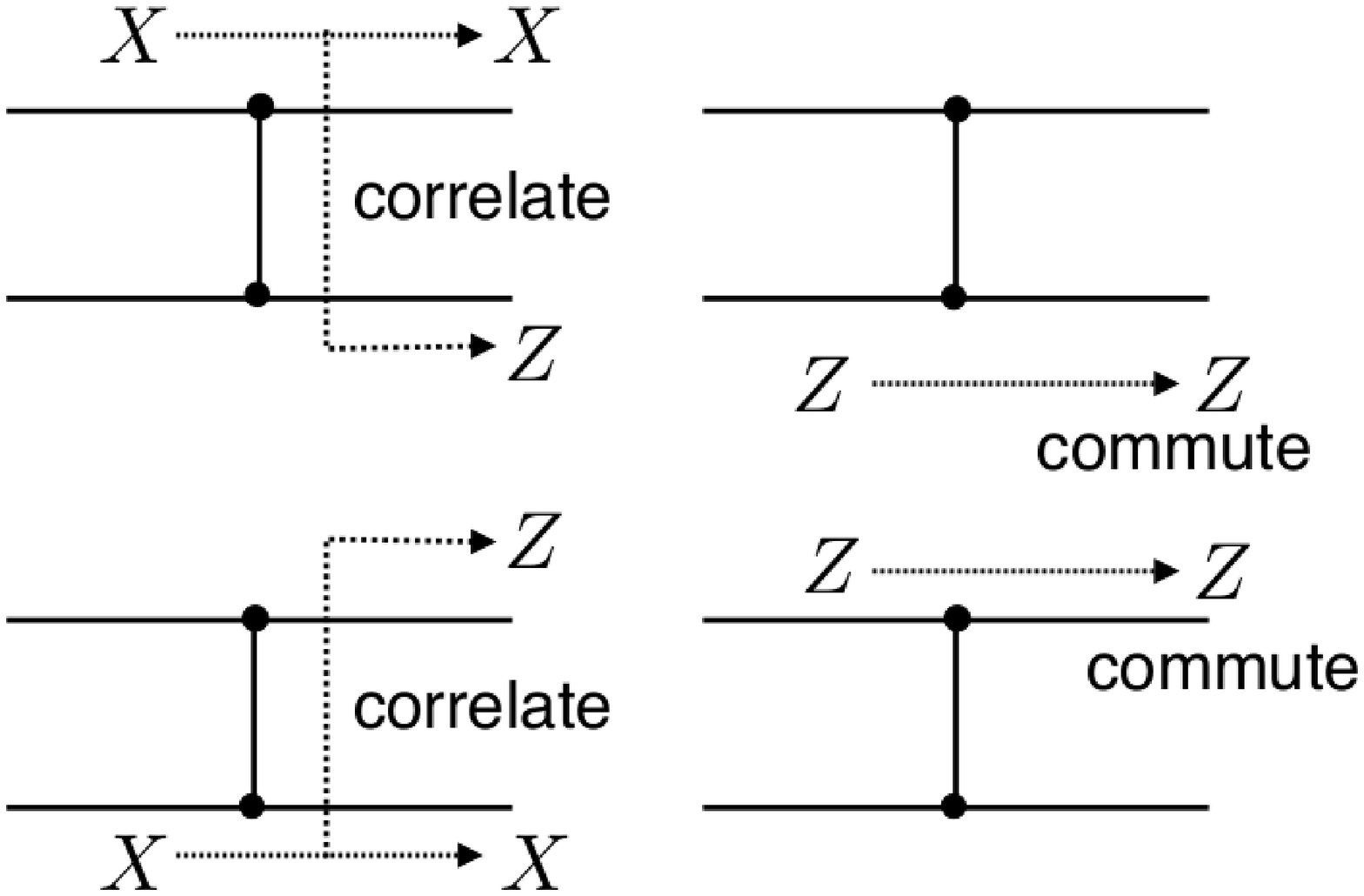}
\end{center}
In this case, note that the Pauli $X$ operation is propagated as the Pauli $Z$ operation.

This graphical understanding allows us to calculate the stabilizer generators of the output of the Clifford circuits.
For example, in the following circuit diagram, the first qubit is stabilized by $X$ before the Clifford operation.
The Pauli $X$ operator is propagated toward the right, and we obtain the stabilizer operator $ZIZIZIZ$ for the output:
\begin{center}
\includegraphics[width=80mm]{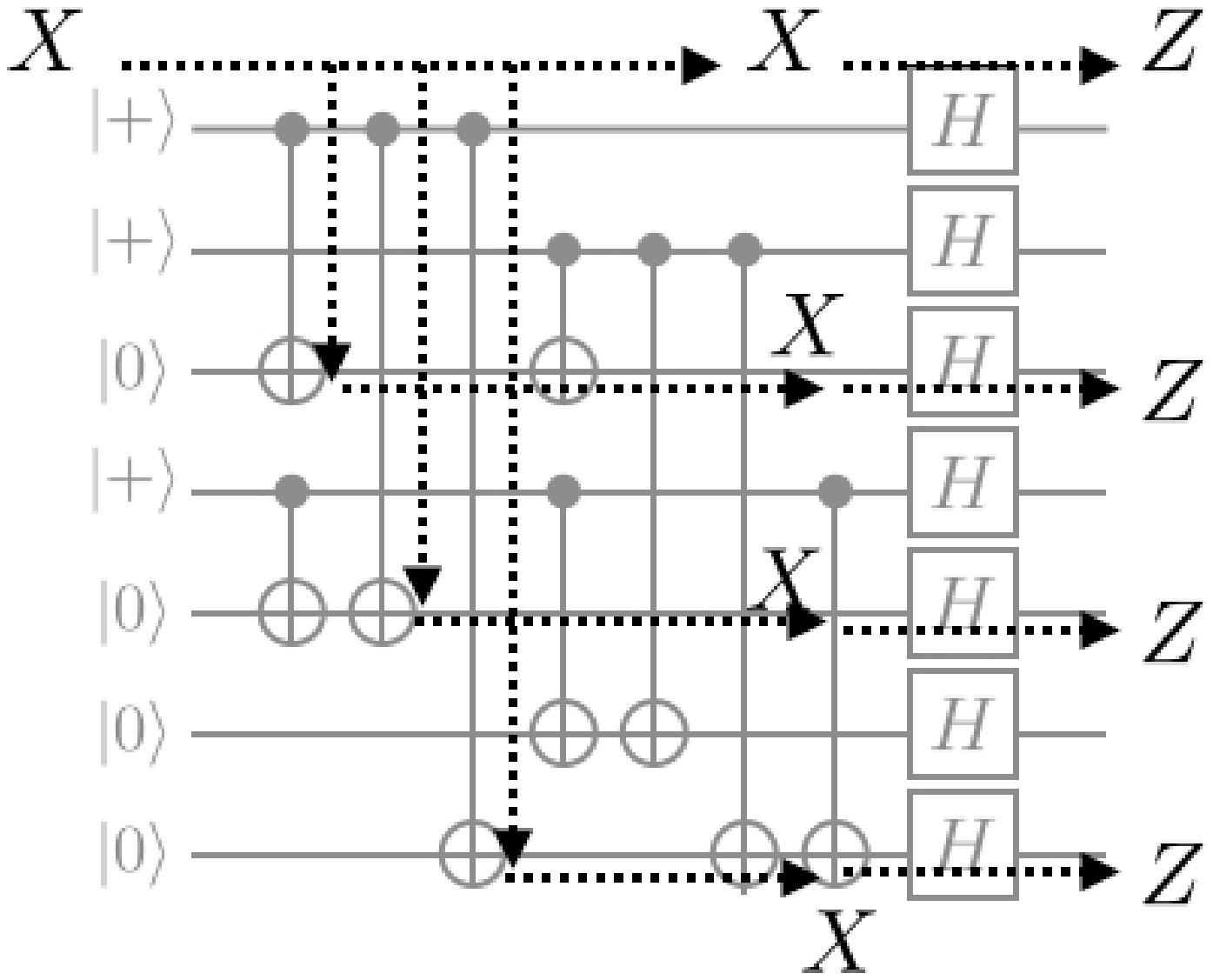}
\end{center}
The reader should use this graphical technique to calculate the other stabilizer generators and verify Eq.\ (\ref{eq:7-qubit}).

\section{Pauli basis measurements}
\label{sec:PauliMeas}
Next, we will see how the Pauli-basis measurements\index{Pauli-basis measurement} on the stabilizer states are described in the stabilizer formalism.
Suppose the $A$-basis ($A=X,Y,Z$) measurement is performed on a stabilizer state $|\psi\rangle$, whose stabilizer group is given by $\langle S_i \rangle$.
(We assume that the number of stabilizer generators is equal to the number of qubits, and hence that the stabilizer state is uniquely defined.)
Depending on the stabilizer group $\langle \{  S_i \} \rangle$ and $A$, there are two possibilities:
\begin{itemize}
	\item[(i)] The Pauli operator $A$ commutes with all stabilizer generators. 
	In that case, either $A$ or $-A$ is an element of the stabilizer group. 
	If $A$ ($-A$) is an element, the eigenvalue $+1$ ($-1$) is obtained with probability 1. 
	The post-measurement state is the same as the stabilizer state before measurement.

	\item[(ii)] At least one stabilizer operator does not commute with $A$. 
	In this case, we can choose another set of generators $\{ S'_i \}$ such that $S'_1$ anti-commutes with $A$ but all other generators commute with $A$. 
	The measurement outcomes $+1$ and $-1$ are obtained with an equal probability of 1/2. The post-measurement state is given by $\langle (-1)^m A, S'_2, ... , S'_k \rangle$ depending on the measurement outcomes $m=0,1$ 
	corresponding to the eigenvalues $(-1)^{m}$.
\end{itemize}

For example, suppose we perform the $Y$-basis measurement on the first qubit of the Bell state stabilized by $\mathcal{S}_{\rm Bell}=\langle XX, ZZ\rangle$.
We can redefine the stabilizer generators by $\{XX, -YY\}$.
Then the stabilizer group after the measurement is given by $\langle YI,-YY \rangle = \langle YI, -IY \rangle$.
Thus, we obtain $|-i\rangle$ as the post-measurement state on the second qubit.

\section{Gottesman-Knill theorem}
\label{sec:GKtheorem}
Because the stabilizer states and Clifford operations are described efficiently in the stabilizer formalism, it implies that such a restricted type of quantum computation can be simulated efficiently on a classical computer.
This is stated by the Gottesman-Knill theorem\index{Gottesman-Knill theorem}~\cite{Gottesman,GottesmanHeisenberg,NielsenChuang}.

\begin{theorem}
Any Clifford operations, applied to the input state $|0\rangle^{\otimes n}$ followed by the $Z$ measurements, can be simulated efficiently in the strong sense.
\label{GKtheorem}
\end{theorem}
Here, an efficient {\it strong} classical simulation of a quantum circuit $C$ is a classical polynomial-time computation that calculates the probability $P_{C}(x)$ 
for a given output $x$ of the circuit $C$,
including an arbitrary marginal distribution $\sum _{x'} P_{C}(x)$. 
(See, for example, Ref.~\cite{IQP} for the definition of a strong simulation.)
Note that this theorem holds true even when the initial state is generalized to an arbitrary stabilizer state, and also any Pauli products are measured, because they are done in the above setup by modifying the Clifford operations appropriately.

{\it Proof:}
The stabilizer group of the input state is $\langle \{ Z_i \} \rangle$
($i=0,1,...,n-1$).
By applying the Clifford operations as mentioned, we obtain the stabilizer generators $\langle \{ S_i \} \rangle$ of the quantum output before the measurements.
Suppose the measurement outcome, the classical output, is given by $\{ m_i=0,1\}$.
Then the probability of obtaining the measurement outcome $\{m_i\}$ can be calculated as follows:
\begin{enumerate}[i)]
	\item Set the stabilizer generators $\mathcal{S}^{(0)} = \langle \{S_i\} \rangle$ and the initial probability $p^{(0)}=1$.

	\item For $k=0,1,...,n-1$, repeat the following procedures.
	\begin{enumerate}[1)]
		\item If $(-1)^{m_k} Z_k \in \mathcal{S}^{(k)}$, update the probability $p^{(k+1)}=p^{(k)}$, because the measurement outcome $m_k$ is obtained with probability $1$. 
		The stabilizer group after the measurement is also updated to $\mathcal{S}^{(k+1)}=\mathcal{S}^{(k)}$.

		\item Else, if $(-1)^{m_k\oplus 1} Z_k \in \mathcal{S}^{(k)}$, update the probability $p^{(k+1)}=0$, because such a measurement outcome does not appear. 
		(You may stop the calculation at this stage, and return the probability 0.)

		\item Else, $\mathcal{S}^{(k)}$ is updated into $\mathcal{S}^{(k+1)}$ by removing an anticommuting generator and adding $(-1)^{m_k} Z_k$ as a new generator. 
		Because the measurement outcome is obtained randomly with probability 1/2, the probability is taken as $p^{(k+1)}=p^{(k)}/2$.
	\end{enumerate}

	\item Return $p^{(n)}$ as the probability of obtaining the measurement outcome $\{ m_i \}$.
\end{enumerate}
Note that, in step (ii), we can efficiently decide which of the three is the case for any $k$ by checking the commutability of $Z_k$ with the stabilizer generators of $\mathcal{S}^{(k)}$.
\hfill $\blacksquare$

The statement of Theorem~\ref{GKtheorem} can be extended by weakening the notion of the classical simulation.
\begin{theorem}
Any Clifford operations, applied to any product states of convex mixtures of the Pauli basis states, followed by $Z$ measurements can be efficiently simulated in the weak sense.
\label{GKtheorem_mix}
\end{theorem}
Here, an efficient {\it weak} classical simulation of a quantum circuit $C$ is a classical polynomial-time randomized computation that samples the output $x$ according to the probability distribution $P_{C}(x)$ of the output of the circuit $C$. 
(See, for example, Ref.~\cite{IQP} for the definition of weak simulation.)
Apparently, a strong simulation includes a weak simulation, because we sample the output by using the marginal distributions~\cite{TerhalDiVincenzo}.

{\it Proof:}
Suppose that the $i$th input qubit is given by
\begin{eqnarray}
\rho _{i} &=& p^{(i)}_{x,+} |+\rangle \langle +| +p^{(i)}_{x,-} |-\rangle \langle -| +p^{(i)}_{y,+} |+i\rangle \langle +i| +p^{(i)}_{y,-} |-i\rangle \langle -i|
\nonumber \\
&&+p^{(i)}_{z,+} |0\rangle\langle0|+ p^{(i)}_{z,-} |1\rangle \langle 1|,
\end{eqnarray}
where $\sum _{\alpha =x,y,z} \sum _{\nu = +,-} p_{\alpha , \nu}^{(i)}=1$.
By using the probability distribution $\{p_{\alpha , \nu}^{(i)} \}$, the input state of each qubit is randomly sampled.
Conditioned by the sampling result, the input state is a product of the Pauli basis states, and hence the output probability distribution can be calculated as shown in Theorem~\ref{GKtheorem}.
Combined with the random sampling of the input state, this provides an efficient weak simulation of the Clifford circuit with noisy input states (convex mixture of the Pauli basis states).
\hfill $\blacksquare$
\\
The input state can be generalized into a classical mixture of stabilizer states, when its polynomial size description of the probability distribution is provided. 
Similarly, the Clifford operations can be extended to stochastic Clifford operations such as the stochastic Pauli error.

\begin{figure}[t]
\centering
\includegraphics[width=100mm]{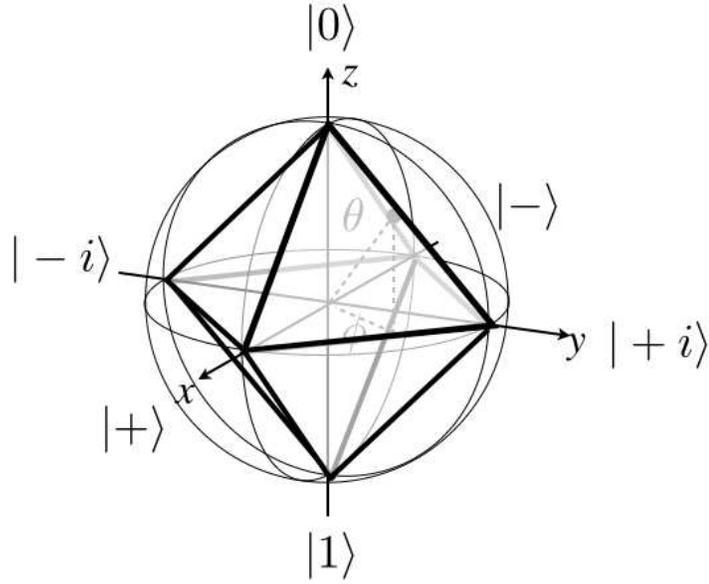}
%
%
\caption{A convex mixture of the Pauli basis states lies inside the octahedron of the Bloch sphere.}
\label{fig36}       
\end{figure}

The convex mixture of the Pauli basis state lies inside the octahedron of the Bloch sphere as shown in Fig.~\ref{fig36}.
It is natural to ask whether or not the Clifford circuit allows universal quantum computation if the input state lies outside the octahedron.
If the input state is a pure non-stabilizer state such as $e^{ - i (\pi/8) Z}|+\rangle$, we can implement a non-Clifford gate $e^{ - i (\pi /8) Z}$ by using gate teleportation, explained in Sec.~\ref{Sec:MBQC}.
Even some mixed states can be converted into a pure non-stabilizer state, the so-called magic state, by using only Clifford operations.
Such a protocol is called magic state distillation~\cite{Magic} and will be explained in Sec.~\ref{sec:magic}.

\section{Graph states}
In this section, we introduce an important class of stabilizer states, the so-called {\it graph states}\index{graph state}~\cite{Graphstate}, whose stabilizer generators are defined on graphs.
The graph states are employed as resource states for MBQC as explained in the next section.
\begin{figure}[t]
\centering
\includegraphics[width=100mm]{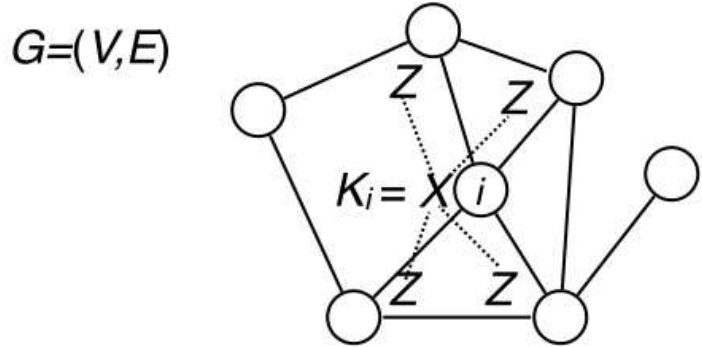}
%
%
\caption{The graph state $|G\rangle$ associated with a graph $G=(V,E)$.
A stabilizer generator $K_i$ is also shown.}
\label{fig28}       
\end{figure}

A graph state is defined by a graph $G=(V,E)$.
Here, $V$ and $E$ are the sets of the vertices and edges, respectively.
A qubit is located on each vertex of the graph.
The stabilizer generator of the graph state $|{G}\rangle$ is defined as
\begin{eqnarray}
K_i =  X_i \prod _{ j \in V_i } Z_j \;\; \textrm {for all } i \in V ,
\end{eqnarray}
where we define a set of vertices $V_i:= \{ j | (i,j) \in E \}$, which are connected to the vertex $i$ by an edge on the graph $G$ (see Fig.~\ref{fig36}).
The graph state $|G\rangle$ is generated from a product state $|+\rangle^{\otimes |V|}$ by applying the CZ gate on each of the graphs:
\begin{eqnarray}
|G\rangle = \prod _{(i,j) \in E} \Lambda (Z)_{i,j} | +\rangle^{\otimes |V|},
\end{eqnarray}
where $|V|$ indicates the number of vertices of the graph $G=(V,E)$.
This can be understood that the stabilizer generator $X_i$ for the state $|+\rangle$ is transformed into $K_i$ by the CZ operations $U\equiv \prod _{(i,j) \in E} \Lambda (Z)_{i,j}$.
Especially, when the graphs are regular lattices such as one-dimensional (1D), square, hexagonal, and cubic lattices, the corresponding graph states tend to be referred to as cluster states\index{cluster states}~\cite{Cluster}.
Any stabilizer state is equivalent to a certain graph state up to local Clifford operations~\cite{LCeq,Graphstate}. 
For example, the cat state is equivalent to the following graph state by applying the Hadamard operation on the $k$th qubit:
\begin{center}
\includegraphics[width=60mm]{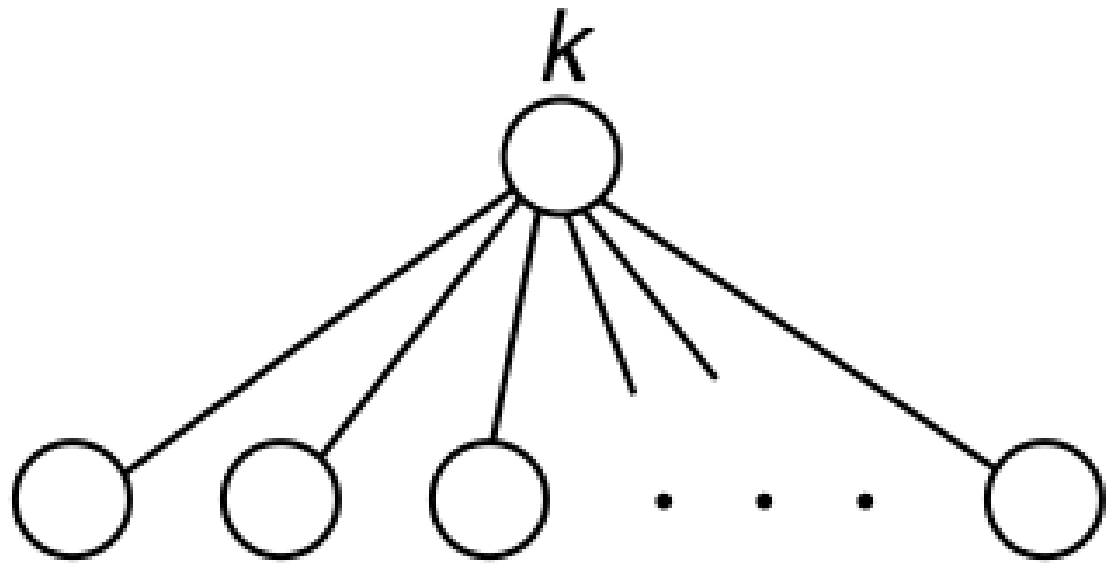}
\end{center}
Unfortunately, the graph associated with a stabilizer state is not uniquely defined, because there are local Clifford operations that change the underlying graph.
This property is called the local complementarity of the graph states~\cite{LCeq,Graphstate}.

Next, we will see how the Pauli basis measurements transform the graph states.
For simplicity, we assume that the state is projected into an eigenstate with eigenvalue $+1$.
Let us consider a 1D graph state as follows:
\begin{center}
\includegraphics[width=70mm]{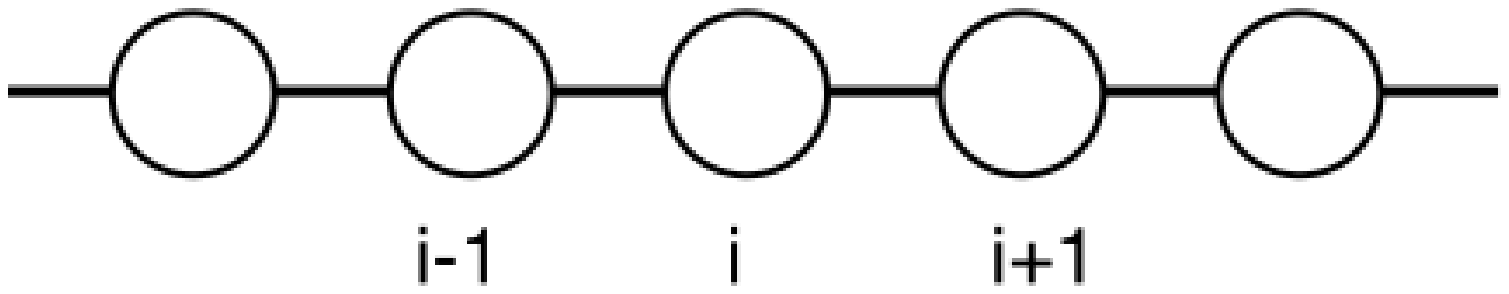}
\end{center}
whose stabilizer generator is given by
\begin{eqnarray}
K_i=Z_{i-1}X_i Z_{{i+1}}.
\end{eqnarray}
We first consider the $Z$ basis measurement (projective measurement of the observable $Z$) on the $i$th qubit. 
Following the procedure seen in Sec.~\ref{sec:PauliMeas}, $K_i$ is removed from the stabilizer generator.
By adding $Z_i$ instead, we obtain the stabilizer group for the post-measurement state
\begin{eqnarray}
\langle ..., K_{i -1}, Z_{i}, K_{i+1} ,...\rangle.
\end{eqnarray}
After the projection, the $i$th qubit is $|0\rangle$ and hence decoupled from the other qubits.
By rewriting the stabilizer generators, we obtain three decoupled stabilizer groups
\begin{eqnarray}
\langle ..., Z_{i-2}X_{i-1}  \rangle  , \langle Z_i \rangle, \langle   X_{i+1} Z_{i+2} ,...\rangle.
\end{eqnarray}
This means that the graph is divided into two parts as follows:
\begin{center}
\includegraphics[width=90mm]{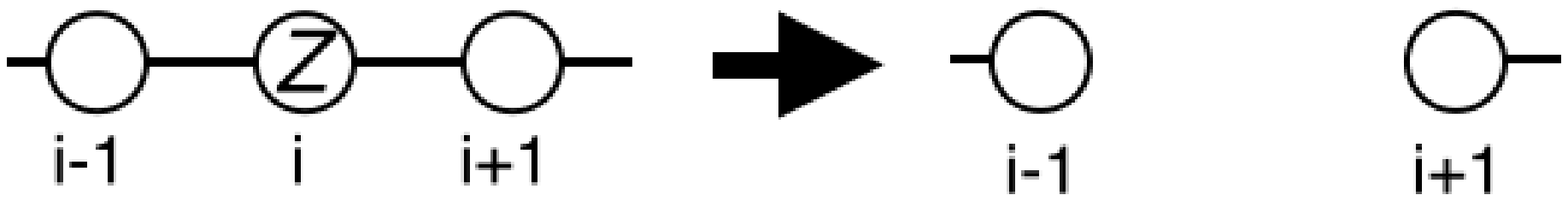}
\end{center}
For any graph, this property of the $Z$-basis measurement holds; the post-measurement state is defined by a modified graph, where the vertex corresponding to the measured qubit and the edges incident to it are removed from the original graph.

Next, we consider the $X$-basis measurement.
The observable $X_i$ does not commute with $K_{i-1}$ and $K_{i+1}$, but does commute with $K_{i-1}K_{i+1}=Z_{i-2} X_{i-1}X_{i+1} Z_{i+2}$.
Following the procedure in Sec.~\ref{sec:PauliMeas}, the stabilizer group for the post-measurement state is calculated to be
\begin{eqnarray}
\langle ...,Z_{i-2} X_{i-1}X_{i+1} Z_{i+2}, Z_{i-1}Z_{i+1}  ,...\rangle, \langle X_i \rangle.
\end{eqnarray}
By performing the Hadamard operation $H$ on the $(i-1)$th qubit, we obtain a new stabilizer group 
\begin{eqnarray}
\langle ...,Z_{i-2} Z_{i-1}X_{i+1} Z_{i+2}, X_{i-1}Z_{i+1}  ,...\rangle, \langle X_i \rangle,
\end{eqnarray}
which indicates that the graph is transformed into the following graph with the Hadamard operation:
\begin{center}
\includegraphics[width=90mm]{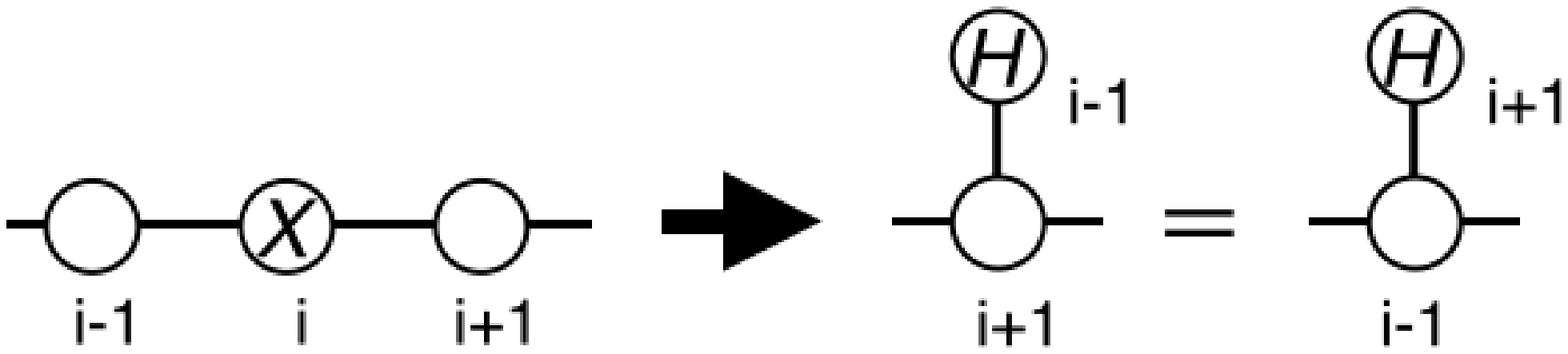}
\end{center}
Instead of the $(i-1)$th qubit, we can obtain a similar result by performing the Hadamard operation on the $(i+1)$th qubit as shown above.

Suppose the $i$th and $(i+1)$th qubits are measured in the $X$-basis on a 1D graph state. 
This is equivalent to measuring the $(i+1)$th qubit of the above post-measurement graph state in the $Z$ basis, because the Hadamard operation is applied on it as a byproduct.
From the previous argument, the $Z$-basis measurement remove the measured qubit from the graph.
Thus, two neighboring $X$-basis measurements remove the measured qubits and connect the left and right hand sides directly:
\begin{center}
\includegraphics[width=90mm]{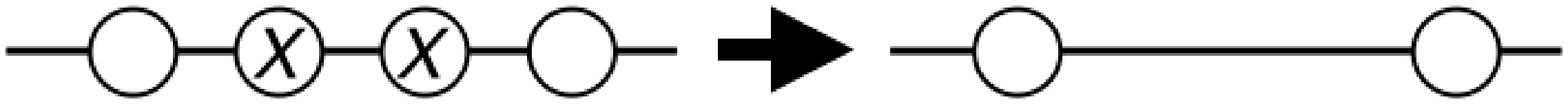}
\end{center}
which we call a contraction.

Finally, we consider the $Y$-basis measurement.
The observable $Y_i$ does not commute with either $K_{i-1}$, $K_i$, or $K_{i+1}$, but does commute with $K_{i-1}K_{i}= Z_{i-2} Y_{i-1} Y_i Z_{i+1} $ and 
$K_{i}K_{i+1}= Z_{i-1} Y_{i} Y_{i+1} Z_{i+2} $.
The stabilizer group for the post-measurement state is calculated to be
\begin{eqnarray}
\langle ...,Z_{i-2} Y_{i-1} Z_{i+1}, Z_{i-1} Y_{i+1} Z_{i+2}  ,...\rangle, \langle Y_i \rangle.
\end{eqnarray}
By performing the phase gates $S$ on the $(i-1)$th and $(i+1)$th qubits, we obtain a new stabilizer group
\begin{eqnarray}
\langle ...,Z_{i-2} X_{i-1} Z_{i+1}, Z_{i-1} X_{i+1} Z_{i+2}  ,...\rangle, \langle Y_i \rangle.
\end{eqnarray}
This indicates that the graph is directly connected up to the phase operation $S$ as a byproduct:
\begin{center}
\includegraphics[width=90mm]{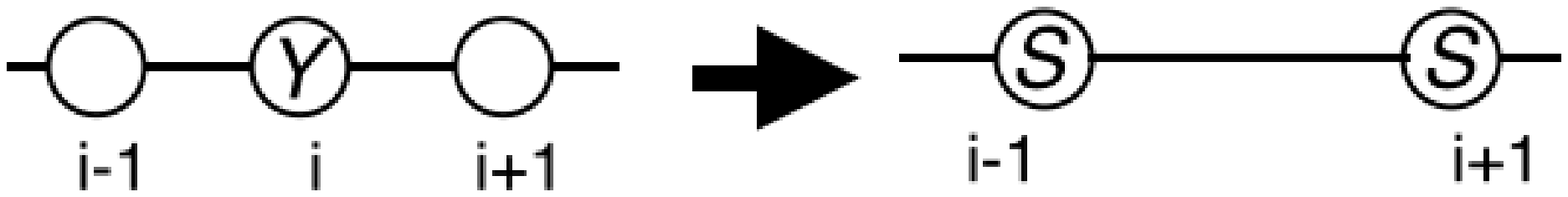}
\end{center}

Suppose three neighboring qubits $(i-1)$, $i$, and $(i+1)$ are measured in the $Y$-basis.
This is equivalent to measuring the $i$th qubit in the $Y$-basis, and then measuring the $(i-1)$th and $(i+1)$th qubits of the post-measurement graph state in the $X$-basis, because there is a phase operation $S$ acting on them as a product.
As seen previously, the $X$-basis measurements on two neighboring qubits result in a contraction of the two qubits on the graph.
Thus, the $Y$-basis measurements on three neighboring qubits contract them from the 1D graph state.
\begin{center}
\includegraphics[width=90mm]{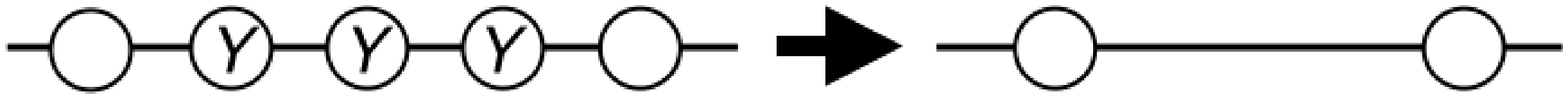}
\end{center}
This property is useful to change even and odd of the length of the 1D graph state.

While we have considered the Pauli-basis measurements only on the 1D graph state, we can generalize these arguments into graph states of general structures.
A graph state is still mapped into another graph state up to some single-qubit Clifford operations as byproducts.

\section{Measurement-based quantum computation}
\label{Sec:MBQC}
Measurement-based quantum computation (MBQC)\index{measurement-based quantum computation} is a model of quantum computation, where quantum gates are implemented by adoptive measurements on a highly entangled resource state~\cite{MBQC,MBQC_PRA,RaussendorfPhD}. 
Specifically, certain graph states, the so-called cluster states, are employed as resource states in MBQC.
Below we will first demonstrate quantum teleportation, a building block of MBQC.
Then, we explain how adoptive measurements on a graph state enable us to emulate universal quantum computation via quantum teleportation. 

Quantum teleportation is a quantum communication protocol, in which Alice sends a quantum state to Bob by using a shared entangled state and classical communication~\cite{teleportation}.
Suppose Alice and Bob share a maximally entangled state, the Bell state, 
\begin{eqnarray}
\frac{|0\rangle _a |0\rangle _b + |1\rangle _a |1\rangle _b}{\sqrt{2}}.
\end{eqnarray}
For an unknown input state $|\psi \rangle_i$ and the half of the Bell state, Alice performs a Bell measurement, which is a projection onto the Bell basis states
\begin{eqnarray}
|\Psi(m_1,m_2)\rangle _{i,a} = X^{m_1}_i Z_i^{m_2}\frac{|0\rangle _i |0\rangle _a + |1\rangle _i |1\rangle _a}{\sqrt{2}},
\end{eqnarray}
where $m_1,m_2 =0,1$ correspond to the measurement outcomes.
A straightforward calculation provides
\begin{eqnarray}
\langle \Psi(m_1,m_2) |_{i,a} \left( |\psi \rangle _i \frac{|0\rangle _a |0\rangle _b + |1\rangle _a |1\rangle _b}{\sqrt{2}} \right)
=Z^{m_2}_b X^{m_1}_b  |\psi \rangle _{b}/2.
\end{eqnarray} 
Hence, the unknown input state is teleported to Bob with a byproduct operator\index{byproduct operator} $X^{m_1} Z^{m_2}$.
If Bob does not know the measurement outcomes $(m_1,m_2)$, the teleported state is a completely mixed state for Bob.
However, if Alice sends the measurement outcome as a classical message, Bob can undo the byproduct and obtain the unknown quantum state at Bob's side.
The circuit diagram of quantum teleportation is:
\begin{center}
\includegraphics[width=100mm]{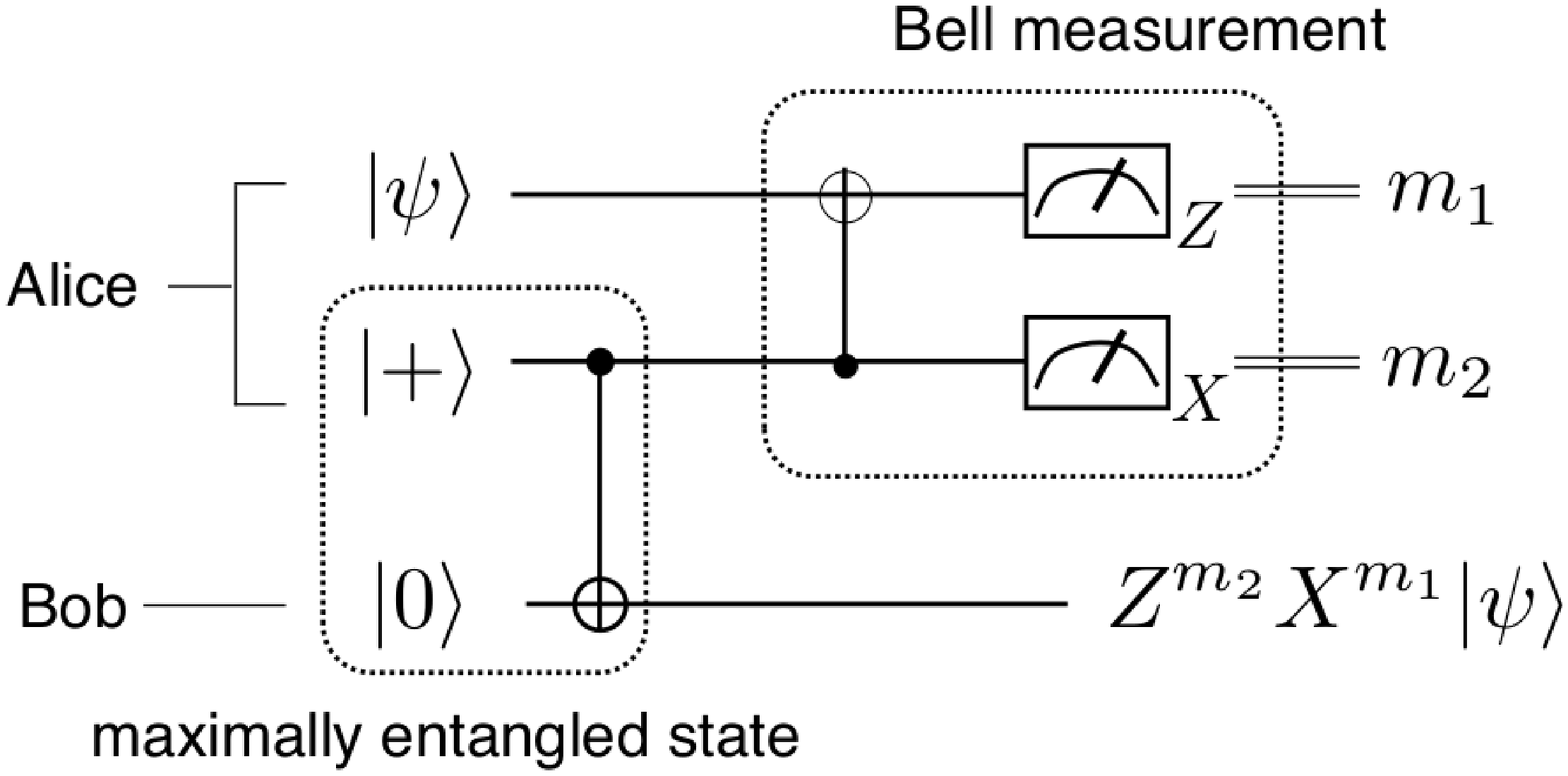}
\end{center}
By using the following circuit equivalence, we can decompose the teleportation circuit into two elementary teleportations, the so-called {\it one-bit teleportations}\index{one-bit teleportation}:
\begin{center}
\includegraphics[width=130mm]{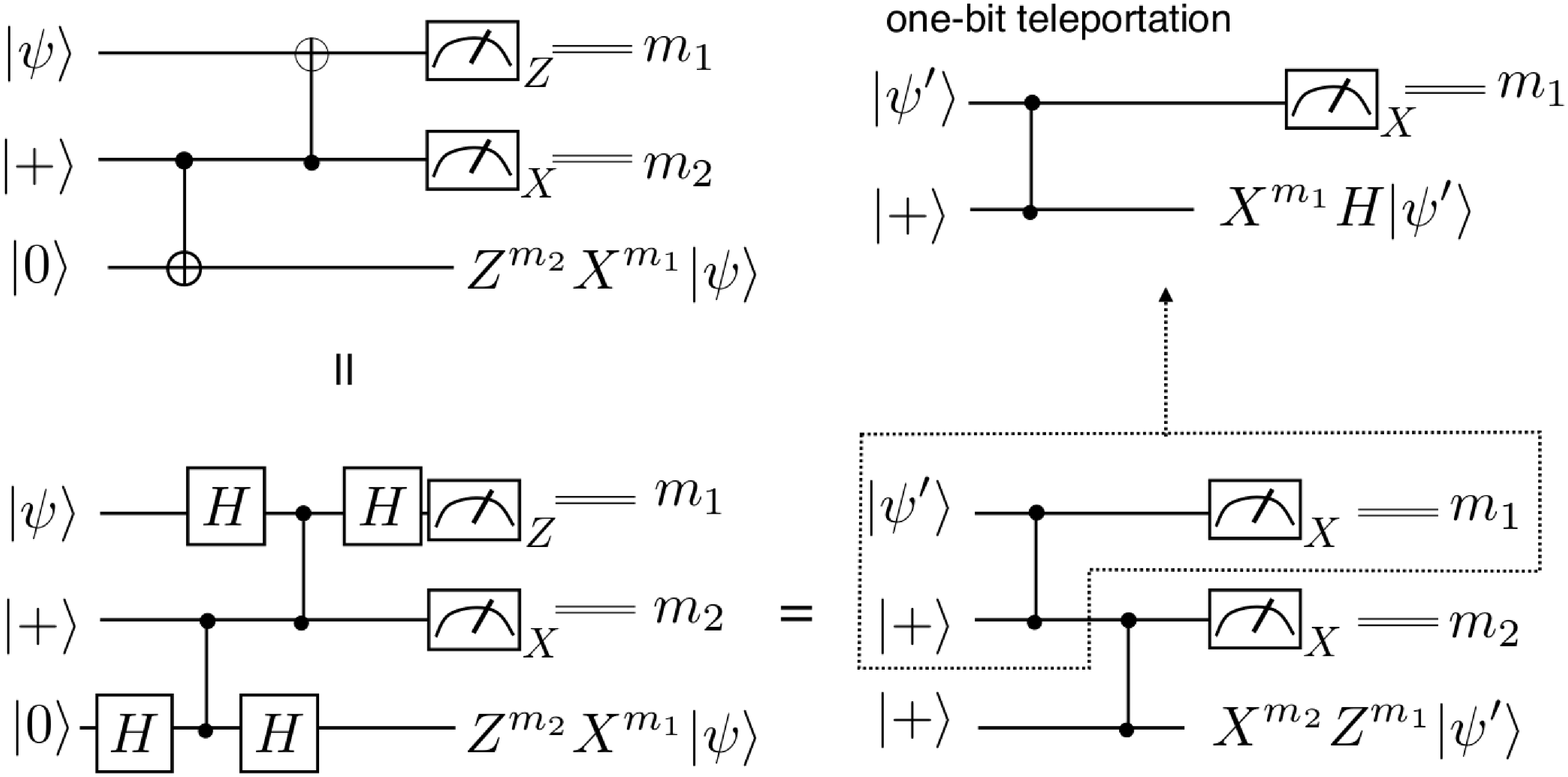}
\end{center}
One-bit teleportation is useful as a building block of the teleportation-based gates employed in MBQC.
A single-qubit $Z$ rotation $e^{i\theta Z}$ can be implemented in a teleportation-based way.
Its action can be understood from the following circuit equivalence:
\begin{center}
\includegraphics[width=80mm]{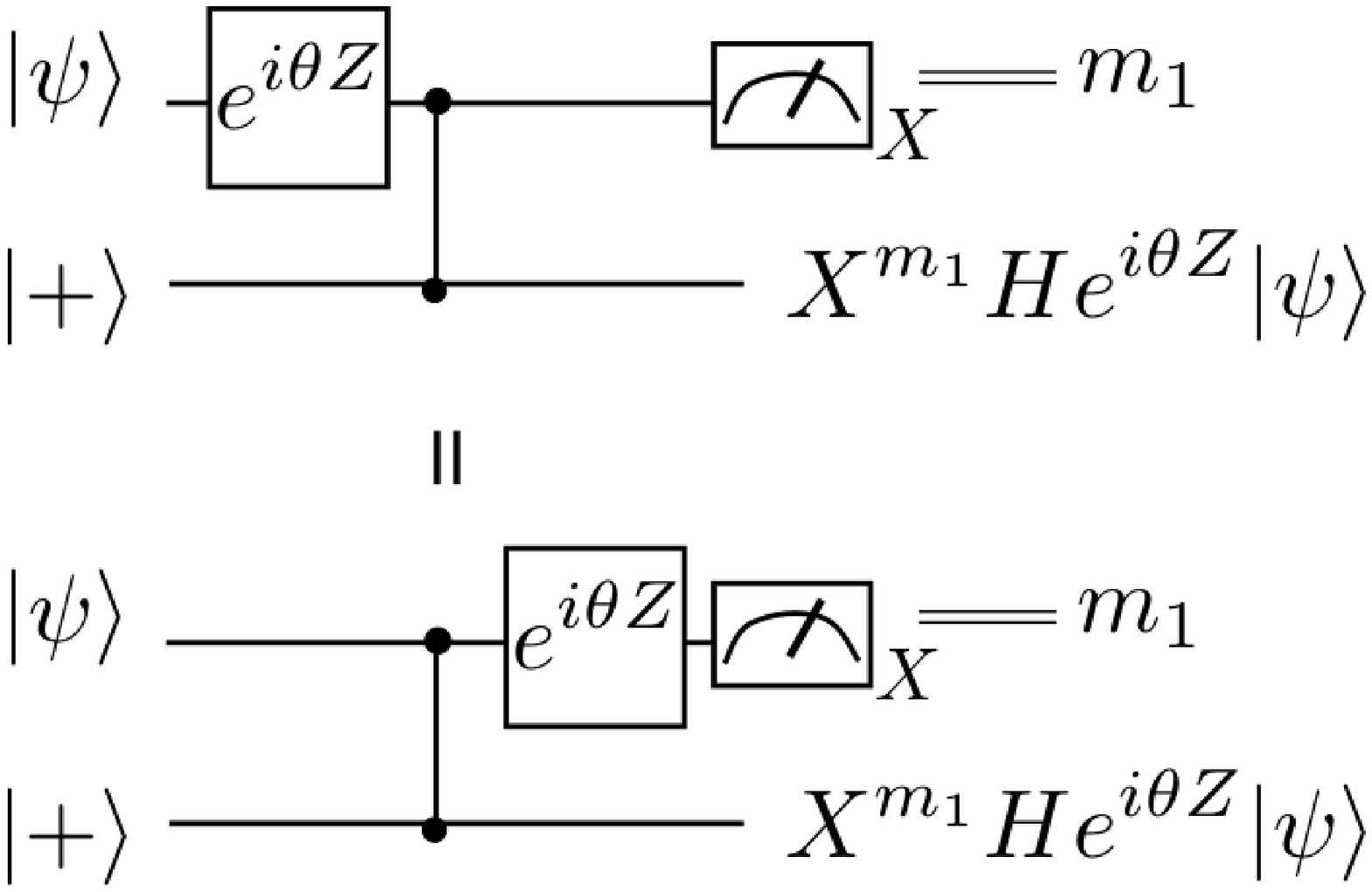}
\end{center}
where we utilized the fact that $e^{i\theta Z}$ and $\Lambda(Z)$ commute.
The controlled-$Z$ operation is also implemented in a teleportation-based way as follows:
\begin{center}
\includegraphics[width=120mm]{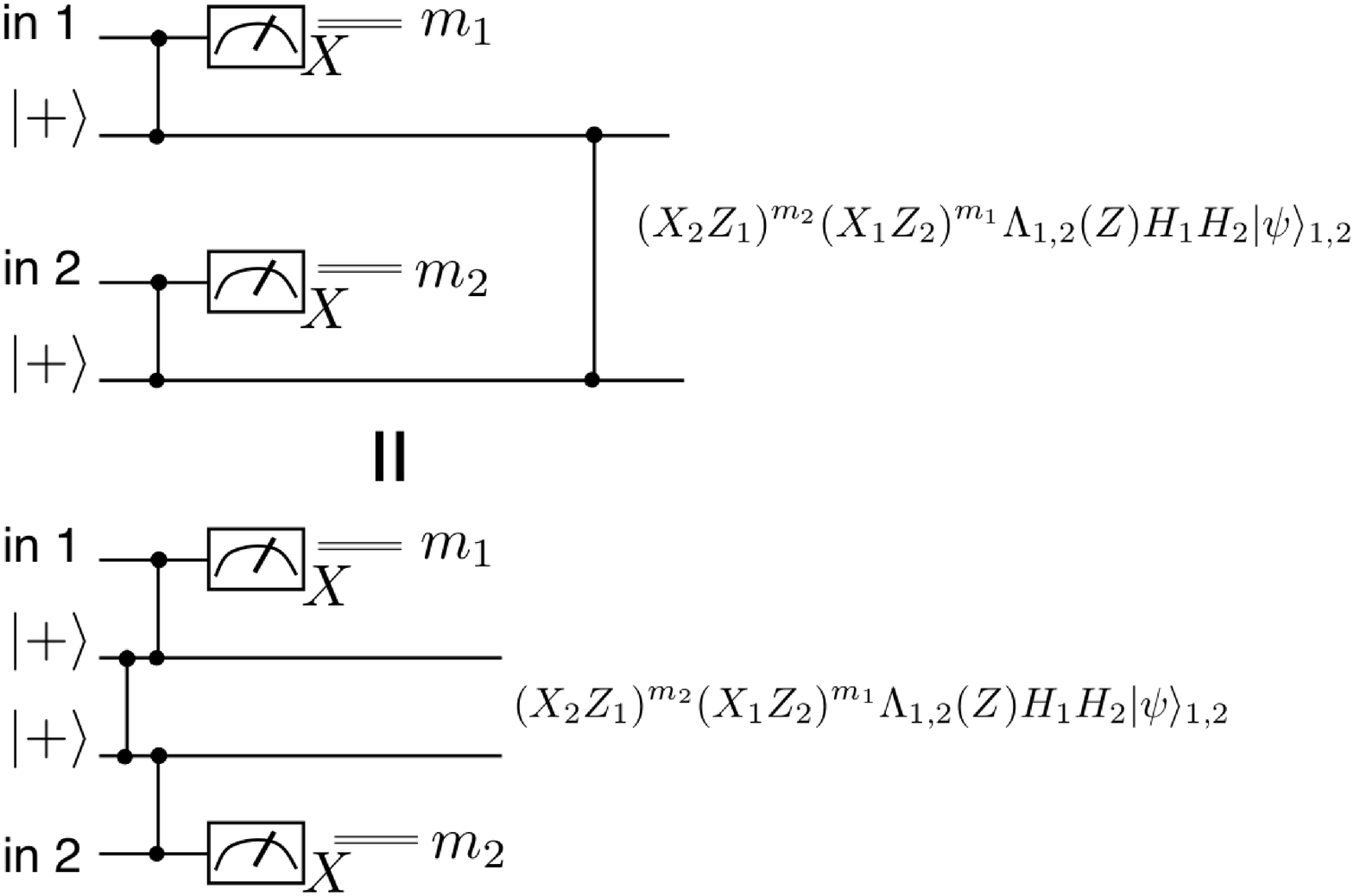}
\end{center}
That is, instead of performing the $\Lambda(Z)$ gate after one-bit teleportations, we can prepare a special resource state, on which the $\Lambda(Z)$ gate is pre-implemented, and the $\Lambda(Z)$ gate is then performed via teleportation.
These quantum operations based on quantum teleportation are called {\it gate teleportation}\index{gate teleportation}~\cite{GottesmanChuang}.

Now we are ready to formulate MBQC.
An arbitrary single-qubit unitary operation $U$ can be decomposed, up to an unimportant global phase, into 
\begin{eqnarray}
U &=& He^{i \phi Z} e^{i \theta X} e^{i \xi Z}
\\
&=&He^{i \phi Z} He^{i \theta Z} He^{i \xi Z}.
\end{eqnarray}
This indicates that we can perform an arbitrary single-qubit unitary operation by a sequence of one-bit teleportations.
The resource state for the sequential one-bit teleportations is a 1D cluster state: 
\begin{center}
\includegraphics[width=50mm]{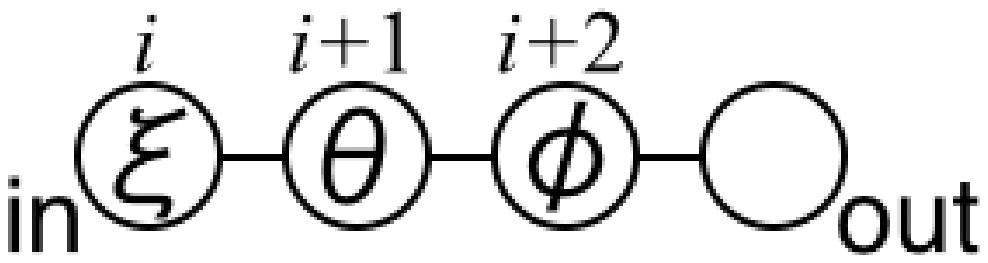}
\end{center}
where the stabilizer generator for the left-most qubit is removed and the input state is encoded.
We have to take care of the byproduct Pauli operators depending on the measurement outcomes.
Fortunately, we can propagate the Pauli byproduct operators\index{byproduct operator} forward as follows:
\begin{eqnarray}
U&=& X^{m_{i+2}}He^{i \phi'  Z} X^{m_{i+1}} He^{i \theta' Z}  X^{m_i} He^{i \xi Z}.
\\
&=& X^{m_{i+2}\oplus m_{i}}Z^{m_{i+1}} He^{i(-1)^{m_{i+1}} \phi'  Z} He^{i (-1)^{m_i}\theta' Z} He^{i \xi Z}.
\end{eqnarray}
\begin{center}
\includegraphics[width=120mm]{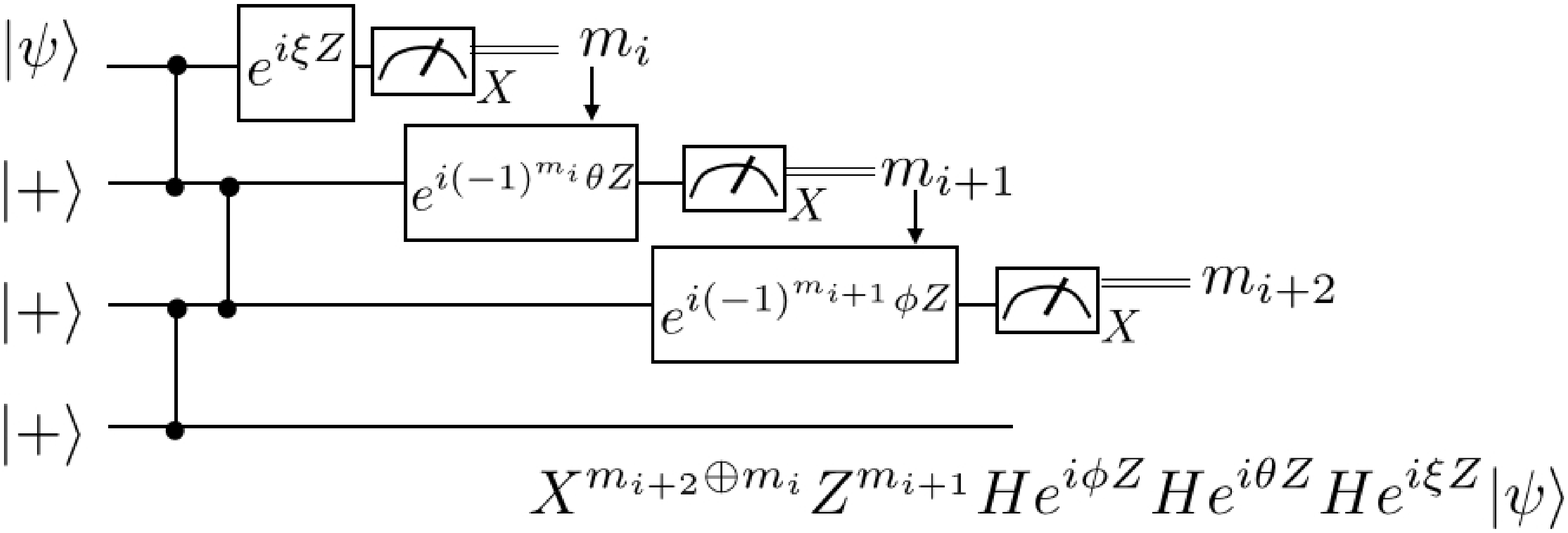}
\end{center}
By choosing $\theta ' = (-1)^{m_i} \theta $ and $\phi' = (-1)^{m_{i+1}} \phi$ adaptively, depending on the previous measurement outcomes, the random nature of the measurements can be managed.
This procedure is called {\it feedforward}\index{feedforward}.
The Pauli byproduct is propagated and updated throughout the computation.
Note that the classical processing required to determine the measurement angle has only XOR (addition modulo two) operations~\cite{MBQC_PRA}. 

Next, we will investigate the measurement-based two-qubit gate operation.
The resource state for the gate teleportation\index{gate teleportation} is the following cluster state:
\begin{center}
\includegraphics[width=120mm]{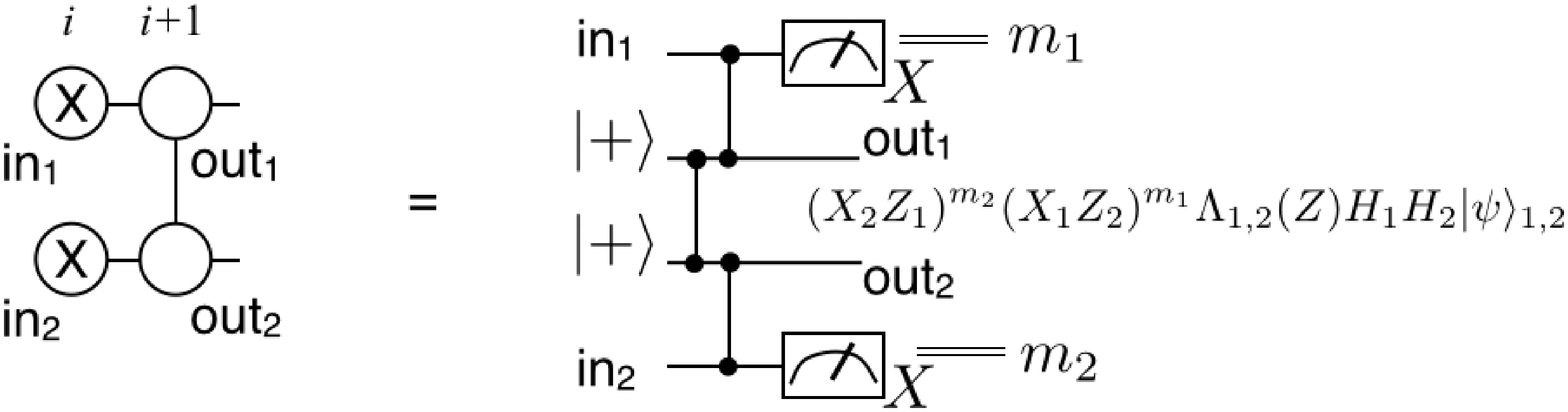}
\end{center}
To adjust the timing of the two-qubit operation, we can insert identity operations depending on the even and odd lengths as follows:
\begin{center}
\includegraphics[width=100mm]{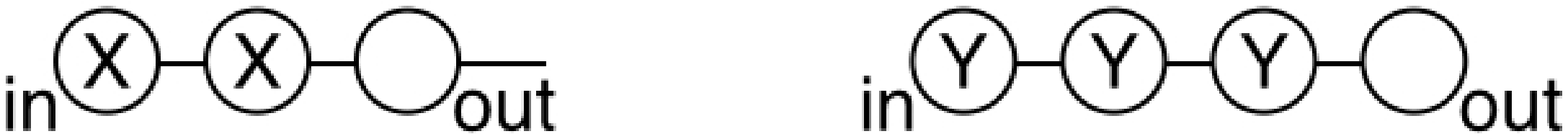}
\end{center}
Without loss of generality, we can assume that all input states of the quantum computation are given by $|+\rangle$, which are automatically encoded by preparing the graph state.
At the end of the computation (on the right-most qubits), measurements are performed to read out the result as follows:
\begin{center}
\includegraphics[width=100mm]{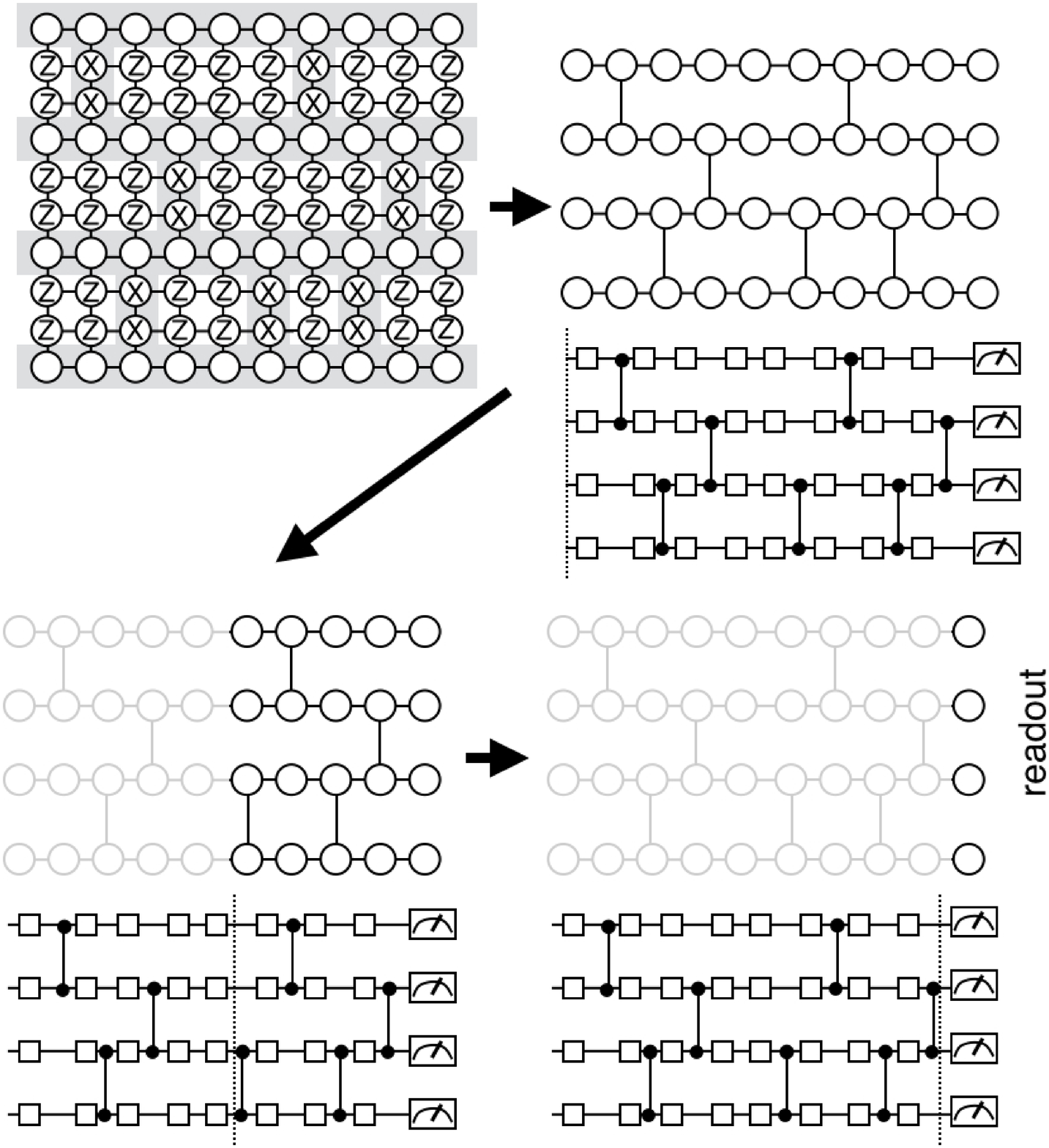}
\end{center}
In this way, universal quantum computation is simulated solely by measurements on a brickwork-like cluster state.
This state can be generated from a cluster state on a square lattice by using the Pauli basis measurements as shown above.
Accordingly, the square lattice cluster states are universal resources for MBQC.

The above circuit-based explanation of MBQC~\cite{NielsenCluster} is very intuitive and straightforward.
However, for a complicated resource state, as will be shown, an operator-based understanding of MBQC~\cite{MBQC_PRA} is quite useful.
Let us reformulate MBQC from an operator viewpoint.
Suppose again an MBQC on a 1D cluster state.
The measurements are executed up to the $(i-1)$th qubits, and hence the operator $K_l$ ($l \leq i$) is removed from the stabilizer generators.
The logical degree of freedom on the remaining resource state can be specified by the $i$th logical operators\index{logical operator}
\begin{eqnarray}
L_X^{(i)} &=&X_i Z_{i+1},
\\
L_Y^{(i)} &=&Y_i Z_{i+1},
\\
L_Z^{(i)} &=&Z_i.
\end{eqnarray}
These logical operators commute with all remaining stabilizer generators $K_l$ ($l \geq i+1$).
Moreover, they anticommute with each other, satisfying the commutation relations for the Pauli operators.
Thus, they specify the state encoded in the graph state.
As seen above, a $Z$-rotation $e^{-i (\theta/2) Z_i}$ is applied before the $X$-basis measurement.
Because $Z_i = L_Z^{(i)}$, this rotation induces a unitary transformation $U$ of the logical operator
\begin{eqnarray}
L_X^{(i)} &\rightarrow& \cos \theta L_X^{(i)} + \sin  \theta L_Y^{(i)},
\\
L_Y^{(i)} &\rightarrow& \cos \theta L_Y^{(i)} - \sin  \theta L_X^{(i)}.
\label{eq:MBQC_ope1}
\end{eqnarray}
Because $L_X^{(i)}=X_i L_Z^{(i+1)}$, the logical $X$ operator after the $X$-basis measurements is given by $(-1)^{m_i} L_Z^{(i+1)}$ depending on the measurement outcome $m_i =0,1$.
On the other hand, the logical operators $L_{Y,Z}^{(i)}$ do not commute with the $X$-basis measurement; they are not relevant logical operators after the measurement.
If two operators are equivalent up to multiplications of the stabilizer operators, their action on the stabilizer state is also the same.
By using this fact, we can replace the logical operators in (\ref{eq:MBQC_ope1}) with
\begin{eqnarray}
L_Z^{(i)} \sim L_Z^{(i)} K_{i+1} = X_{i+1}Z_{i+2} \equiv L_X^{(i+1)},
\\
L_Y^{(i)} \sim K_{i+1} = X_i Y_{i+1}Z_{i+2} \equiv X_i L_Y^{(i+1)},
\end{eqnarray}
where $\sim$ indicates that two operators are equivalent up to stabilizer operators.
After the $X$-basis measurement, $X_i$ can be replaced by its eigenvalue $(-1)^{m_i}$.
Then the logical operator of the post-measurement state is given by
\begin{eqnarray}
L_X^{(i)} &\rightarrow& (-1)^{m_i} (\cos \theta L_Z^{(i+1)} + \sin  \theta L_Y^{(i+1)}) = U L_X^{(i+1)} U^{\dag},
\\
L_Y^{(i)} &\rightarrow&  (-1)^{m_i}(\cos \theta L_Y^{(i+1)} + \sin  \theta L_Z^{(i+1)})= UL_Y^{(i+1)}U^{\dag} ,
\\
L_Z^{(i)} &\rightarrow &  L_X^{(i+1)} = U L_Z^{(i+1)}U^{\dag} .
\end{eqnarray}
We now realize that the logical operators for the $i$th step are transformed into those for the $(i+1)$th step rotated by $U \equiv X^{m_i}He^{-i (\theta/2) Z}$.

Similarly, a two-qubit gate in MBQC can also be regarded as a propagation of a correlation by a projection on the stabilizer state.
Consider the following graph state.
\begin{center}
\includegraphics[width=110mm]{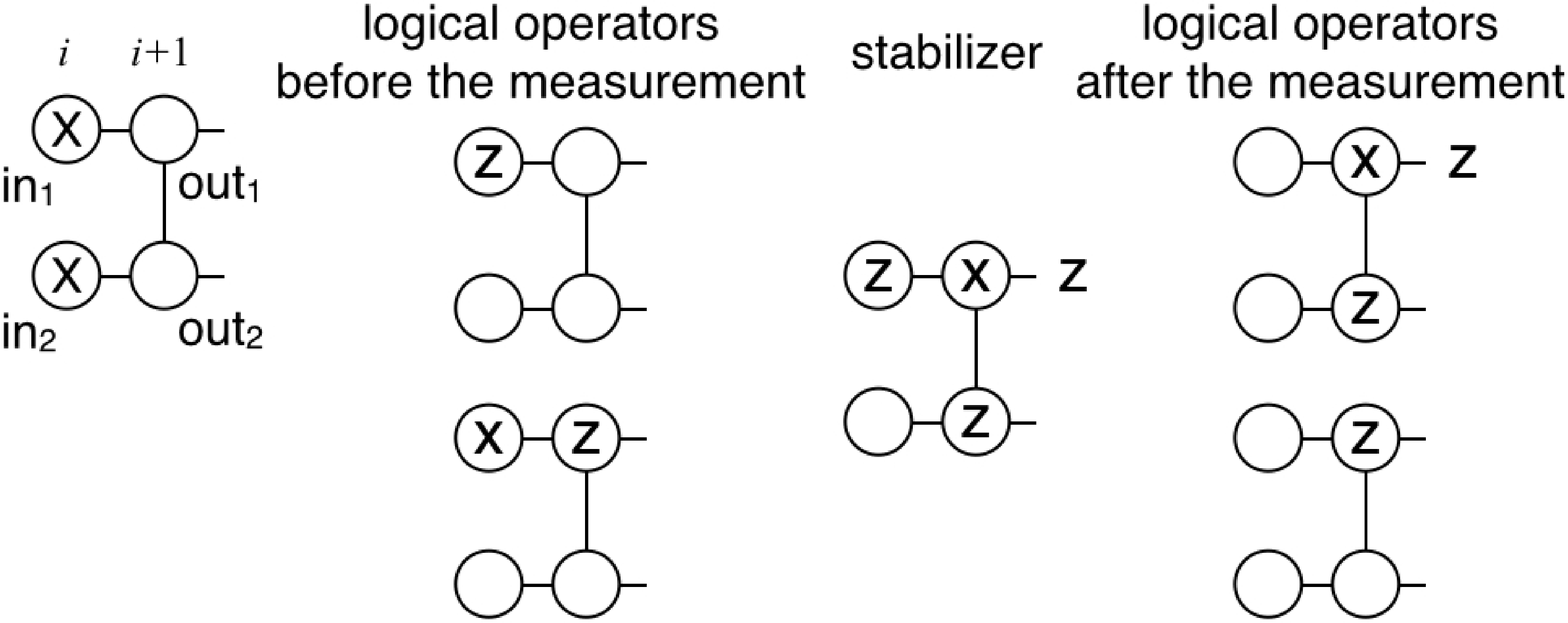}
\end{center}
The logical operators for the $i$th step are given by $\{ L_{X1}^{(i)}, L_{Z1}^{(i)} \}$ and $\{ L_{X2}^{(i)}, L_{Z2}^{(i)} \}$.
By multiplying the stabilizer operator, we obtain
\begin{eqnarray}
L_{Z1}^{(i)} &\sim& X_{1,i+1} Z_{1,i+1} Z_{2,i+1} = L_{X1}^{(i+1)}L_{Z2}^{(i+1)}
,
\\
L_{Z2}^{(i)} &\sim& X_{2,i+1} Z_{2,i+1} Z_{1,i+1}= L_{X2}^{(i+1)}L_{Z1}^{(i+1)}
.
\end{eqnarray}
The logical operators for the $(i+1)$th step after the projections are calculated to be
\begin{eqnarray}
\{ L_{X1}^{(i)}, L_{Z1}^{(i)} \} \rightarrow \{ (-1)^{m_1} L_{Z1}^{(i+1)}, L_{X1}^{(i+1)}L_{Z2}^{(i+1)} \} = \{ VL_{X1}^{(i+1)}V^{\dag},VL_{Z1}^{(i+1)}V^{\dag} \},
\\
\{ L_{X2}^{(i)}, L_{Z2}^{(i)} \} \rightarrow \{ (-1)^{m_2} L_{Z2}^{(i+1)}, L_{X2}^{(i+1)}L_{Z1}^{(i+1)} \} = \{ VL_{X2}^{(i+1)}V^{\dag},VL_{Z2}^{(i+1)}V^{\dag} \}.
\end{eqnarray}
Again, we realize that the logical operators for the $i$th step are transformed into those for the $(i+1)$th step with a two-qubit unitary operation
\begin{eqnarray} 
V\equiv (X_1 Z_2)^{m_1} (X_2 Z_1)^{m_2} \Lambda_{1,2} (Z)H_1 H_2.
\end{eqnarray}
By combining single-qubit rotations $X^{m}He^{i (\theta/2) Z}$ and the two-qubit operation $ (X_1 Z_2)^{m_1} (X_2 Z_1)^{m_2} \Lambda_{1,2} (Z)H_1 H_2$ as seen above, we can perform a universal quantum computation. 
In this way, MBQC can be understood in the Heisenberg picture.

\begin{figure}[b]
\centering
\includegraphics[width=120mm]{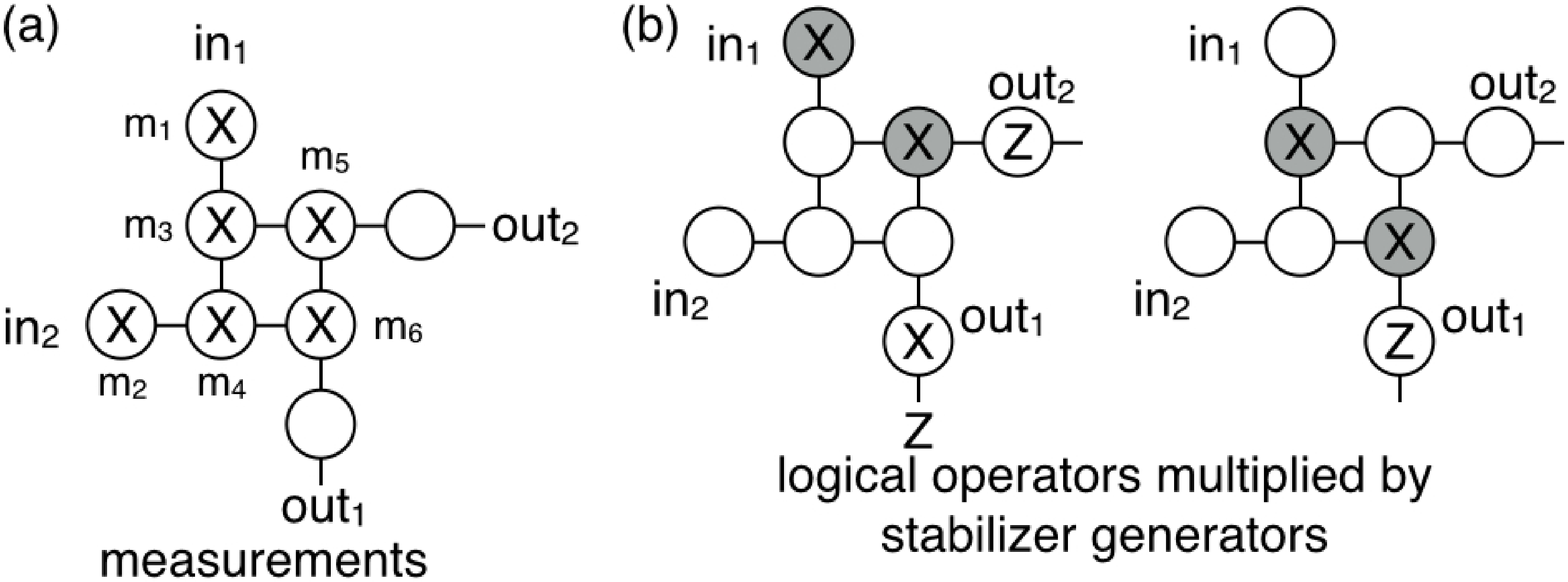}
%
%
\caption{(a) A graph state and a measurement pattern.
(b) The logical $X$ operator of input 1 is multiplied by the stabilizer generators and we obtain a correlated operator on outputs 1 and 2 (left). 
The logical $Z$ operator of input 1 is multiplied by the stabilizer generators and we obtain the logical $Z$ operator of output 1.
The gray colored $X$ operators are replaced by $\pm1$ depending on the measurement outcomes.}
\label{fig98}       
\end{figure}

Suppose the logical Pauli operators of the $k$th input and output qubits are related by the measurements as follows:
\begin{eqnarray}
\{ L_{X,k}^{({\rm In})}, L_{Z,k}^{(\rm In)} \}
\rightarrow 
\{ U L_{X,k}^{({\rm Out})} U^{\dag}, UL_{Z,k}^{(\rm Out)}U^{\dag} \}.
\end{eqnarray}
The unitary operator $U$ is performed on the input qubits.
Here a Pauli byproduct, depending on the measurement outcomes, is also included in $U$.
Moreover, if two graph states, which perform $U$ and $V$, are concatenated with the appropriate feedforwarding of the Pauli byproducts, then $VU$ is performed:
\begin{center}
\includegraphics[width=70mm]{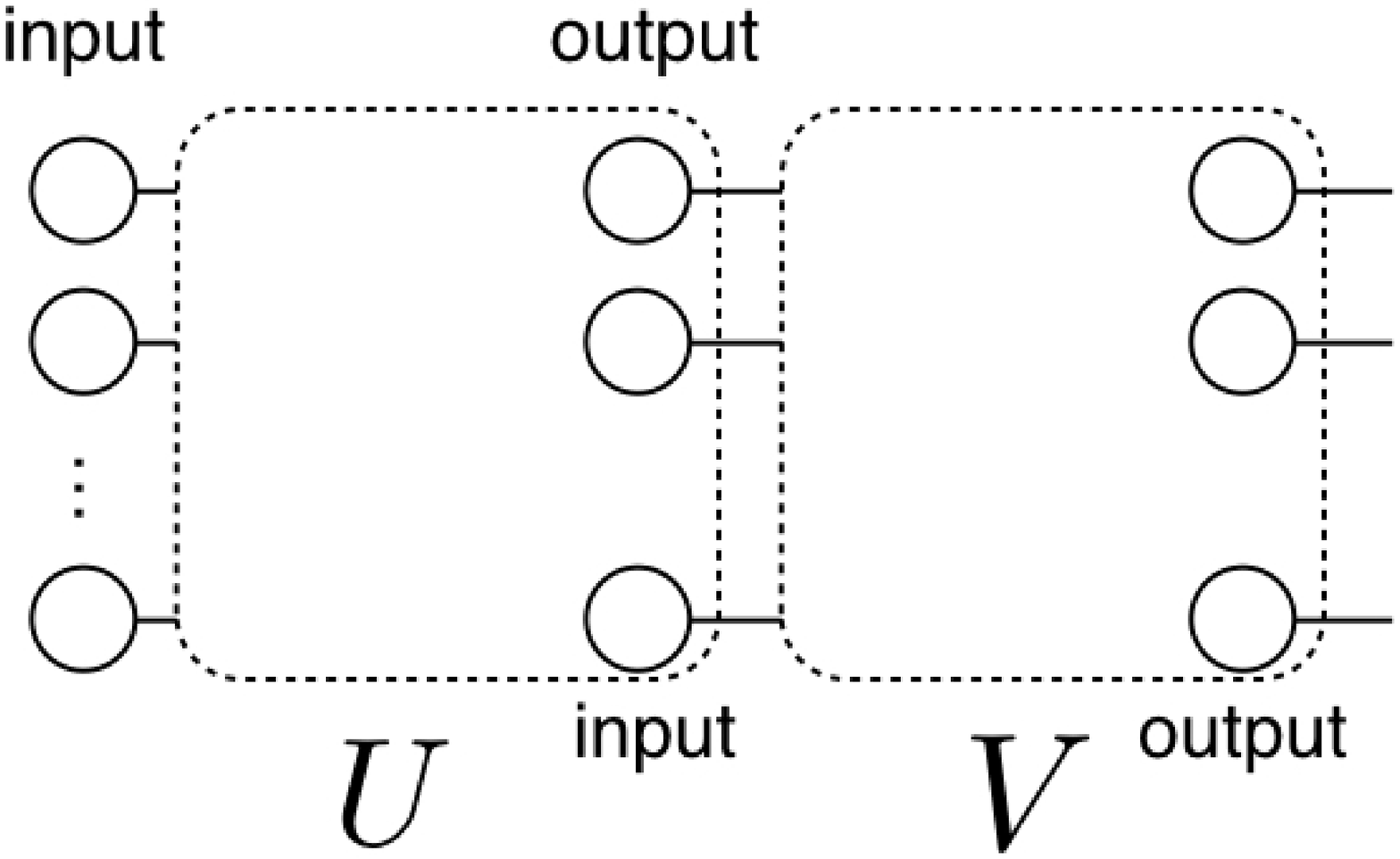}
\end{center}

Let us consider the example shown in Fig.~\ref{fig98} (a).
The logical operators on the inputs are replaced by multiplying stabilizer generators so that they commute with the $X$-basis measurements.
Then the $X$ operators on the measured qubits are replaced by $\pm1$.
The measurements transform the input logical operators as follows:
\begin{eqnarray}
&&\{ L_{X,1}^{({\rm In})}, L_{Z,1}^{(\rm In)}, 
L_{X,2}^{({\rm In})}, L_{Z,2}^{(\rm In)} \}
\rightarrow 
\nonumber \\
&&\{ (-1)^{m_1 \oplus m_5} L_{X,1}^{({\rm Out})}L_{Z,2}^{({\rm Out})}, (-1)^{m_3 \oplus m_6}L_{Z,1}^{(\rm Out)}, 
(-1)^{m_2 \oplus m_6}L_{X,2}^{({\rm Out})} L_{Z,1}^{({\rm Out})} ,(-1) ^{m_4 \oplus m_5}L_{Z,2}^{(\rm Out)} \}.
\nonumber \\
\end{eqnarray}
Thus, the $\Lambda_{1,2}(Z)$ gate is implemented up to a Pauli byproduct.

Using this fact and concatenation of the input-output relations, we can construct a measurement-based CNOT gate between the separated two-qubit as follows~\cite{MBQC_PRA}:
\begin{center}
\includegraphics[width=90mm]{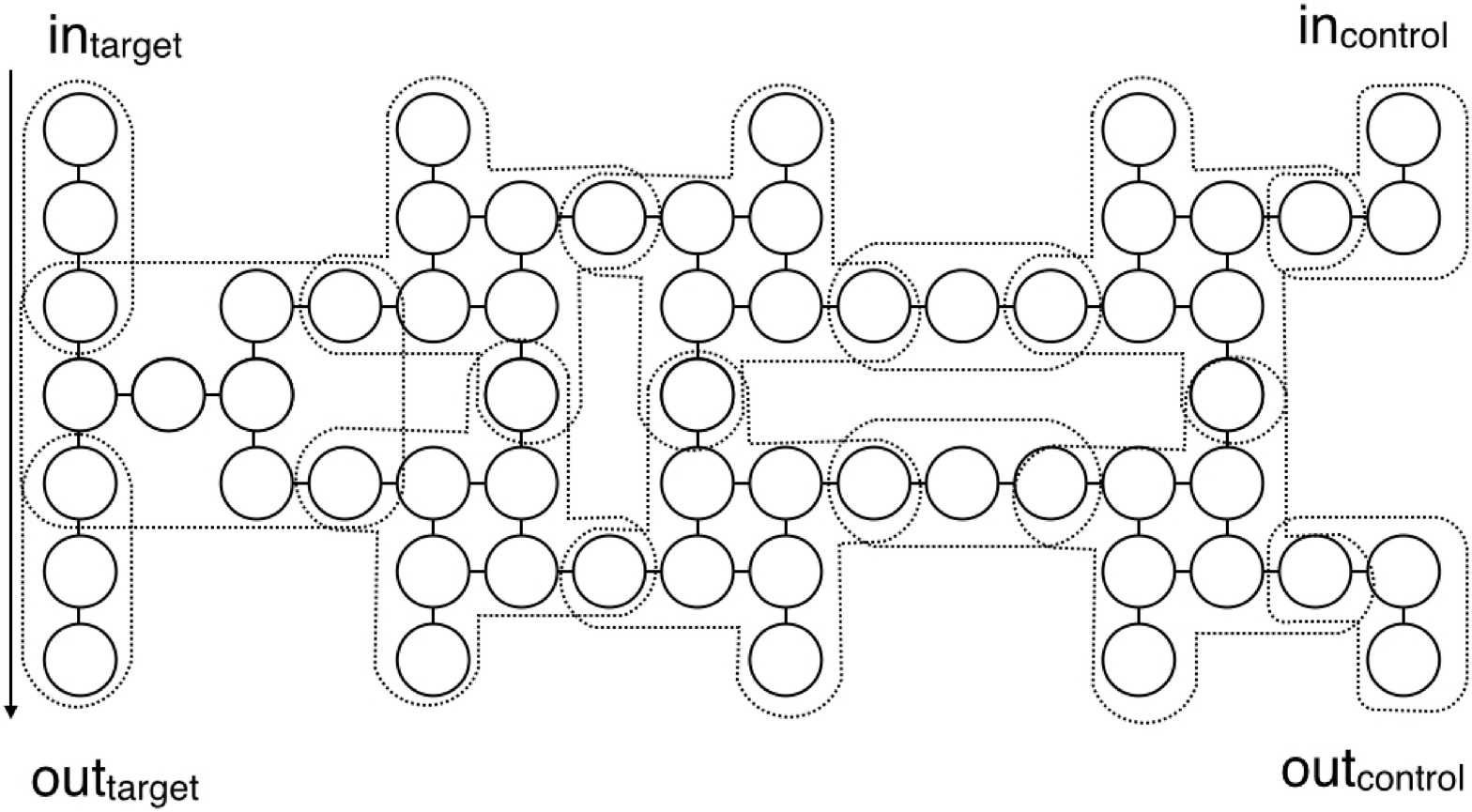}
\end{center}
The input-output relation of the above graph state is equivalent to that for the following circuit:
\begin{center}
\includegraphics[width=90mm]{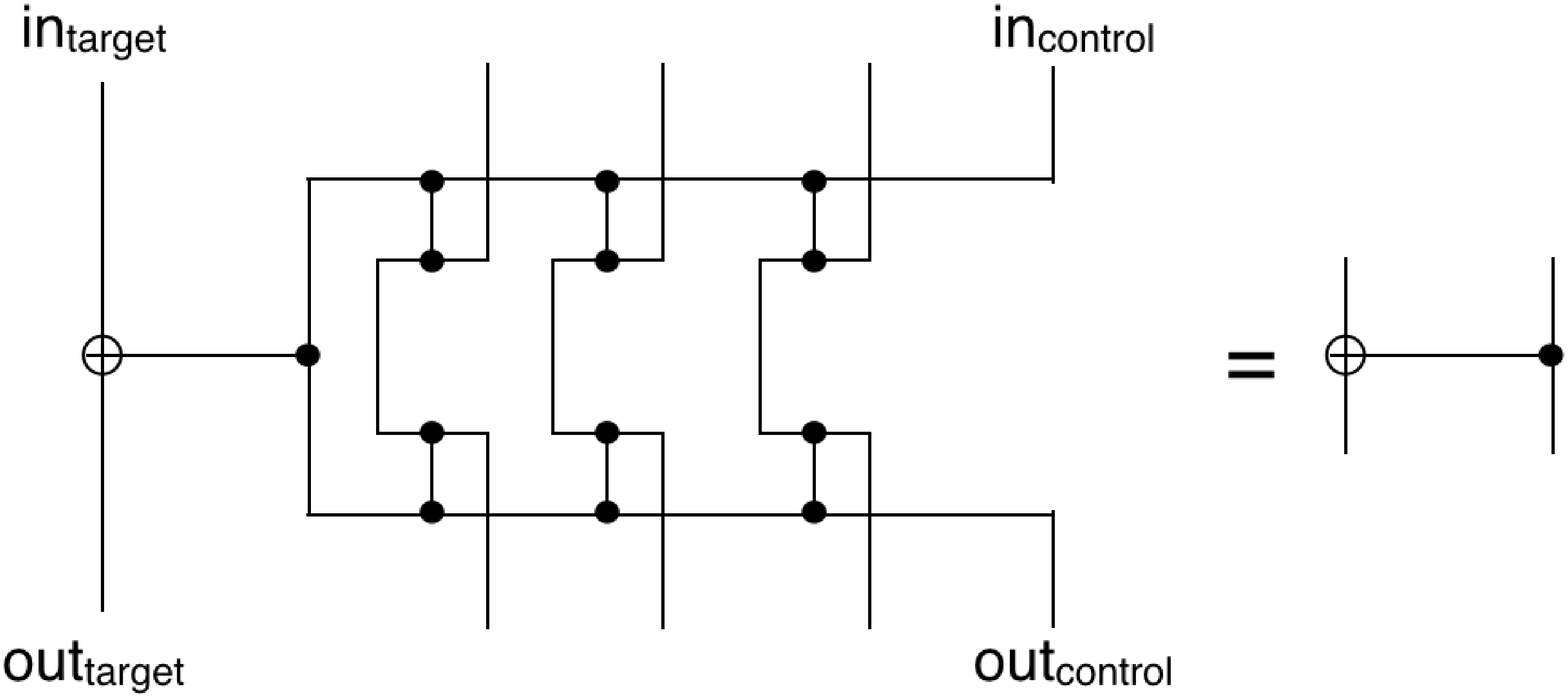}
\end{center}
In this way, a CNOT gate between two arbitrary separated qubits can be implemented using a constant depth (constant width) resource state.

Topologically protected quantum computation was first proposed as a MBQC on a 3D cluster state~\cite{RaussendorfAnn,RaussendorfNJP}
as we will see in Chapter~\ref{Chap:TMBQC}.
It was further applied to the circuit-based model in two dimensions (2D)~\cite{RaussendorfPRL,FowlerHigh},
which will be explained in detail in Chapter~\ref{Chap:TQC}.
Specifically, in the formulations, the operator-based understanding is quite useful for tracking how correlation is propagated in a topologically protected way.
In Chapter~\ref{Chap:TMBQC}, we will formulate a topologically protected MBQC from the operator viewpoint and see how it is related to the circuit-based understanding explained in Chapter~\ref{Chap:TQC}.

To conclude, we summarize the properties of MBQC and recent progress in this area.
A unique feature of MBQC is that the resource state for universal quantum computation is prepared offline. 
Entangling operations, which would be one of the most difficult tasks in experiments, are employed only in this stage. 
Quantum computation is executed solely by adaptive measurements.
This property is useful for experimental realization in certain physical systems.
For example, a deterministic entangling operation is difficult to achieve in an optical quantum computation.
In such a case, we can utilize linear optics and measurement-induced nonlinearity to generate a cluster state~\cite{Yoran03,NielsenOpt,Browne05}.
Importantly, the entangling operation can be nondeterministic, as long as the successful or non-successful outcome is heralded.
By using such a probabilistic entangling operation, we can gradually expand the cluster state.
After successful cluster state generation, we can start the measurements for quantum computation.
Note that the probability of successful cluster state generation is not exponentially small by using a divide and conquer approach~\cite{NielsenOpt,Browne05,DuanRaussendorf,Matsuzaki10,LiBenjamin,FTProb}.

The clear separation between the quantum stage requiring entangling operations and the measurement stage is useful, not only for the physical implementation, but also in a quantum cryptographic scenario.
Suppose that Bob (server) possesses a fully fledged quantum computer and
that Alice (client), who has a less advanced quantum device, such as a single-qubit state generator, wants to delegate quantum computation to Bob.
By using the idea of MBQC, such a delegated quantum computation can be made unconditionally secure.
This is called a blind quantum computation and was proposed by Broadbent, Fitzsimons, and Kashefi (BFK)~\cite{Blind} (see also the related earlier works~\cite{ChildsBlind,ArrighiBlind}).
In the BFK protocol, Alice sends randomly rotated qubits $\{ e^{i \theta _j Z } |+\rangle\}$ to Bob, where the angle is chosen to be $\theta _j = k_j \pi /4$ ($k_j=0,1,...,7$).
Bob generates a cluster state by using the randomly rotated qubits.
In the computation phase, Alice sends a classical message $\delta _j = \phi _j + \theta _j + r_j \pi $.
Here, $\phi _j$ is the measurement angle with which Alice want to perform a measurement. The angle $\theta _j$ is for the randomly rotated state (which is secret to Bob). The random bit $r\in \{0,1\}$ makes the measurement angle completely random for Bob.
Then Bob performs the measurement in the $\{ e^{i\delta_j} |\pm \rangle\}$ basis.
Because the initial state is pre-rotated by $\theta_j$ (from Alice's viewpoint), Bob performs the measurement in the $\{ e^{i(\phi _j + r_j \pi) } |\pm \rangle\}$ basis, which is what Alice wants to do.
However, from Bob's viewpoint, the state is a completely mixed state with no information about $\{\phi _j\}$.
Thus, Bob is blind to any information about the input, the algorithm, and the output.
Instead of the state generation, Alice, who has a measurement device, can also perform a blind quantum computation, whose security is guaranteed by the no-signaling principle~\cite{MoriBlind}.
A fault-tolerant blind quantum computation has been proposed, based on topologically protected MBQC~\cite{BlindMoriFuji}.

\section{Quantum error correction codes}
In this section, we introduce stabilizer codes, which are a class of quantum error correction (QEC)\index{quantum error correction, QEC} codes.

\subsection*{Three-qubit bit-flip code}
The QEC codes can be  described elegantly in the stabilizer formalism.
Let us first consider the simplest one, the three-qubit bit flip code\index{three-qubit bit flip code}, whose stabilizer generators are given by
\begin{eqnarray}
S_{1} = Z_{1}Z_{2} , \;\;\; S_{2} = Z_{2}Z_{3}.
\end{eqnarray}
The stabilizer subspace is spanned by the following two logical states:
\begin{eqnarray}
|0_{L} \rangle = |000\rangle , \;\;\; |1_{L} \rangle=|111\rangle.
\end{eqnarray}
The logical Pauli-$X$ operator is given by $L_X \equiv X_{1}X_{2}X_{3}$.
The logical Pauli-$Z$ operator is defined as $L_Z\equiv Z_{1}$.
We may, equivalently, choose the logical Pauli $Z$ operator to be $Z_2$ or $Z_3$, because their actions on the code space are equivalent.
The present code is a quantum analogue of the classical three-bit repetition code.
Consider a bit flip error with an error probability $p$:
\begin{eqnarray}
\mathcal{E}_{i} \rho = (1-p) \rho + p X_{i} \rho X_{i}.
\end{eqnarray}
If the initial state $|\psi _{L} \rangle = \alpha|0_{L} \rangle +\beta |1_{L}\rangle $ undergoes the bit flip error independently, the output state is transformed in leading order as
\begin{eqnarray}
\mathcal{E} _{1} \circ \mathcal{E} _{2} \circ \mathcal{E} _{3}| \psi _{L} \rangle \langle \psi _{L} |
=  (1-p) ^3 | \psi _{L} \rangle \langle \psi _{L} |+p (1-p)^2 \sum_{i} X_{i} | \psi _{L}\rangle \langle \psi _{L} | X_{i} + O (p^2).
\nonumber \\
\label{eq:noise}
\end{eqnarray}
The error $X_i$ maps the code space to an orthogonal space.
We perform a projective measurement onto the orthogonal subspaces, $P_{k}^{\pm} = (I \pm S_{k})/2$, which we call a syndrome measurement, to know in which orthogonal space the state lies.
Note that the encoded quantum information is not destroyed by the syndrome measurement, because it commutes with the logical operators.
According to the measurement outcomes, the logical state can recover from the error as follows:
\begin{eqnarray}
\mathcal{R} \circ \mathcal{E} _{1} \circ \mathcal{E} _{2} \circ \mathcal{E} _{3}| \psi _{L} \rangle \langle \psi _{L} |
&=& [(1-p)^3 + 3p(1-p)^2 ] | \psi _{L} \rangle \langle \psi _{L} | + O(p^2),
\label{eq:recovery}
\end{eqnarray}
where the recovery operator is given by
\begin{eqnarray}
\mathcal{R} \rho &=& P_{1}^{+}P_{2}^{+} \rho P_{2}^{+}P_{1}^{+}+X_{1} P_{1}^{-}P_{2}^{+} \rho P_{2}^{+}P_{1}^{-} X_{1}+X_{2} P_{1}^{-}P_{2}^{-} \rho P_{2}^{-}P_{1}^{-} X_{2}
\nonumber \\
&&+X_{3} P_{1}^{+}P_{2}^{-} \rho P_{2}^{-}P_{1}^{+} X_{3}.
\end{eqnarray}
The four terms in $\mathcal{R}\rho$ correspond to the measurement outcomes (eigenvalues) $(+1,+1)$, $(-1,+1)$, $(-1,-1)$, and $(+1,-1)$
of the stabilizer generators, respectively.
By comparing Eqs.\ (\ref{eq:noise}) and (\ref{eq:recovery}), one can understand that if $p$ is sufficiently small, the fidelity of the logical state is improved.

Similarly, we can construct a three-qubit phase flip code\index{three-qubit phase flip code}, which can correct a phase flip error, by changing the basis with the Hadamard transformation:
\begin{eqnarray}
\langle Z_1 Z_2, Z_2 Z_3 \rangle \rightarrow \langle X_1 X_2 , X_2 X_3 \rangle.  
\end{eqnarray}

\subsection*{9-qubit Shor code}
The three-qubit bit-flip code cannot correct $Z$ errors, which commute with the stabilizer generators.
A QEC code that can correct all $X$, $Y$, and $Z$ errors was developed by Shor based on a concatenation of three-qubit bit-flip and phase-flip codes~\cite{Shor95}.
The stabilizer generators of the 9-qubit Shor code\index{9-qubit Shor code} are given as follows:
\begin{equation}
\begin{array}{ccccccccc}
\hline \hline
X & X & X& X& X& X & I & I & I
\\
I&I&I&X&X&X&X&X&X
\\
Z&Z&I&I&I&I&I&I&I
\\
I&Z&Z&I&I&I&I&I&I
\\
I & I & I & Z & Z & I & I & I & I
\\
I&I&I&I&Z&Z&I&I&I
\\
I&I&I&I&I&I&Z&Z&I
\\
I&I&I&I&I&I&I&Z&Z
\\
\hline \hline
\end{array}
\end{equation}
The code space is spanned by
\begin{eqnarray}
\frac{|0 _{L}\rangle + |1_{L} \rangle}{\sqrt{2}}
&=& 
\frac{\left(|000\rangle + |111 \rangle \right) \left(|000\rangle + |111 \rangle \right) \left(|000\rangle + |111 \rangle\right)}{2\sqrt{2}} ,  
\\
\frac{|0 _{L}\rangle - |1_{L} \rangle}{\sqrt{2}}
&=& 
\frac{\left(|000\rangle -|111 \rangle \right) \left(|000\rangle -|111 \rangle \right) \left(|000\rangle -|111 \rangle\right)}{2\sqrt{2}}.
\end{eqnarray}
The logical Pauli operators are given by $X_{L}=X^{\otimes 9}$ and $Z_{L}=Z^{\otimes 9}$, which are bitwise tensor products of physical Pauli operators.
If the logical $A$ operator is given by a bitwise tensor product of the physical $A$ operators on the QEC code, we say that the operation $A$ has {\it transversality}\index{transversality}.
The 9-qubit code is capable of correcting all $X$, $Y$, and $Z$ errors for each qubit, which can be understood because the three-qubit phase flip code 
$\{ |+++\rangle , |---\rangle \}$ is constructed by using the three logical qubits of the three-qubit bit flip codes $\{ |000\rangle, |111\rangle \}$.

Note that any single-qubit noise $\mathcal{E}$ can be described by using the Kraus operators $\{ K_j \}$:
\begin{eqnarray}
\mathcal{E} \rho = \sum _j K_j \rho K_j^{\dag}.
\end{eqnarray}
Any operator $K_j$ can be decomposed into the Pauli operators $\sigma _0 = I$, $\sigma _1=X$, $\sigma _2 = Y$, and $\sigma _3 =Z$:
\begin{eqnarray}
K_j = \sum _{l} c_{jl} \sigma _l.
\end{eqnarray}
Thus, if the $X$ and $Z$ errors on a single qubit are both corrected appropriately, we can correct any single-qubit noise automatically.
Specifically, because noise contains a superposition of the Pauli errors, it can be collapsed by the syndrome measurements.

\subsection*{Stabilizer codes}
To summarize the above examples, let us formalize the stabilizer quantum error correction codes and their properties. \index{stabilizer QEC code}
The code space of a stabilizer QEC code is defined by a stabilizer group $\langle \{ S_i \}\rangle$.
The encoded degree of freedom is specified by the mutually independent logical operators\index{logical operator} $\{L^{Z}_j\}$, which commute with all stabilizer generators and are independent of the stabilizer generators.
The computational basis state of the code state is completely determined by the stabilizer group $\langle \{ S_i\}, \{(-1)^{m_j} L^{Z}_j \} \rangle$.
We can always find another set of the logical operators $\{L^{X}_j\}$ being subject to
\begin{eqnarray}
L^X_j L^Z_i = (-1)^{\delta _{ij}} L^Z_i L^X_j ,
\end{eqnarray}
where $\delta _{ij}$ is the Kronecker delta.
Hence, the pair of logical operators $L_i^{Z}$ and $L_i^{X}$ represents the $i$th logical qubit\index{logical qubit}.
In terms of the numbers $n$ and $(n-k)$ of qubits and stabilizer generators, respectively, the number of pairs of logical operators is $k$. 

Let us define the {\it weight}\index{weight} ${\rm wt}(S_i)$ of a Pauli product $S_i$ as the number of qubits on which a Pauli operator (except for the identity $I$) is acting. 
The minimum weight of the logical operator over all possible logical operators is called the {\it code distance}\index{code distance} $d$. 
This implies that all Pauli products whose weights are smaller than $d$ are elements of the stabilizer group or anticommute with the stabilizer generators.
Thus, they act trivially on the code state	 or 
map the code state into an orthogonal subspace.
If the weight of a Pauli product as an error is less than $(d-1)/2$, we can find a unique recovery operator that returns the erroneous state into the code space.
Thus, we can correct weight-$\lfloor (d-1)/2 \rfloor$ errors.
Such a stabilizer QEC code is called a $[[n,k,d]]$ stabilizer code.
For example, the code distance of the 9-qubit code is a $[[9,1,3]]$ stabilizer code correcting weight-one errors.

The nine-qubit code is not the smallest QEC code that can correct all weight-one $X$, $Y$, and $Z$ errors.
The smallest code is the five-qubit code\index{five-qubit code}, found independently by Laflamme {\it et al.}~\cite{Laflamme96} and Bennett 
{\it et al.}~\cite{BCDSW96}.
The stabilizer generators and the logical Pauli operators are given as follows:
\begin{equation}
\begin{array}{cccccc}
\hline \hline
S_{1}=&X&Z&Z&X&I
\\
S_{2}=&I&X&Z&Z&X
\\
S_{3}=&X&I&X&Z&Z
\\
S_{4}=&Z&X&I&X&Z
\\
X_{L}=&X&X&X&X&X
\\
Z_{L}=&Z&Z&Z&Z&Z
\\
\hline \hline
\end{array}
\end{equation}
We see that the code distance is three, and hence an arbitrary single-qubit error can be corrected.

\subsection*{Calderbank-Shor-Steane codes}
The readers who are familiar with classical coding theory might already be aware of the correspondence between stabilizer codes and classical linear codes.
Let us recall the 9-qubit code.
The $X$ and $Z$ errors are detected independently through the $Z$-type and $X$-type stabilizer generators, respectively.
This implies that $X$ and $Z$ error corrections are described by classical coding theory, where two classical error corrections are subject to a certain constraint to appropriately form a stabilizer group.

To formulate this, we briefly review classical linear codes\index{classical linear code}.
A $[[n,k]]$ classical linear code $C$ is defined as a $k$-dimensional space $V_C$ over $GF(2^{n})$ by using an $n \times k$ generator matrix
\begin{eqnarray}
G=( \mathbf{b}_1 , ..., \mathbf{b}_k),
\end{eqnarray}
where the column vectors $\{ \mathbf{b}_i\}$ are the basis vectors of $V_C$. 
A $k$-bit classical information $\mathbf{y}$ is encoded into the code $\mathbf{c}$ as
\begin{eqnarray}
\mathbf{a}= G \mathbf{c}.
\end{eqnarray}
To detect and analyze the errors, we define an $(n-k)\times n$ parity check matrix\index{parity check matrix} $H$ such that $H \mathbf{b}_k=0$ for all basis vectors 
$\{ \mathbf{b}_k\}$.
Suppose an error $\mathbf{e}$ occurs on the code state, $\mathbf{a}' = \mathbf{a}\oplus \mathbf{e}$, where $\oplus$ indicates a bitwise addition modulo two.
By using the parity check matrix $H$, we can detect the error
\begin{eqnarray}
H \mathbf{a}'  = H(\mathbf{a} \oplus \mathbf{e})=H \mathbf{e}\equiv \mathbf{s},
\end{eqnarray}
where $\mathbf{s}$ is called an error syndrome.

For example, the three-bit repetition code is defined by the generator
\begin{eqnarray}
G = \left(\begin{array}{c}1 \\ 1 \\ 1 \end{array} \right).
\end{eqnarray}
A classical bit $0$ and $1$ is encoded into $(0,0,0)^{\rm T}$ and $(1,1,1)^{\rm T}$, respectively.
The parity check matrix is defined to be
\begin{eqnarray}
H= \left(
\begin{array}{ccc}
1 & 1 & 0
\\
0 & 1 & 1
\end{array}
\right).
\end{eqnarray}

Now, we realize that the positions of the 1s of the parity check matrix are exactly the same as those of the $Z$s in the stabilizer generators of the three-qubit bit flip code.
This suggests to use the parity check matrices $H_x$ and $H_z$ of the two classical linear codes $C_x$ and $C_z$, respectively, 
in the definition of the $X$-type and $Z$-type stabilizer generators:
\begin{eqnarray}
S_{X}^{(i)} = \prod _{j} X_j^{(H_x)_{ij}},
\;\;\;\;
S_{Z}^{(i)} = \prod _{j} Z_j^{(H_z)_{ij}}.
\end{eqnarray}
For these operators to commute with each other, the two parity check matrices have to satisfy
\begin{eqnarray}
H_x H_z^{\rm T} = \mathbf{0},
\end{eqnarray}
where $\mathbf{0}$ indicates a matrix with all elements $=0$.
To define the logical $Z$ operators, we define a quotient space ${\rm Ker}(H_x)/V_{H_z}:=\{ \mathbf{c}| H_x \mathbf{c} =0, \mathbf{c} \in V_{C_z} \}/V_{H_z}$, where $V_{H_z}$ is the space spanned by the column vectors of $H_{z}^{\rm T}$.
Denoting the basis vectors of the quotient space ${\rm Ker}(H_x)/V_{H_z}$ by $\{ [ \mathbf{b}_k^{z} ]\}$, we define the logical $Z$ operators
\begin{eqnarray}
L_Z^{(k)}=  \prod _{i} Z_i^{(\mathbf{b}^{z}_k)_i}.
\end{eqnarray}
Similarly, we can define the logical $X$ operators
\begin{eqnarray}
L_X^{(k)}=  \prod _{i} X_i^{(\mathbf{b}^{x}_k)_i},
\end{eqnarray}
using the basis vectors $\{ [\mathbf{b}_k^{x}] \}$ of a quotient space ${\rm Ker}(H_z)/V_{H_x}$, where $\mathbf{b}_k^{x}$ is chosen such that 
$L_Z^{(i)} L_X^{(j)} = (-1)^{\delta _{ij}}L_X^{(j)} L_Z^{(i)}$.
Note that dimensions of these kernel subspaces are the same, and we can easily find such pairs of anticommuting logical operators.
The above stabilizer code constructed from two classical linear codes is called a Calderbank-Shor-Steane (CSS) code\index{Calderbank-Shor-Steane code, CSS code}.

Let us see an important example of CSS codes, the 7-qubit code\index{7-qubit Steane code} introduced by Steane~\cite{Steane96}.
Specifically, we utilize a classical linear code, the $[[7,4,3]]$ Hamming code, whose generator and parity check matrices are given by
\begin{eqnarray}
G= \left( 
\begin{array}{cccc}
1 & 0 & 0& 0
\\
0 & 1 & 0 & 0
\\
0 & 0& 1 & 0
\\
0 & 0& 0& 1
\\
0 & 1 & 1 & 1 
\\
1 & 0 & 1 & 1 
\\
1 & 1 & 0 & 1
\end{array}
\right),
\;\;\;
H= \left( 
\begin{array}{ccccccc}
1 & 0 & 1 & 0 & 1 & 0 & 1
\\
0 & 1 & 1 & 0 & 0 & 1 & 1
\\
0 & 0 & 0 & 1 & 1 & 1 & 1
\end{array}
\right).
\end{eqnarray}
Because $H H^{\rm T}=\mathbf{0}$, we can employ the Hamming code to define both $X$- and $Z$-type stabilizer generators:
\begin{equation}
\begin{array}{cccccccccc}
\hline \hline
S_{1}&=& I&I&I&X&X&X&X
\\
S_{2}&=& I&X&X&I&I&X&X
\\
S_{3}&=& X&I&X&I&X&I&X
\\
S_{4}&=& I&I&I&Z&Z&Z&Z
\\
S_{5}&=& I&Z&Z&I&I&Z&Z
\\
S_{6}&=& Z&I&Z&I&Z&I&Z
\\
\hline
\hline
\end{array}
\end{equation}
There is an element $(1,1,1,1,1,1,1)^{\rm T}$ in the quotient space ${\rm Ker}(H)/V_H$.
The logical operators are given by
\begin{eqnarray}
\begin{array}{cccccccccc}
L_X&=& X&X&X&X&X&X&X,
\\
L_Z&=& Z&Z&Z&Z&Z&Z&Z.
\end{array}
\end{eqnarray}

The 7-qubit code is quite useful for fault-tolerant quantum computation.
Both the $X$- and $Z$-type stabilizer generators are defined from the Hamming code, and the stabilizer group is invariant under the transversal Hadamard operation 
$\bar H \equiv H^{\otimes 7}$.
Moreover, the logical $X$ operator is mapped into the logical $Z$ operator, $\bar H L_X \bar H = L_Z$.
Thus, the transversal Hadamard operation acts as the logical Hadamard operation for the encoded degree of freedom.
Similarly, a transversal phase operation $\bar S \equiv (ZS)^{\otimes 7}$ acts as a logical phase operation, $\bar S L_X \bar S^{\dag} =  L_X L_Y$.
Furthermore, a transversal CNOT operation $\bar \Lambda (X) = \Lambda(X)^{\otimes 7}$ keeps the stabilizer group of two logical qubits invariant:
\begin{eqnarray}
&&\langle \{ S_X^{(i)} \otimes I^{\otimes 7} \},
\{ S_Z^{(i)} \otimes I^{\otimes 7} \} ,
\{  I^{\otimes 7} \otimes S_X^{(i)} \},
\{  I^{\otimes 7} \otimes S_Z^{(i)} \} \rangle
\\
&&=
\langle \{ S_X^{(i)} \otimes S_X^{(i)} \},
\{ S_Z^{(i)} \otimes I^{\otimes 7} \} ,
\{  I^{\otimes 7} \otimes S_X^{(i)} \},
\{  S_Z^{(i)} \otimes S_Z^{(i)} \} \rangle.
\end{eqnarray}
The logical Pauli operators are subject to the transformation rule of the CNOT gate.
Accordingly, the transversal CNOT operation $\bar \Lambda (X)$ acts as a logical CNOT operation for the encoded degree of freedom.
Because the Hadamard, phase, and CNOT operations are implemented transversally, whole Clifford group elements can be implemented by transversal operations.

The transversal implementation is fault-tolerant because the operations apparently do not increase the number of errors on a code block and there is no internal interaction between the qubits in the same code block.
Combined with a fault-tolerant gadget for measuring the error syndrome as explained in Appendix~\ref{Ap:FT_syndrome_meas}, we can implement Clifford operations fault-tolerantly.

Unfortunately, the non-Clifford operation does not transform a Pauli operator into another Pauli operator.
For example, the $\pi/8$ operation $e^{ -i (\pi /8) Z}$ transforms the Pauli $X$ operator into a Clifford operator:
\begin{eqnarray}
e^{-i (\pi/8) Z} X e^{i (\pi/8) Z} = (X+Y)/\sqrt{2}.
\end{eqnarray}
This implies that a transversal non-Clifford operation hardly results in a logical non-Clifford operation.
Thus, a fault-tolerant non-Clifford gate operation is not so straightforward.
To settle this, we can utilize magic state distillation, consisting of noisy non-Clifford resource states and ideal (fault-tolerant) Clifford operations, as explained in the next section.

\section{Magic state distillation}
\label{sec:magic}
\subsection{Knill-Laflamme-Zurek protocol}
A fault-tolerant implementation of a non-Clifford gate was first proposed in an earlier paper by Knill, Laflamme, and Zurek~\cite{Knill98a,Knill98b}.
Instead of implementing the non-Clifford gate directly, we consider a fault-tolerant preparation of the non-stabilizer state, the so-called magic state\index{magic state},
\begin{eqnarray}
|A \rangle \equiv e^{ - i (\pi /8 ) Y }| +\rangle.
\end{eqnarray}
The magic state can be utilized to implement a non-Clifford gate $A \equiv e^{- i (\pi /8) Y}$ by using one-bit teleportation consisting of Clifford gates and the Pauli basis state preparations and measurements.
Thus, if we can prepare a clean magic state, we can create a fault-tolerant non-Clifford gate by using fault-tolerant Clifford gates.

The Knill-Laflamme-Zurek construction of the fault-tolerant preparation of the magic state was based on the fact that $|A\rangle$ is an eigenstate of $H$.
The Hadamard operation has transversality, e.g., on the 7-qubit code.
Hence, if we perform a projective measurement of $H^{\otimes 7}$, we obtain a clean magic state.
The circuit is given as follow
\begin{equation}
\includegraphics[width=120mm]{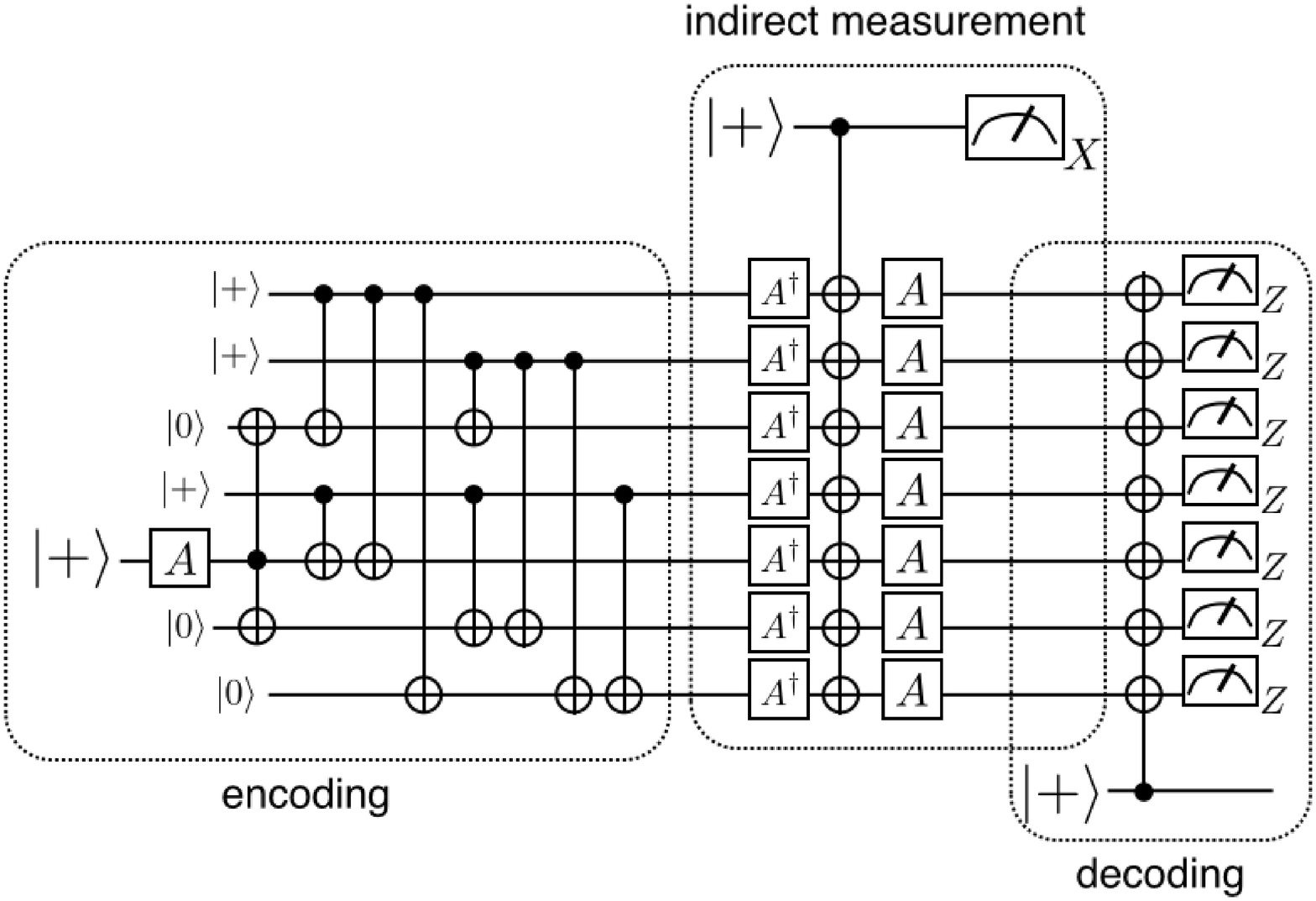}
\end{equation}
where $A=e^{-i (\pi/8)Y}$ and we used the fact that $A_t \Lambda _{c,t}(X)A_t^{\dag}= \Lambda_{c,t}(H)$.
The above circuit consists of three parts, encoding of the logical magic state into the 7-qubit code, indirect measurement of $H^{\otimes 7}$, and decoding by one-bit teleportation.
Note that all Clifford operations are assumed to be ideal, because they are easily made fault-tolerant by using a stabilizer code.
With an appropriate randomization operation 
made up by Clifford operations, a noisy magic state can be transformed into a mixture of 
$|A\rangle$ and $|A^{\perp}\rangle = Y|A\rangle$:
\begin{eqnarray}
\rho _{A} = (1-p) |A\rangle \langle A| + p |A^{\perp}\rangle \langle A|
\end{eqnarray}
Hence, the noise on the magic state can be expressed as a $Y$ error.
The $Y$ error can be detected by transversal $Z$ measurements for the decoding (see also Knill's gadget for a fault-tolerant syndrome measurement in Appendix~\ref{Ap:FT_syndrome_meas}).
Assuming an ideal Clifford operation, the error probability $p$ decreases as $O(p^{3})$, when we employ the 7-qubit code with distance 3.
Including the $A$ gate for the initial encoding, we need 15 noisy magic states $|A\rangle$ to obtain a clean magic state.
This distillation protocol works for all self-dual CSS codes, which are symmetric between the $X$- and $Z$-type stabilizer generators and has transversality for the Hadamard gate.
Recently, improved protocols have been proposed based on this approach~\cite{Meier12,JonesMagic}.
 
 \subsection{Bravyi-Kitaev protocol}
Bravyi and Kitaev proposed another magic state distillation protocol based on a 15-qubit code~\cite{Magic}.
While their and the Knill-Laflamme-Zurek protocols seem quite different, they are, interestingly, known to be equivalent~\cite{Reichardt05}.
In the Bravyi-Kitaev protocol, a [[15,1,3]] quantum code is defined by the [[15,7,3]] classical Reed-Muller code\index{Reed-Muller code}, whose parity check matrix is given by
\begin{eqnarray}
H_x= \left( 
\begin{array}{ccccccccccccccc}
1 & 0 &0 &0 &0 &1 & 1& 0& 0& 1& 1 & 1& 1& 0 & 1
\\
0 & 1 & 0 & 0& 1 & 0 & 1 & 0 & 1 & 0 & 1 & 1 & 0 & 1 & 1 
\\
0 & 0 & 1 & 1 & 0 & 0 &1 & 0 & 1 & 1 & 0 & 0& 1 & 1 & 1 
\\
0 & 0& 1 & 0 & 1 & 1 & 0 & 1 & 0 & 0& 1 & 0& 1 & 1& 1
\end{array}
\right)
\end{eqnarray}
The $X$-type stabilizer generators are defined by $S_X^{(i)}= \prod _j X_j^{(H)_{ij}}$.
Then, we choose a parity check matrix of another classical code
\begin{eqnarray}
H_z= \left( 
\begin{array}{ccccccccccccccc}
0 & 0 &1 &1 &0 &0 & 0& 1& 0& 0& 0 & 0& 0& 0 & 1
\\
0 & 1 &0 &0 &1 &0 & 0& 1& 0& 0& 0 & 0& 0& 0 & 1
\\
1 & 0 &0 &0 &0 &1 & 0& 1& 0& 0& 0 & 0& 0& 0 & 1
\\
1 & 1 &0 &1 &0 &0 & 1& 0& 0& 0& 0 & 0& 0& 0 & 0
\\
0 & 1 &0 &1 &0 &0 & 0& 0& 1& 0& 0 & 0& 0& 0 & 1
\\
1 & 0 &0 &1 &0 &0 & 0& 0& 0& 1& 0 & 0& 0& 0 & 1
\\
1 & 1 &0 &0 &0 &0 & 0& 1& 0& 0& 1 & 0& 0& 0 & 0
\\
1 & 1 &0 &0 &0 &0 & 0& 0& 0& 0& 0 & 1& 0& 0 & 1
\\
1 & 0 &0 &1 &0 &0 & 0& 1& 0& 0& 0 & 0& 1& 0 & 0
\\
0 & 1 &0 &1 &0 &0 & 0& 1& 0& 0& 0 & 0& 0& 1 & 0
\end{array}
\right)
\end{eqnarray}
so that $H_x H_z^{\rm T}=\mathbf{0}$ and the $Z$-type stabilizer generators are defined similarly.
The logical operators are given by $L_X=X^{\otimes 15}$ and $L_Z=Z^{\otimes 15}$.
The logical states are written down explicitly as 
\begin{eqnarray}
|0_L\rangle 
&=& \prod _{i=1}^{4}(I+S_X^{(i)}) |00...0\rangle,
\\
|1_L\rangle 
&=& \prod _{i=1}^{4}(I+S_X^{(i)}) |11...1\rangle.
\end{eqnarray}
The number of 1s in each term of $|0_L\rangle$ and $|1_L\rangle$ is 8 and 7, respectively.
By applying the $T=e^{-i (\pi /8 )Z}$ gate transversally, we obtain
\begin{eqnarray}
T^{\otimes 15} |0 _{L}\rangle &=& e^{i \pi /8} |0 _{L}\rangle,
\\
T^{\otimes 15} |1 _{L}\rangle &=& e^{-i\pi/8} |0 _{L}\rangle.
\end{eqnarray}
Thus, the transversal $T$ gate acts as a logical $T^{\dag}$ gate.
Note that this transversality does not hold in the orthogonal (erroneous) subspace, e.g., spanned by $\{ X_k|0_L\rangle, X_k|1_L\rangle \}$.
However, we can show that this is enough to perform a fault-tolerant logical $T$ gate.

Instead of applying the $T$ gate directly, we implement it using a one-bit teleportation:
\begin{equation}
\includegraphics[width=110mm]{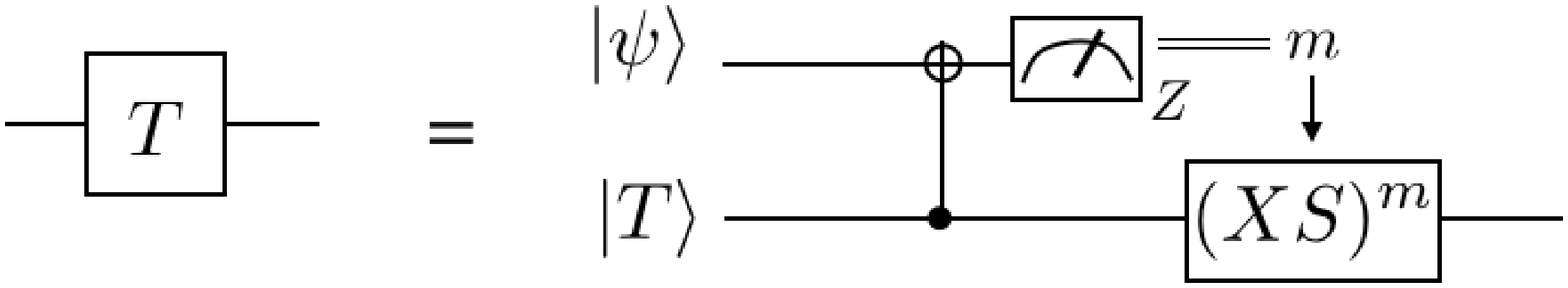}
\end{equation}
The $Z$-basis measurement and CNOT operation are both implemented transversally on a CSS code.
Thus, if the preparation of the non-Clifford ancilla state $|T\rangle = (|0 _{L} \rangle + e^{-i \pi /4} |1_{L}\rangle)/\sqrt{2}$, called a magic state, is done fault-tolerantly, one can ensure fault-tolerance of the logical $T$ gate.

By using an appropriate randomization process, we can prepare a noisy magic state as follows:
\begin{eqnarray}
\rho _{T} = (1-p) |T\rangle \langle T| + p Z|T\rangle \langle T|Z.
\end{eqnarray}
Thus, a phase error $Z$ is located on the ideal magic state with probability $p$.
This phase error causes a $Z$ error after the $T$ gate by one-bit teleportation.
Because the code space is invariant under the transversal $T$ gate, we can detect such a $Z$ error by the following circuit.
\begin{center}
\includegraphics[width=90mm]{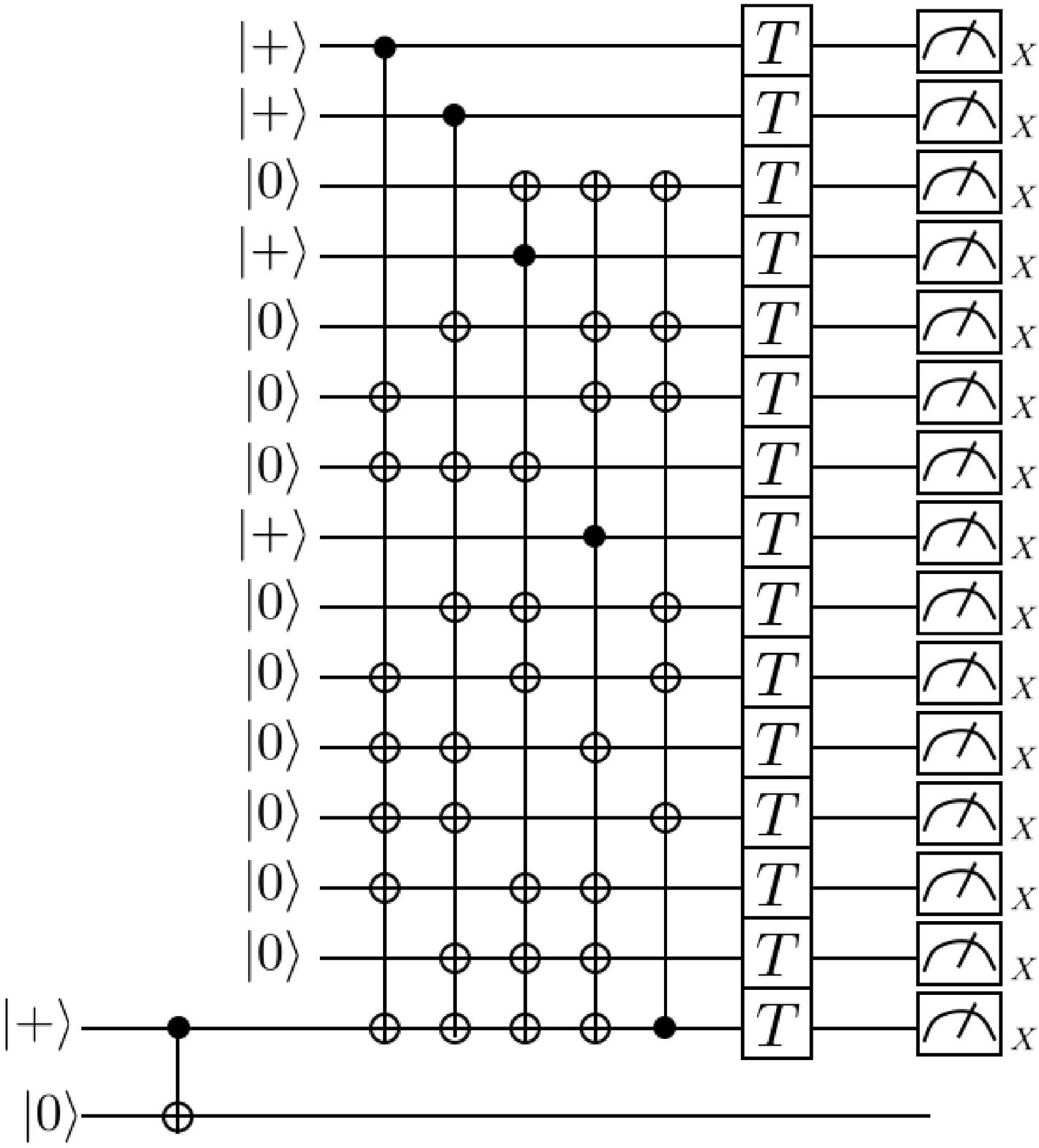}
\end{center}
The first part of the above circuit consisting of CNOT operations is an encoding circuit for the quantum Reed-Muller code.
The transversal $T$ gate is applied by using one-bit teleportation.
The logical $|T\rangle$ state is measured in the $X$-basis transversally, which detects $Z$ errors on the code state, projecting the code state on the local $X$-basis.
The input state in the second lowest wire is entangled with the ancilla qubit in the lowest wire, where the distilled magic state is teleported.

Let $\mathbf{c}$ be a 15-bit string specifying the location of the $Z$ errors, $ E(\mathbf{c})\equiv \prod_i Z_i^{(\mathbf{c})_i}$.
If $E(\mathbf{c})$ commutes with the $X$-type stabilizer generators, the state passes through the distillation circuit.
To calculate this probability, we define a weight enumerator of a subspace $V \in GF(2^n)$,
\begin{eqnarray}
W_V(x,y)= \sum _{\mathbf{c} \in V} x^{n-{\rm wt}(\mathbf{c})} y^{{\rm wt}(\mathbf{c})}.
\end{eqnarray}
The probability of passing the distillation circuit is calculated to be
\begin{eqnarray}
p_{\rm pass}=W_{V_{H_x}^{\perp}}(1-p,p) = \frac{1}{|V_{x}|} W_{V_{H_x}}(1,1-2p) = \frac{1+15(1-2p)^8}{16} ,
\end{eqnarray}
where the orthogonal subspace $V_{H_x}^{\perp}$ is equivalent to the kernel of $H_x$, ${\rm Ker}(H_x)$. 
We also used the MacWilliams identity~\cite{macwilliams}:
\begin{eqnarray}
W_{V}(x,y)= \frac{1}{|V|} W_{V^{\perp}}(x+y,x-y).
\end{eqnarray}
Similarly, the error probability of the output can be calculated to be 
\begin{eqnarray}
W_{V_{H_z}}(p,1-p)&=&\frac{1}{|V_{H_z}^{\perp}|}W_{V_{H_z}^{\perp}}(1,2p-1) 
\\
&=& \frac{1+15(2p-1)^8+15(2p-1)^{7}+(2p-1)^{15}}{32}.
\end{eqnarray}
Accordingly, the error probability, under the condition of passing the distillation circuit, is given by
\begin{eqnarray}
p' = \frac{1+15(2p-1)^8+15(2p-1)^{7}+(2p-1)^{15}}{2[1+15(1-2p)^8]} = 35p^3+O(p^4).
\end{eqnarray}
If $p'>0.141$, we can reduce the error probability on the magic state via the distillation circuit.
After $l$ rounds of distillation, the error probability decreases to $(\sqrt{35} p)^{3^l}/\sqrt{35}$.
At each round, we need 15 noisy magic states. 
Because the probability of successfully passing the distillation circuit converges rapidly to 1, the average number of noisy magic states consumed after $l$ rounds becomes $15^{l}$.
Accordingly, the average number of noisy magic states required to achieve an error probability $\epsilon$ of the magic state scales like
\begin{eqnarray}
[\log (\sqrt{35} \epsilon)/ \log (\sqrt{35} p)]^{\log(15)/\log(3)} =O(\log ^{2.5} \epsilon).
\end{eqnarray}

\begin{figure}[t]
\includegraphics[width=120mm]{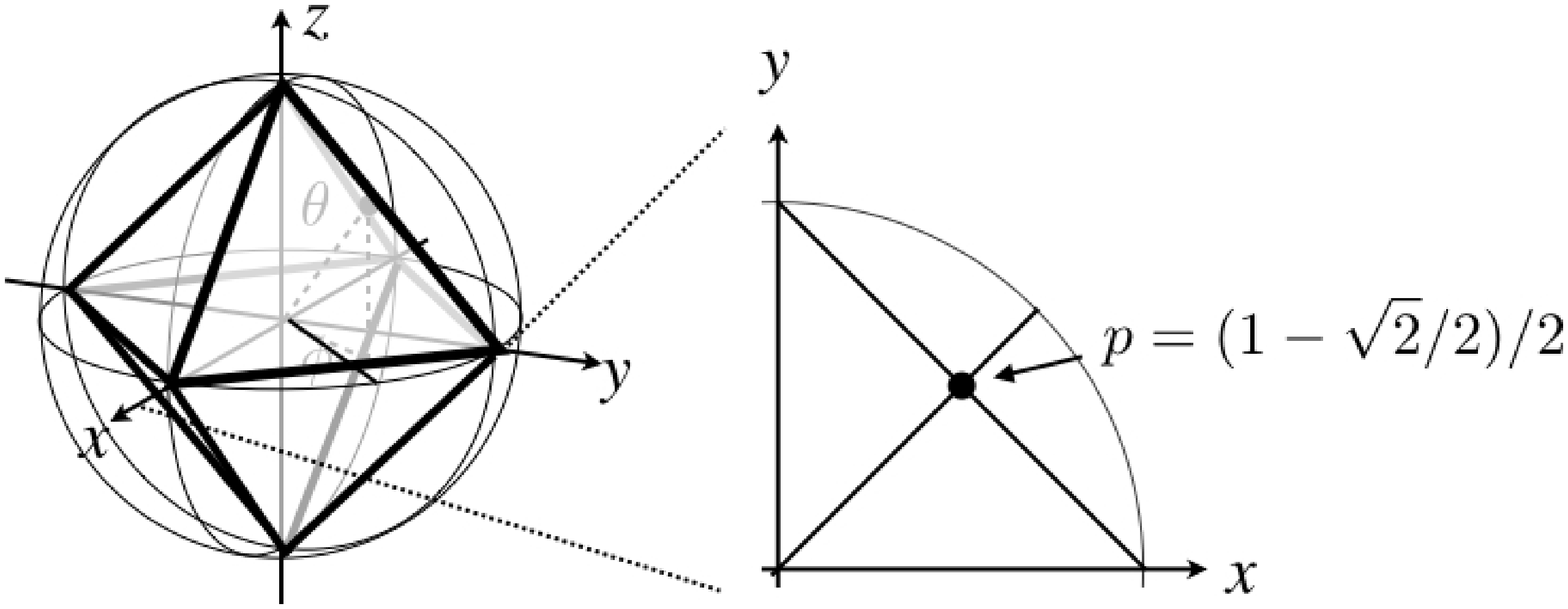}
%
%
\caption{The noisy magic state in the Bloch sphere.}
\label{fig53}       
\end{figure}

In Sec.~\ref{sec:GKtheorem}, we saw that if an input state is a convex mixture of the Pauli basis states followed by Clifford operations and Pauli basis measurements, the measurement outcomes can be simulated classically in the weak sense.
The noisy magic state $\rho _{T} = (1-p) |T\rangle \langle T| + p |T\rangle \langle T|$ lies on a line in the $x$--$y$ plane, as shown in Fig.~\ref{fig53}.
If $p > (1-\sqrt{2}/2)/2=0.146$, $\rho _{\pi/8}$ lies inside the octahedron, and the Gottesman-Knill theorem is applicable. 
On the other hand, if $p< 0.141$, magic state distillation allows us to implement universal quantum computation with an arbitrary accuracy as seen above.
Unfortunately, magic state distillation based on the Reed-Muller code does not provide a tight distillation threshold against the classically simulatable region.
In Ref.~\cite{Reichardt05}, a distillation protocol using the 7-qubit code was proposed and achieved a tight threshold $p=(1-\sqrt{2}/2)/2$.
In this sense, the classically simulatable and quantum universal regions 
are divided sharply on the $x$-$y$ plane.

By combining the magic state distillation and fault-tolerant Clifford operations on the CSS code, we can perform universal quantum computation fault-tolerantly.
In order to make the error probability arbitrarily small, we can employ concatenated quantum computation, in which logical qubits of a lower concatenation level are utilized as the physical qubits at a higher level.
At the higher level, all operations, including logical qubit preparations and syndrome measurements for QEC, have to be done fault-tolerantly.
If the error probability is smaller than a certain constant value, which we call the noise threshold, the logical error probability at the highest concatenation level decreases super-exponentially.
On the other hand, the overhead increases exponentially.
Thus, we can make the logical error probability small enough to maintain a quantum computation of size $N$ with a polylogarithmic overhead ${\rm polylog} (N)$.
This implies that we can obtain a quantum benefit even for a quantum algorithm with a quadratic speedup, such as the Grover algorithm~\cite{Grover}.
In Appendix~\ref{Ap:FT_syndrome_meas}, we briefly review fault-tolerant syndrome measurements, concatenated quantum computation, and the threshold theorem.

While the resource increment for protecting quantum computation scales polylogarithmically, its constant factor is quite huge.
Almost all overheads for the fault-tolerant quantum computation are employed for the magic state distillation~\cite{VanMeter,Jones12}. 
Thus, much effort has been spent recently on developing resource-efficient magic state distillation~\cite{Meier12,BravyiHaah,Eastin13,Jones13}. 
\chapter{Topological stabilizer codes}
\label{Chap:TpologicalCodes}
Protecting quantum information from decoherence is of prime importance to realize quantum information processing.
Several approaches have been proposed toward reliable quantum information processing, ranging from passive to active protections, such as decoherence-free subspaces
\cite{DFS}, dynamical decoupling \cite{DD}, and quantum error correction \cite{Shor95}.
Among these, the most comprehensive approach is fault-tolerant quantum computation based on quantum error correction \cite{DiViShor}.
Quantum fault-tolerance theory ensures scalable quantum computation with noisy quantum devices as long as the error probability of such devices is smaller than a threshold value (see Appendix~\ref{Ap:FT_syndrome_meas}). 
The first threshold values were obtained, at almost the same time (1996), independently by Aharonov {\it et al.}, who used a polynomial code of distance five \cite{Aharonov97,Aharonov08}; Knill {\it et al.}, who used the Steane seven-qubit code \cite{Knill98a,Knill98b}; and Kitaev, who used toric (surface) codes \cite{Kitaev97}.
All of these studies were based on a concatenated quantum computation and achieved similar threshold values $\sim 10^{-6}$. 
Since then, fault-tolerant quantum computing has been studied as one of the most important issues in quantum information science.
In 2005, Knill proposed a novel scheme based on the $C_4/C_6$ error-detecting code, namely the Fibonacci scheme \cite{KnillNature}, and achieved a considerably higher threshold, a few \%. 
Recently, Fujii {\it et al.} have further improved this approach by making use of measurement-based quantum computation (MBQC) \cite{RaussendorfAnn,RaussendorfNJP} on logical cluster states, which gives the highest threshold value so far, $\sim 5\%$ \cite{FYCluster,FYClusterTopo}.

All these approaches rely on the availability of two-qubit gates between arbitrarily separated qubits.
However, all physically available interactions are local in space. 
Moreover, if we consider manufacturing convenience and the selective addressability of individual qubits, 2D nearest-neighbor architectures are necessary.
It was thought that, if we restrict the two-qubit gates to nearest-neighbor ones in 2D, then the threshold value would decrease significantly \cite{Svore}.
The situation changed completely by the proposal made by Raussendorf {\it et al.}, i.e., topologically protected quantum computation~\cite{RaussendorfPRL,RaussendorfNJP,RaussendorfAnn}. 
They utilized a surface code, a kind of topological quantum code, originally proposed by Kitaev as a toric code on a torus~\cite{Kitaev} and investigated by Dennis {\it et al.} in detail~\cite{Dennis}.
All the stabilizer generators of the surface code are spatially local, as will be seen later, and hence syndrome measurements are done by using nearest-neighbor two-qubit gates.
More importantly, Raussendorf {\it et al.} developed a novel way to implement a fault-tolerant logical gate operation by braiding defects on the surface, which can be done by using only single-qubit measurements and nearest-neighbor two-qubit gates.
Nevertheless, its threshold value is very high $\sim 1\%$~\cite{RaussendorfPRL,RaussendorfNJP,RaussendorfAnn,WangFowler}.
Based on these analyses, physical implementations and quantum architecture designs have recently been suggested~\cite{VanMeter,LiBenjamin,FTProb,Jones12,FujiiDis,Benjamin1,Benjamin2,Fowler1,Fowler2}, which clarifies the experimental requirements for building a fault-tolerant quantum computer.
On the other side, extensive experimental resources have been used to achieve these requirements, and very encouraging results have already been obtained in several experiments 
in superconducting systems \cite{Chow12IBM,IBM13,UCSB14,IBM14}.

Another important motivation for studying topological quantum codes are their connection with topologically ordered systems in condensed matter physics~\cite{Kitaev}.
Topological stabilizer codes, whose stabilizer generators are local and translation invariant, provide toy models for topologically ordered systems.
Using quantum coding theory with a geometrical constraint such as locality and translation invariance, we can understand the nature of topologically ordered condensed matter systems.
One of the most important issues in this direction is to answer whether or not thermally stable topological order exists in three or lower dimensions~\cite{BravyiTerhal,BeniThermal3D}.

When we consider the decoding problems of topological quantum codes, we encounter classical statistical physics~\cite{Dennis}.
Specifically, the posterior probabilities for the error correction correspond to partition functions of classical spin glass models~\cite{Dennis}, such as the random-bond Ising model.
If the corresponding random-bond Ising model lies inside a ferromagnetic phase, the error correction on a surface code succeeds, and the logical error probability decreases exponentially.

In this chapter, we introduce a representative example of the topological stabilizer codes\index{topological stabilizer code}, the surface codes. 
Before that, we also briefly review the ${\bf Z}_2$ chain complex, which is a useful tool to define the surface codes. 
We also apply it to a classical repetition code as an exercise.
After introducing the surface codes, we explain how topological error correction is performed.
We also mention the relation between topological error correction and the classical spin glass model.
Another example of topological stabilizer codes, the topological color codes, is also presented.
Finally, we explain the connection between topological stabilizer codes and topologically ordered systems studied in condensed matter physics.

\section{${\bf Z}_2$ chain complex}
\label{sec:chain_complex}
Before defining the surface codes, we introduce a mathematical tool, the $\mathbf{Z}_2$ chain complex\index{$\mathbf{Z}_2$ chain complex}, which is useful for describing the surface codes. 

Consider a surface $G=(V,E,F)$ that consists of the vertices $V=\{v_k\}$, edges $E=\{e_l\}$, and faces $F=\{f_m\}$.
(Here we consider a graph $G=(V,E)$ embedded on a 2D manifold, the {\it surface}, on which the faces $F$ are defined.)
We define a vector space $C_0$ over $\mathbf{Z}_2$, using each vertex $v_k \in V$ as a basis $B(C_0)=\{ v_k\}$.
A vector in $C_0$
\begin{eqnarray}
c_0 &=& \sum _k z_k v_k, 
\end{eqnarray}
where $z_k =\{ 0, 1\}$, is referred to as a 0-chain.
The vector space $C_0$ is an Abelian group under component-wise addition modulo 2.
Similarly, we can define the Abelian groups $C_1$ and $C_2$ using the edges $E=\{e_l\}$ and faces $F=\{ f_m \}$ as base vectors $B(C_1)=\{ e_l\}$ and $B(C_2)=\{ f_m\}$, respectively.
The elements
\begin{eqnarray}
c_1 &=& \sum _l z_l e_l, \;\;\;
\\
c_2 &=& \sum _m z_m f_m, \;\;\;
\end{eqnarray}
where $z_l, z_m=\{ 0,1\}$, are called the 1-chain and 2-chain, respectively.
With a mild abuse of the notation, a set of $i$-dimensional elements (i.e., vertices, edges, and faces) is also specified by a $i$-chain $c_i \in C_i$ as a set of elements having $z_j=1$.

We can define a homomorphism $\partial_i : C_i \rightarrow C_{i-1}$ such that 
\begin{eqnarray}
\partial _i \circ \partial _{i-1}=0.
\label{eq:boundary_op}
\end{eqnarray} 
Specifically, $\partial _i c_i $ is defined as an $(i-1)$-chain that is the boundary of $c_i$. 
For example, $\partial e_l$ is a 1-chain for which $z_k=1$ if and only if (iff) $v_k$ is an endpoint of $e_l$.
Thus, the homomorphism $\partial _i$ is called a boundary operator.
When there is no risk for confusion, we will denote $\partial _i$ simply by $\partial$.
A chain $c_i$ is called a cycle, if it is in a kernel of the boundary operator $\partial _i$, i.e., $\partial c_i=0$.
For example, the boundary $\partial_2 f_m$ of a face $f_m$ is a cycle, because $\partial _2 \circ \partial _1 =0$. 
Such a sequence of Abelian groups $C_i$, connected by the boundary operators $\partial _i$ with $\partial _i \circ \partial _{i-1}=0$, is called a $\mathbf{Z}_2$ chain complex.
We define a trivial $i$-cycle $c_i$ if there exists an $(i+1)$-chain such that $c_i = \partial c_{i+1}$, i.e., $c_i \in {\rm Img} (\partial _{i+1})$.
By regarding the trivial cycles\index{trivial cycle} as generators of the continuous deformation, we may define an equivalent class having the same topology.
More precisely, a homology group $h_i$ is defined as a quotient group\index{quotient group} formed by the cycles and the trivial cycles:
\begin{eqnarray}
h_i = \ker (\partial _i) / {\rm Img} (\partial _{i+1}).
\end{eqnarray}
An element of the quotient group $h_i$ is called a homology class\index{homology class}.
If two $i$-chains $c_i$ and $c'_i$ belong to the same homology class, there exists an $(i+1)$-chain $c_{i+1}$ such that $c_i = c'_i + \partial c_{i+1}$.
Two such $i$-chains $c_i$ and $c'_i$ are said to be homologically equivalent.

We also define the dual surface $\bar G=(\bar V,\bar E,\bar F)$, where we identify the elements of the dual and original (primal) lattices such that $\bar V=F$, $\bar E=E$ and $\bar F=V$.
Specifically, the dual lattice is constructed such that the two vertices $\bar v, {\bar v}' \in \bar V$ are connected by an edge $\bar e$, if the corresponding two faces $f$ and $f'$ share the same edge $e$.
We can also define a $\mathbf{Z}_2$ chain complex on $\bar G$ using the dual bases $\bar B(C_i)$, a dual $i$-chain $\bar c_i \in \bar C_i$ and a boundary operator 
$\bar \partial _i: \bar C_i \rightarrow \bar C_{i-1}$.
When there is no risk of confusion, $\bar \partial _i$ will be denoted by $\partial$.
 
In the construction of the surface codes, a qubit is defined on each edge $e_l \in E$ of the surface $G$ (or equivalently each dual edge $\bar e_l$ of the dual surface $\bar G$).
The Pauli product is defined using a 1-chain $c_1 = \sum _l z_l e_l$ (or dual 1-chain $\bar c_1$) such that
\begin{eqnarray}
W(c_1) = \prod _{l} W_l^{z_l},
\end{eqnarray}
where $W_l \in \{X_l , Y_l,Z_l \}$ is a Pauli operator acting on the qubit on edge $e_l$.
Specifically, we have 
\begin{eqnarray}
W(c_1)W(c'_1)=W(c_1 + c'_1).
\end{eqnarray}
Consider the two operators $X(c_1)$ and $Z(c'_1)$, defined by the two 1-chains $c_1$ and $c'_1$, respectively.
The commutability of these two operators is determined by the inner product of the two chains (vectors) $c_1\cdot c'_1 \equiv \sum _{l}z_l z'_l$, where the addition is taken modulo 2:
\begin{eqnarray}
c_1\cdot c'_1 &=&0, \;\; {\rm iff} \;\; X(c_1)Z(c'_1)=Z(c'_1)X(c_1) \;\; {\rm (commute)},
\\
c_1\cdot c'_1&=&1, \;\; {\rm iff} \;\; X(c_1)Z(c'_1)=-Z(c'_1)X(c_1) \;\; {\rm (anticommute)}.
\end{eqnarray}

Let $M(\partial _i)$ be a matrix representation of $\partial _i$ with respect to the basis vectors $B(C_i)$ and $B(C_{i-1})$.
We have
\begin{eqnarray}
\bigl(M(\partial _i) c_i \bigr) \cdot c_{i-1} =  c_i \cdot \bigl( M(\partial _i)^{T}  c_{i-1} \bigr).
\label{eq:adjoint}
\end{eqnarray}
By identifying the primal and dual bases $B(C_0)=\bar B(C_2)$, $B(C_1)=\bar B(C_1)$, and $B(C_2)=\bar B(C_0)$, the duality relation between primal and dual lattices can be expressed by
\begin{eqnarray}
M(\partial _1) = M(\bar \partial _2)^{T},
\;\;\;
M(\partial _2) = M(\bar \partial _1)^{T},
\label{eq:dual_rep}
\end{eqnarray}
where $[\cdot]^{T}$ indicates the matrix transpose.

Using Eq.\ (\ref{eq:boundary_op}) $\partial _1 \circ \partial _2=0$, we obtain $ M(\bar \partial _2)^{T} M(\partial _2)=0$.
Thus, for any primal and dual 2-chains, $c_2$ and $\bar c_2$, we have 
\begin{eqnarray}
\bar \partial \bar c_2 \cdot \partial c_2=0,
\label{eq:commutability}
\end{eqnarray}
i.e., $X(\partial c_2)$ and $Z(\bar \partial \bar c_2)$ always commute.
This property is useful to construct a stabilizer group, because stabilizer generators have to commute with each other.
Using Eqs.\ (\ref{eq:adjoint}) and (\ref{eq:dual_rep}), we also have
\begin{eqnarray}
c_1 \cdot \bar \partial _{2} \bar c_2 = \partial c_1 \cdot \bar c_2.
\label{eq:comm_rep}
\end{eqnarray}

\section{A bit-flip code: exercise}
\label{sec:bit-flip}
\begin{figure}[t]
\centering
\includegraphics[width=130mm]{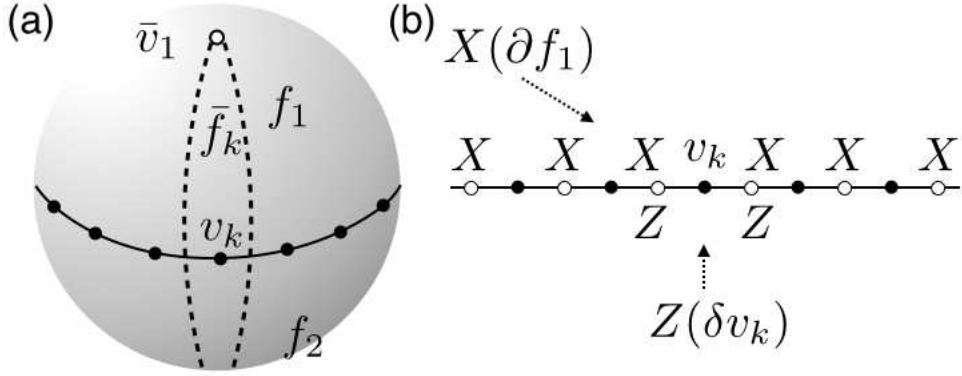}
%
%
\caption{(a) A regular polygon embedded on a sphere. (b) A stabilizer operator defined as $A_k=Z(\partial \bar f_k)=\prod _{\delta v_k} Z_k$. 
The logical $X$ operator is given by $L_X=X(\partial f_1)$. Because primal and dual 1-chains share an even number of qubits, $L_X$ and $A_k$ commute with each other. }
\label{fig54}       
\end{figure}
As an exercise, we define a bit-flip code (classical repetition code) using the $\mathbf{Z}_2$ chain complex.
(Readers who are familiar with this topic may skip this section.)
Consider a regular polygon $G(V,E,F)$ on a sphere consisting of $n=|E|$ edges and two faces $F=\{ f_1 , f_2\}$ (corresponding to the top and bottom hemispheres) as shown in Fig.~\ref{fig54}.
The number of qubits is $n$.

We define a stabilizer generator for each dual face $\bar f _k= v_k$ as follows:
\begin{eqnarray}
A_{k}=Z( \partial \bar f_k)= \prod _{e_l \in \delta v_k} Z_l,
\end{eqnarray}
where, for convenience, $\delta v_k \equiv \partial \bar f $ is defined as a set of edges that are incident to the vertex $v_k$.
(Note that both the dual and primal objects are identified.)

Because $ \prod _{v_k \in V} A_k = I$, there are $n-1$ independent stabilizer generators.
The dimension of the stabilizer subspace is two.
The code subspace is described by a one-cycle $c_1= \partial f_1=\partial f_2$ surrounding the sphere. 
We define the logical $X$ operator
\begin{eqnarray}
L_X=X(\partial f_1)= X(c_1).
\end{eqnarray}
Note that $L_X=X(\partial f_1)$ and $A_{k}=Z( \partial \bar f_k)$ commute, due to Eq.\ (\ref{eq:commutability}).
The logical Pauli $Z$ operator, which commutes with $\{ A_k\}$ and anticommutes with $L_X$, is defined as $L_Z= Z_l$.
(We may choose any of the $Z_l$s on the edge $e_l$, because they act equivalently on the code space.)
The code is a bit-flip code, which protects quantum information against bit flip errors.

\begin{figure}[t]
\centering
\includegraphics[width=110mm]{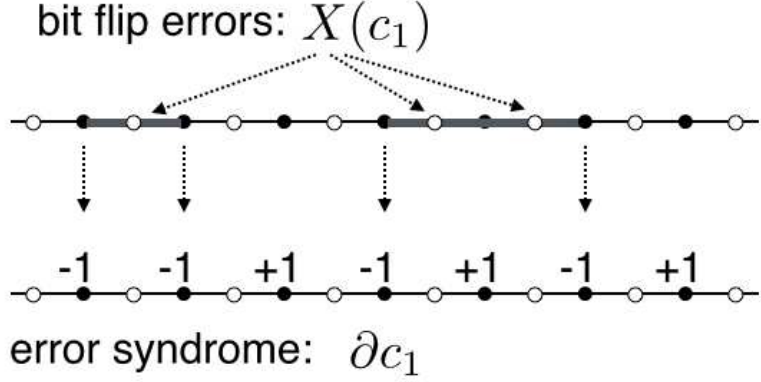}
%
%
\caption{An $X$ error chain $X(c_1)$ and the corresponding error syndrome $\partial c_1$.
The eigenvalue of the stabilizer generator becomes $-1$ on vertices in $\partial c_1$.}
\label{fig55}       
\end{figure}
A string of bit errors $X(c_1)$ is defined using a 1-chain $c_1$, which we call an error chain.
The error is detected by measuring the eigenvalues of the stabilizer generators $A_k=Z(\delta v_k)$, i.e., through syndrome measurements.
From Eq.\ (\ref{eq:comm_rep}), we have $c_1 \cdot \delta v_k = \partial c_1 \cdot v_k$.
Thus, $X(c_1)$ anticommutes with the stabilizer generator $A_k$ if $\partial c_1 \cdot v_k =1$. 
Hence, $X(c_1)$ anticommutes with $A_k$ on the vertex $v_k$ that is a boundary $v_ k \in \partial c_1$ of the error chain $c_1$.
The eigenvalue of the stabilizer generator becomes -1.
Error correction is the task of finding a recovery chain $c'_1$ such that $\partial (c_1 + c'_1)=0$.
In the case of the bit-flip code, there are only two possibilities $c_1+c'_1=0$ or $c_1+c'_1=\partial f_1$.
In the former case, error correction succeeds, while the latter case results in a logical error $L_X$.

The bit-flip code cannot protect quantum information against phase errors. 
A natural extension of the classical repetition code to handle both bit and phase errors is the surface codes introduced in the next section.

\section{Definition of surface codes}
\subsection{Surface code on a torus: toric code}
\begin{figure}[t]
\centering
\includegraphics[width=100mm]{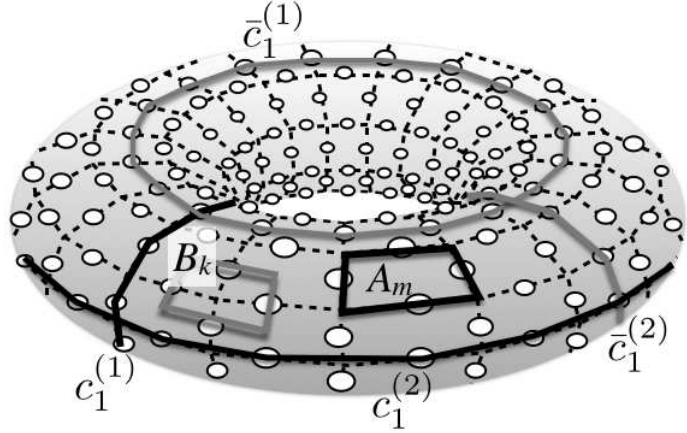}
%
%
\caption{A square lattice on a torus with a periodic boundary condition. 
The $Z$ and $X$ stabilizer generators are defined by being associated with each face and vertex (dual face). 
The logical $Z$ and $X$ operators are defined by non-trivial cycles.}
\label{fig56}       
\end{figure}
Now we are ready to define the surface code.
Let us consider a square lattice $G=(V,E,F)$ on a torus with a periodic boundary condition as shown in Fig.~\ref{fig56}.
A dual square lattice $\bar G=(\bar V, \bar E, \bar F)$ is also defined on the torus.
We define stabilizer generators of the $Z$- and $X$-types for each face $f_m$ and vertex $v_k$ as follows:
\begin{eqnarray}
A_m=Z(\partial f_m), \;\;\;  B_k=X(\delta v_k)=X(\partial \bar f_k).
\end{eqnarray}
Because $\partial f_m \cdot \partial \bar f_k =0$ from Eq.\ (\ref{eq:commutability}), $A_m$ and $B_k$ commute.
The stabilizer generators $A_m$ and $B_k$ are called plaquette and star operators \index{plaquette and star operators}, respectively.
By the definition of the stabilizer code, the code state $|\Psi\rangle$ satisfies 
\begin{eqnarray}
A_m |\Psi \rangle = |\Psi \rangle , \;\;\; B_k |\Psi \rangle = |\Psi \rangle,
\end{eqnarray}
for all $f_m \in F$ and $v_k \in V$.

If two 1-chains, $c_1$ and $c'_1$, are homologically equivalent, the actions of $Z(c_1)$ and $Z(c'_1)$ on the code state are the same, because there exists a 2-chain $c_2$ such that 
\begin{eqnarray}
Z(c'_1)=Z(c_1)Z(\partial c_2)=Z(c_1) \left(\prod _{f_m \in c_2}A_m\right).
\label{eq:homo_eq}
\end{eqnarray}
We will denote this simply by $Z(c'_1) \sim Z(c_1)$.
This is also the case for homologically equivalent dual 1-chains, $\bar c_1$ and $\bar c'_1$, and the actions of $X(\bar c_1)$ and $X(\bar c'_1)$ on the code state.
Hence, the homology classes, which are equivalent classes over trivial cycles, of the primal and dual 1-chains correspond to the actions of $Z$- and $X$-type operators on the code state, because their actions are equal up to stabilizer operators.

Let us define the logical operators $Z(c_1)$ and $X(\bar c_1)$.
The logical operators have to commute with all stabilizer generators and also be independent of them.
The former condition implies that $\partial c_1 =0$ and $\partial \bar c_1=0$ from Eq.\ (\ref{eq:comm_rep}).
This is because the commutability imposes that $c_1 \cdot \delta v_k = \partial c_1 \cdot v_k =0$ for all vertices $k$ and 
$\bar c_1 \cdot \partial f_m = \partial \bar c_1 \cdot \bar v_m=0$ for all faces (dual vertices) $m$.
The latter condition implies that $c_1$ and $\bar c_1$ are nontrivial cycles, because for two homologically equivalent cycles $c_1$ and $c'_1$, the actions of $Z(c_1)$ and $Z(c'_1)$ on the code state are the same
as seen in Eq.(\ref{eq:homo_eq}).
We can find two non-trivial cycles, which belong to different homology classes, for each primal and dual 1-chain, as shown in Fig.~\ref{fig56}.
(This is a natural consequence, because the homology group $h_1$ on the torus is $\mathbf{Z}_2 \times \mathbf{Z}_2$.)
Then, we define two pairs of logical Pauli operators
\begin{eqnarray}
\{ L_Z^{(1)}=Z(c_1^{(1)}), L_X^{(1)}=X(\bar c_1^{(1)})\}
\;\; {\rm and} \;\;
\{ L_Z^{(2)}=Z(c_1^{(2)}), L_X^{(2)}=X(\bar c_1^{(2)})\}.
\end{eqnarray}
Note that $L_Z^{(i)}$ and $L_X^{(j)}$ anticommute with each other if $i=j$.
Otherwise, they commute:
\begin{eqnarray}
L_Z^{(i)}L_X^{(j)} = (-1)^{\delta _{ij}} L_X^{(j)}L_Z^{(i)}.
\end{eqnarray}
That is, they satisfy a commutation relation equivalent to that of the Pauli operators for two qubits.
The logical Pauli basis states are defined as follows:
\begin{eqnarray}
L_Z^{(i)} | \Psi _Z (s_1 ,s_2)\rangle &=& (-1)^{s_i}| \Psi _Z (s_1 ,s_2)\rangle ,
\\
L_X^{(i)} | \Psi _X (s_1 ,s_2)\rangle &=& (-1)^{s_i}| \Psi _X (s_1 ,s_2)\rangle .
\end{eqnarray}
The number of stabilizer generators of the $n \times n$ square lattice on the torus is given by 
\begin{eqnarray}
|F|+|\bar F|-2=|F|+|V|-2=2n^2-2,
\end{eqnarray}
where $-2$ comes from the fact that $\prod_{f_m \in F} A_m=I$ and $\prod_{v_k \in V} B_k=I$, and hence there are two non-independent operators.
On the other hand, the number of qubits is given by 
\begin{eqnarray}
|E| = 2 n^2.
\end{eqnarray}
Thus, we have a $2^{|E|-(|F|+|V|-2)}=2^{2}$-dimensional stabilizer subspace.
The above two pairs of logical operators appropriately describe the degrees of freedom in the code space.
The code distance, the minimum weight of the logical operators, is determined by the linear length of the square lattice.
Thus, the surface code on the torus is an $[[n^2,2,n]]$ stabilizer code.

In short, an operator that commutes with the stabilizer generators corresponds to a cycle, i.e., ${\rm ker}(\partial _1)$.
The stabilizer operators correspond to the boundaries of the 2-chains $c_2 \in C_2$.
The logical operators, which commute with and are independent of the stabilizer generators, correspond to the homology class $h_1 = {\rm ker}(\partial _1)/{\rm Img}(\partial_2)$.
We summarize the correspondence between the surface code and the $\mathbf{Z}_2$ chain complex in Tab.~\ref{tab:ChainStabCode}.
\begin{table}[htdp]
\caption{The correspondence between $\mathbf{Z}_2$ chain complex and stabilizer codes.}
\begin{center}
\begin{tabular}{c|c}
\hline
\hline
stabilizer code & chain complex
\\
\hline
$Z$-type stabilizer generator & face $f_m$
\\
\hline
$X$-type stabilizer generator & vertex $v_k$ (dual face $\bar f_k$)
\\
\hline
$Z$- and $X$-type stabilizer operators & boundaries of 2-chains, 
\\
& ${\rm Img}(\partial _2)$ and  ${\rm Img}(\bar{\partial} _2)$
\\
\hline
commutability & cycle conditions: $\partial c_1 =0$ and $\partial \bar c_1=0$
\\
\hline
the operators commuting with stabilizer generators & ${\rm ker} (\partial _1)$ and  ${\rm ker} (\bar{\partial} _1)$
\\
\hline
the logical $Z$ and $X$ operators & nontrivial cycles $c_1$ and $\bar c_1$
\\
 & (homology classes $[c_1] = {\rm ker}(\partial _1) /{\rm Img}(\partial _2)$ 
 \\
 &
 and $[\bar c_1]={\rm ker}(\bar \partial _1) /{\rm Img}(\bar \partial _2)$)

\\
\hline
$Z$ and $X$ errors & 1-chains $c_1$ and $\bar c_1$
\\
\hline
$Z$ and $X$ error syndromes & $\partial c_1$ and $\partial \bar c_1$
\\
\hline \hline
\end{tabular}
\end{center}
\label{tab:ChainStabCode}
\end{table}%

All properties so far hold for a general tilling $G=(V,E,F)$, and hence we can define a surface code on general, e.g., triangular and hexagonal, lattices~\cite{FujiiTokunaga12}.
Moreover, the numbers of edges, faces, and vertices are subject to the Euler characteristic formula 
\begin{eqnarray}
|F|+|V|-|E| = 2- 2 g,
\end{eqnarray}
where $g$ is the genus, the number of ``handles'' of the surface. 
Thus, the dimension of the stabilizer subspace is calculated to be $2^{|E|-(|F|+|V|-2)}=2^{2g}$.
These are the degrees of freedom equivalent to $2g$ qubits in the stabilizer subspace.

Because almost all arguments are made from the operator viewpoint (Heisenberg picture), there is no need to write down $|\Psi_A(s_1,s_2) \rangle$ explicitly.
However, for the readers who worry about such things, we will give an explicit description from the state viewpoint:
\begin{eqnarray}
|\Psi _Z (s_1,s_2)\rangle &=& \mathcal{N}  Z(c_1^{(1)})^{s_1} Z(c_1^{(2)})^{s_2} \left( \prod _{v_k \in V} \frac{I+B_k}{2} \right) |0\rangle ^{\otimes n^2}
\\
|\Psi _X (s_1,s_2)\rangle &=&\mathcal{N}  X(\bar c_1^{(1)})^{s_1} X(\bar c_1^{(2)})^{s_2}\left( \prod _{f_m \in F} \frac{I+A_m}{2} \right) |+\rangle ^{\otimes n^2}
\end{eqnarray}
where $\mathcal{N}$ is a normalization factor.
The code state can be viewed as an equal weight superposition of all cycles belonging to the same homology class.

\subsection{Planar surface code}
\label{subsec:planar}
\begin{figure}[t]
\centering
\includegraphics[width=90mm]{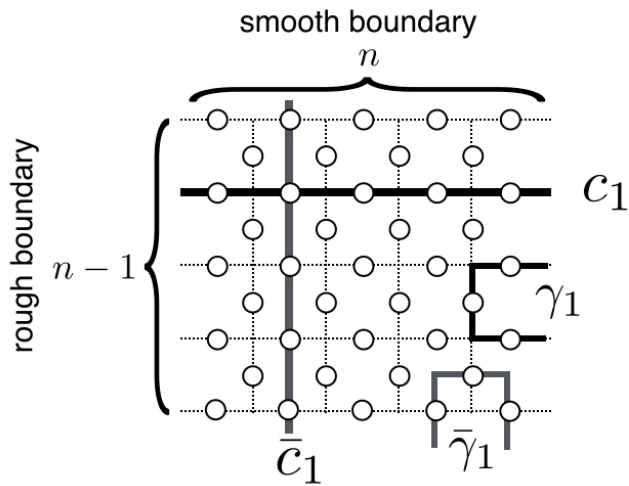}
%
%
\caption{A planar surface code. 
Top and bottom are smooth boundaries consisting of three-qubit star operators. 
Left and right are rough boundaries consisting of three-qubit plaquette operators. 
At the boundary, we can define sets of 1-chains $\Gamma _1 = \{ \gamma _1 \} \in C_1$ and $\bar \Gamma _1 = \{ \bar \gamma _1 \} \in \bar C_1$ for the primal and dual lattices, respectively, by which the relative homology is defined.}
\label{fig58}       
\end{figure}
The periodic boundary condition might be hard to implement in experiments.
We can also define a surface code on a planar $n\times (n-1)$ square lattice with an appropriate boundary condition as shown in Fig.~\ref{fig58}.
The top and bottom boundaries consist of three-qubit star operators, which we call smooth boundaries, because they are complete plaquette operators.
On the other hand, the left and right boundaries consist of three-qubit plaquette operators, which we call rough boundaries.
The $X$ operators on a dual 1-chain can terminate at the smooth boundary, while the $Z$ operators on a 1-chain can terminate at the rough boundary.
Thus, we should use a relative homology to define logical operators, where two chains $c_i$ and $c'_i$ are said to be (relative) homologically equivalent iff 
$c'_i = c_i + \partial c_{i+1} + \gamma_i$ with $\gamma _i \in \Gamma _i \subset C_i$.
In this case, $\Gamma _1$ is chosen, specifically, to be the vector space
spanned by the set of 1-chains each of which consists of three edges corresponding to the three-qubit plaquette operator at the top and bottom smooth boundaries.
Similarly, we also define $\bar \Gamma _1$ at the left and right rough boundaries of the dual lattice.
Because $Z(\gamma _1)$ and $X(\bar \gamma _1)$ are both stabilizer operators, if the shapes of two logical operators are homologically equivalent, their actions on the code state are the same.
The number of edges, faces, and vertices are now $|E|=2n^2 -2n +1$, $|V|=n^2-n$, and $|F|=n^2-n$.
Thus, the stabilizer subspace is a 2D subspace.
We can define the logical operators $L_Z=Z(c_1)$ and $L_X=X(\bar c_1)$ using horizontal and vertical 1-chains $c_1$ and $\bar c_1$, as shown in Fig.~\ref{fig58}.

\begin{figure}[t]
\centering
\includegraphics[width=90mm]{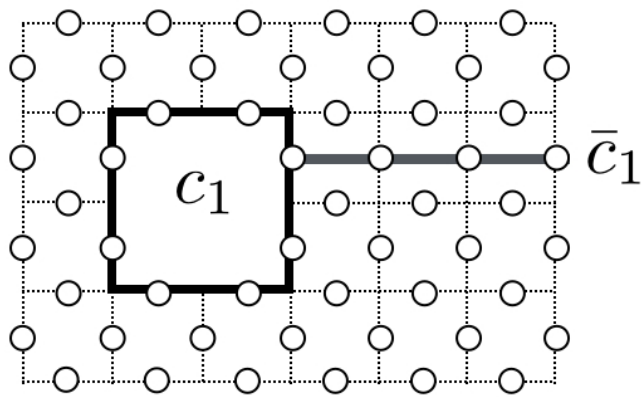}
%
%
\caption{A planar surface code with a defect. 
The stabilizer generators are not defined  inside a defect, which provides a nontrivial cycle wrapping around it.}
\label{fig59}       
\end{figure}
If one chooses all boundaries smooth, the stabilizer state is uniquely defined, and hence there is no logical degree of freedom.
By punching a hole, we can introduce a defect on the surface, whereby we can define a nontrivial cycle for a logical $Z$ operator, as shown in Fig.~\ref{fig59}.
The logical $X$ operator can be chosen as a dual 1-chain that connects the defect and the smooth boundary, as shown in Fig.~\ref{fig59}.
As mentioned previously, the properties of the logical operators are the same if the corresponding 1-chains are homologically equivalent.

\section{Topological quantum error correction}
\label{Sec:TQEC}
\begin{figure}[t]
\centering
\includegraphics[width=110mm]{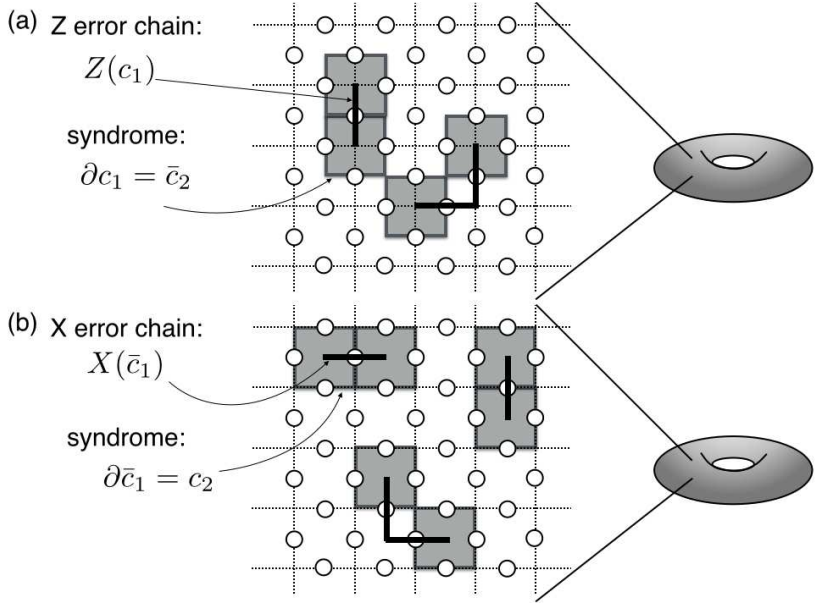}
%
%
\caption{(a) A $Z$ error chain $Z(c_1)$ is detected at the boundary $\partial c_1$.
The eigenvalue of the stabilizer generator $B_k$ on the face $\bar f_k \in \bar c_2 =\partial c_1$ becomes -1. 
(b)  An $X$ error chain $X(\bar c_1)$ is detected at the boundary $\partial \bar c_1$.
The eigenvalue of the stabilizer generator $A_m$ on the face $f_m \in  c_2 =\partial \bar c_1$ becomes -1. }
\label{fig57}       
\end{figure}
Let us return to the surface code on the torus to explain how errors are corrected.
There are $2^{2n^2-2}$ orthogonal subspaces, the so-called syndrome subspaces, each of which is an eigenspace of the stabilizer generators and has the same structure as the code space.
These orthogonal subspaces are utilized to identify the location of errors and to infer a recovery operation.
Suppose the $X$ and $Z$ errors $X(\bar c_1^{e})$ and $Z(c_1^{e})$, defined by error 1-chains $\bar c_1^{e}$ and $c_1^{e}$, respectively, occur on the surface code, as shown in Fig.~\ref{fig57}.
The code state is now mapped into one of the orthogonal subspaces.
From Eq.\ (\ref{eq:comm_rep}), $X(\bar c_1^{e})$ and $Z(c_1^{e})$ anticommute with $A_m$ and $B_k$ on the face $f_m \in \partial \bar c_1^e$ and vertex $v_k \in \partial c_1^e$.
Thus, the orthogonal subspace is specified by error syndromes 
\begin{eqnarray}
\partial \bar c_1^{e} \equiv c_2^{s}, \;\;\; \partial c_1^{e} \equiv c_0^{s}.
\end{eqnarray}
More precisely, the eigenvalues with respect to the stabilizer generators $A_m$ and $B_k$ are given by $(-1)^{z^s_m}$ and $(-1)^{z^s_k}$, where $c_2^{s}=\sum _m {z^s_m} f_m$ and $c_0^{s}=\sum _k z_k^s v_k$ (see Fig.~\ref{fig57}), respectively.
Error correction is the task of finding recovery 1-chains $\bar c_1^r$ and $c_1^{r}$ such that $\partial (\bar c_1^e + \bar c_1^r)=0$ and $\partial (c_1^e +  c_1^r)=0$,
meaning that the state is returned into the code space by applying $X(\bar c_1^r)$ and $Z(c_1^r)$, respectively.
Below we will, for simplicity, explain only how to correct the $Z$ errors, but the extension to the $X$ errors is straightforward.

\begin{figure}[t]
\centering
\includegraphics[width=150mm]{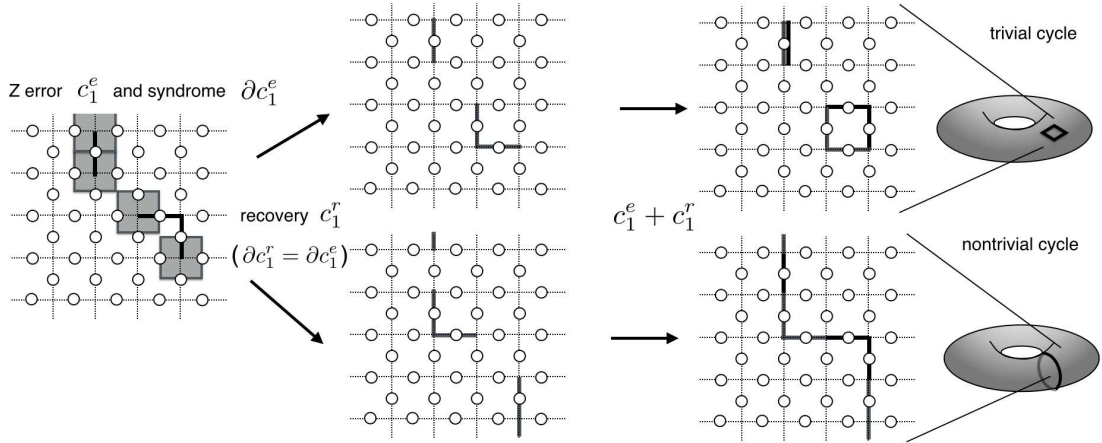}
%
%
\caption{ From left to right, error chain $Z(c_1^e)$, recovery chain $Z(c_1^r)$ with $c_1^r=c_0^s$, and net actions $Z(c_1^r + c_1^e)$.
If $c_1^r + c_1^e$ is a trivial cycle (top), error correction succeeds. 
If $c_1^r + c_1^e$ is a non-trivial cycle (bottom), the error correction results in a logical error. }
\label{fig60}       
\end{figure}

Suppose each $Z$ error occurs with an independent and identical probability $p$.
Conditioned on the error syndrome $c_0^s=\partial c_1^e$, the posterior probability of an error $Z(c_1)$ occurring with $c_1= \sum _l z_l e_l$ is written as
\begin{eqnarray}
P( c_1 | c_0^{s}) &=&\mathcal{N} \prod _{l }  \left( \frac{p}{1-p} \right) ^{z_l} \big |  _{ \partial c_1 =c_0^s }
\label{eq:posterior}
\end{eqnarray}
where $\mathcal{N}$ is a normalization factor.
One way to find a recovery chain is by maximizing the posterior probability 
\begin{eqnarray}
c_1^r \equiv {\rm arg} \max _{c_1} P( c_1 | c_0^{s}) ={\rm arg} \min_{c_1} \left( \sum _{l} z_l \right) |_{\partial c_1 = c_0^s},
\end{eqnarray}
where $c_1=\sum _l z_l e_l$.
As seen in the l.h.s., this corresponds to minimizing the number of errors $\sum _{l}z^r_l$ such that $ \partial c_1^r = c_0^s$.
Hence, this is called minimum distance decoding.
If $c_1^r +c_1^e$ is a trivial cycle, the error correction succeeds, because the net action $Z(c_1^r +c_1^e)$ after the recovery operation is identity on the code state as shown in Fig~\ref{fig60}.
If $c_1^r +c_1^e$ is a nontrivial cycle, the recovery operation results in the logical operation $Z(c_1^r+c_1^e)\sim L_Z$, i.e., a logical error.

\begin{figure}[t]
\centering
\includegraphics[width=130mm]{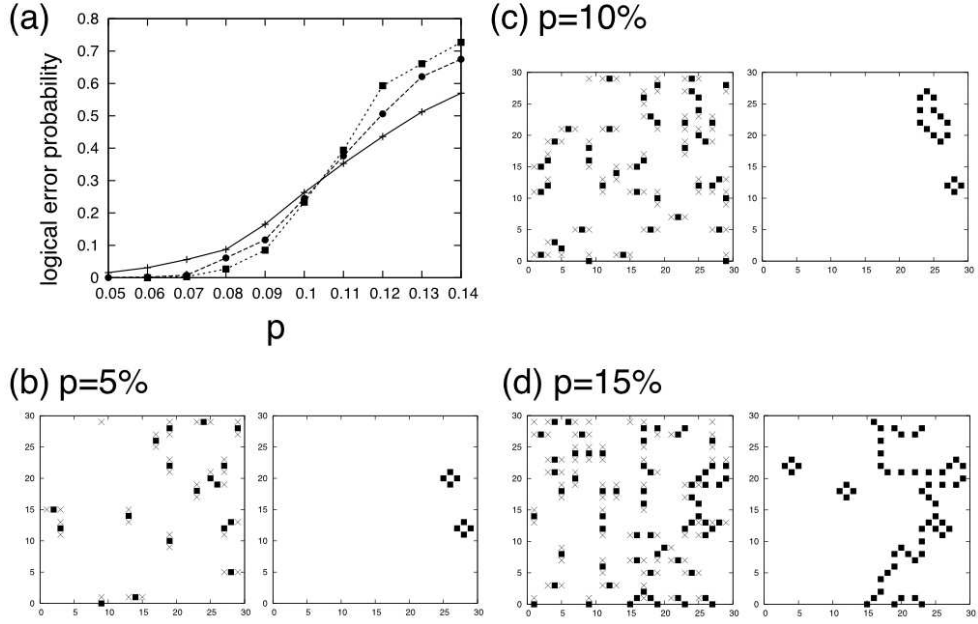}
%
%
\caption{(a) The logical error probability is plotted as a function of the physical error probability $p$. 
Error chains $c_1^e$ and error plus recovery chains $c_1^e+c_1^r$ for $p=0.05$ (b), $p=0.10$ (c), and $p=0.15$ (d).}
\label{fig66}       
\end{figure}

Minimum distance decoding is hard in general, because it can be mapped into an integer programing problem, which is NP-hard~\cite{DecodingNP,DecodingNP2,DecodingSharpP}. 
However, in the present case, there is a nice geometrical property that makes the decoding problem feasible.
The condition $ \partial c_1 = c_0^s$ and $\min _{c_1}$ read 
that it is sufficient to find a 1-chain that connects pairs of two vertices in $c_0^s$ with a minimum Manhattan length.
There is a classical polynomial time algorithm to do such a task, the so-called minimum-weight perfect matching (MWPM)\index{minimum-weight perfect matching, MWPM} algorithm of Edmonds~\cite{Edmonds,Barahona,Cook99}.
The algorithm scales like $O(n^6)$, with $n$ being the linear length of the lattice and with the fixed error probability $p$.
Typical examples of the error chain $c^e_1$ and the error plus recovery chain $c^e_1+e^r_1$ are shown in Fig.~\ref{fig66} for each $p=0.05$ (b), $p=0.10$ (c), and $p=0.15$ (d).
Here, we employ an implementation of MWPM, blossom V~\cite{Kolmogorov}.
The higher the physical error probability is, the longer the error plus recovery chain becomes.
For a high physical error probability $p=0.15$, the error plus recovery chain becomes a large cycle, and unfortunately results in a logical error. 
Such a logical error probability is plotted as a function of the physical error probability $p$ in Fig.~\ref{fig66} (a) for each $n=10$ (solid line), $n=20$ (dashed line), and $n=30$ (dotted line).
If the error probability is sufficiently smaller than a threshold value, the logical error probability decreases for increasing $n$.
The threshold value for decoding by the MWPM algorithm has been estimated to be 10.3\% (MWPM)~\cite{Dennis,Wang}.

\begin{figure}[t]
\centering
\includegraphics[width=100mm]{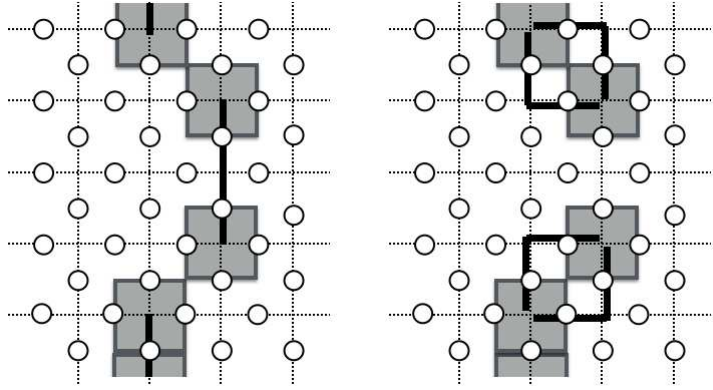}
%
%
\caption{Suppose the top and bottom boundaries are connected by a periodic boundary condition. 
The minimum-weight error is shown in the left panel. 
The error shown in the right panel is not minimum-weight but has a fourfold degeneracy. 
The recovery chain should be chosen by comparing the total error probabilities $p^3$ and $4 \times p^4$.   }
\label{fig61}       
\end{figure}

The minimum distance decoding with MWPM is not optimal for our purpose, i.e., making the logical error probability as small as possible.
An error correction with another recovery chain $c_1^{r'}$, for which $c^e_1+c^{r'}_1$ belongs to the same homology class as $c^e_1+c^r_1$, provides exactly the same result.
We may use such a recovery chain $c_1^{r'}$ to correct the error.
Thus, we should maximize, not each posterior probability $p(c_1^{r}|c_0^s)$, but a summation of it over the same homology class.
This problem originated from the degeneracy of the surface code, where each syndrome is assigned not uniquely, but for many error instances.
A prototypical example of an error syndrome,
for which we should consider not olny the weight of errors,
but also combinatorics (an entropic effect) of the error configurations,
is shown in Fig.~\ref{fig61}.

Denoting the homology class by $h_i$, the posterior probability for a homology class $h_i$ is given by
\begin{eqnarray}
p_i =\sum _{c_1^{r'}| c_1^r+c_1^{r'}\in h_i} P( c_1^{r'} | c_0^{s}),
\label{eq:prob_homo}
\end{eqnarray}
where $c_1^{r}$ is a recovery chain satisfying $\partial c_1^r = c_0^s$ and chosen arbitrarily as a reference frame, and the summation is taken over all 1-chains $c_1^{r'}$ such that $c_1^r + c_1^{r'}$ belongs to the homology class $h_i$.

The posterior probability may be rephrased by using the stabilizer language as follows (see also Appendix~\ref{Ap:Decode}).
Let $\mathcal{G}$ and $\mathcal{L}$ be the stabilizer and logical operator groups, respectively.
For a given error syndrome $c_0^{s}$, we define the recovery operator $Z(c_1^{r})$ a priori, such that the erroneous state is returned into the code space. 
We can decompose an arbitrary error operator $Z(c_1^e)$, providing the syndrome $c_0^{s}$, into 
\begin{eqnarray}
Z(c_1^e) = Z(c_1^{r}) G L_i,
\label{eq:decomposition}
\end{eqnarray}
where $G \in \mathcal{G}$ and $L_i \in \mathcal{L}$ are a stabilizer and logical operator, respectively.
The posterior probability of the logical operator $L_i$ is calculated by summing over all stabilizer operators $G \in \mathcal{G}$
\begin{eqnarray}
p_i = P (L_i| c_0^{s}) = \frac{1}{\mathcal{N}} \sum _{G \in \mathcal{G}} P[Z(c_1^{r}) G L_i ]. 
\end{eqnarray}
We choose the most likely homology class $h_i$ or, equivalently, the most likely logical operator $L_i$ that maximizes the probability $p_i =  P(L_i)$:
\begin{eqnarray}
L_{\bar i} \equiv {\rm arg}\max _{L_i}  P (L_i| c_0^{s}).
\end{eqnarray}
The error correction is completed by applying 
$L_{\bar i} Z(c^{r}_1)$, and the logical error probability 
is given by $1-p_{\bar i}$.

In the next section, we relate the posterior probability summed over the same homology class to a partition function of the random-bond Ising model.
The partition function of the Ising model on a planar graph with general coupling strengths (without magnetic fields) can be calculated in polynomial time using the Kasteleyn-Barahona algorithm with the Pfaffian method~\cite{Kasteleyn,Barahona,BravyiRaussendorf}.
Thus, the optimal decoding is also implemented by a polynomial time classical processing, though it takes more overhead than MWPM.
The threshold value for optimal decoding is also discussed in the next section.

Efficient decoding has been one of the most important issues for the realization of fault-tolerant quantum computing, because the coherence time of quantum information would be very short; a fast classical processing is essential.
Fowler {\it et al.} proposed an efficient decoding method based on MWPM~\cite{FowlerDecode}. 
Because a long error chain is exponentially suppressed, we can assign weights between each pairs of vertices having $-1$ eigenvalues according to their Manhattan length.
This allows us to reduce the number of edges from $O(n^4)$ to $O(n^2)$.
Moreover, the exponential suppression of longer error chains allows us to search the pairs within a local small region; matching a long-distance pair is exponentially rare.
Because the matching process employs almost exclusively local information, this algorithm can be parallelized to an O(1) average time per round. 

\begin{figure}[t]
\centering
\includegraphics[width=150mm]{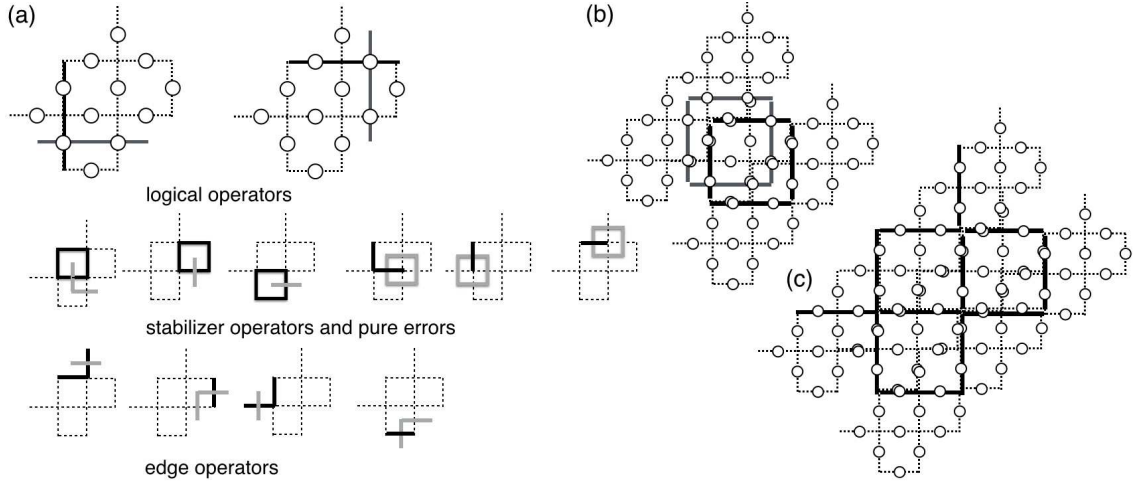}
%
%
\caption{(a) A unit cell of the level-1 logical qubit. 
Two pairs of logical Pauli operators (top). 
The 6 stabilizer operators and the pure error operators,
each of which anticommutes with an stabilizer operator (middle).
The 4 pairs of edge operators.
These 12 pairs of mutually anti-commuting operators 
generate the Pauli group of the 12 qubits.
(b) Using the level-1 logical operators, the level-2 stabilizer generators are defined. 
(c) Using 12 level-1 logical qubits, a unit cell of the level-2 logical qubit is defined.}
\label{fig62}       
\end{figure}
Another decoding method is using a renormalization technique~\cite{DuclosPoulin1,DuclosPoulin2}.
As explained in Appendix~\ref{Ap:Decode}, we can efficiently execute an optimal decoding on a concatenated quantum code by using a brief-propagation on a tree factor graph~\cite{Poulin06}.
Although the surface code itself does not have such a hierarchal structure, a renormalization technique is employed to introduce a hierarchal structure on the surface code~\cite{DuclosPoulin1,DuclosPoulin2}.

Using 12 qubits as a unit cell, we define a couple of level-1 logical qubits, as shown in Fig.~\ref{fig62} (a),
which include
2 pairs of logical operators $\mathcal{L}^{(1)}$, 
6 stabilizer generators $\mathcal{G}^{(1)}$. 
We define 6 pure error operators $\bar{\mathcal{G}}^{(1)}$
each of which anticommutes with a stabilizer generator
as shown in Fig.~\ref{fig62} (a) (middle).
Moreover, we define 4 pairs of anticommuting operators as shown in Fig.~\ref{fig62} (a) (bottom),
which we call edge operators $\mathcal{E}^{(1)}$, to generate the Pauli group of the 12 qubits.
Any Pauli operator $A$ on the unit cell
can be decomposed in terms of these operator,
\begin{eqnarray}
A = L^{(1)} G^{(1)} \bar G^{(1)} E^{(1)},
\end{eqnarray}
where $B^{(1)} \in \mathcal{B}^{(1)}$ for $B=L,G,\bar G, E$.
The pure error operator $\bar G^{(1)}$ is 
chosen uniquely according to the error syndrome $S^{(1)}$
to return the state into the code space.
(If a stabilizer generator $G_1$ has an eigenvalue $-1$,
then we employ $\bar G_1$, which anticommutes with $G_1$.)
From the error distribution $P(A)$, 
the posterior probability of the level-1 logical operator
is calculated by taking a marginal over $\mathcal{G}^{(1)}$ and $\mathcal{E}^{(1)}$,
\begin{eqnarray}
P(L^{(1)} | S^{(1)})=\sum _{G^{(1)} \in \mathcal{G}^{(1)}, E^{(1)} \in \mathcal{E}^{(1)}} P(A=L^{(1)} G^{(1)} \bar G^{(1)} E^{(1)}|S^{(1)}), 
\end{eqnarray}
which are utilized to model the error distribution at the level 2.

Similarly, a level-$k$ unit cell is defined 
by using 8 level-$(k-1)$ unit cells and the 12 pairs of 
the level-$(k-1)$ logical operators on them.
Similarly to the level-1 case,
we define the level-$k$ logical, stabilizer, pure error, edge operators.
At the highest level $k=l$,
we obtain the logical operators of the surface code.
For example, the level-2 stabilizer generators are shown in Fig.~\ref{fig62} (b).
A unit cell of the level-2 logical qubit is shown in Fig.~\ref{fig62} (c).

The posterior probability of the level-$k$
logical operator $P(L^{(k)} | S^{(k)})$ is calculated
using the posterior probabilities at the level $(k-1)$
by assuming that they are independent for each level-$(k-1)$ unit cell.
Under this assumption,
we can calculate the posterior probability $P(L^{(l)}|S^{(l)})$
conditioned on all error syndrome $S^{(l)}$ 
by using the belief propagation~\cite{Poulin06,DuclosPoulin1,DuclosPoulin2} 
(see Appendix ~\ref{Ap:Decode} for 
decoding the concatenated codes by using the belief propagation).
Then the maximization procedure over all logical operators
provide us a most likely logical operator,
\begin{eqnarray}
L^*= {\rm arg}\max _{L^{(l)}} P(L^{(l)} | S^{(l)}).
\end{eqnarray}
The level-$(k-1)$ unit cells share physical qubits with each other, and hence the conditional logical error probability is not independent 
for level-$(k-1)$ unit cells. 
If we employ a message passing in order to reweigh
the level-$(k-1)$ logical error probability as the error model 
for the level-$k$ unit cell,
we can further improve the approximation of $P(L^{(l)} | S^{(l)})$.
If the decoding process is done in parallel, the belief propagation takes $O(\log_2 (n))$ time, similar to the case of the concatenated code as explained in Appendix~\ref{Ap:Decode}.
Because the posterior probability is approximated, the decoding based on the renormalization is not optimal, but results in a reasonable threshold value,
$\sim 9\%$ for the independent $X$ and $Z$ error
and $\sim 15.2\%$ for the depolarizing error
(against $10.3\%$ and $15.5\%$ obtained by MWPM, respectively)
~\cite{DuclosPoulin1,DuclosPoulin2}.
The renormalization decoder can also be applied for an arbitrary topological code~\cite{BombinTopoStab}.

\section{Error correction and spin glass model}
\label{Sec:TQECandSpinGlass}
The behavior of the logical error probability in Fig.~\ref{fig66} suggests the existence of a critical phenomenon behind the error correction problem.
Indeed, there is a beautiful correspondence between quantum error correction on the surface code and a spin glass model, the so-called random-bond Ising model (RBIM)\index{random-bond Ising model, RBIM}~\cite{Dennis}.
More precisely, the posterior probability Eq.\ (\ref{eq:prob_homo}) of the logical operator is mapped into a partition function of the RBIM as seen below.

To solve the condition $\partial c_1 = c_0^s $ in Eq.\ (\ref{eq:posterior}), we rewrite the recovery chain as $c_1 = c_1^{r_k}+c_1^{t}$, where $c_1^{t}\in {\rm Img}(\partial _2)$ is a trivial cycle.
$c_1^{r_k}$ is further decomposed into $c_1^{r_k} = c_1^e+ c_1^{(k)}$, where $c_1^e$ determines the actual location of errors, and $c_1^{(k)}$ is a (nontrivial) cycle belonging to the homology class $h_k$.
Note that this decomposition corresponds to Eq.\ (\ref{eq:decomposition}).
The posterior probability is rewritten as
\begin{eqnarray}
P( c_1  | c_0^{s}) &=&\mathcal{N} \prod _{l }  \left( \frac{p}{1-p} \right) ^{z_l^{t} \oplus z_l^{r_k}} \Bigl|_{c_1^{t} \in {\rm Img}(\partial _2)},
\end{eqnarray}
where $c_1^{\alpha}=z_l^{\alpha} e_l$, with $z_l^{\alpha} \in \{ 0,1\}$. 
In order to take the condition $c_1^{t} \in {\rm Img}(\partial _2)$
automatically, 
we introduce a gauge valuable $z_m^g \in \{ 0,1\}$ on each dual vertex $\bar v_m$
and a dual 0-chain $\bar c_0^{g} = \sum _m z_m^g \bar v^g_m$.
Any trivial cycle $c_1^{t}$ is replaced by the gauge valuables using the following relation
\begin{eqnarray}
z_l^t = \bigoplus _{\bar v_m \in \partial \bar e_l} z_m^g.
\end{eqnarray}
For example, if we choose $\bar c_0^{g} = \bar v_m$, we obtain $c_1^{t}= \partial f_m$.
(This corresponds to a multiplication of the face stabilizer generator defined on the face $f_m$.)
There is a one-to-one correspondence between a trivial cycle and a gauge dual 0-chain.
Using $c_0^g$ we can formally solve the condition 
$c_1^{t} \in {\rm Img}(\partial _2)$ in Eq.\ (\ref{eq:posterior}),
\begin{eqnarray}
P( c_1  | c_0^{s}) &=&\mathcal{N} \prod _{l }  \left( \frac{p}{1-p} \right) ^{ z_l^{r_k}\bigoplus _{\bar v_m \in\partial \bar e_l} z_m^g}.
\label{eq:gauge_trans}
\end{eqnarray}
The binary valuables $z_i\in \{ 0,1\}$ are transformed into spin variables $\sigma _i \in \{ +1,-1\}$ by $\sigma _i= (-1)^{z_i}$.
Moreover, we define a coupling constant $e^{-J} = \sqrt{p/(1-p)}$.
Then Eq.\ (\ref{eq:gauge_trans}) is rewritten as 
\begin{eqnarray}
P( c_1  | c_0^{s}) &=&\mathcal{N}' e ^{J \sum _l \sigma_l^{r_k}\sigma^g _{m(l)}\sigma^g _{m'(l)}},
\end{eqnarray}
where $\mathcal{N}'$ is a normalization factor and $m(l)$ and $m'(l)$ are end points of the edge $e_l$.
By changing the notation $l \rightarrow ij$ and $m(l), m'(l) \rightarrow i,j$, the posterior probability is reformulated as a Boltzmann factor of the $\pm J$ RBIM:
\begin{eqnarray}
P( c_1  | c_0^{s}) &=&\mathcal{N}' e ^{ \sum _{ij} J_{ij}^{(k)}\sigma^g _{i}\sigma^g _{j}},
\end{eqnarray}
where $J_{ij}^{(k)} = J\sigma _{ij}^{r_k}$.
On the edges where the errors are located, anti-ferromagnetic interactions are assigned.
The posterior probability of the logical operator Eq.\ (\ref{eq:prob_homo}) is calculated by taking summation over all gauge spin configurations (this corresponds to the summation over all stabilizer operators in Eq.\ (\ref{eq:prob_homo})):
\begin{eqnarray}
p_{k}&=& \sum _{c_1^{r'}| c_1^{r_k}+c_1^{r'} \in h_k} P( c_1^{r'}  | c_0^{s})
\\
&=& \sum _{c_1^{t} |  c_1^{t} \in {\rm Img}(\partial _2)} P( c_1^{r_k}+ c_1^{t}  | c_0^{s})
\\
&=&
\sum _{\{ \sigma _i^{g}\}} \mathcal{N}' e ^{ \sum _{ij} J_{ij}^{(k)}\sigma^g _{i}\sigma^g _{j}}.
\\
&=& \mathcal{N}' \mathcal{Z}(\{ J_{ij}^{(k)}\}),
\label{eq:logi_prob1}
\end{eqnarray}
where $\mathcal{Z}(\{ J_{ij}^{(k)}\})$ is the partition function of the $\pm J$ RBIM.

Let us consider the performance under this decoding by taking an average of the logarithm of $p_k$ over the error distribution, which corresponds to a sample average with respect to quenched randomness:
\begin{eqnarray}
\bar p^{\rm ln}_k \equiv [\ln (p_k)] &=& \sum _{c_1^{e}} P(c_1^{e}) p_k
\\
&=&
\sum _{\{ J_{ij}\}} P(\{ J_{ij}^{e} \}) \ln \mathcal{Z}(\{ J_{ij}^{(k)}\}) +\ln \mathcal{N}'
\\
&\equiv & -F_k + \ln \mathcal{N}', 
\label{eq:sample_ave}
\end{eqnarray}
where $[\cdot]$ indicates the sample average and $\bar p^{\rm ln}_k$ is the sample average of the logarithm of the logical error probability $p_k$.
The quenched randomness is determined by the error distribution
\begin{eqnarray}
P( J_{ij}^{e} ) =(1-p') \delta ( J_{ ij}^{e} -1 )+ p'\delta ( J_{ i  j}^{e} +1 ).
\end{eqnarray}
Note that $p$ is a parameter in the posterior probability and could be different from the actual error probability $p'$.
Of course, optimal decoding,
in the sense of Bayesian inference, is achieved by maximizing the posterior probability with a true error probability $p=p'$.
$F_0=-[\ln \mathcal{Z}(\{ J_{ij}^{(0)}\})]$ is the free-energy of the RBIM, while $F_k =-[\ln \mathcal{Z}(\{ J_{ij}^{(k)}\})]$ ($k=1,2,3$) is the free-energy with respect to the interactions $\{ J_{ij}^{(k)}\}$, where a domain wall of anti-ferromagnetic interactions corresponding to a cycle $c_1^{(k)}$ of a homology class $h_k$ is inserted.

\begin{figure}[t]
\centering
\includegraphics[width=130mm]{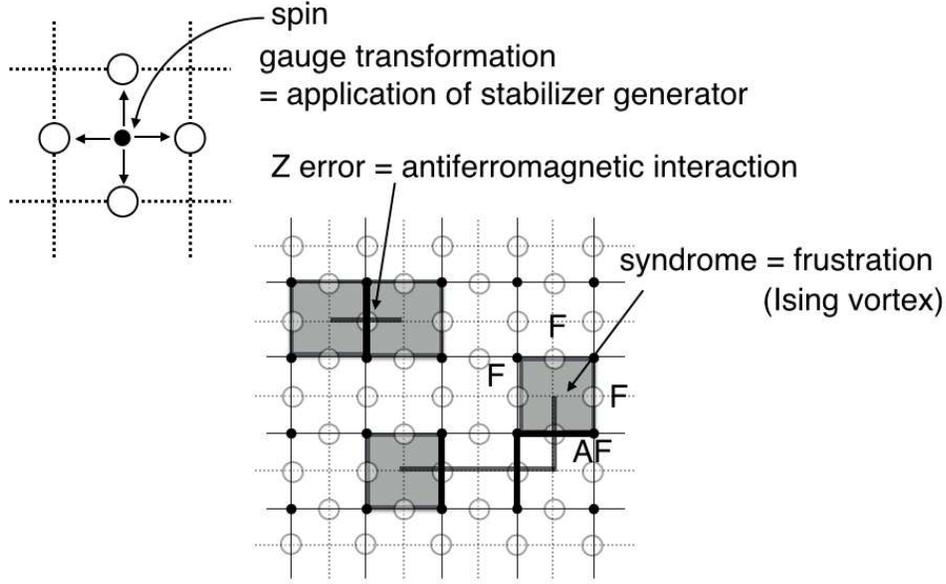}
%
%
\caption{A gauge spin $\sigma ^g_f$ located at the face-center of a plaquette.
The distribution of anti-ferromagnetic interactions corresponds to the error chain $c_1^e$.
The error syndrome $c_0^s=\partial c_1^e$ corresponds to the distribution of frustrations. 
The ground state configuration is determined by a domain-wall consisting of excited domain-wall and anti-ferromagnetic bonds, both of which have frustrations at their end points.}
\label{fig64}       
\end{figure}
From Eqs.\ (\ref{eq:logi_prob1}) and (\ref{eq:sample_ave}), the relation between quantum error correction and a spin glass model becomes apparent; the posterior probability of the logical operator is proportional to the partition function of RBIM, whose Hamiltonian is given by $H= - \sum _{ij} J_{ij}^{(k)} \sigma^g _i \sigma^g _j$.
The location of the $Z$ errors, represented by $J_{ij}^{(k)} = -1$, corresponds to the anti-ferromagnetic interaction due to disorder as shown in Fig.~\ref{fig64}
(see also Tab.~\ref{tab:ChainStabCode2}).
The error probability and the coupling strength (inverse temperature)
are related by $e^{-J} = \sqrt{p/(1-p)}$.
The error syndrome $c_0^s=\partial c_1^e$ corresponds to the distribution of frustrations of the Ising interactions and also the end points (Ising vortex) of the excited domain walls. 
\begin{figure}[t]
\centering
\includegraphics[width=120mm]{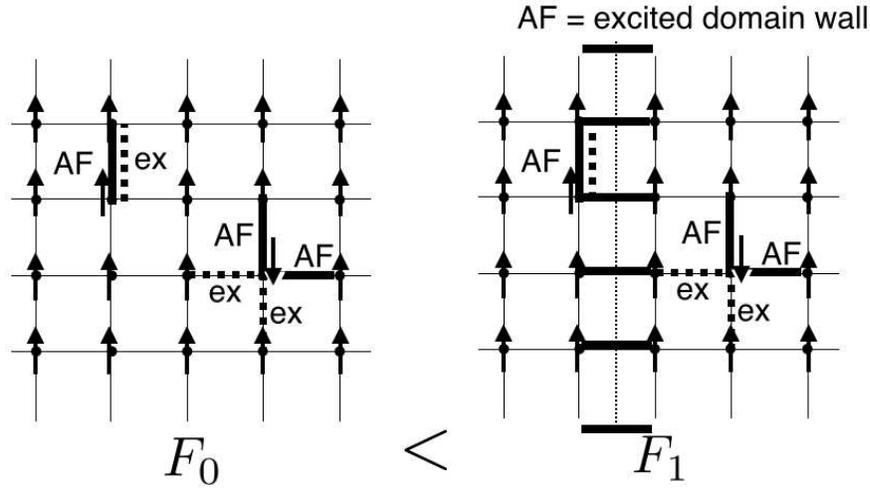}
%
%
\caption{Examples of ground states in a ferromagnetic phase are shown for $\{ J_{ij}^{(k)}\}$ with $k=0,1$.
In the ferromagnetic case, the insertion of the anti-ferromagnetic bonds results in an excited domain-wall and hence increases the free energy (ground state energy at zero temperature).} 
\label{fig63}       
\end{figure}
The probability of the homology class is expressed by the domain-wall free energy:
\begin{eqnarray} 
-(\bar p_k^{\ln} - \bar p_0^{\ln}) = {F_k-F_0}.
\label{eq:exponent}
\end{eqnarray}
If the physical error probability $p$ is smaller than the threshold value, $-\bar p _k^{\ln}$ ($k=1,2,3$) diverges in the large $n$ limit.
On the other hand, if $p$ is higher than the threshold value, $-\bar p _k^{\ln}$ ($k=0,1,2,3$) converges to $2\ln 2$ meaning the stored information becomes completely destroyed.
This non-analytical behavior also appeared in the r.h.s.\ of Eq.\ (\ref{eq:sample_ave}), i.e., the free-energy of RBIM.
Indeed, the difference between the free energies $\Delta =F_k - F_0$ is an order parameter, the so-called domain-wall free energy~\cite{Hukushima99}, for a ferromagnetic ordered phase.
The phase diagram of RBIM with respect to $J=(1/2)\ln [p/(1-p)] $) and $p'$ is shown in Fig.~\ref{fig65}.
\begin{figure}[t]
\centering
\includegraphics[width=90mm]{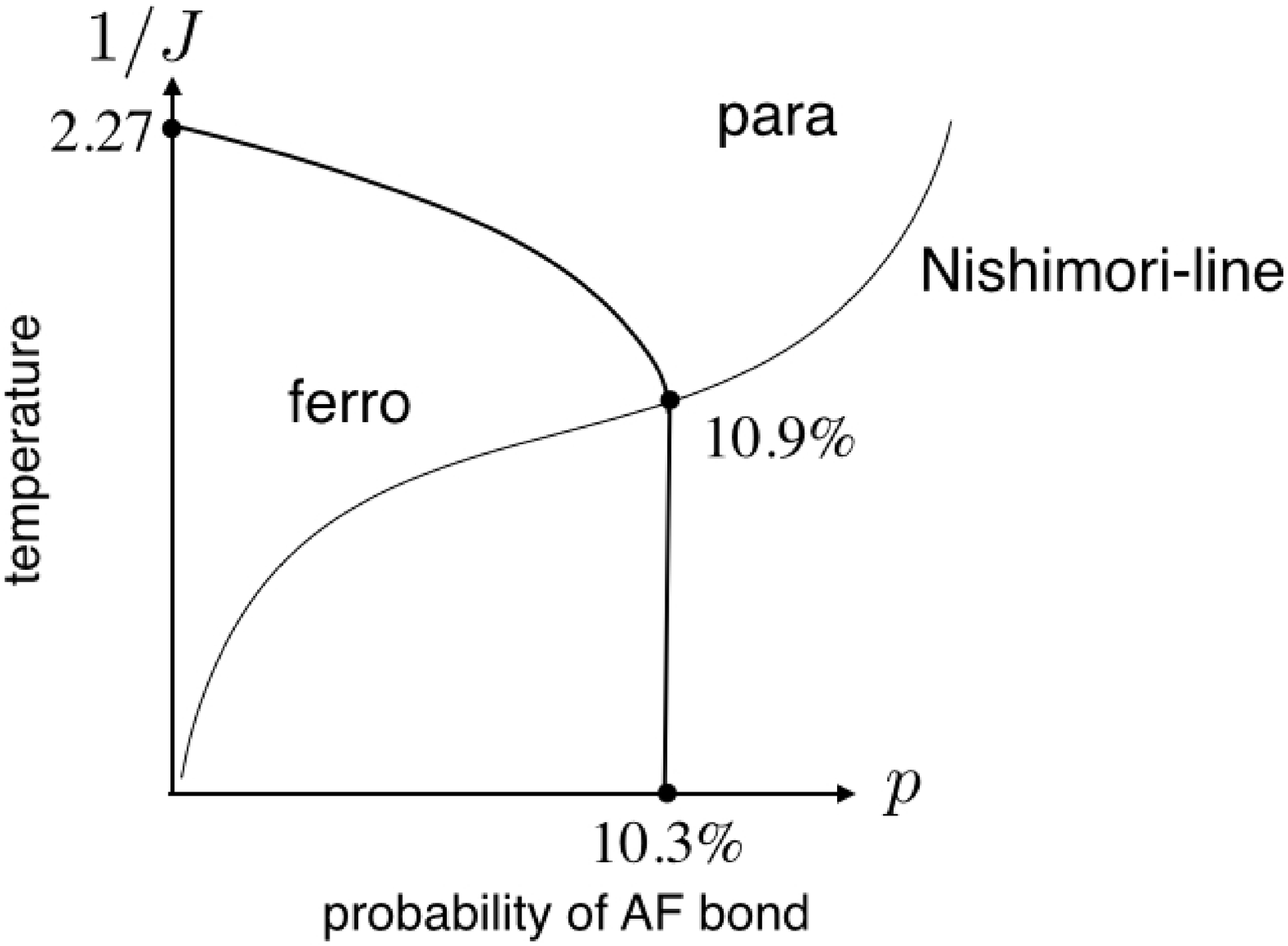}
%
%
\caption{A phase diagram of RBIM with respect to the coupling strength $J=(1/2)\ln [p/(1-p)]$ and the probability of anti-ferromagnetic interaction $p'$.
The probability $p$ in $j$ corresponds to the parameter in the posterior probability. 
When $p=p'$, called the Nishimori line\index{Nishimori line}, an optimal decoding is achieved. }
\label{fig65}       
\end{figure}
As mentioned before, optimal decoding is achieved with $p=p'$, 
i.e., $e^{-J}=\sqrt{p'/(1-p')}$, 
which is called the Nishimori line~\cite{Nishimori81}.
The critical point on the Nishimori line, which is called a multi-critical point, has been numerically calculated to be $10.94\pm 0.02\%$ by Honecker {\it et al.} and 
Merz {\it et al.}~\cite{Reis99,Honecker01,Merz02}.
The optimal threshold value of the surface code is good in the following sense:
The existence of CSS codes with asymptotic rate $R\equiv k/n$ is guaranteed if
\begin{eqnarray}
R=1 - 2 H_2 (p),
\end{eqnarray}
by the quantum Gilbert-Varshamov bound\index{quantum Gilbert-Varshamov bound} under independent $X$ and $Z$ errors with probability $p$~\cite{CalderbankShor}.
The rate becomes zero with $p=11.00\%$.
The optimal threshold of the surface code, which consists only of local stabilizer generators, achieves a value very close to this.
Indeed, Nishimori conjectured that the multi-critical point of the RBIM is determined by 
\begin{eqnarray}
 H_2 (p)= 1/2,
\end{eqnarray}
arguing from the self-duality in RBIM with a replica method~\cite{NishimoriNemoto,NishimoriConj,KramersWannier} (if the reader is interested this derivation, please see a review in Ref.~\cite{OhzekiKindaiBook}).
Ohzeki has evaluated the multi-critical point precisely to be $10.92\%$ by using a real-space renormalization~\cite{Ohzeki09}, which is in a good agreement with the numerical result~\cite{Hasenbusch08}.

The minimum distance decoding with MWPM is achieved in the limit $p \rightarrow 0$, which is the low temperature limit $J \rightarrow \infty$ where the entropic effect is suppressed.
The threshold of MWPM corresponds to the critical point at zero temperature~\cite{Barahona}, which has been investigated numerically and determined to be $10.4\pm 0.1$ by Kawashima {\it et al.}~\cite{Kawashima,Wang,Dennis}.

\begin{table}[htdp]
\caption{The correspondence among 
random-bond Ising model, $\mathbf{Z}_2$ chain complex and stabilizer codes.}
\begin{center}
\begin{tabular}{c|c|c}
\hline
\hline
RBIM & stabilizer code & chain complex 
\\
\hline
dual lattice & $Z$ error correction & primal lattice
\\
primal lattice & $X$ error correction & dual lattice
\\
Ising interactions & qubits & edge $e_l$
\\
gauge spin & $Z$-type stabilizer generators & face $f_m$
\\
gauge spin & $X$-type stabilizer generators & vertex $v_k$ (dual face $\bar f_k$)
\\
domain wall & logical $X$ and $Z$ operators & nontrivial cycles $c_1$ and $\bar c_1$
\\
anti-ferromagnetic interactions & $X$ and $Z$ errors & 1-chains $c_1$ and $\bar c_1$
\\
frustration & $X$ and $Z$ error syndromes & $\partial c_1$ and $\partial \bar c_1$
\\
\hline \hline
\end{tabular}
\end{center}
\label{tab:ChainStabCode2}
\end{table}%

\section{Other topological codes}
\begin{figure}[t]
\centering
\includegraphics[width=150mm]{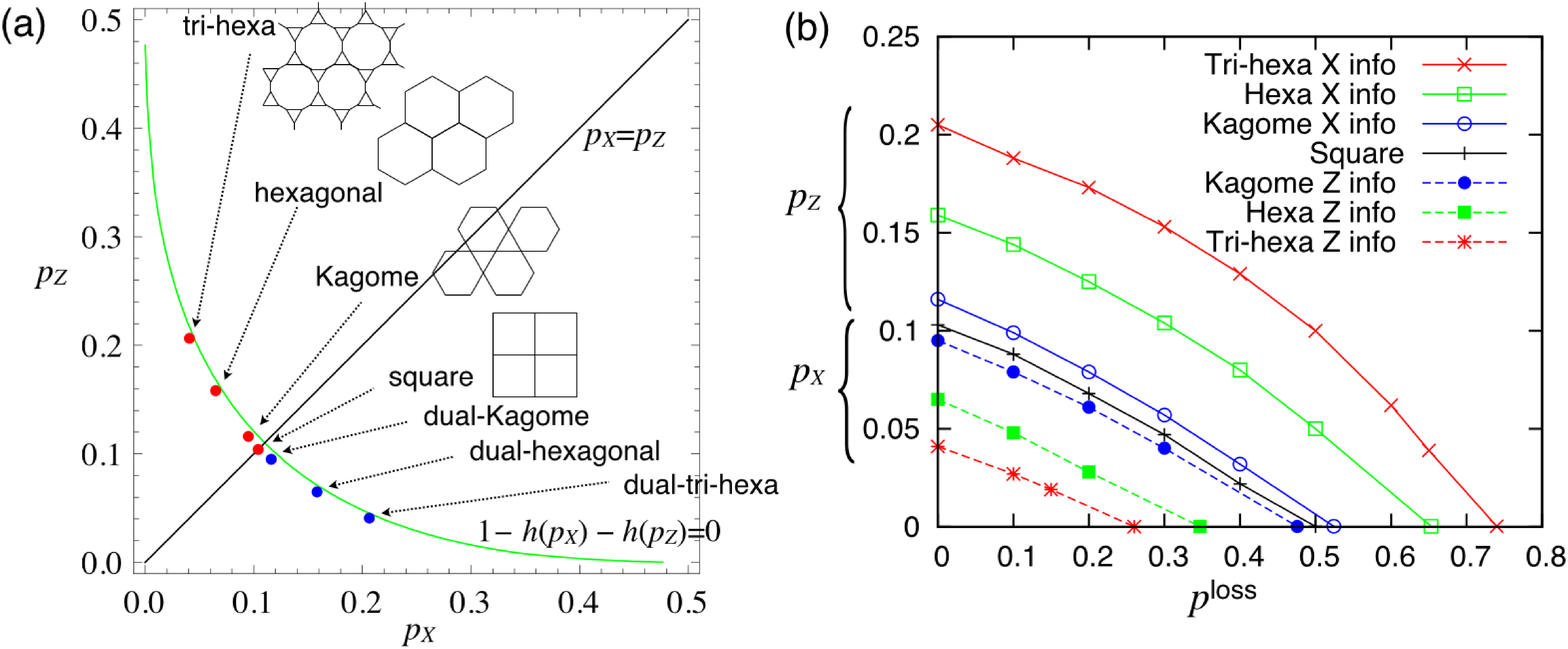}
%
%
\caption{(a) The thresholds $(p_x,p_z)$ of the surface codes on square, Kagome, hexagonal, and triangle-hexagonal lattices, as well as their duals. 
The curve is the quantum Gilbert-Varshamov bound hitting zero asymptotic rate under a bit and phase flip channel with probabilities $p_x$ and $p_z$, respectively.
(b) The trade-off curves between qubit loss rate $p^{\rm loss}$ and unheralded error rates $p_x$ and $p_z$ are shown for square, Kagome, hexagonal, and triangle-hexagonal lattices. 
If the loss and error rates are below these curves, the logical information is protected. }
\label{fig72}       
\end{figure}

The surface codes have been also studied on general lattice tillings, such as, triangle, hexagonal, and random lattices~\cite{Ohzeki09,Roth,FujiiTokunaga12,OhzekiFujii,Al-Shimary,Kay14}.
Suppose a surface code is defined on a lattice $G=(V,E,F)$.
As seen in the previous section, the $X$ error correction is mapped into a RBIM on a lattice $G=(V,E)$.
On the other hand, the $Z$ error correction is mapped into a RBIM on its dual lattice $\bar G=(\bar V, \bar E)$.
The mutual duality relation\index{mutual duality relation}~\cite{NishimoriNemoto} between $G$ and $\bar G$ allows us to predict the relation between the optimal threshold values 
for $X$ and $Z$ errors:
\begin{eqnarray}
H(p_x)+H(p_z)=1,
\label{eq:mutual_duality}
\end{eqnarray}
where $p_x$ and $p_z$ are the $X$ and $Z$ error probabilities, respectively.
This equality is the same as the quantum Gilbert-Varshamov bound evaluated for the independent $X$ and $Z$ errors with probabilities $p_x$ and $p_z$, respectively.
The precise locations of the optimal thresholds for regular lattices have been investigated by Ohzeki using a real-space renormalization technique~\cite{Ohzeki09}.
The thresholds have been investigated using the MWPM algorithm by Fujii {\it et al.}~\cite{FujiiTokunaga12}.
The thresholds with MWPM (i.e., the critical points of RBIMs at zero temperature) even approaches Eq.\ (\ref{eq:mutual_duality}), as shown in Fig.~\ref{fig72}.
These codes with an asymmetry between the $X$ and $Z$ error tolerances would be useful to correct a biased error~\cite{BiasedNoiseNJP}.
In Refs~\cite{Roth,OhzekiFujii}, the asymmetry is continuously controlled by changing a lattice parameter.

\begin{figure}[t]
\centering
\includegraphics[width=120mm]{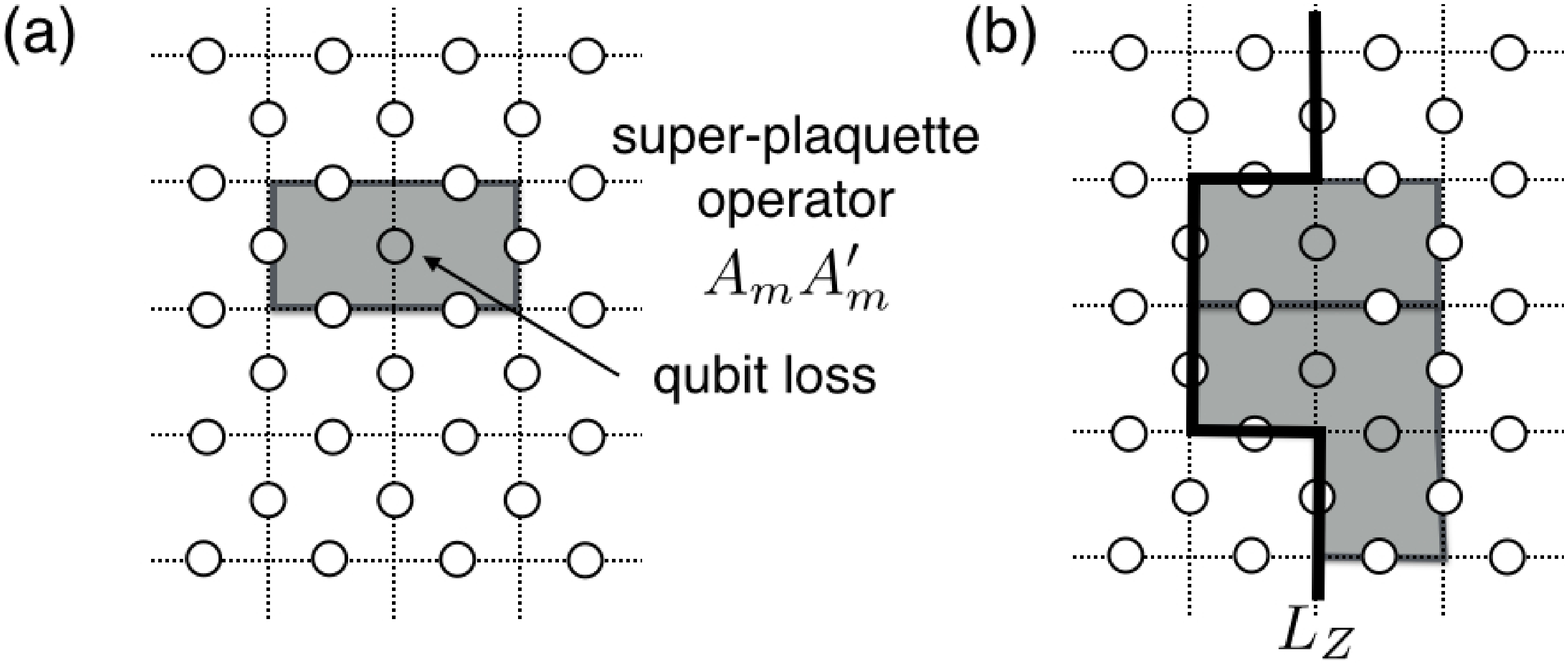}
%
%
\caption{(a) A super-plaquette defined as a product of two plaquette operators. 
The lost qubit is not contained in the super-plaquette operator. 
(b) The logical operator is chosen appropriately by avoiding the lost qubits. 
The lost qubits (bonds) are not percolated, so we can find such a logical operator.}
\label{fig67}       
\end{figure}
A leakage process or qubit loss is an important source of noise.
Unlike the $X$ and $Z$ errors discussed so far, the qubit loss is detectable (heralded), and hence we can tolerate much more loss rate than for the (undetectable) error rate.
Stace {\it et al.} proposed to cope with the qubit loss on the surface code~\cite{StacePRL,StacePRA}. 
Suppose a qubit is lost on the surface as shown in Fig.~\ref{fig67} (a).
The plaquette operators containing the lost qubit are undetermined. 
However, we can construct a {\it super-plaquette} multiplying two neighboring plaquette so that the super-plaquette does not have the lost qubit.
By using the super-plaquette as a stabilizer generator, we can perform MWPM.
If two super-plaquettes are neighbors, the weight between these super-plaquettes has to be modified appropriately in MWPM, because the plaquettes share two physical qubits and the error probability is effectively doubled.
The logical operator is defined by avoiding the lost qubits.
Unless the lost qubits, which are located on edges, are percolated throughout the lattice, we can find such a logical operator.
Thus, the threshold for qubit loss in the large lattice limit without any (undetected) error is determined by the bond percolation threshold.
The trade-off curves between qubit loss and (unheralded) error rates with MWPM for various lattice tillings are shown in Fig.~\ref{fig72} (b).

On the other hand,
when the error correction problem is mapped into the RBIM,
the qubit loss corresponds to a bond-dilution.
The duality relation in the presence of the bond-dilution on the square lattice is given by~\cite{OhzekiLoss}
\begin{eqnarray}
(1-q)h(p)+q=1/2.
\label{eq:duality_loss}
\end{eqnarray}
From the numerical data in Refs.~\cite{FujiiTokunaga12}, we could expect a more general equality,
\begin{eqnarray}
(1-q_x) h(p_x)+(1-q_z)h(p_z)+q_x+q_z=1,
\label{eq:m_duality_loss}
\end{eqnarray}
where $q_x$ and $q_z$ are the probabilities of the heralded $X$ and $Z$ errors, respectively.
With $p_x=p_z=0$, this is reduced to $q_x+q_z=1$, which corresponds to Kesten's duality relation of the bond percolation thresholds between mutually dual lattices.
With $q_x=q_z=0$, Eq.\ (\ref{eq:m_duality_loss}) is reduced to Eq.\ (\ref{eq:mutual_duality}).
With $q_x=q_z=q$ and $p_z=p_x=p$, Eq.\ (\ref{eq:m_duality_loss}) is reduced to Eq.\ (\ref{eq:duality_loss}).

Another important class of local stabilizer codes is the topological color code\index{topological color code} proposed by Bombin and Martin-Delgado~\cite{Bombin06,Bombin07}.
\begin{figure}[t]
\centering
\includegraphics[width=100mm]{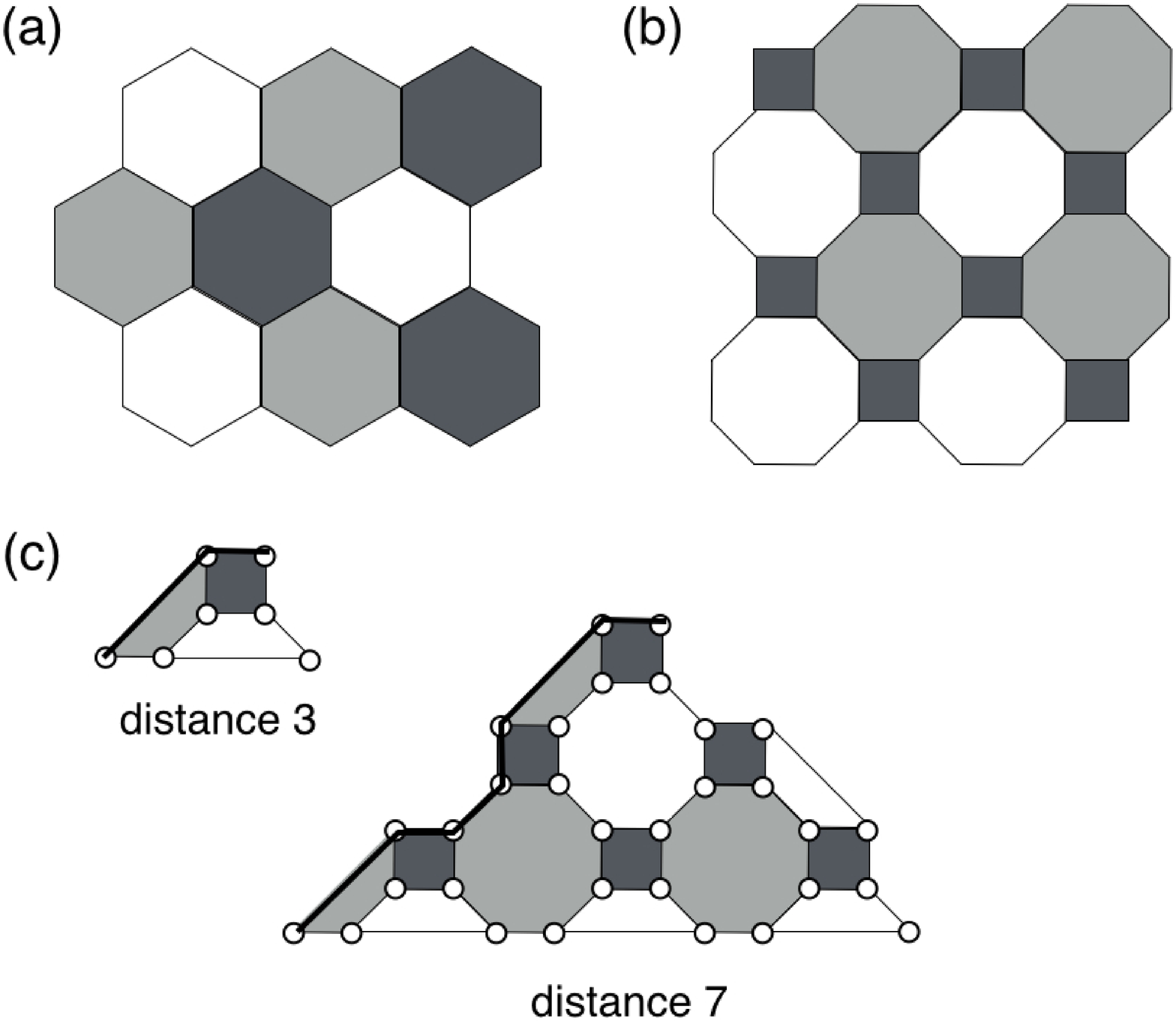}
%
%
\caption{(a) and (b) Trivalent lattices that can be colored with three distinct colors. 
(c) Color codes defined on (4.8.8) lattices with a triangle open boundary condition. 
The logical operators are shown by solid lines. 
Any string at the boundaries can be employed as a logical operator.}
\label{fig69}       
\end{figure}
The topological color codes are defined on trivalent graphs with faces that can be colored with three colors, such as
hexagonal (6.6.6) and (4.8.8) lattices, as shown in Fig.~\ref{fig68} (a) and (b), respectively.
A qubit is defined on each vertex $v$ and stabilizer generators are defined on each face $f$:
\begin{eqnarray}
B^{x}= \prod _{v \in f} X_v , \;\;\;B^{z}= \prod _{v \in f} Z_v.
\label{eq:color_stab}
\end{eqnarray}
Specifically, the color code on a (4.8.8) lattice, shown in Fig.~\ref{fig69} (c), allows all single-qubit Clifford gates transversally.
The distance-3 topological color code on the (4.8.8) lattice is equivalent to Steane's 7-qubit code~\cite{Steane96}.
The extension to a 3D lattice also enables a transversal non-Clifford gate on the code space~\cite{Bombin07}, whose distance-3 version corresponds to the Reed-Mullar 15-qubit code~\cite{Magic}.

The topological color codes are also described by a $Z_2$ chain complex on hyper-graphs~\cite{Delfosse14}.
Consider a trivalent graph $G=(V,E)$, on which a topological color code is defined.
We define a hyper-graph $\mathcal{G}=(\mathcal{V}, \mathcal{E},\mathcal{F})$ consisting of the sets of hyper-vertices $\mathcal{V}$, hyper-edges $\mathcal{E}$, and hyper-faces $\mathcal{F}$ as follows (see Fig.~\ref{fig68}):
\begin{figure}[t]
\centering
\includegraphics[width=120mm]{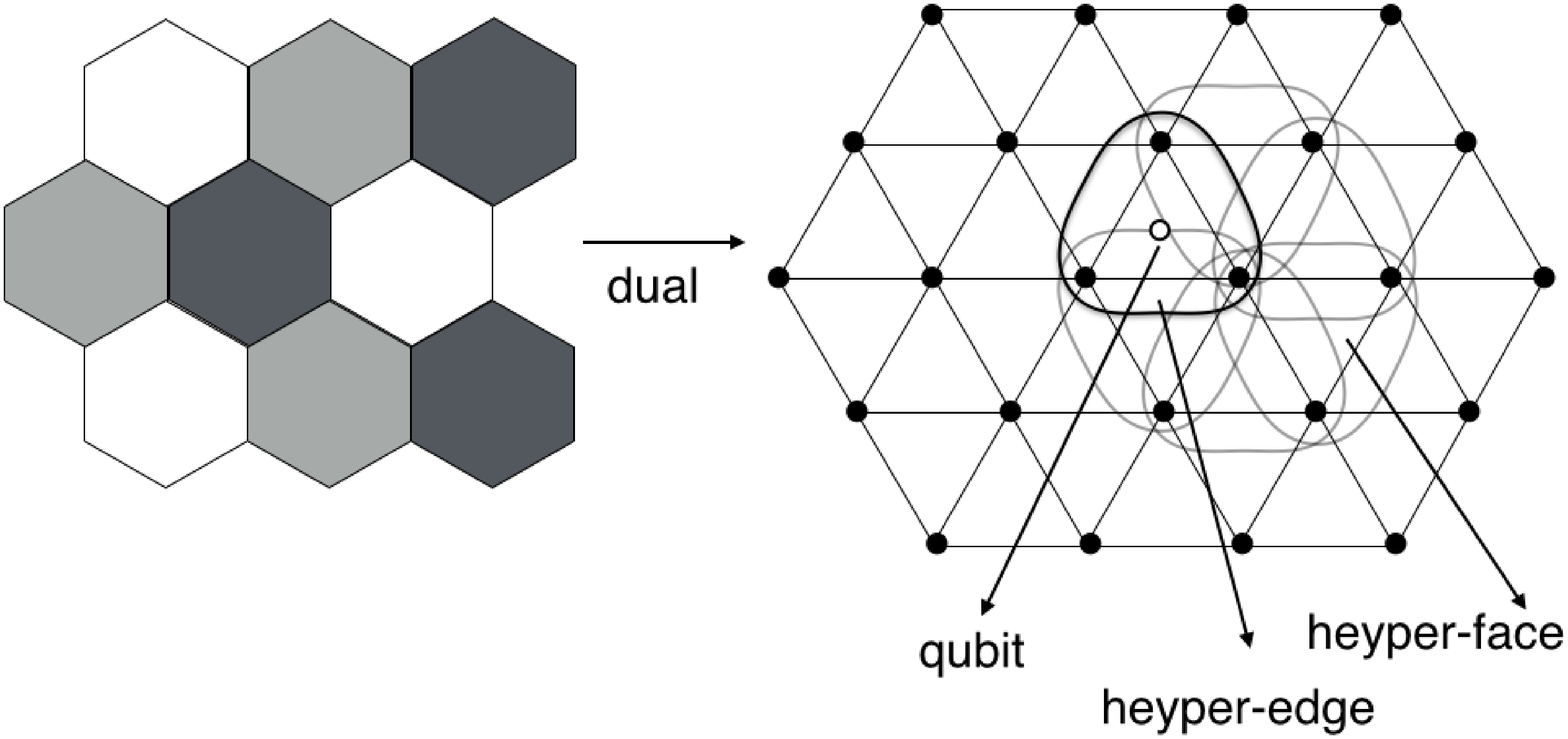}
%
%
\caption{(left) A hexagonal lattice with three-coloring. 
(right) A triangular lattice, the dual of the hexagonal lattice. 
A hyper-graph is defined on a triangular lattice. The color code on the hexagonal lattice corresponds to a surface code defined on the hyper-graph.}
\label{fig68}       
\end{figure}
The hyper-vertices $\mathcal{V}$ are vertices $\bar V$ of the dual graph $\bar G$.
A hyper-edge $ \tilde{e} \in \mathcal{E}$ is defined as a triplet of (hyper-)vertices $\bar v \in \bar V$ on a face $\bar f$ of the dual graph $\bar G$.
A hyper-face $\tilde f \in \mathcal{F}$ is defined as a set of hyper-edges $\tilde{e}$ that are incident to the vertex $\bar v$ of the dual graph $\bar G$.
As defined in Sec.~\ref{sec:chain_complex}, we can define both $\mathbf{Z}_2$ valued vector spaces and Abelian groups $C_{1,2,3}$ with the hyper-graph elements as their bases.
Because the hyper-edge and hyper-face are defined as sets of hyper-vertices and hyper-edges, respectively, we can define boundary maps $\partial _i : C_i \rightarrow C_{i-1}$ 
naturally by such sets.
The plaquette and star stabilizer generators of the surface code defined on such a hyper-graph are as follows:
\begin{eqnarray} 
A_{m} = \prod _{\tilde e \in \partial \tilde f} Z_{\tilde e},
\;\;\;
B_k = \prod _{\tilde e \in \delta \tilde v} X_{\tilde e}.
\end{eqnarray}
This definition is equivalent to the previous definition, Eq.\ (\ref{eq:color_stab}).
In Ref.~\cite{Delfosse14}, the authors defined the projections from the $\mathbf{Z}_2$ chain complex on the hyper-graph into chain complexes on a dual graph $\bar G$.
Specifically, a certain subset of vertices is removed in each projection.
This procedure corresponds to the removal of stabilizer generators colored by one of the three colors.
This allows us to utilize MWPM to decode the topological color codes efficiently on the projected surface codes~\cite{Delfosse14}.

The decoding problem of topological color codes is mapped into random three-body Ising models using the mapping between quantum error correction and spin glass models. 
This can be understood as follows. 
For each face center a gauge spin valuable 
is located associated with each stabilizer generator;
three gauge spin valuables (stabilizer generators) 
that share the same qubit (vertex)
interact with each other.
The locations of the optimal thresholds have also been investigated via the spin glass theory and Monte Carlo simulations~\cite{OhzekiColor,BombinOhzeki}.

\section{Connection to topological order in condensed matter physics}
Topological order\index{topological order, topological ordered system} is an exotic quantum phase of matter, which cannot be characterized by the Landau-Ginzburg theory of symmetry breaking, in which an ordered phase can be characterized by local order parameters.
The ground state of a topologically ordered system is degenerate, but its degeneracy cannot be destroyed by any local perturbation. 
Thus, no local order parameter can succeed to capture the topologically ordered phase.
Understanding the nature of topological order is one of the most important goals of modern condensed matter physics.
Moreover, the ground state of a topologically ordered system is, by definition, robust against any local perturbations, and hence it is also useful for storing quantum information.
There is a beautiful correspondence between topological quantum codes and topologically ordered systems, which provides a promising way to understand condensed matter physics
via quantum information.

Let us first consider the bit-flip code defined in Sec.~\ref{sec:bit-flip}.
The Hamiltonian, which we call a stabilizer Hamiltonian\index{stabilizer Hamiltonian}, is defined as follows:
\begin{eqnarray}
H_{\rm Ising} = - J \sum _k A_k = -J \sum _{i=1}^{n-1} Z_iZ_{i+1}.
\label{eq:Hami_Ising}
\end{eqnarray}
Here, the summation is taken over all stabilizer generators, but one stabilizer operator, which is not independent, is removed.
The Hamiltonian Eq.\ (\ref{eq:Hami_Ising}) corresponds to the Ising model in 1D with an open boundary condition.
By its construction, the Hamiltonian is diagonalizable, and the stabilizer state becomes the ground state.
The bit-flip code has a 2D stabilizer subspace spanned by $\{ |00...0\rangle, |11...1\rangle\}$.
This means that the ground state is degenerate.
The bit-flip errors occurring on the code space excite the ground state to an excited state.
Thus, the states in the orthogonal subspaces of the bit-flip code correspond to excited states.

To address topological order, let us consider the robustness of the ground state against perturbations from transversal fields $h_x\sum _i X_i$.
By using standard perturbation theory~\cite{Kato}, we can show that the degeneracy of the ground states is not lifted up to the $(n-1)$th order of perturbation.
This happens because the code distance of the $n$-qubit bit flip code against $X$ errors is $n$, and any $X$ errors of weight up to $n-1$ map the code state into an orthogonal space.
Accordingly, the ground state degeneracy is robust against the transversal fields.

On the other hand, longitudinal fields $h_z \sum _{i}Z_i$ spoil the ground state degeneracy in the large $n$ limit, even if $h_z$ is small. 
More precisely, the energy between $|00..0\rangle$ and $|11..1\rangle$ is shifted by $n h_z$.
Thus, a superposition $\alpha |00..0\rangle + \beta |11..1\rangle$ in the ground subspace is easily destroyed by the longitudinal fields.
In this sense, the stabilizer Hamiltonian constructed by the bit-flip code is not topologically ordered.
However, if any perturbation with respect to the longitudinal fields is prohibited due to some symmetry of nature, the ground state degeneracy is robust under that symmetry.
This type of robustness of ground-state degeneracies is called {\it symmetry protected topological order}\index{symmetry protected topological order}~\cite{SymmetGuWen,SymmetPollMann}.
(Note that in this case the ground-state degeneracy is 
not related to the geometrical property of the underlying manifold
in contrast to the genuine topological order in 2D.)

The symmetry prohibiting the longitudinal fields seems to be somewhat artificial.
We can, however, impose the symmetry by transforming the Ising Hamiltonian to a free-fermion model using the following Jordan-Wigner transformation\index{Jordan-Wigner transformation}~\cite{JWtrans}:
\begin{eqnarray}
a_{2i-1} =\prod _{k=1}^{i-1} X_k Z_i,
\\
a_{2i} = \prod _{k=1}^{i-1} X_k Y_i.
\end{eqnarray}
The operators, called Majorana fermion\index{Majorana fermion} operators, are hermitian $a_k= a_k^{\dag}$ and satisfy the fermion commutation relation 
$\{ a_{k}, a_{k'} \} = \delta _{k,k'}I$.
The Ising stabilizer Hamiltonian is reformulated in terms of $a_k$:
\begin{eqnarray}
H_{\rm Isng}= -J \sum_{i=1}^{n-1} (-i) a_{2i} a_{2i+1}.
\end{eqnarray}
The logical operators acting on the degenerated ground states are given by
\begin{eqnarray}
L_Z=a_1=Z_1, \;\; L_Y=a_{1} a_{2n}= Y_1 \left(\prod _{k=2}^{n-1}X_k \right) Y_n.
\end{eqnarray}
The degree of freedom in the degenerated ground states is called the Majorana zero mode or the unpaired Majorana fermion~\cite{KitaevMajorana}.
If the parity of the number of fermions is preserved, the fermion operators would appear with a quadratic form $a_k a_k'$.
Under such a symmetry, there is no perturbation that lifts the ground state degeneracy. 
Thus, the ground state of the unpaired Majorana fermion is symmetry protected.

\begin{figure}[t]
\centering
\includegraphics[width=110mm]{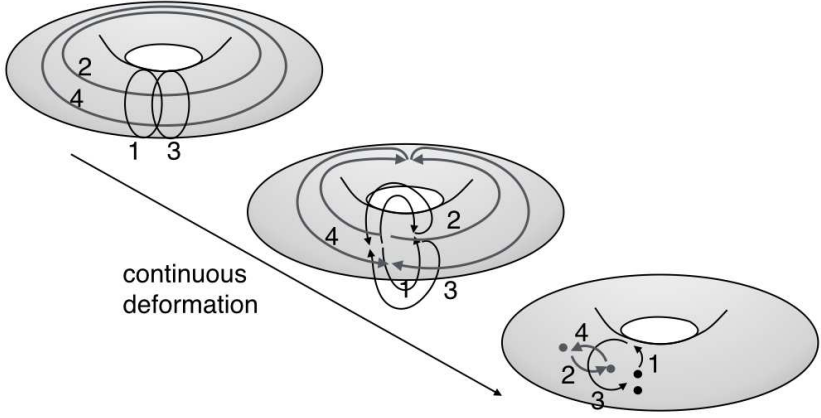}
%
%
\caption{(left top) (1) A pair of $X$-type excitations are created, moved around the torus, and annihilated. 
(2) A pair of $Z$-type excitations are created, moved around the torus, and annihilated. 
(3) Do the process (1) again. 
(4) Do the process (2) again. 
(middle) The creation, movement, and annihilation process is continuously deformed. 
(right bottom) A braiding operation of an $X$-type excitation around a $Z$-type excitation.}
\label{fig70}       
\end{figure}
Next, we will provide a genuine topologically ordered system based on the surface code (Kitaev's toric code).
The stabilizer Hamiltonian, the so-called Kitaev's toric code Hamiltonian \index{Kitaev's toric code Hamiltonian}~\cite{Kitaev97}, is given as a summation of all plaquette and star operators:
\begin{eqnarray}
H_{\rm Kitaev} = -J \sum _{m}A_m - J \sum _{k} B_k.
\end{eqnarray}
The ground state has a fourfold degeneracy corresponding to the code space.
Errors on the code state correspond to excitations.
Specifically, there are two types of excitations, corresponding to the $Z$ error $Z( c_1)$ and $X$ error $X(c_1)$.
Excitations appear at the boundaries of the error chains $\partial c_1$ and $\partial \bar c_1$, because the local energy changed from $-J$ to $+J$ there.
Such excitations, i.e., at the end points of the error chains, are always created as pairs, can be viewed as a pair creation process on the ground state.

Suppose these two types of the excitations were created on the system and moved as shown in Fig.~\ref{fig70} (left top).
This process can be described by
\begin{eqnarray}
Z(c_1^{(2)})X(\bar c_1^{(2)})Z(c_1^{(1)})X(\bar c_1^{(1)}) |\Psi \rangle =- |\Psi \rangle.
\label{eq:topo_braid}
\end{eqnarray}
On the other hand, by continuously changing the trajectory of the particles, as shown in Fig.~\ref{fig70}, this process can also be regarded as a braiding process of an $X$-type excitation around a $Z$-type excitation.
After the braiding operations, a phase factor is applied to the state
as shown in the r.h.s. of Eq. (\ref{eq:topo_braid}).
Thus, the excitations are neither bosonic nor fermionic, which are invariant under the braiding operation, i.e., the swapping operation twice.
In this sense, the excitations on the surface code are referred to as anyons.
Specifically, since the phase factors is $\mathbf{Z}_2$, 
they are called $\mathbf{Z}_2$ Abelian anyons\index{Abelian anyon}.
By using the generalized Pauli operators
on for the qudit,
we can also define $\mathcal{Z}_d$ Kitaev's toric code~\cite{Kitaev},
on which excitations are $\mathcal{Z}_d$ Abelian anyons.
More generally, using a finite group $G$,
we can define a quantum state $|g\rangle$ ($g \in G$)
in a $|G|$-dimensional Hilbert space.
Then we define four types of operators for each $g \in G$:
\begin{eqnarray}
L^g_+ = \sum _{h \in G} |gh \rangle \langle h|,
L^g_- = \sum _{h \in G} |hg^{-1} \rangle \langle h|,
T_+^h= |h\rangle \langle h|,
T_-^h= |h^{-1}\rangle \langle h^{-1}|.
\end{eqnarray}
The non-Abelian Kitaev's toric code model,
which is called the quantum double model~\cite{Kitaev},
is defined as
\begin{eqnarray}
H=-J\sum _{m} A(f_m) -J \sum _{k} B(v_k),
\end{eqnarray}
in terms of
the following plaquette and star operators
\begin{eqnarray}
A(f_m) &=& \sum _{g_1 g_2 g_3 g_4 = I} T_-^{g_1}(e^m_{l_1})T_-^{g_2}(e^m_{l_2})
T_+^{g_3}(e^m_{l_3})T_+^{g_4}(e^m_{l_4})
\\
B(v_k) &=& \frac{1}{|G|}\sum _{g \in G} L_+^{g} (\bar e^k_{l_1})
L_+^{g}(\bar e^k_{l_2}) L_-^{g} (\bar e^k_{l_3}) L_-^{g} (\bar e^k_{l_4})
\end{eqnarray}
where the four edges $e^m_{l_{1,2,3,4}} \in \partial f_m$ 
and $\bar e^k_{l_{1,2,3,4}} \in 
\delta v_k = \partial \bar f _k$
are labeled clock wise.
The quantum double model
supports non-Abelian anyonic excitations~\cite{Kitaev,Pachos2012},
which allows us to implement universal quantum computation 
solely by braiding them.

Let us return to the $\mathbb{Z}_2$ Kitaev's toric code Hamiltonian.
The code distance of the surface code on the $n\times n$ torus is $n$.
Thus, neither $X$-type (transverse) nor $Z$-type (longitudinal) fields can lift the ground state degeneracy up to the $(n-1)$th order of perturbation.
More generally, no local perturbation can lift the ground state degeneracy in the large $n$ limit.
Thus, the ground state of the Kitaev's toric code Hamiltonian is topologically ordered.
For any properly defined stabilizer code, we can define a stabilizer Hamiltonian, whose ground state is topologically ordered.
However, in condensed matter physics, the local interactions are of central importance.
Thus, it is natural to restrict the stabilizer generators to be spatially local, i.e., topological stabilizer codes.

\begin{figure}[t]
\centering
\includegraphics[width=120mm]{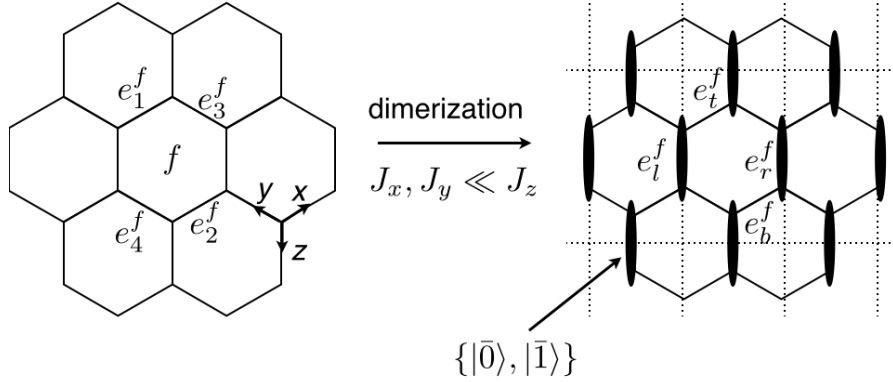}
%
%
\caption{A hexagonal lattice on which Kitaev's compass model is defined.
In the large $J_z$ limit, the two spins on each vertical edge are confined into a 2D subspace forming a dimmer.
The dimmer is located on each edge of a square lattice shown by the dotted lines.}
\label{fig71}       
\end{figure}

Only the nearest-neighbor two-body interactions are attained in physically natural systems.
It has been known that the Kitaev's toric code Hamiltonian 
is obtained as an effective low energy model of 
another model consisting only of two-body nearest-neighbor interactions~\cite{Kitaev06}.
Let us consider the following two-body nearest-neighbor model, called Kitaev's compass model\index{Kitaev's compass model}:
\begin{eqnarray}
H_{\rm comp} = - J_x \sum _{(i,j) \in E_x} X_i X_j  - J_y \sum _{(i,j) \in E_y} Y_i Y_j - J_z \sum _{(i,j) \in E_Z} Z_i Z_j ,
\end{eqnarray}
where $E_x$, $E_y$, and $E_z$ are sets of right-up, left-up, and vertical bonds, respectively, of a hexagonal lattice (see Fig.~\ref{fig71} (left)).
If we take the large $J_z$ limit, each vertical bond favors the two-dimensional subspace spanned by $\{ |00\rangle , |11\rangle\}$, because it is stabilized by $Z_iZ_j$.
Thus, in the large $J_z$ limit, we can derive an effective low energy Hamiltonian, which commutes with the $Z_i Z_j$ interactions, by using perturbation theory:
\begin{eqnarray}
H_{\rm eff} = - \frac{J_x^2 J_y^2}{16|J_z|^3} \sum _{ f} \tilde X_{e_1^{f}} \tilde X_{e_2^f} \tilde Y_{e_3^f} \tilde Y_{e_4^f} - J_Z \sum _{e \in E_Z} \tilde Z_{e},
\end{eqnarray}
where $\tilde A_e = A_i A_j$ with $e=(i,j)$ and $A=X,Y,Z$, depending on $e \in E_x, E_y , E_z$.
The edges $\{ e_i^f\}$ are left-up and right-up edges on a hexagonal face $f$ and the summation $\sum_f$ is taken over all faces.

Let us define a qubit $\{ |\bar 0 \rangle =|00\rangle , |\bar 1\rangle =|11\rangle \}$ for each dimerized edge and Pauli operators $\bar X= XX$ or $=\bar YY$ and 
$\bar Z= Z \otimes I$ or $=I \otimes Z$.
Now a qubit is assigned to each edge of a square lattice, which is an vertical edge in the hexagonal lattice as shown in Fig.~\ref{fig71} (right).
Using this definition, the effective Hamiltonian can be reformulated as
\begin{eqnarray}
\bar H_{\rm eff} = - \frac{J_x^2 J_y^2}{16|J_z|^3} \sum _{f} \bar X _{e_l^f} \bar X_{e_r^f} \bar Z_{e_t^f} \bar Z_{e_b^f},
\end{eqnarray}
where $\{ e_l^f, e_r^f, e_t^f, e_b^f\}$ are the left, right, top, and bottom edges, respectively, on a square face $f$.
If we apply the Hadamard transformation on all horizontal edges, we obtain the Kitaev's toric code Hamiltonian.
In this way, the stabilizer Hamiltonian can be obtained as a low-energy effective model of a two-body nearest-neighbor system~\cite{Kitaev06}.
This shows the validity of employing topological stabilizer codes and quantum coding theory to understand the quantum phase of matter in condensed matter physics.

A complete classification of the topological stabilizer codes in 2D has been obtained by Yoshida~\cite{BeniTopoStab} (see also the classification by 
Bombin {\it et al.}\cite{BombinTopoStab}).
Specifically, the quantum phases of the 2D stabilizer Hamiltonians are classified by the geometric shapes of the logical operators.
The thermal stability of the topological order in stabilizer Hamiltonian systems at finite temperatures has also been studied via quantum coding theory by Bravyi and Terhal~\cite{BravyiTerhal} for 2D, and Yoshida for 3D~\cite{BeniThermal3D}.

If there is a thermally stable topological order, we can store quantum information reliably even at a finite temperature without any active error correction, i.e., we have a  self-correcting quantum memory.
Of course, if a fault-tolerant quantum computer were realized, we could store quantum information reliably with a repetitively performing quantum error correction, which, however, requires selective addressing of each individual qubits.
There are also several intermediate approaches for a reliable quantum storage without selective addressing using global dissipative dynamics~\cite{MFTP}, an interaction with an engineered environment~\cite{ToricBoson,ToricBoson2,Kapit14}, and decoding by cellular automata with local update rules~\cite{AutomatonDec14,Harrington}.
However, a genuine topologically ordered self-correcting quantum memory
seems to be hard to achieve in 2D
even in the presence of effective long-range interactions
~\cite{Landon-Cardinal15}.

\chapter{Topological quantum computation}
\label{Chap:TQC}
In this chapter, we explain how to perform topological fault-tolerant quantum computation on the surface code.
All operations employed are nearest-neighbor (at most) two-qubit gates and signle-qubit measurements on a 2D array of qubits.
This property is quite favorable for the fabrication of qubits and their control lines on a chip.
Fault-tolerance ensures that no local noise during any sort of operations ever spoil the quantum computation, if the noise strength is smaller than a certain threshold value, as seen below.

We first introduce the defect pair qubit, a logical qubit using a pair of defects on the surface, which allows us to encode many logical qubits on the surface.
Then, we explain the elementary operations of defects by local quantum information processing.
A braiding operation, based on the elementary operations, of the defects is further utilized to implement the logical CNOT gate between the defect pair qubits.
Combining with the topologically protected operations, the magic states are injected 
and distilled for fault-tolerant universal quantum computation.
Based on these understandings, we introduce a topological diagram and the topological calculus on it, as a set of the transformations that preserve the logical actions on the code space.
These provide us with a deep understanding of topological quantum computation on the surface. 
Finally, we return to the microscopic viewpoint to explain how topological quantum error correction is executed and analyzed.

\section{Defect pair qubits}
\label{Sec:DPQ}
In order to perform quantum computation of $N$ qubits, we have to arrange $N$ logical qubits in the code subspace while keeping their code distance long enough.
A way of doing this is to use a surface of higher genus. 
Specifically, we use a large planar surface and punch holes on it, which we call defects on the surface.

In Sec.~\ref{subsec:planar}, we have seen that we can define a logical qubit by introducing a defect on a planar surface, where the stabilizer generators inside the defect region 
are removed from the stabilizer group.
The logical operators are defined by a nontrivial cycle surrounding the defect and a chain connecting the defect to the boundary.
If we introduced many degrees of freedom, i.e., many defects
based on this strategy, the logical operators defined by chains connecting each defect to the boundary would become complicated.
To avoid this, we use a pair of defects to define a logical qubit, which we call a {\it defect pair qubit}\index{defect pair qubit}.

\begin{figure}[t]
\centering
\includegraphics[width=120mm]{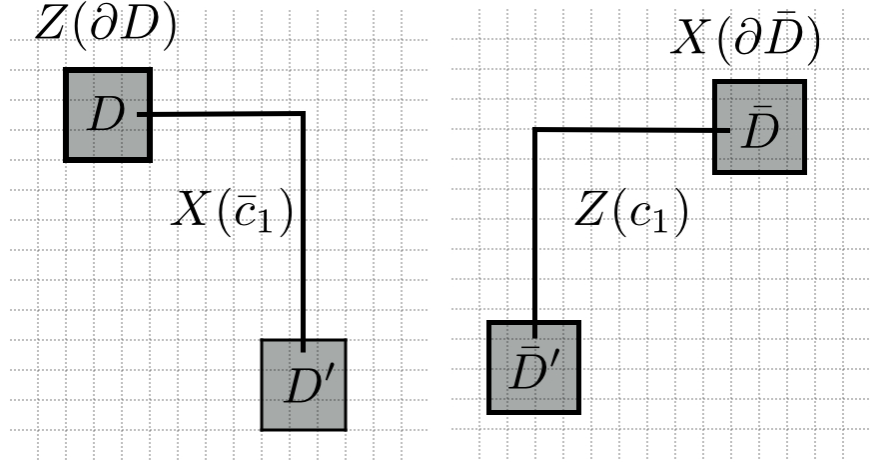}
%
%
\caption{Primal (left) and dual (right) defect pair qubits. 
The primal defect pair qubit is defined by removing the plaquette operators inside the defect regions $D$ and $D'$. 
Then the cycle $\partial D$ around the defect becomes a nontrivial cycle, by which the logical $Z$ operator $Z(\partial D)$ is defined. 
The logical $X$ operator $X(\bar c_1)$ is defined as a dual 1-chain $\bar c_1$ connecting two defects. 
Any homologically equivalent logical operators acts the same way on the code space.
The dual defect pair is defined similarly, but the basis is changed by the Hadamard transformation.
  }
\label{fig73}       
\end{figure}
We define two defect regions $D \in C_2$ and $D'\in C_2$, as shown in Fig.~\ref{fig73}.
We remove all $Z$-type stabilizer generators inside these regions, i.e., $\{ A_m\} _{f_m \in D \cup D'}$.
Thus, the operators $Z(\partial D)$ and $Z(\partial \bar D')$ are not stabilizer operators.
Instead, we append a $Z$-type stabilizer operator $Z(\partial D+\partial \bar D)$ as a stabilizer generator.
Moreover, we also append the Pauli $X$ operator on all edge qubits inside (not including the boundary) the regions $D \cup D'$, i.e., 
$\{ X_l \}_{e_l \in (D \cup D') \backslash (\partial D \cup \partial D')}$.
By this definition, the star ($X$-type) stabilizer generators in the defect region are still in the stabilizer group. 
We choose $Z(\partial D)$ as a logical operator, because it commutes with all stabilizer operators, but does not belong to the stabilizer group.
Because $Z(\partial D + \partial \bar D)$ is a stabilizer operator, $Z(\partial D')$ also acts the same way, i.e., $Z(\partial D) \sim Z(\partial D')$.
Moreover, 
the actions of operators represented by 
the homologically equivalent cycles are the same.
The logical $X$ operator is given by $X(\bar c_1)$, with a dual 1-chain $\bar c_1$ connecting two defects, as shown in Fig.~\ref{fig73}.
The actions of the operators represented by any homologically equivalent chains in the sense of relative homology are the same.
That is, we may choose any dual 1-chain $\bar c_1$
which connects two defect regions $D$ and $D'$.
The code distance is given by the circumference of the defect or the distance between two defects.
Because the plaquette operators on the primal lattice are removed, we call these defects and the logical qubit {\it primal defects} and {\it primal defect pair qubit}, respectively.

Similarly, we can also define a logical qubit by removing the star ($X$-type) stabilizer operators on the dual defect regions $\bar D$ and $\bar D'$ defined on the dual lattice as shown in Fig.~\ref{fig74}.
We call such defects and the logical qubit {\it dual defects} and {\it dual defect pair qubit}, respectively.
Hereafter, the planar surface code state, on which the defects are introduced, is referred to as {\it vacuum}.
Below we will explain how defect qubits are created, deformed, and moved in the vacuum. 

\section{Defect creation, annihilation, and movement}
\index{defect creation, annihilation, and movement}
\label{Sec:DefectOperation}
\begin{figure}[t]
\centering
\includegraphics[width=130mm]{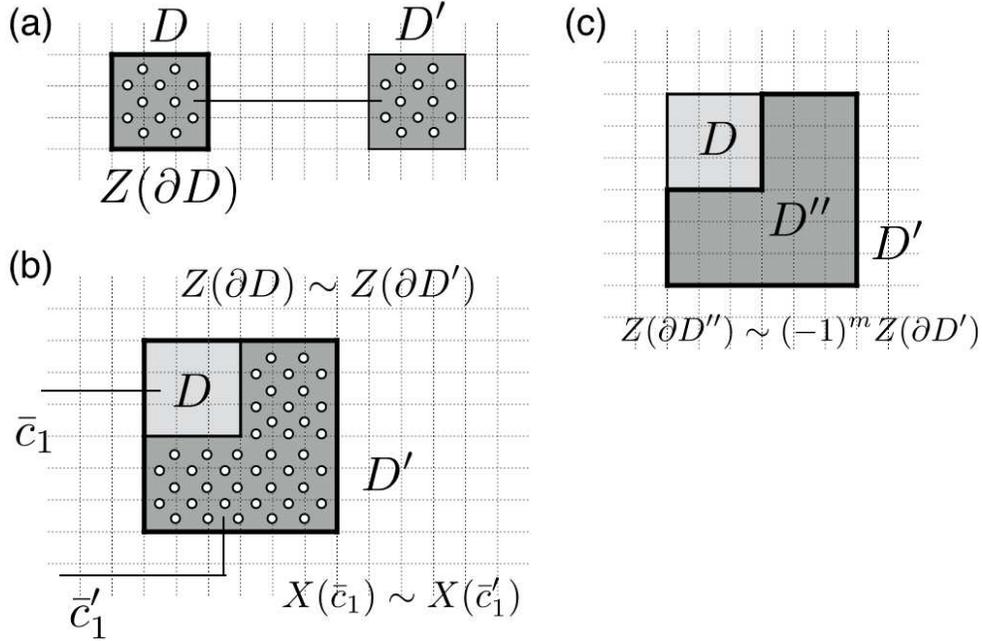}
%
%
\caption{(a) A defect creation. 
The qubits inside the defect regions (excluding the qubits on the boundary), denoted by circles, are measured in the $X$-basis. 
(b) The defect region $D$ is expanded into $D'$ by measuring the qubits inside the region $D'$ in the $X$-basis, where the encoded quantum information is preserved. 
(c) A defect region $D'$ is contracted into $D$ by measuring the plaquette operators in $D'' = D' \backslash D$. 
The logical operator $Z$ after the contraction is defined depending on the measurement outcome $m$. }
\label{fig74}       
\end{figure}
The defect creation is accomplished by measuring the qubits inside the defect region $D$, not including the qubits on its boundary, in the $X$-basis [see Fig.~\ref{fig74} (a)].
These measurements remove the plaquette operators inside $D$ from the stabilizer group, because these $X$-basis measurements do not commute with them.
On the other hand, the measurements do commute with the star operators, and hence a parity of four measurement outcomes corresponding to a star operator is always even (if there is no measurement error).
While the measurement outcomes are random, we can prepare all measured qubits to be in the $|+\rangle$ state by applying $Z$ operators according to the measurement outcomes.
(In practice, there is no need to apply the $Z$ operations. 
It is enough to keep the information of the byproduct Pauli operator dependent on the measurement outcomes.)

As the qubits on the boundary are not measured, the post-measurement state is stabilized by $Z(\partial D)$.
In order to create a pair of defects, we do the same thing for another defect $D'$.
Apparently, the resultant state is stabilized by $Z(\partial D + \partial D')$ and satisfies the definition of the defect pair qubit. 
Moreover, the logical qubit is stabilized by $Z(\partial D)$ and hence a logical $Z$-basis state is prepared.

The defect region can be expanded by creating a larger defect $D'$, which includes the original defect region $D$, i.e., $D \subset D'$, 
as shown in Fig.~\ref{fig74} (b).
We can choose $Z(\partial D')$ as a logical $Z$ operator of the defect pair qubit, so that the information with respect to $Z(\partial D')$ is preserved during this operation.
(Recall that logical operators act the same if the corresponding cycles are homologically equivalent. Thus we can choose a large enough cycle 
surrounding the defect in advance.)
We can expand the defect region step-by-step such that, at each step, a logical $X$ operator is untouched as shown in Fig.~\ref{fig74} (b).
(Or equivalently, we can compensate for the difference in the logical $X$ operator according the $X$-basis measurement outcomes inside the defect region.)
Thus, the information with respect to the logical $X$ operator is also preserved.
Accordingly, the logical information of the defect pair qubit is stored, but now the defect region $D$ is expanded into $D+D'$.

The defect annihilation is executed by measuring the plaquette operators inside the region $D$ to restore them to the stabilizer group.
The surface with a defect $D$ is rewritten by
\begin{eqnarray}
|D\rangle \propto \prod _{e_l \in D} \left(\frac{I+X_{l}}{2} \right) |v\rangle,
\end{eqnarray}
where $|v\rangle$ indicates the surface code state without defect, i.e., vacuum.
Thus, $|D\rangle$ is a superposition of all possible applications of $X$ operators on the vacuum $|v\rangle$.
The measurements of the plaquette operators collapse the superposition.
By applying a recovery operation inside the region $D$ such that all eigenvalues becomes $+1$, the defect is annihilated.
(Note that there is no need to actually apply the recovery operation. It is enough to keep the record of the eigenvalues.)
The parity of all measurement outcomes of the plaquette operators inside $D$ corresponds to the eigenvalue of $Z(\partial D)$, because we have 
\begin{eqnarray}
Z(\partial D) = \prod _{f_m \in D} Z(\partial f_m)= \prod _{f_m \in D} A_m.
\end{eqnarray}
This indicates that a logical $Z$-basis measurement can be done by annihilating the defect completely.

Suppose a defect region $D$ inside the defect region $D'$ (i.e., $D \in D'$) is annihilated by measuring the qubits inside $D$, as shown in Fig.~\ref{fig74} (c).
We can then obtain the eigenvalue $(-1)^m$ ($m=0,1$) of $ Z (\partial D)$.
Let $D''$ be the complement of $D$ in $D'$.
Because we have 
\begin{eqnarray}
Z( \partial D')  = Z(\partial D'' ) Z( \partial D),
\end{eqnarray}
depending on the eigenvalue $(-1)^m$, the operator $(-1)^{m }Z(\partial D'')$ acts as the logical operator of the defect $D''$.
Thus, the defect $D'$, consisting of the defect pair qubit, is contracted into $D''$ without changing the stored quantum information (up to the logical Pauli $X$ flip).

\begin{figure}[t]
\centering
\includegraphics[width=100mm]{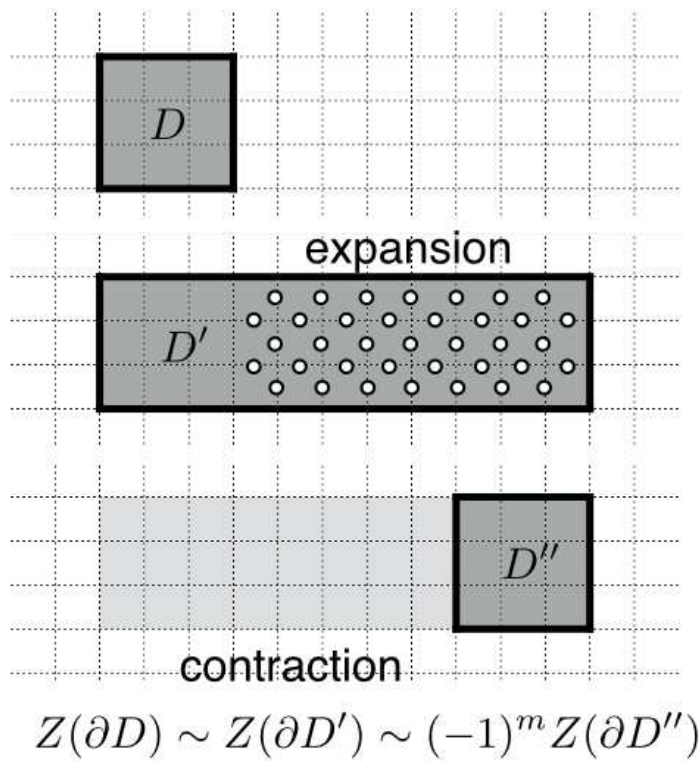}
%
%
\caption{A defect movement by expansion and contraction.}
\label{fig78}       
\end{figure}
The defect movement on the surface is implemented by combining the defect expansion and contraction, as shown in Fig.~\ref{fig78}.
At first, we expand a defect $D$ into $D'$ ($D \in D'$) by the previously mentioned procedure.
Second, the defect region $D$ is annihilated, and we obtain the eigenvalue $(-1)^m$ of $ Z (\partial D)$.
As we already pointed out, the stored logical information is unchanged under these procedures up to the logical Pauli $X$ operator depending on the measurement outcome $m$.

\begin{figure}[t]
\centering
\includegraphics[width=100mm]{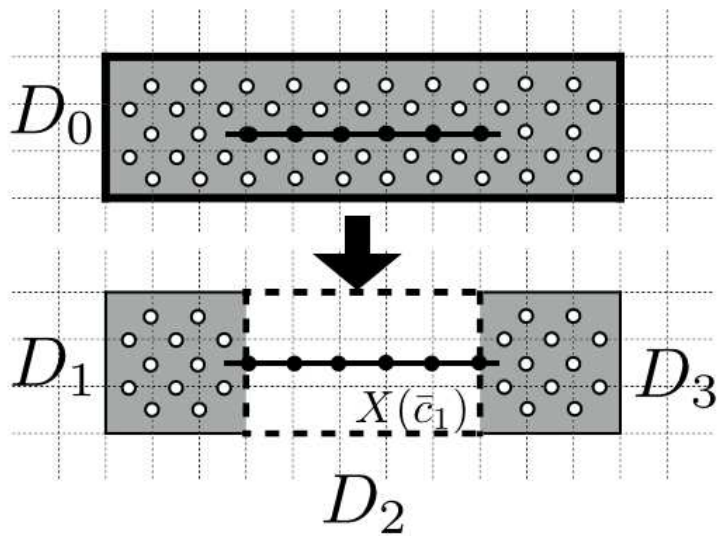}
%
%
\caption{A preparation of the logical $X$-basis state by creation of a defect region $D_0$ and annihilation of a defect region $D_2$ between two defect regions $D_1$ and $D_3$. Note that after the first defect creation, the state is stabilized by the logical $X$ operator $X(\bar c_1)$, which is untouched during the following defect annihilation.}
\label{fig79}       
\end{figure}
The logical $X$-basis state preparation is executed by combining the defect creation and annihilation, as shown in Fig.~\ref{fig79}.
We first create a defect region $D_0=D_1+D_2+D_3$, which consists of three adjacent defect regions $D_1$, $D_2$, and $D_3$.
The two defects $D_1$ and $D_3$ are utilized as a defect pair.
We can define a logical $X$ operator by employing the qubits inside the defect region $D_2$.
Because all qubits inside the defect region are in the $|+\rangle$ state, the eigenvalue of the logical $X$ operator at this time is $+1$.
Then the defect region $D_2$ in-between $D_1$ and $D_3$ is annihilated by measuring the star operators.
These measurements commute with the logical $X$ operator, and hence its eigenvalue is still $+1$ after the annihilation.
Now we have a defect pair qubit, which is the eigenstate of the logical $X$ operator.
A measurement in the logical $X$ basis can be implemented by doing the $X$-basis state preparation in an inverse way.
More precisely, two defects $D_1$ and $D_3$ are connected by making a larger defect $D_0=D_1+D_2+D_3$.
We can choose a logical $X$ operator such that all constituent qubits belong to the defect region $D_0$.
Then, we obtain the eigenvalue of the logical $X$ operator.

The dual defect creation, annihilation, expansion, contraction, and propagation can be done in the same way on the dual lattice with the Hadamard transformation.
In Sec.~\ref{Sec:TopoCalc}, these elementary operations of the primal and dual defects are depicted by a topological diagram.

\section{Logical CNOT gate by braiding}
\label{Sec:LogicalCNOT}
Using the elementary operations explained in the previous section, we can perform logical gate operations on the defect pair qubits.
Indeed, the defects created from the vacuum behave like anyonic particles, i.e., braiding a defect around another defect results in a nontrivial operation.
(It is better to say that we change the manifold continuously to deform the ground state, in contrast to the anyons appeared as an excitation on the toric code Hamiltonian.)
All operations employed in the elementary operations are single-qubit or local stabilizer measurements, which can be done by nearest-neighbor two-qubit gates and single-qubit measurements.
This contrasts with concatenated quantum computation, where the logical operations are implemented as transversal operations and hence, essentially, non-local two-qubit gates are employed.

\begin{figure}[t]
\includegraphics[width=150mm]{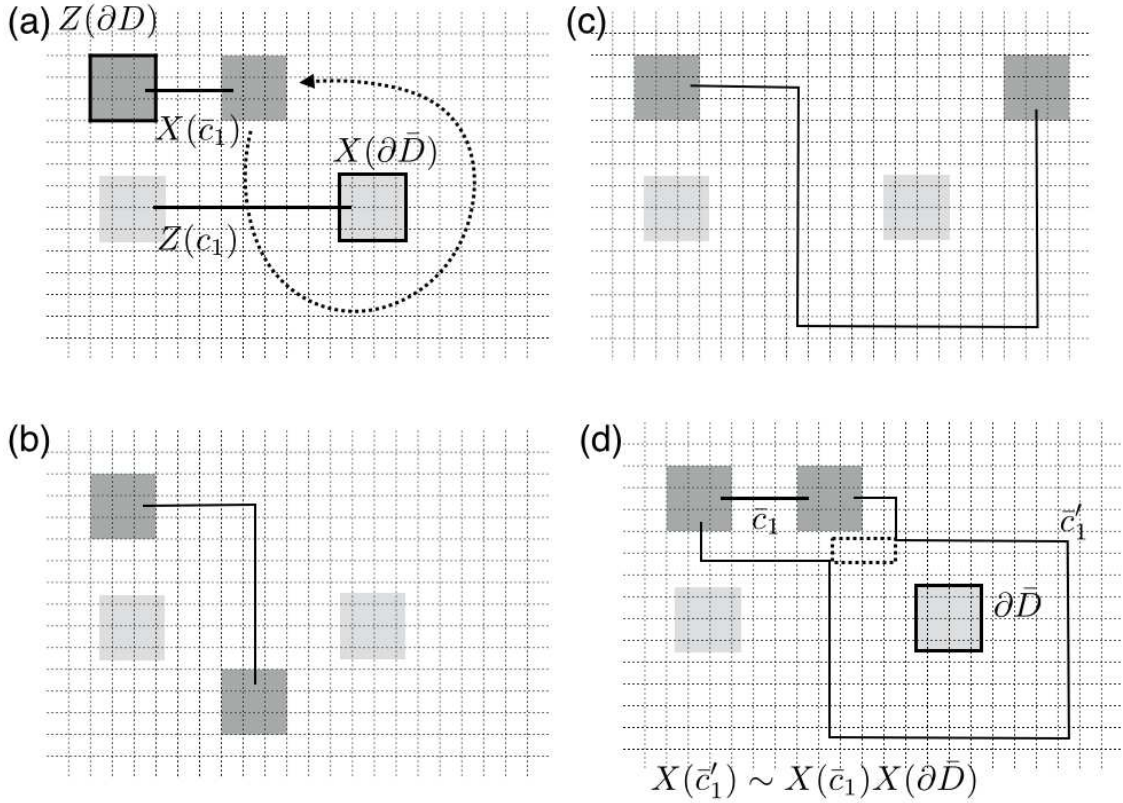}
%
%
\caption{A logical CNOT gate created by braiding. 
(a) A primal defect is braided around a dual defect. 
(b,c) The logical $X$ operator is deformed continuously during the braiding operations. 
(d) After the braiding, the logical $X$ operator is represented by the dual chain $\bar c_1'$ winding around the primal defect.
By applying star operators inside the rectangle denoted by dotted lines, $\bar c_1'$ is decoupled into a chain connecting two primal defects and a chain surrounding the dual defect. 
The corresponding operators are equivalent to the products of the logical $X$ operators on the primal and dual qubits.
}
\label{fig75}       
\end{figure}
We first consider the CNOT gate between primal (control) and dual (target) defect pair qubits.
Suppose there are primal and dual qubits on the surface code as shown in Fig.~\ref{fig75} (a), where logical operators are specified by the chains $\{ \partial D, \bar c_1\}$ 
and $\{ \partial \bar D,  c_1\}$ for the primal and dual qubits, respectively.
We can freely move the defect everywhere we want by the defect expansion and contraction.
Let us braid the primal defect around the dual defect, as shown in Fig.~\ref{fig75} (a)-(d).
After the braiding operation, the operator $X(\bar c_1')$ in Fig.~\ref{fig75} (d) 
has the same information as the operator $X(\bar c_1)$ before the braiding.
Using the equivalence relation,
\begin{eqnarray}
X(\bar c_1') \sim  X(\bar c_1) X(\partial D),
\end{eqnarray}
we understand that a correlation is made by the braiding operation between the logical $X$ operators $X(\bar c_1)$ and $X(\partial D)$ of the primal and dual defect pair qubits.
\begin{figure}[t]
\centering
\includegraphics[width=90mm]{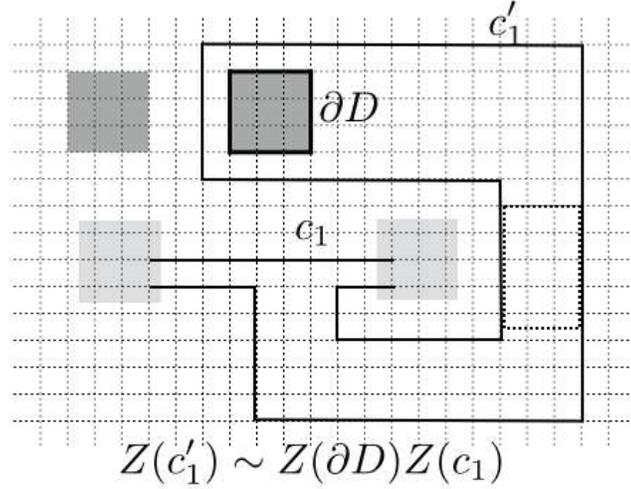}
%
%
\caption{The time evolution of the logical $Z$ operator caused by braiding. 
The logical $Z$ operator on the dual qubit is transformed into the product of the logical $Z$ operators on the primal and dual qubits.}
\label{fig76}       
\end{figure} 
A similar observation holds for the logical operator $Z(c_1)$, which is transformed into $Z(c_1') \sim Z(c_1) Z(\partial D)$ (see Fig.~\ref{fig76}).
On the other hand, $Z(\partial D)$ and $X(\partial \bar D)$ are invariant under this operation.
In short, the braiding operation transforms the logical Pauli operators as follows:
\begin{eqnarray}
Z(\partial D) &\rightarrow& Z(\partial D),
\label{eq:braid_CNOT1}
\\
Z(c_1) &\rightarrow& Z(c_1) Z(\partial D),
\label{eq:braid_CNOT2}
\\
X(\partial \bar D) &\rightarrow & X(\partial \bar D),
\label{eq:braid_CNOT3}
\\
X(\bar c_1) &\rightarrow& X(\bar c_1) X(\partial \bar D).
\label{eq:braid_CNOT4}
\end{eqnarray}
This transformation is equivalent to that for the CNOT gate, Eqs.\ (\ref{eq:CNOT1}-\ref{eq:CNOT4}).
Thus, braiding the primal defect around the dual defect results in a logical CNOT gate between the primal and dual defect pair qubits.

\begin{figure}[t]
\centering
\includegraphics[width=140mm]{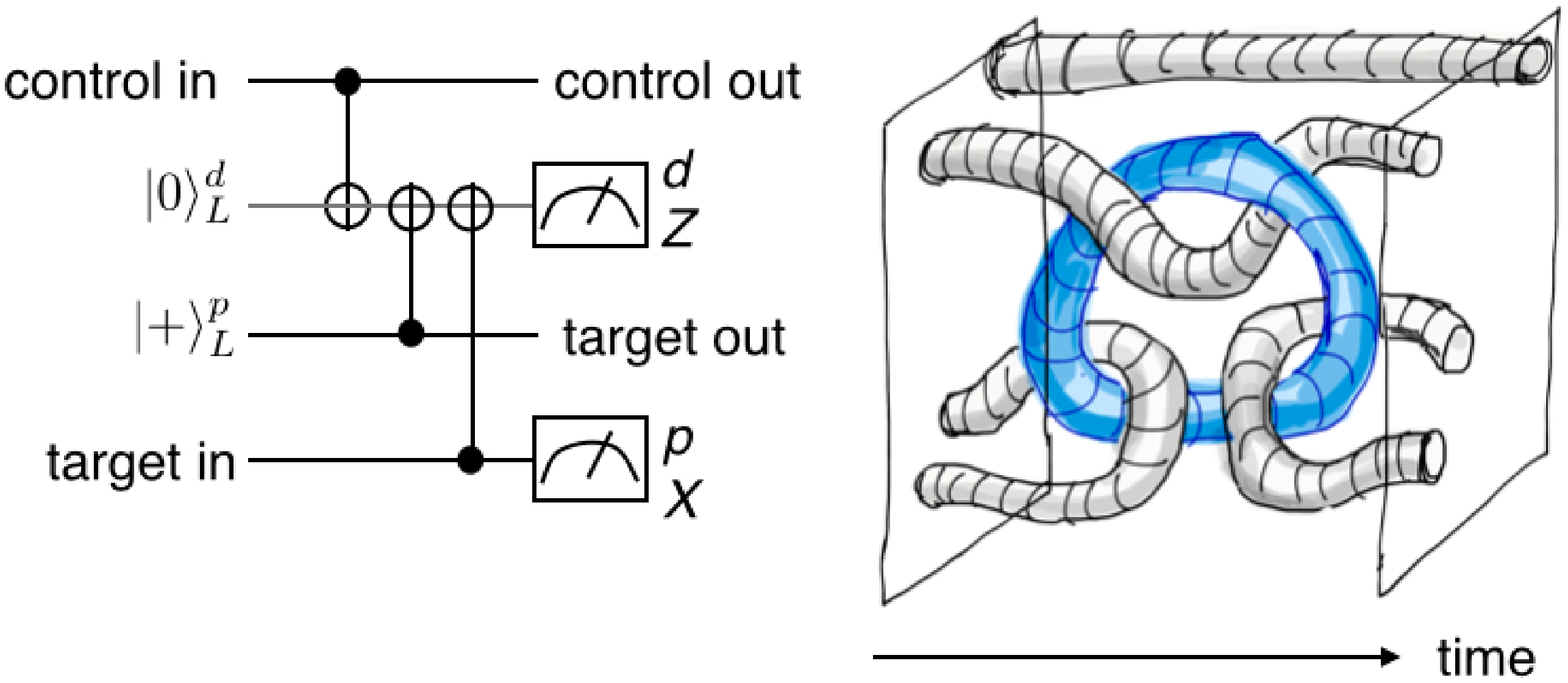}
%
%
\caption{The circuit diagram of the teleportation-based CNOT gate between primal qubits using a dual qubit as ancilla (left). 
The control and target qubits of the CNOT gates are always primal and dual qubits, respectively. The corresponding braiding operation (right).
}
\label{fig77}       
\end{figure}
Unfortunately, in the above CNOT gate, the primal (dual) defect is always a control (target) qubit.
Such CNOT gates always commute with each other, which is a natural consequence of the fact that the defect qubits on the surface code are Abelian.
In order to realize a genuine CNOT gate between the same type of qubits (and hence noncommuting gates), we utilize a teleportation-based gate \cite{GottesmanChuang}, as shown in Fig. \ref{fig77}.
Only CNOT gates between primal (control) and dual (target) qubits are employed.
The Pauli basis measurements are also done as mentioned in the previous section.
Accordingly, the CNOT gate between primal qubits is realized by braiding the primal defects around {\it virtual} dual defects, which are created and annihilated as ancillae.

The above braiding operations are implemented with topological quantum error correction at each elementary step, which will be explained later in detail.
All operations considered so far can be executed keeping the defect size (circumference) and defect distance larger than a length $d$, which provides a code distance of the logical qubits.
Thus, the logical error probability for the logical CNOT gates decreases exponentially by increasing the characteristic length $d$ of the system.
In other words, the CNOT gates are topologically protected.

We should mention that 
the braiding operations explained above are not the only way to 
perform fault-tolerant operations on the surface code.
There are another approaches to perform logical operations 
fault-tolerantly for the encoded degrees of freedom of the surface code.
One is the lattice surgery scheme~\cite{Horsman12},
where the boundary conditions of two planar surface codes 
are engineered to perform a logical operation.
Another is to employ twists,
which are topological defects introduced by
point defects on lattices~\cite{BombinTwist,Bombin11}.
All Clifford gates can be implemented by the
twist creation, braiding, and annihilation
similarly to the defect pair qubits explained above.
All these different approaches 
can be view as the logical operations by the code deformations~\cite{Dennis,RaussendorfAnn,Bombin09}
and seem to be a unique feature 
for the topological codes 
contrasting to the transversal logical gate for the concatenated codes.

\section{Magic state injections and distillation}
\label{Sec:MagicInjection}
Unfortunately, the topologically protected CNOT gates do not allow universal quantum computation, because Clifford circuits can efficiently be simulated classically due to the Gottesman-Knill theorem~\cite{GottesmanKnill}.
Here, we explain how to perform single-qubit rotations on the defect pair qubit, while, unfortunately, they are not topologically protected.

\begin{figure}[t]
\centering
\includegraphics[width=130mm]{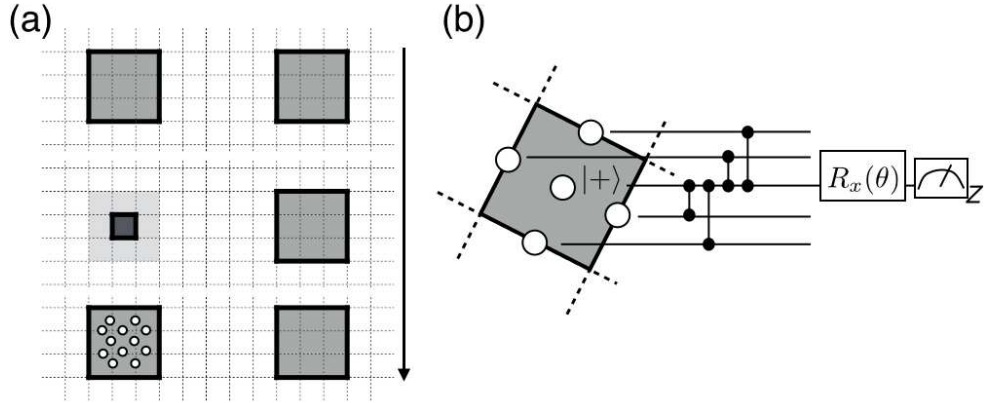}
%
%
\caption{(a) A magic state injection by contracting the defect into 
an elementary cell. 
(b) A circuit diagram for the logical $Z$ rotation on the defect of an elementary cell.}
\label{fig80}       
\end{figure}
Suppose we have a defect pair qubit, whose logical operators are given by $Z(\partial D)$ and $X(c_1)$.
We first consider a logical $Z$ rotation $e^{-i (\theta/2) Z(\partial D)}$ on the defect pair qubit.
To this end, the defect region $D$ is contracted to a single face $f_l$ by using the annihilation process, as shown in Fig.~\ref{fig80} (a).
Now the logical operator $Z(f_l)$ is a four-body operator.
(This implies that the code distance becomes 4 at this stage.)
A rotation with respect to the logical operator $Z(f_l)$, $e^{-i (\theta/2) Z(f_l)}$, can be implemented 
indirectly by using an acilla qubit located at the center of the face
shown in Fig.~\ref{fig80} (b).
First, the four CZ gates are applied 
between the ancilla and the four edge qubits on the face.
Second, after applying a $X$-rotation $R_x(\theta )= e^{-i (\theta/2) X}$,
the ancilla qubit is measured in the $Z$-basis. 
According to the measurement outcome $m=0,1$,
a logical rotation $e^{-i(-1)^{m} (\theta/2) Z(f_l)}$ is implemented.
After the above procedure, the defect, as a single face $f_l$, is expanded again into the defect region $D$ to restore the code distance of the defect pair qubit.

\begin{figure}[t]
\centering
\includegraphics[width=90mm]{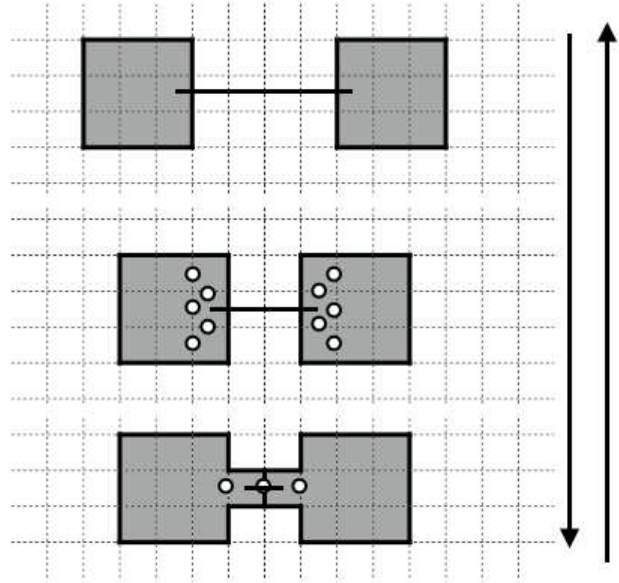}
%
%
\caption{A magic state injection by connecting two defects.
If two defects regions are adjacent, the logical $X$ operator is a physical Pauli $X$ operator between two defects.}
\label{fig81}       
\end{figure}
Next we consider a logical $X$ rotation, $e^{-i (\theta/2) X(c_1)}$.
In this case, the two defects are moved and deformed near each other such that $X(c_1')$ becomes a one-body operator as shown in Fig.~\ref{fig81}.
The rotation $e^{-i (\theta/2) X(c_1')}$ is now easily implemented by a single-qubit gate.
After that, the distance between two defects is restored.

\begin{figure}[t]
\centering
\includegraphics[width=110mm]{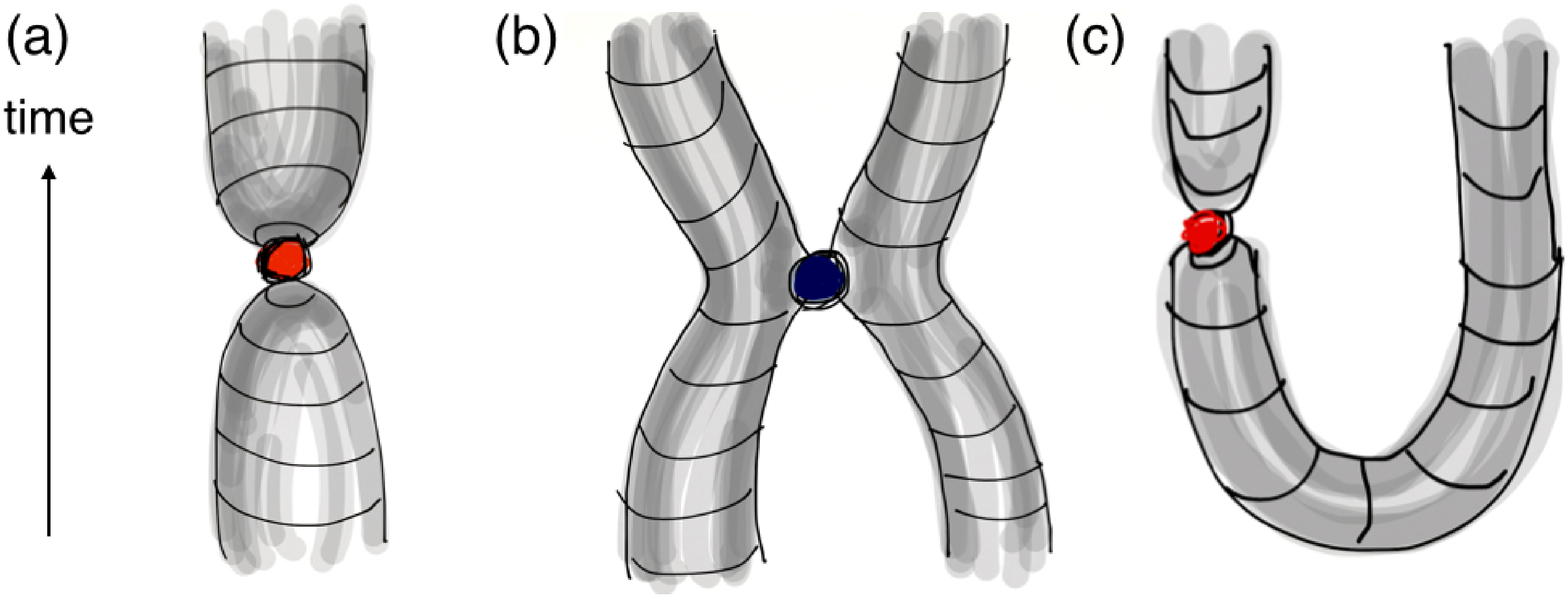}
%
%
\caption{(a) The logical $Z$ rotation. (b) The logical $X$ rotation. (c) The logical 
$e^{-i (\theta /2)Z}|+\rangle$ state injection.}
\label{fig125}       
\end{figure}
In this way, we can perform an arbitrary single-qubit unitary operation on the defect pair qubit by continuously deforming the defect pair qubit.
These deformations and the operations for the logical rotations
are depicted in Fig.~\ref{fig125}.
Unfortunately, during the deformation, the code distance inevitably becomes relatively small, so that we can perform logical rotations directly with nearest-neighbor operations.
Thus, these processes are not topologically protected.

\begin{figure}[t]
\centering
\includegraphics[width=110mm]{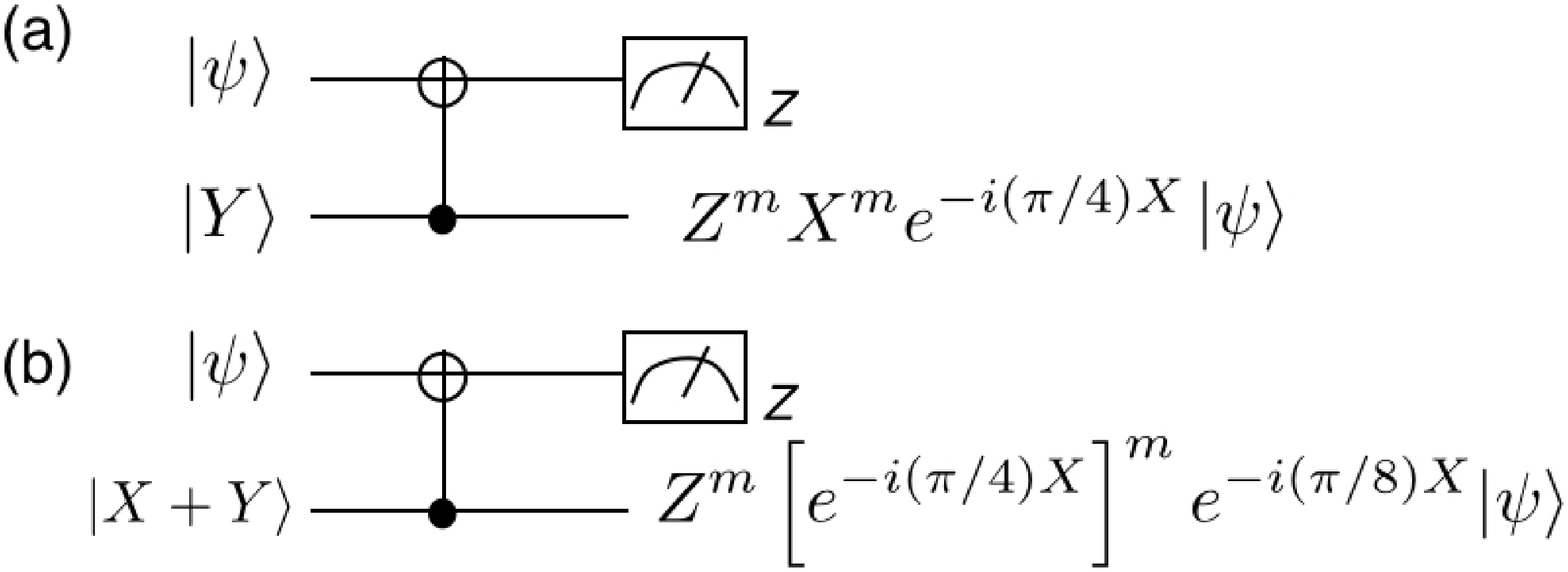}
%
%
\caption{(a,b) Single-qubit rotations by one-bit teleportations~\cite{Leung}.}
\label{fig82}       
\end{figure}

These non-topological operations can be utilized to inject noisy magic states on the surface.
Specifically, the logical $e^{-i(\theta/2)Z}|+\rangle$ state
is injected as shown in Fig.~\ref{fig125} (c).
Then, these noisy magic states are distilled by topologically protected operations 
to clean magic states, which have a fidelity high enough for reliable quantum computation.
Note that we are allowed to use only the CNOT gates, and there is no topologically protected single-qubit Clifford gate.
To manage this, we distill two types of magic states.
One is the eigenstate of the Pauli-$Y$ operator, which is used to implement the $S$ gate via gate teleportation~\cite{GottesmanChuang,Leung}, as shown in Fig.~\ref{fig82} (a).
Another is the eigenstate of the $(X+Y)/\sqrt{2}$ operator, which is used to implement the $\pi/8$ gate, a non-Clifford gate necessary for universal quantum computation.
The magic state distillation for the $Y$-basis state is executed solely by the topologically protected CNOT gates using the Steane 7-qubit code,
similarly to the method introduced in Sec.~\ref{sec:magic}.
The topologically protected CNOT gates and the $S$ gates with the distilled $Y$-basis states are further employed to distill the $(X+Y)/\sqrt{2}$-states through the Reed-Muller 15-qubit code, as explained in Sec.~\ref{sec:magic}.

Using these distilled magic states and the CNOT gates, we can implement the single-qubit gates shown in Fig.~\ref{fig82} (a) and (b), which together with the CNOT gate constitute a universal set of gates.
Accordingly, universal quantum computation is performed reliably on the surface code.
Note that all operations employed are single-qubit gates, two-qubit nearest-neighbor gates, and single-qubit measurements on the 2D array of qubits.

\section{Topological calculus}
\label{Sec:TopoCalc}
Based on the microscopic understanding of the topological operations, we can introduce a {\it topological diagram}\index{topological diagram} and a {\it topological calculus}\index{topological calculus}, 
which allows us a diagrammatic description 
of topological quantum computation on the surface codes.

In this diagram, the 3D space-time trajectory of the defects 
is depicted by projecting it onto a 2D plane
like link diagrams mentioned in Sec.~\ref{sec:JonesPoly}.
The trajectory of the primal and dual defects
as tubes in 3D space-time are denoted 
by pairs of solid and gray-colored lines:
\begin{center}
\includegraphics[width=90mm]{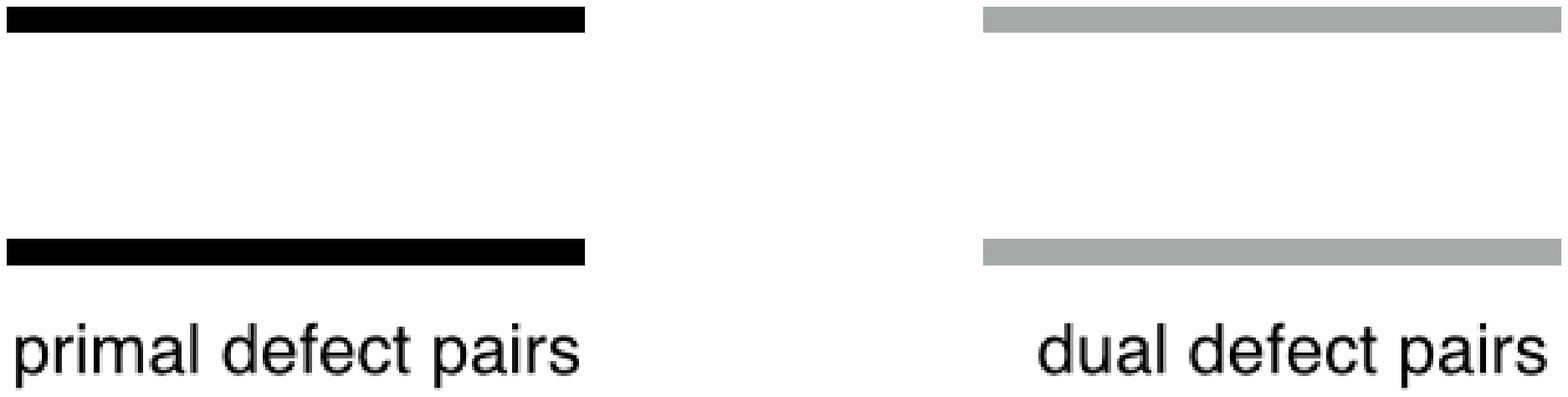}
\end{center}
The braiding operation is denoted by a double crossing as follows:
\begin{center}
\includegraphics[width=70mm]{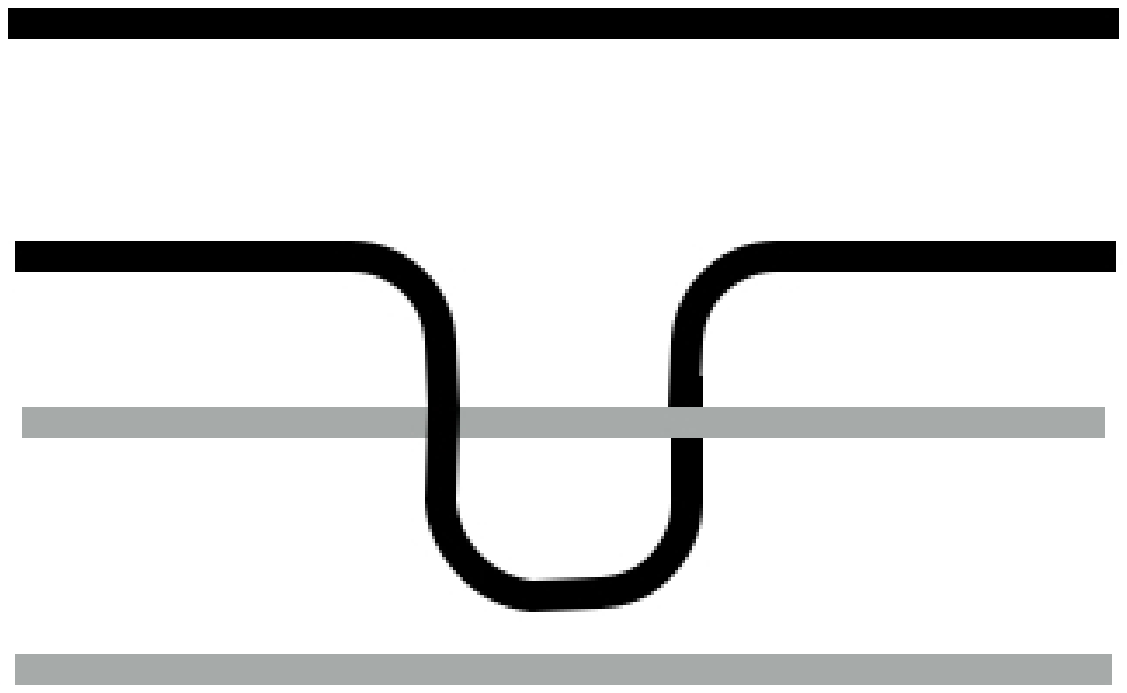}
\end{center}
The logical $X$ state preparation and $X$-basis measurement of the primal defect are denoted by closures:
\begin{center}
\includegraphics[width=70mm]{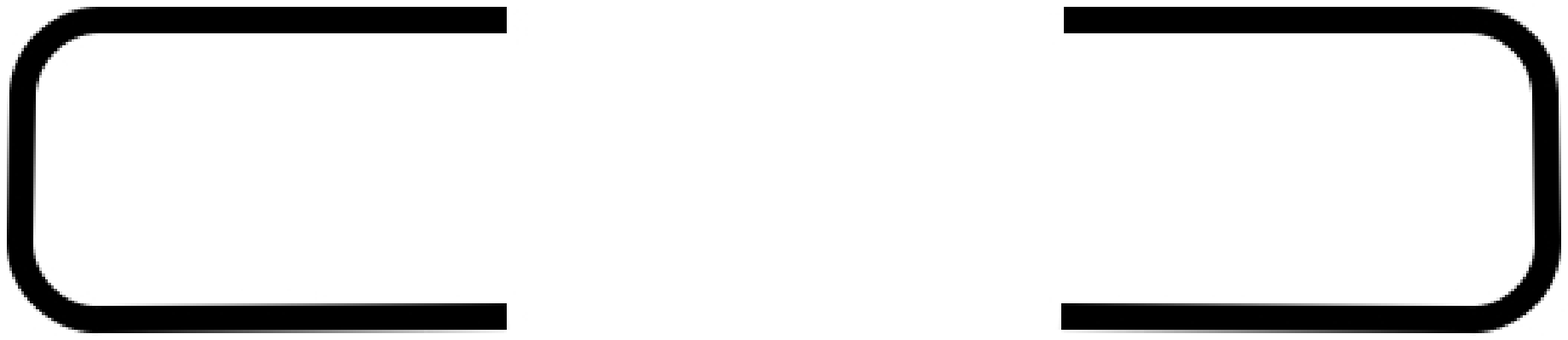}
\end{center}
The logical $Z$ state preparation and $Z$-basis measurement of the primal defect are denoted by endpoints of the tubes:
\begin{center}
\includegraphics[width=70mm]{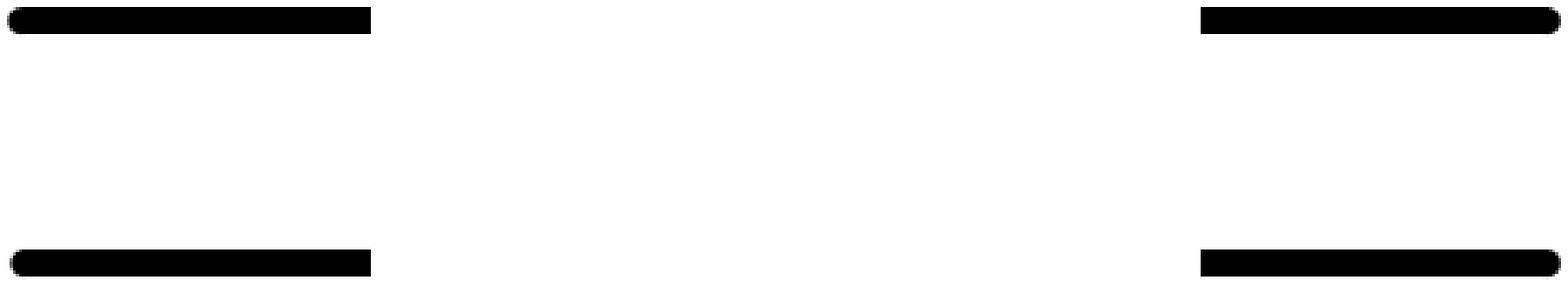}
\end{center}
The state preparations and measurements for the dual defect pair qubit are depicted in a similar way with gray (curved) lines, except for the basis change by the Hadamard transformation.

The trajectories of the logical operators are represented by surfaces
in 3D space-time like a Seifert surface, 
which we call {\it correlation surfaces}\index{correlation surface}.
Specifically, the logical $Z$ operator for the primal defect corresponds to a surface wrapping around the primal defect.
The logical $X$ operator for the primal defect corresponds to a surface whose boundary is the primal defect tubes.
(This is also the case for the dual defect, except for the Hadamard transformation.)
For example, the time evolution of the logical $X$ and $Z$ operators by the braiding operation can be represented by the following surfaces
\begin{center}
\includegraphics[width=100mm]{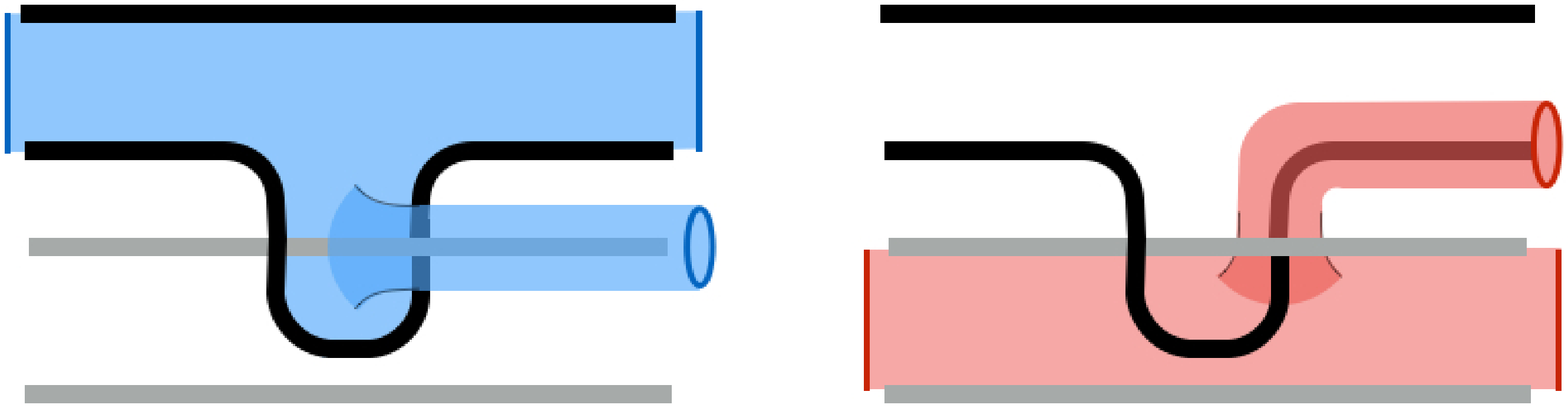}
\end{center}
The logical operators before and after the operation correspond to the left and right boundaries of these surfaces, respectively.
We can confirm Eqs.~(\ref{eq:braid_CNOT1})-(\ref{eq:braid_CNOT4})
from the left and right boundaries the correlation surfaces of the above diagrams.

The CNOT gate between the primal defect pair qubits is depicted as follows:
\begin{center}
\includegraphics[width=120mm]{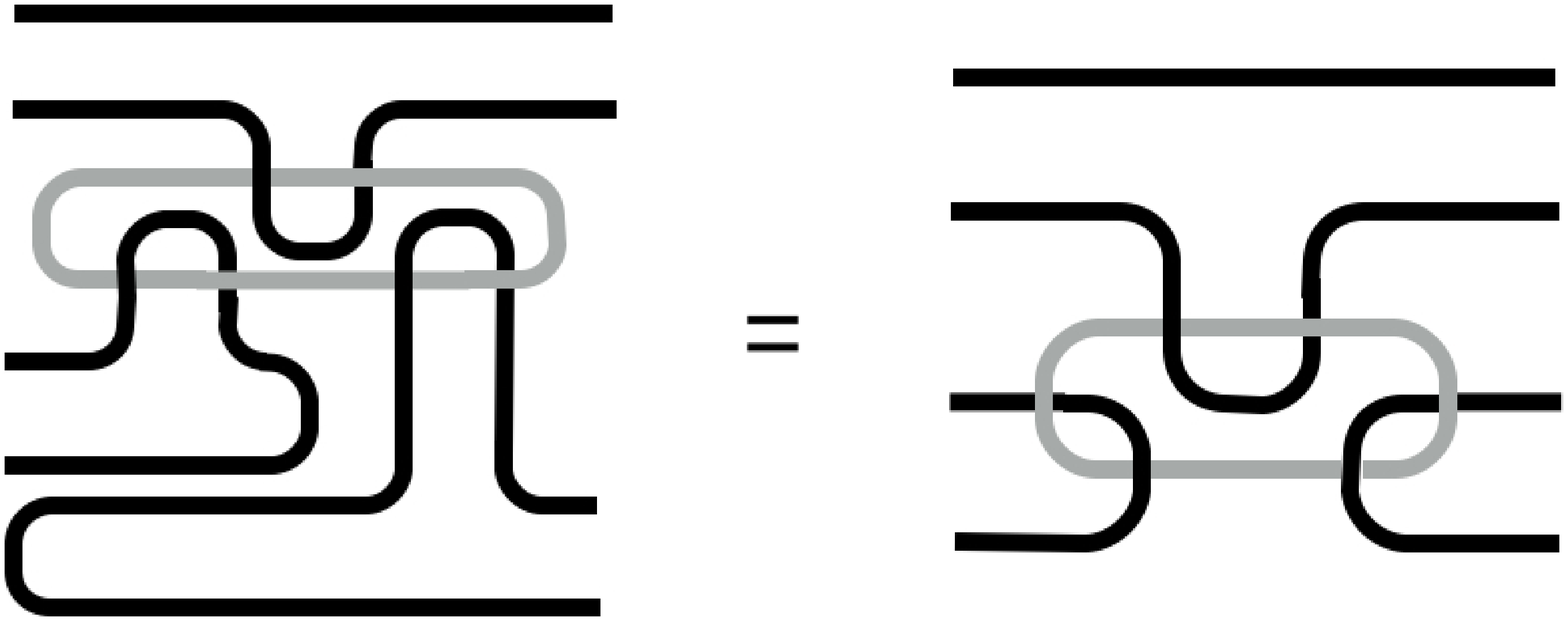}
\end{center}
In this diagram, we can easily confirm that the braiding operations transform the logical operators according to the rule for the CNOT gate, i.e., 
$X\otimes I \rightarrow X\otimes X$ and $I \otimes Z \rightarrow Z \otimes Z$:
\begin{center}
\includegraphics[width=120mm]{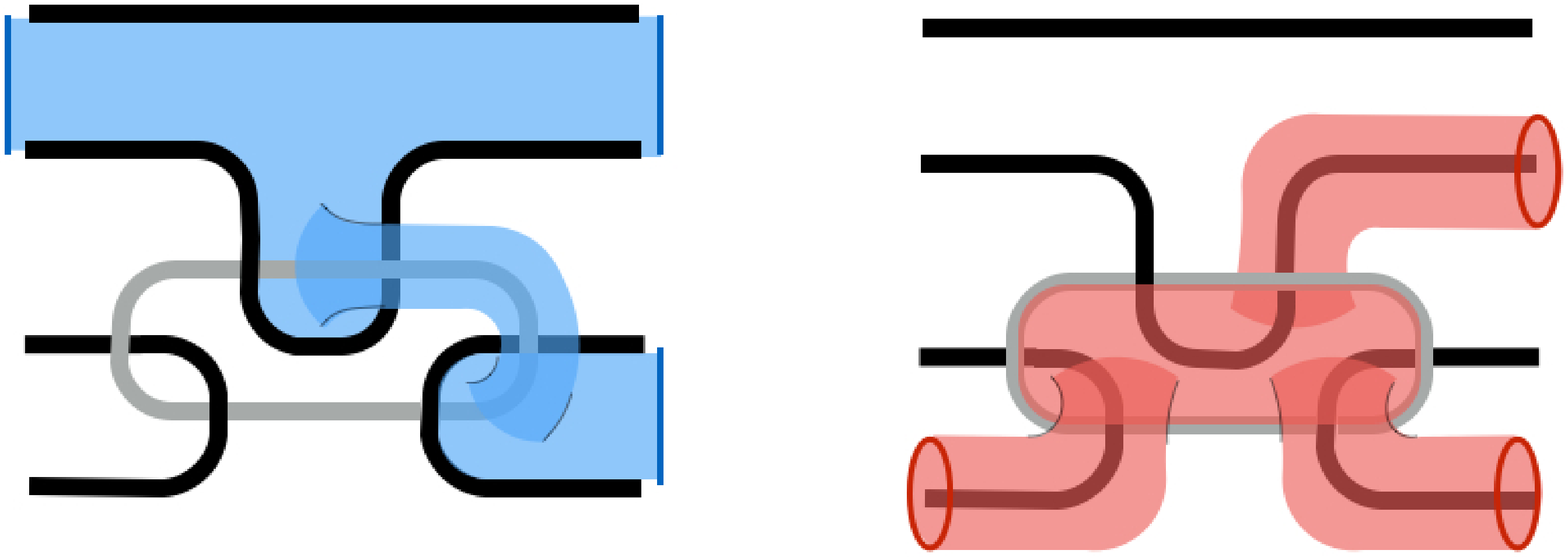}
\end{center}

As seen above, the action on the code space is defined by the left and right boundaries of the correlation surfaces.
Thus, the logical action of the topological operation is invariant under transformations that do not change the topology of the left and right boundaries of the correlation surfaces. 
Similarly to link diagrams, the boundary topology is invariant under the Reidemeister moves\index{Reidemeister move}:
\begin{center}
\includegraphics[width=140mm]{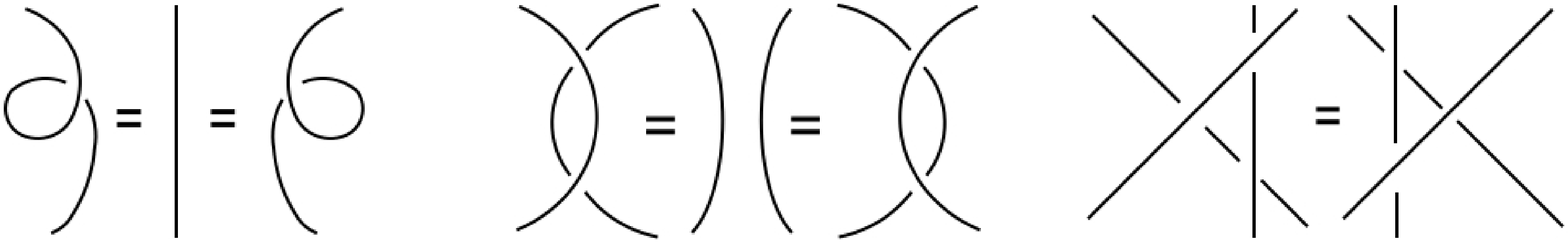}
\end{center}
In the present case, we have further transformations under which the logical action on the code space is invariant due to the properties of the defect pair qubits.
At first, the following two crossings of tubes of the same type are equivalent:
\begin{eqnarray}
\includegraphics[width=80mm]{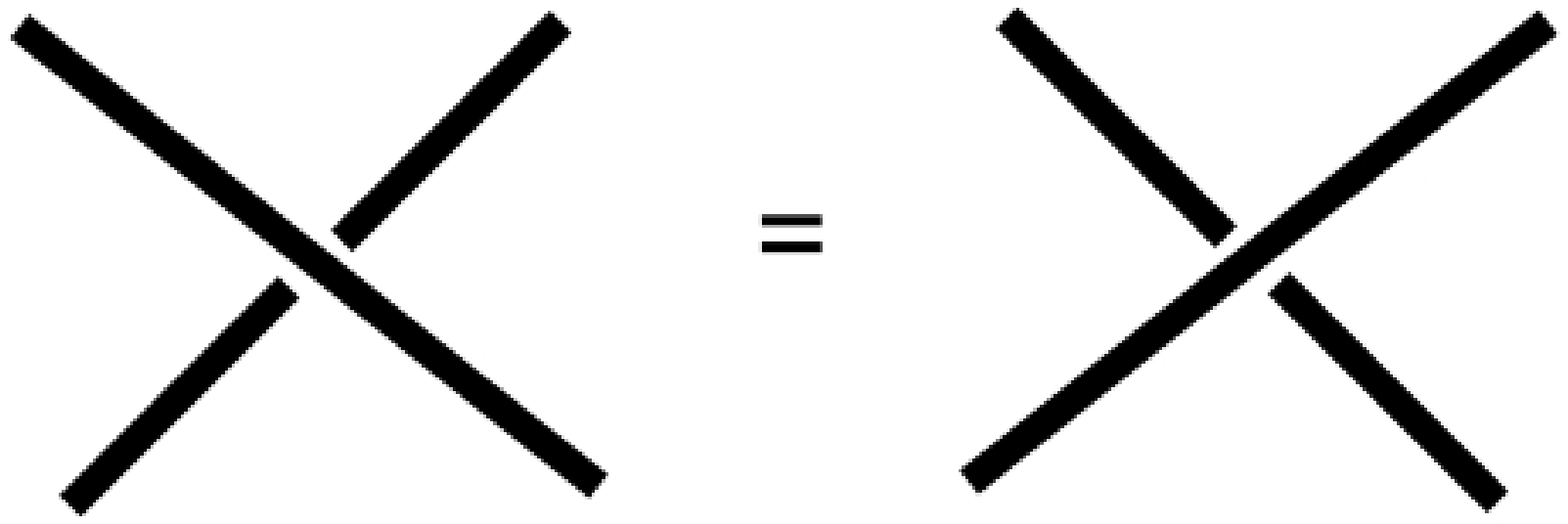}
\label{TopoRule1}
\end{eqnarray}
Second, if two defect tubes are 
a defect pair qubit, a dual loop wrapping around a defect pair can be removed:
\begin{eqnarray}
\includegraphics[width=80mm]{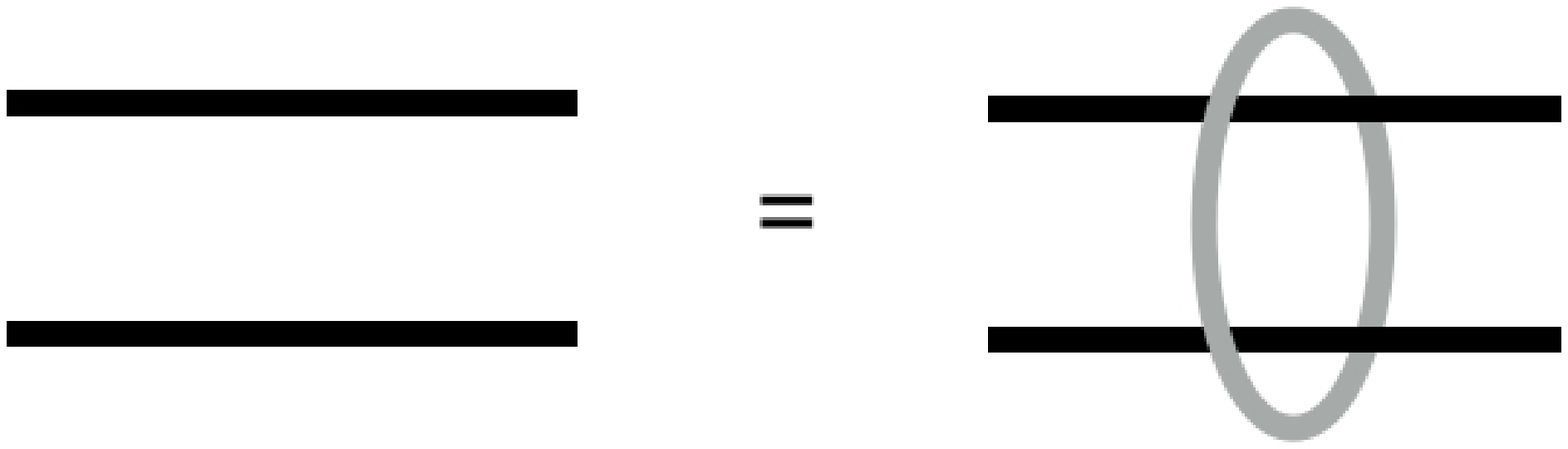}
\label{TopoRule2}
\end{eqnarray}
This is because the primal defect pair qubit is stabilized by a $Z$-type loop operator surrounding the defect pair, and hence the dual loop is nothing but a trivial measurement of the stabilizer operator.
Third, because the defect pair qubit is a $\mathbf{Z}_2$ Abelian anyon, if we braid a primal defect around a dual defect twice, it results in an identity operation:
\begin{eqnarray}
\includegraphics[width=100mm]{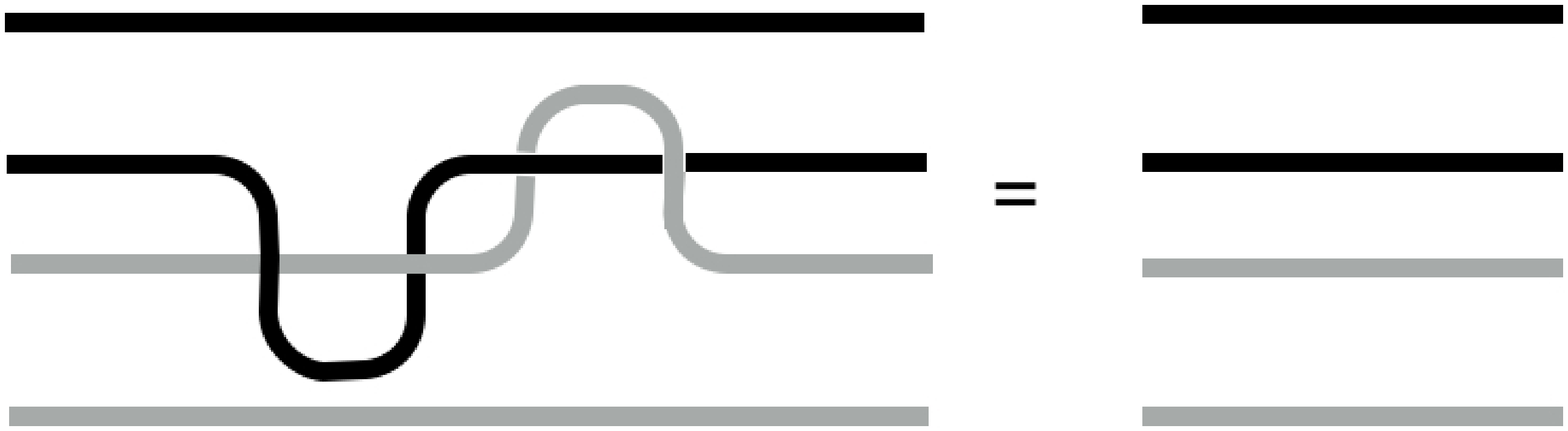}
\label{TopoRule3}
\end{eqnarray}
Fourth, and most importantly, two tubes can be connected or disconnected by the following procedure, which we call a $\Phi$-transformation\index{$\Phi$-transformation}:
\begin{eqnarray}
\includegraphics[width=90mm]{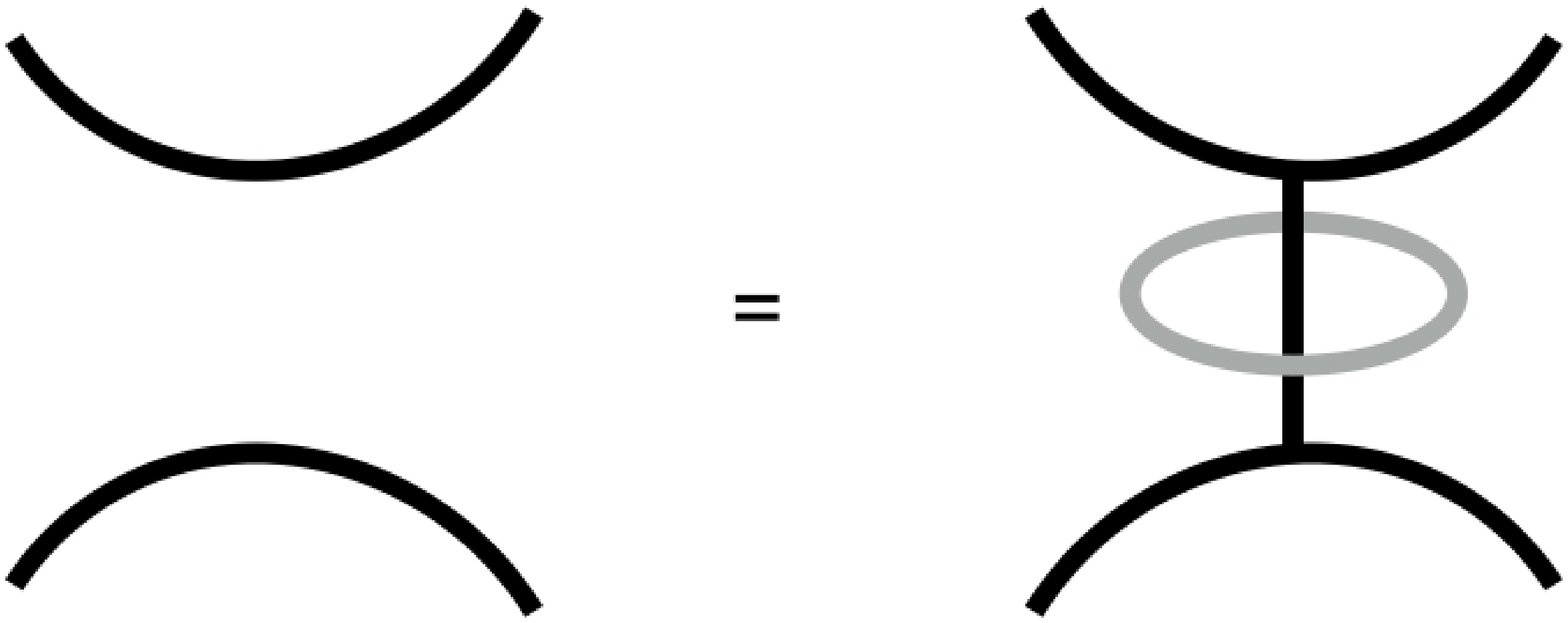}
\label{TopoRule4}
\end{eqnarray}
We can easily confirm that the logical operators are invariant under the $\Phi$-transformation:
\begin{center}
\includegraphics[width=120mm]{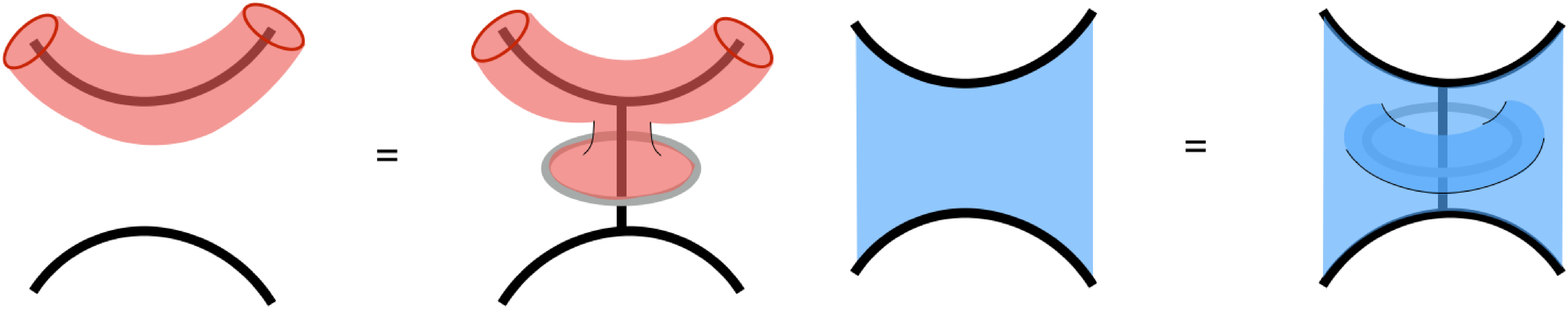}
\end{center}
where we note that the dual ring in the middle serves to stop the $Z$-type correlation surface from propagating toward the bottom and also serves to mediate the $X$-type correlation surface toward the right.
Specifically, if two defect pair qubits are connected by the $\Phi$-transformation between their closures, we can remove the dual ring by using the rule (\ref{TopoRule3}):
\begin{center}
\includegraphics[width=120mm]{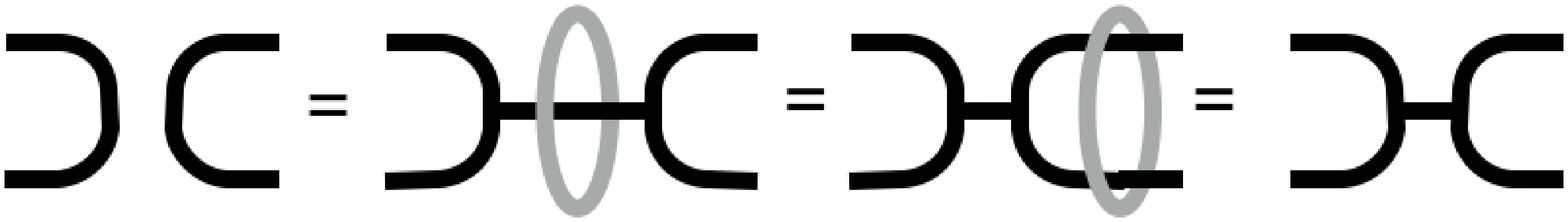}
\end{center}
By using the $\Phi$-transformation and rules (\ref{TopoRule3}) and (\ref{TopoRule2}), we can transform the defect pair qubit as follows:
\begin{center}
\includegraphics[width=120mm]{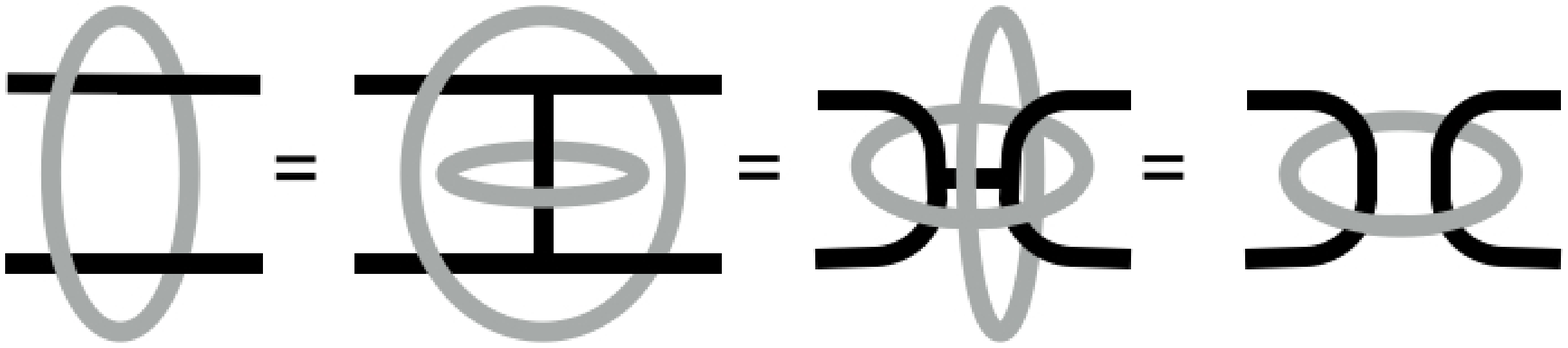}
\end{center}
We can also easily confirm that the logical operators are invariant under this transformation as follows:
\begin{center}
\includegraphics[width=120mm]{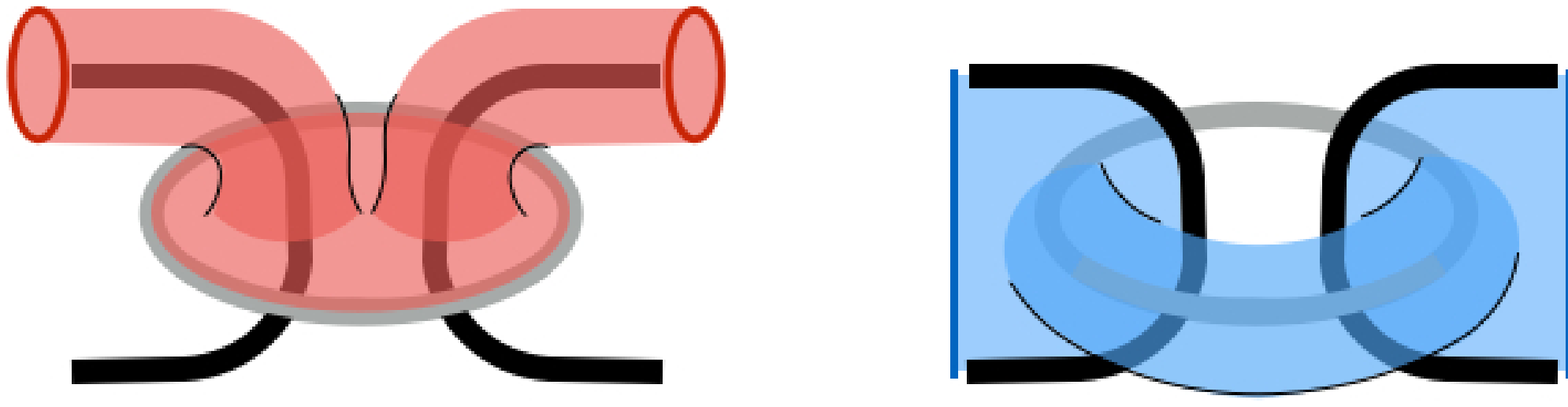}
\end{center}

The $\Phi$-transformation is very useful to simplify a complicated topological operation on the surface by joining the tubes as follows:
\begin{center}
\includegraphics[width=140mm]{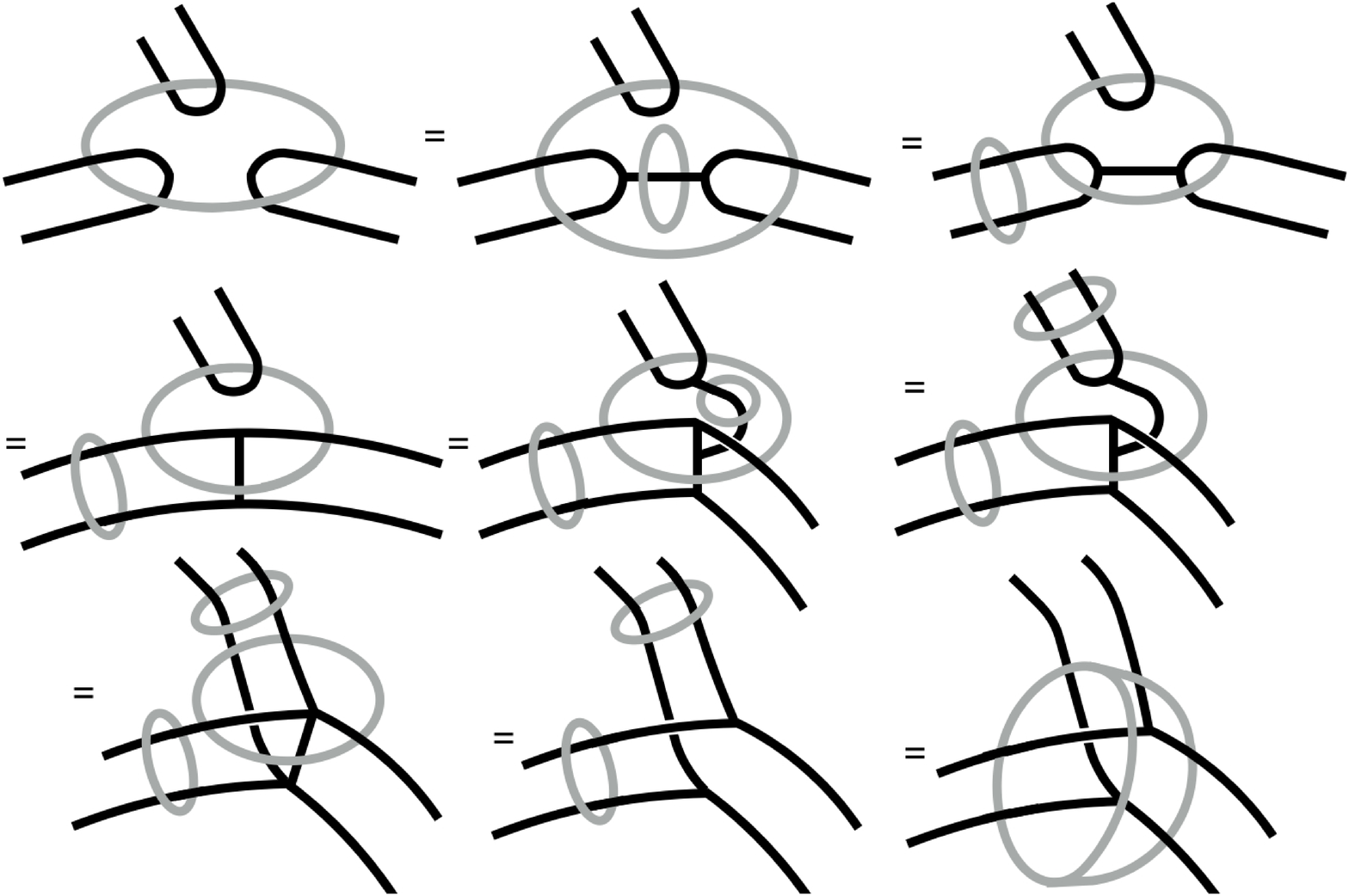}
\end{center}
where we frequently used the $\Phi$-transformation and rule (\ref{TopoRule2}).
Here, the dual triple-ring wrapping around each of the two defect tubes serves to reflect the logical $Z$ operator from the upper to the lower tubes as follows:
\begin{center}
\includegraphics[width=50mm]{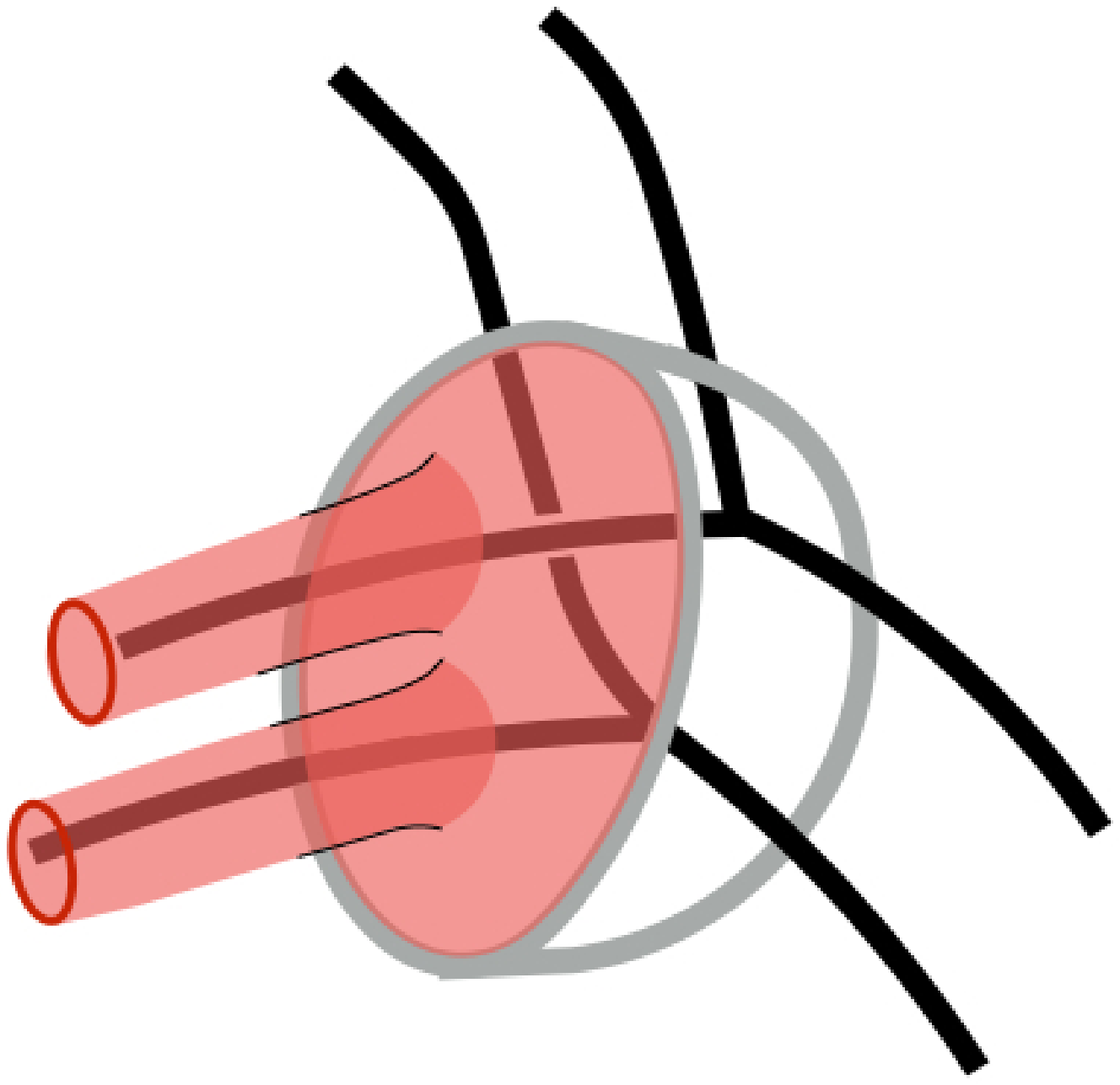}
\end{center}

For example, if we apply this transformation to the CNOT gate between the primal defect pair qubits, we obtain a much simpler diagram as follows:
\begin{center}
\includegraphics[width=100mm]{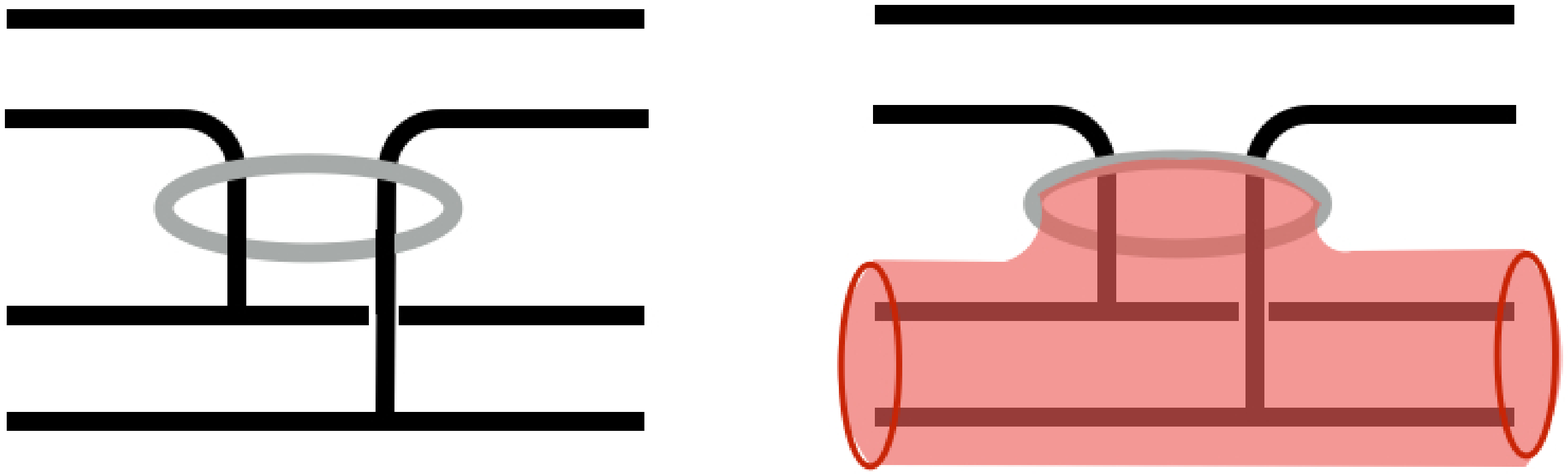}
\end{center}
where the dual ring in the middle serves to keep the code space of the defect pair qubit as shown in the left above.
It is straightforward to check the transformations of the logical operators.
In general, by joining the defect tubes directly, we can transform the logical operators under multiple CNOT gates. 
However, if two tubes are joined, the definition of the defect pair is broken. 
In order to keep the defect pair qubit encoding, we need the dual ring wrapping around the two tubes at the joint.

The magic state injection is denoted by the following diagram:
\begin{center}
\includegraphics[width=80mm]{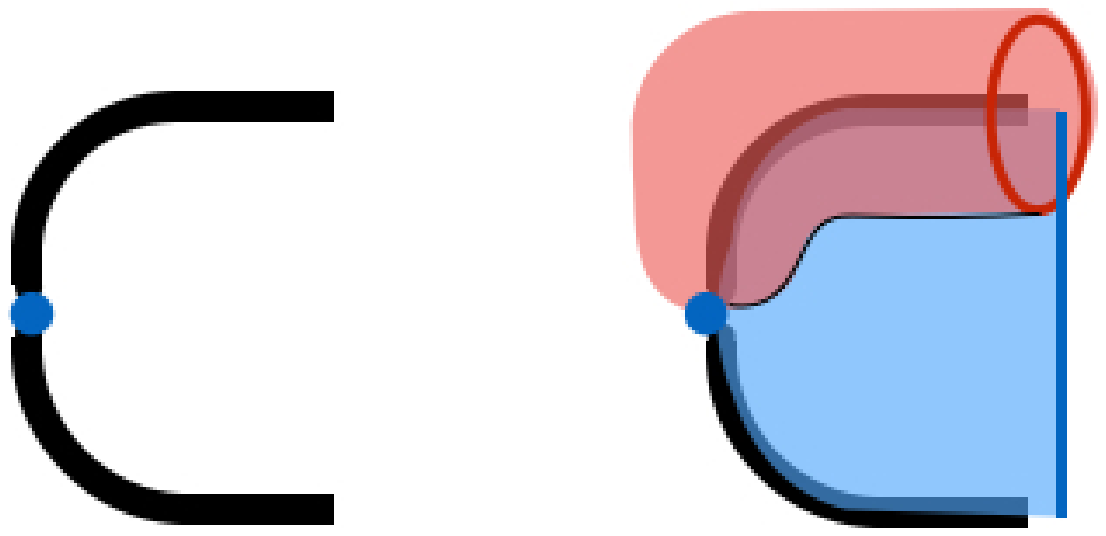}
\end{center}
which can be viewed as a superposition of two correlation surfaces for the two anti-commuting logical operators.
The non-Clifford gate by one-bit teleportation can be described in the topological diagram as follows:
\begin{center}
\includegraphics[width=100mm]{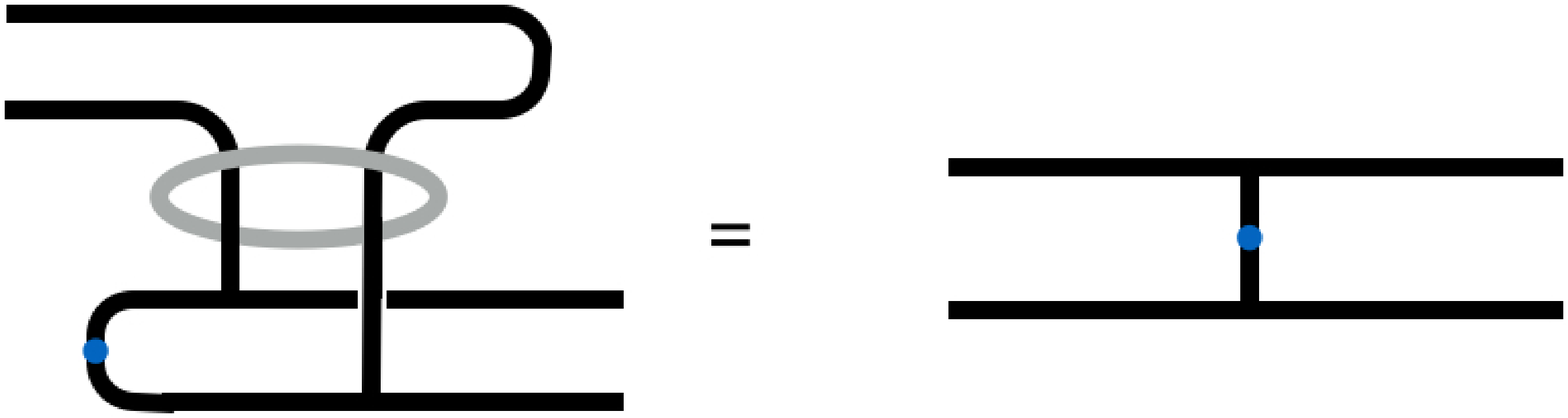}
\end{center}
The right diagram, which is topologically equivalent to the left one, indicates that the non-Clifford gate by one-bit teleportation with the magic state injection is equivalent to simply performing the non-Clifford gate by the method mentioned in the previous section.

The present diagrammatic description of the topological operations and correlation surfaces on them provide us an intuitive understanding of how topological quantum computation on the surface code is performed.
These transformation rules will be useful to optimize the complexity (space-time volume required) of the braiding operations.

\section{Faulty syndrome measurements and noise thresholds}
\label{Sec:TopoFaultTolerance}
We have seen how universal quantum computation is executed on the surface code.
At each step, we have to perform topological quantum error correction to protect the quantum information encoded by the defects.
\begin{figure}[t]
\centering
\includegraphics[width=120mm]{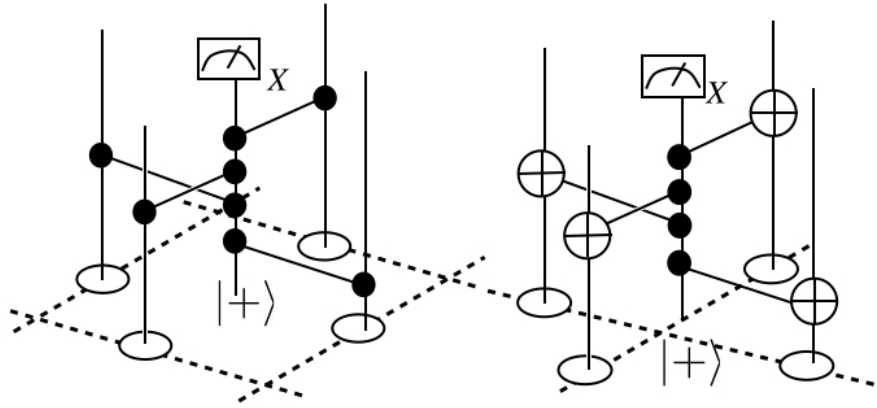}
%
%
\caption{Syndrome measurements for plaquette (left) and star (right) operators, where qubits on a face center and on a vertex are employed as ancillae.
The alternative syndrome measurements for the plaquette and star operators are done repeatedly.}
\label{fig83}       
\end{figure}
In Sec.~\ref{Sec:TQEC}, we analyzed topological quantum error correction on the surface code.
However, at that time, we assumed ideal syndrome measurements.
However, in a fault-tolerant quantum computation, we have to take into account all reasonable sources of noise, including faulty syndrome measurements, and the quantum computation has to tolerate them as well.
Here, we explain how faulty syndrome measurements are handled during topological quantum computation, which completes the big picture of fault-tolerant topological quantum computation.

During the topological operations, the star and plaquette operators are measured in the vacuum region to obtain the error syndrome.
These measurements are implemented using ancillae located on each vertex and face center for the star and plaquette operators, respectively (see Fig.~\ref{fig83}).
The ancilla state is prepared to be $|+\rangle$ and the $\Lambda (X)$ or $\Lambda(Z)$ gates are performed from the ancilla qubit as a control to the four qubits on $\partial f_l$ or $\delta v_k$.
By measuring the ancilla qubit in the $X$-basis, we obtain the eigenvalue of the star or plaquette operator. 
If an error is introduced during the measurement, we obtain an incorrect syndrome value, which has to be dealt with in the topological quantum error correction.

Again, we assume that the error is given as a Pauli error for simplicity.
We also assume that the $X$ and $Z$ errors, which might be correlated in general, are corrected independently.
The former assumption could be justified as follows.
Any Kraus operator of the noise map can be decomposed into a superposition of Pauli operators. 
Such a superposition is collapsed into Pauli errors by the syndrome measurements, which map different Pauli errors (of low weight) into orthogonal subspaces.
Note that, in this case, we have to model the error per gate carefully. 
The latter assumption makes the analysis very simple, but only results in an underestimation of the noise threshold.
Below we only consider correction of the $Z$ errors, but it can be applied straightforwardly to the $X$ errors.

Let $c_1^{e}(t) = \sum _l z^e_l(t) e_l$ be a 1-chain specifying the space-time $Z$ error location, where if $z_l(t)=1$, an $Z$ error occurs on qubit $e_l$ at time step $t$.  
The $Z$ operators on the code state at time $t$ are denoted by $c_1(t) = \sum _l z_l(t)$ (i.e., $Z[c_1(t)]$).
 The state at time $t$ satisfies the following equation of the motion:
\begin{eqnarray}
c_1(t+1) = c_1(t) + c_1^{e} (t+1).
\end{eqnarray}
At each time step $t$, we measure the star operator $B_k(t)$ and obtain the measurement outcome
\begin{eqnarray}
m_k (t) = \left( \bigoplus _{e_l \in \delta v_k} z_l (t)  \right) \oplus z^e_k (t),
\end{eqnarray}
where the bit $z^e_k (t) \in \{ 0,1\}$ indicates an error on the measured syndrome, which we call a measurement error.
To obtain an equation consisting only of errors, we calculate the parity of the syndrome at time steps $t$ and $t-1$,
\begin{eqnarray}
s_k(t) = m_k (t) \oplus m_k(t-1) = \left( \bigoplus _{e_l \in \delta v_k} z^e_l (t) \right) \oplus z^e_k (t) \oplus z^e_k (t-1). 
\end{eqnarray}
Hence, errors on the measured syndrome can be detected by the parity (difference) of the syndrome values at time $t$ and $t-1$.
We redefine $\{ s_k (t) \}$ as an error syndrome of the space-time errors including the measurement error.
\begin{figure}[t]
\centering
\includegraphics[width=130mm]{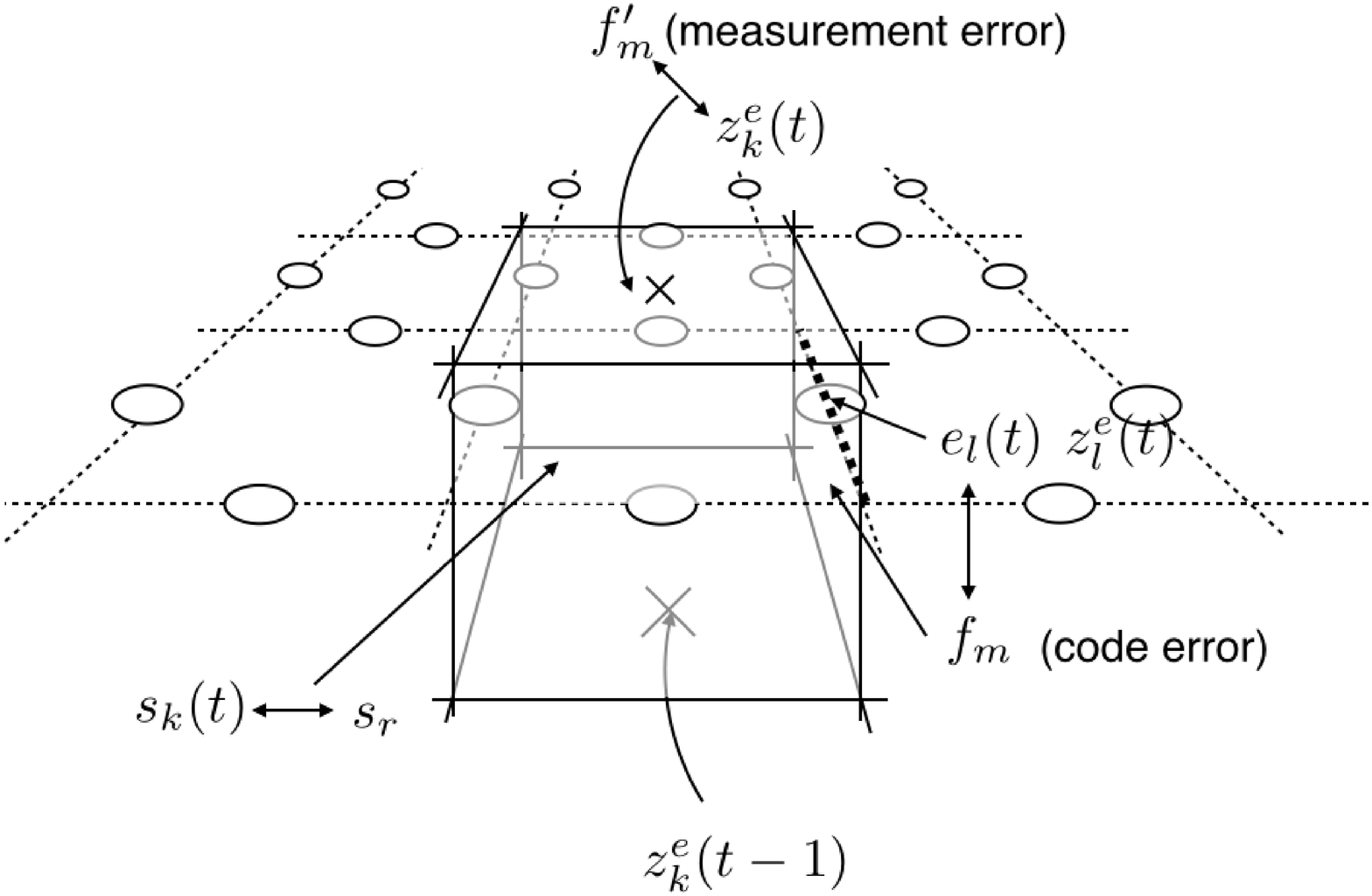}
%
%
\caption{A mapping from the chain complex in 2D with time evolution to another chain complex in 3D.}
\label{fig84}       
\end{figure}

In the case of the perfect syndrome measurement, the error syndrome is given by the boundary of errors $\partial c_1^e$, which allows us to use the MWPM algorithm for the error correction.
What about the case with the faulty syndrome measurement?
Using a chain complex in a 3D manifold, we can again reformulate the error syndrome $\{ s_k (t) \}$ as the boundary of a space-time error chain as shown in Fig.~\ref{fig84}.
We consider a chain complex $\{ C_0, C_1, C_2, C_3\}$ on a cubic lattice.
The basis for the 3-chain is given by a set of cubes $\{ q_r \}$. 
The 3-chain $c_3 = \sum _r z_r q_r \in C_3$ ($z_r\in \{ 0,1\}$) is a linear combination over $\mathbf{Z}_2$.
The dual chain complex $\{ \bar C_0, \bar C_1 , \bar C_2 , \bar C_3\}$ is also defined.
Specifically, $C_i$ and $C_{3-i}$ are identified via the duality transformation.

Let us explain how to embed the space-time error location to a 1-chain in the chain complex in 3D.
The edge $e_l(t)$ and its coefficient $z_l^e(t)$ at time $t$ are mapped into a vertical face $f_m$ and its coefficient $z_m^e$ in the 3D chain complex,
respectively.
The measurement error $z^e(t)$ is mapped to the coefficient $z_{m'}^e$ of a horizontal face $f_{m'}$.
Then, the space-time error location including the measurement errors is described by a 2-chain or a dual 1-chain in the 3D chain complex:
\begin{eqnarray}
c_2 ^e = \sum _m z_m^e f_m, \;\; \leftrightarrow \;\; \bar c_1 ^e = \sum _m z_m^e \bar e_m.
\end{eqnarray}
A syndrome $s_k (t)$ is assigned on each cube $q_r$, or equivalently each dual vertex $\bar v_ r$, in the 3D chain complex, and denoted by $s_r$.
Now we realize that 
\begin{eqnarray}
s_r = \bigoplus _{f_m \in q_r} z_m^e.
\end{eqnarray}
By using the same argument for the 2D case, $s_r =1$ iff $\bar v_r$ belongs to the boundary $\partial \bar c_1^e$ of the dual 1-chain $\bar c_1^e$, specifying the space-time error location.
Thus, we can apply the MWPM algorithm on the cubic lattice to perform topological quantum error correction in space-time. 

To calculate the noise thresholds, we have to model the noise distribution $P(\bar c_1^e)$.
First, we consider the simplest case, where the errors are located on each dual edge $\bar e_l$ of the dual cubic lattice with probability $p$.
This means that $Z$ errors occur on each qubit independently with probability $p$
at each time step.
Moreover, the measured syndrome is flipped with probability $p$.
Such a noise model is called a {\it phenomenological noise model}.
Using the MWPM algorithm, the noise threshold has been estimated to be $2.93\%$~\cite{Wang}.
The error correction problem under the phenomenological noise model can be mapped into a phase transition of the random-plaquette $\mathbf{Z}_2$ gauge model by using the same argument as in Sec.~\ref{Sec:TQECandSpinGlass}.
Specifically, the loop condition for the dual 1-chain (primal 2-chain) in the 3D model can be solved by introducing a gauge spin on each primal edge $\sigma _l$
and defining the dual trivial 1-cycle
$\bar c_1 = \sum _{m} z_m \bar e_m \in {\rm Img}(\bar \partial _2)$
via $(-1)^{z_m} = \prod _{e_l \in \partial f_m} \sigma _l$,
where $\bar e_m$ and $f_m$ are related through the duality relation.
In this way,
the variable on the face center
is provided as a product of the gauge spins on the boundary of the face, which leads to the random-plaquette $\mathbf{Z}_2$ gauge model.
The ordered Higgs and disordered confinement phases correspond to fault-tolerant and non-fault-tolerant regions, respectively~\cite{Dennis}.
The threshold value $2.93\%$ corresponds to the critical point at zero temperature.
The optimal threshold is provided on the Nishimori line and has been estimated to be $3.3\%$ using the Monte Carlo simulation~\cite{Ohno}.

In fault-tolerant quantum computation, we have to consider any source of noise, including the gate operations, during the syndrome measurement. 
As a standard way to model a realistic situation,
suppose that each elementary gate is followed by a depolarizing channel.
This is called a {\it circuit-based noise model}.
More precisely, 
an ideal single-qubit gate is followed by single-qubit depolarizing noise,
\begin{eqnarray}
(1-p_1)\rho + \sum _{A \in \{X,Y,Z\} } \frac{p_1}{3} A \rho  A.
\end{eqnarray}
An ideal two-qubit gate is followed by two-qubit depolarizing noise,
\begin{eqnarray}
(1-p_2)\rho + \sum _{A,B \in \{I, X,Y,Z\} \backslash I\otimes I} \frac{p_2}{15} (A\otimes B) \rho   (A\otimes B).
\end{eqnarray}
A faulty Pauli-basis state preparation is modeled by the state
\begin{eqnarray}
P_A^{+}(p_p) = (1-p_p) \frac{I+A}{2} + p_p \frac{I-A}{2},
\end{eqnarray}
where $A=X,Y,Z$.
A faulty Pauli-basis measurement is modeled by a POVM measurement with POVM elements
\begin{eqnarray}
\{ P_A^{+}(p_m), P_A^{-}(p_m)\},
\end{eqnarray}
where 
\begin{eqnarray}
P_A^{-}(p_m)= (1-p_m) \frac{I-A}{2} + p_m \frac{I+A}{2}.
\end{eqnarray}
Thus, the measurement outcome is flipped with probability $p_m$.

In Refs.~\cite{RaussendorfAnn,RaussendorfNJP,RaussendorfPRL}, the error probabilities were parameterized as $p_1 = p_2 = (3/2)p_p= (3/2)p_m$.
The noise threshold with the MWPM algorithm was obtained by numerical simulations to be $p_2 = 0.75\%$.
In Ref.~\cite{WangFowler}, the threshold value was further improved to $\sim 1\%$ by assigning a weight for each edge in MWPM appropriately, according to the amount of possible errors.
Roughly speaking, the threshold is located where the expectation value of each syndrome becomes $0.7$:
\begin{eqnarray}
\langle (-1)^{s_r} \rangle = 0.7.
\end{eqnarray}
In the phenomenological noise model, the expectation value is provided by
\begin{eqnarray}
\langle (-1)^{s_r} \rangle = (1-2p)^6,
\end{eqnarray}
which hits $0.7$ for $p=2.89\%$; thus being in good agreement with $2.93\%$.
Moreover, in the case of the circuit-based noise model, $p$ is roughly given by $p_p+p_m+4p_2$ for the measurement error, which means that one state preparation, one measurement, and four two-qubit gates are employed.
The error probability $p$ for the code state is given by $4p_2$, meaning that four two-qubit gates are employed in the syndrome measurements of the star and plaquette operators.
This yields
\begin{eqnarray}
\langle (-1)^{s_r} \rangle = (1-2p_p-2p_m-8p_2)^2 (1-8p_2)^4,
\end{eqnarray}
which hits 0.70 for $p_2 = 0.63\%$ with $p_1 = p_2 = (3/2)p_p= (3/2)p_m$.
This is again in a good agreement with the numerical result $0.75\%$.
This simple calculation provides rough estimates of the threshold values, but there is no validity.
If we need a more accurate threshold value, we should perform a numerical simulation by taking noise propagation and correlation into account~\cite{RaussendorfNJP,WangFowler}.

\section{Experimental progress}
Topologically protected quantum computation in 2D has been utilized as a platform to design fault-tolerant architectures for quantum computation.
One promising approach is on-chip monolithic architectures, such as quantum dots \cite{VanMeter,Jones12}, silicon-based nuclear spins~\cite{SiliconSpin}, and superconducting qubits \cite{Fowler1,Fowler2,IBM13,IBM14}, where huge number of qubits are integrated on a single chip and each individual qubit and the interactions between the qubits are manipulated by multiple lasers or electrodes.
Among them, the superconducting qubits system is one of the most promising candidates for implementing topological quantum computation on the surface code, because they can be fabricated on a 2D chip, and all elementary operations have already been demonstrated experimentally~\cite{ChargeQubit,ChargeCNOT,FluxQubit,DispersiveMeas}.
The coherence time and gate fidelity of the superconducting systems have improved rapidly.
An important breakthrough was made by Martinis's group at University of California Santa Barbara in 2014, where single-qubit gates with a fidelity of $99.92\%$ and two-qubit gates with a fidelity of $99.4\%$ were demonstrated on a 1D array of five superconducting transmon qubits~\cite{MartinisLegend,QuantumQuest}.
In addition, repetitive quantum non-demolition measurements were demonstrated on a 1D array of 9 qubits, which improved the fidelity of a state preparation by a factor of 8.5, even when using 9 qubits and faulty two-qubit gates and measurements~\cite{Martinis9qubit}.
This is an important building block of the fault-tolerant quantum computation on the surface code.

Another approach is the distributed modular architecture, where few-qubit local modules are connected with quantum channels mediating interactions between separate modules~\cite{LiBenjamin,FTProb,FujiiDis,Benjamin1,Benjamin2,Monroe}.
The few-qubit quantum module has already been experimentally realized in various physical systems, such as nitrogen-vacancy centers in diamond and trapped ions.
Furthermore, entangling operations between separate local modules have been experimentally demonstrated.
This experimental and theoretical progress will gradually lead us to large-scale quantum computation.
\chapter{Topologically protected MBQC}
\label{Chap:TMBQC}
In this chapter, 
we reformulate topological fault-tolerant quantum computation
explained in the previous chapter
in terms of meausrement-based quantum computation.

\section{Topological cluster state in 3D}
Consider a (primal) cubic lattice $\mathcal{L}$ 
and $\mathbf{Z}_2$ chain complex on it, $\{ C_0, C_1, C_2, C_3\}$,
where
\begin{eqnarray}
c_0 &=& \sum _k z_k v_k \in C_0, \;\;\;
c_1 = \sum _l z_l e_l \in C_1, 
\\
c_2 &=& \sum _m z_m f_m \in C_2, \;\;\;
c_3 = \sum _n z_n q_n \in C_3, \;\;\;
\end{eqnarray}
with $z_k,z_l, z_m, z_n \in \mathbf{Z}_2$.
We also consider a dual cubic lattice $\bar{\mathcal{L}}$
through the relations $v_k \leftrightarrow \bar q_k$,
$e_l \leftrightarrow \bar f_l$, $f_m \leftrightarrow \bar e_m$,
and $q_n \leftrightarrow \bar v_n$.

\begin{figure}[t]
\centering
\includegraphics[width=120mm]{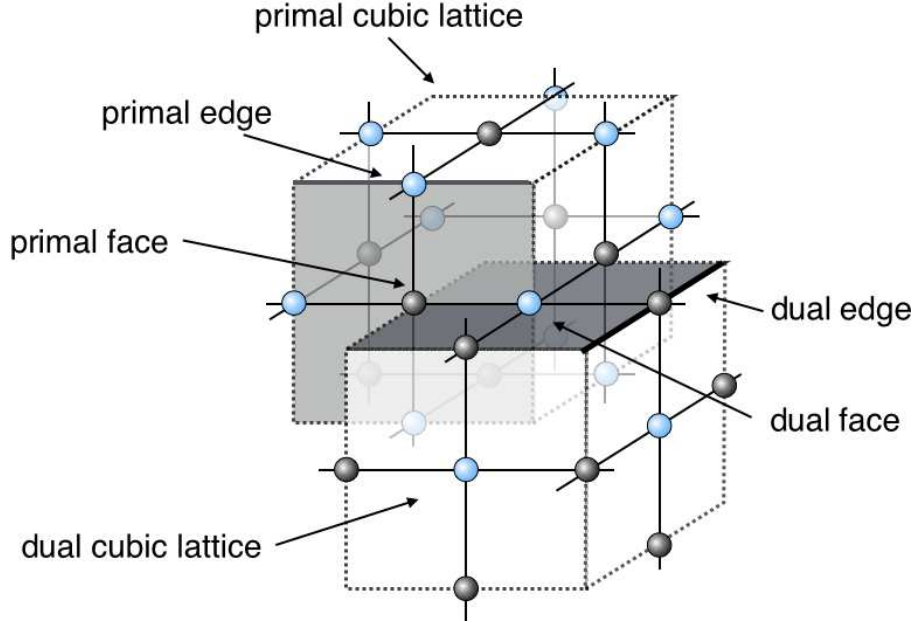}
\caption{A unit cell of the cluster state for topological MBQC. 
The primal and dual cubes, faces, and edges are also shown.
}
\label{figAp01}
\end{figure}

\begin{figure}[t]
\centering
\includegraphics[width=120mm]{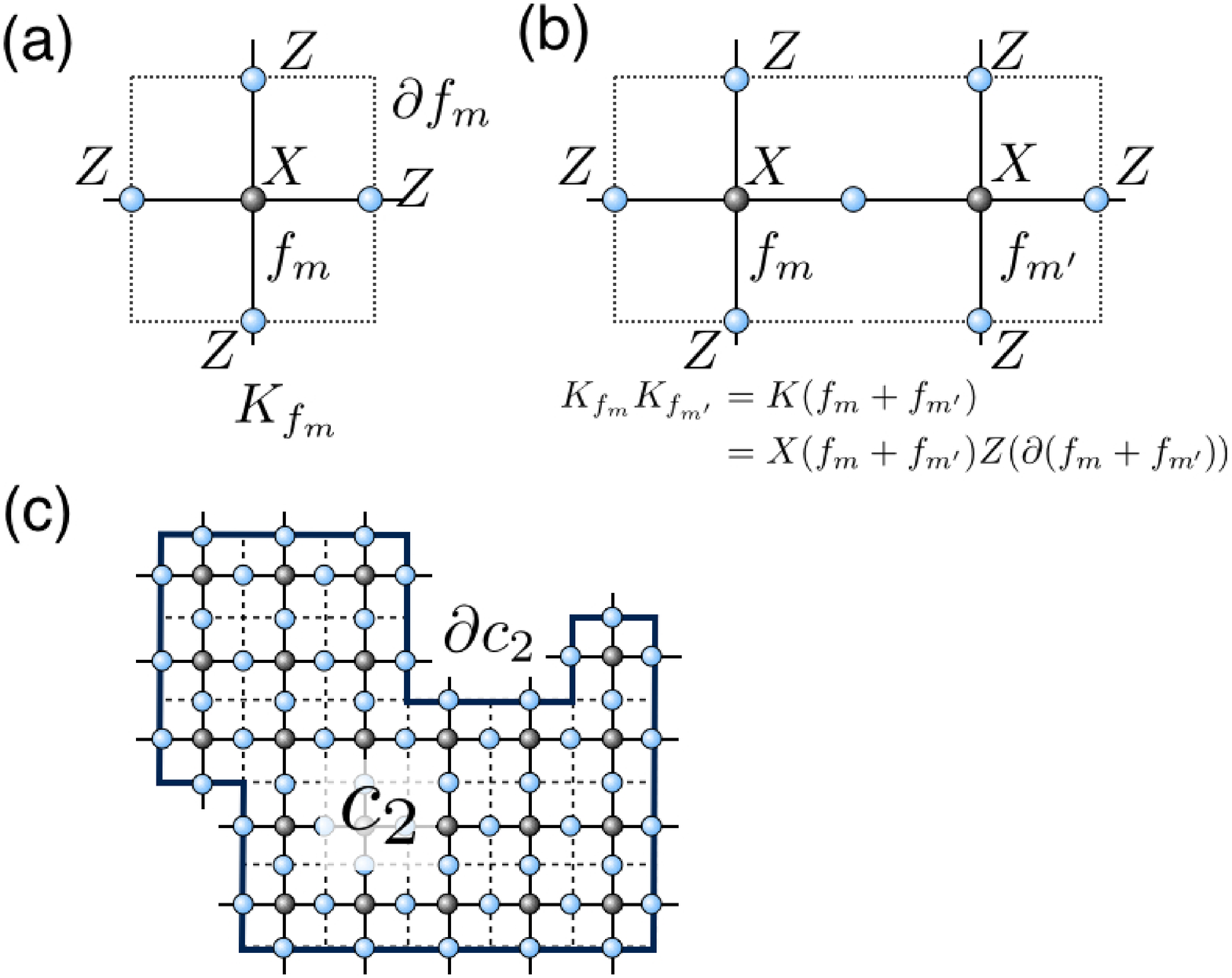}
\caption{(a) A stabilizer operator defined on a primal face.
(b) $K_{f_m} K_{f_{m'}} = X(f_m+f_{m'}) Z(\partial (f_m+f_{m'}))$. 
(c) $K(c_2) \equiv \prod _m K_{f_m}^{z_m} = X(c_2) Z(\partial c_2)$.}
\label{figAp02}
\end{figure}

Qubits are defined on the edges and faces of the primal lattice $\mathcal{L}$ (or equivalently primal and dual edges), as shown in Fig.~\ref{figAp01}.
We define an operator $A(c_i)$ ($i=1,2$) 
in terms of a 1-chain $c_1 = \sum _j z_j e_j$ or a 2-chain $c_2=\sum _j z_j f_j$
as 
\begin{eqnarray}
A(c_i) = \prod _{j} A^{z_j}.
\end{eqnarray}
The stabilizer generators of a 3D cluster state for topologically protected MBQC are defined on the primal and dual elementary faces $f_m$, $\bar f_{m'}$ (see Fig.~\ref{figAp02} (a)): 
\begin{eqnarray}
K_{f_m} &=& X_{f_m} Z(\partial f_m),
\\
K_{\bar f_{m'}} &=& X_{\bar f _ {m'}} Z(\partial \bar f_{m'}).
\end{eqnarray}
A unit cell of the 3D cluster state is shown in Fig.~\ref{figAp01}.
This notion of stabilizer generators of the cluster state is quite useful; it provides a connection between the operators and the chain complex as follows.
By multiplying the two stabilizer operators $K_{f_m}$ and $K_{f_{m'}}$, we have 
\begin{eqnarray}
K_{f_m} K_{f_{m'}} = X(f_m+f_{m'}) Z(\partial (f_m+f_{m'})),
\end{eqnarray}
(see also Fig.~\ref{figAp02} (b)).
By using this property, we can define a stabilizer operator on a 2-chain $c_2$,
\begin{eqnarray}
K(c_2) \equiv \prod _m K_{f_m}^{z_m} = X(c_2) Z(\partial c_2),
\end{eqnarray}
(see also Fig.~\ref{figAp02} (c)).
Furthermore, for the two 2-chains, $c_2$ and $c'_2$, we have
\begin{eqnarray}
K(c_2+c'_2)= K(c_2)K(c'_2).
\end{eqnarray}

Let us see how the 3D cluster state is related to 
topological quantum computation on the surface code 
explained in Chapter~\ref{Chap:TQC}.
Recall the circuit diagrams for the 
syndrome measurements of the plaqette and star operators
in Fig.~\ref{fig83}.
The measurement for the plaquette operator is done by 
applying the CZ gates between 
the ancilla qubit on the face center and the four qubits on the edges.
This operation is the same as generation of the cluster state
stabilized by $K(f_m)$ with a horizontal face $f_m$.
The syndrome measurement for the star operator 
can be done by the CZ gates with the basis change by 
the Hadamard gates.
This corresponds to generation of the cluster state
stabilized by $K(\bar f_l)$ with a horizontal dual face $\bar f_l$.
Moreover, the horizontal edge qubits,
which constitute the surface code, are connected 
by applying the CZ gates vertically
in order to perform the Hadamard gates for the basis change.
In this way, we recover the 3D cluster state
stabilized by $K(f_m)$ and $K(\bar f_l)$
for all primal and dual faces $f_m$ and $\bar f_l$.
Two of three dimensions are employed for the 
spatial degrees of freedom, constituting the surface code.
One is for the time evolution of measurement-based quantum computation.
The measurements are done along the time-like axis,
where even and odd layers,
corresponding to the syndrome measurements 
of the plaquette and star operators respectively, 
together with constitute an elementary time step of the topologically 
protected MBQC.
Below we will see how the topological operations on 
the surface code are translated into a measurement pattern 
of the MBQC on the 3D cluster state.

\section{Vacuum, defect, and singular qubit regions}
The cubic lattice is divided into three regions: the vacuum $\mathcal{V}$, defect $\mathcal{D}$, and singular qubits $\mathcal{S}$ (the detailed definitions are provided later).
In the vacuum region, the topological quantum computation is protected
through topological quantum error correction.
The defect regions are utilized to implement topological quantum computation by braiding defects. 
We have two types of defects: the primal ($D$) and dual ($\bar D$) defects.
For simplicity, we only consider the primal defect.
The extension to the dual case is straightforward by replacing primal by dual in the derivation.
The primal defect $D$ is defined as a set of primal cubes.
The primal face qubits inside the primal defect (except for those on the boundary $\partial D$) are measured in the $Z$-basis to remove the corresponding bonds of the cluster state (or, equivalently, we can prepare the cluster state without those bonds from the beginning).
On the boundary $\partial D$, the primal face qubits are measured in the $X$-basis.
The primal edge qubits belonging to the primal defect (including its boundary) are measured in the $X$-basis.
The measurement pattern for the dual defect is defined similarly.

Unfortunately, only Clifford circuits such as Pauli-basis preparations, measurements, and CNOT gates, are implemented in a topologically protected way.
For universal quantum computation, magic states for the non-Clifford gates are injected on the singular qubits, 
which are always located in-between two defects.
The injections are executed by
measuring the singular qubits in the $Y$- and $(X+Y)/\sqrt{2}$-bases.
These measurements correspond to injections of $(|0\rangle + e^{-i\pi/2}|1\rangle)/\sqrt{2}$ and $(|0\rangle + e^{-i \pi /4 }|1\rangle)/\sqrt{2}$ (up to global phases), 
which are utilized to implement the 
$S=e^{-i (\pi/4) Z}$ and $T= e^{-i (\pi /8) Z}$ gates via gate teleportation,
respectively.
The singular qubits are not topologically protected,
because two defects are made close to each other 
resulting in shortening the code distance.
However, we can obtain clean magic states 
with topologically protected Clifford gates through 
the magic state distillation protocols~\cite{Magic}.
In this way, an arbitrary quantum computation is executed fault-tolerantly.
Below, we will define these three regions more precisely 
and see how topological quantum computation 
is executed in a measurement-based way.

\begin{figure}[t]
\centering
\includegraphics[width=80mm]{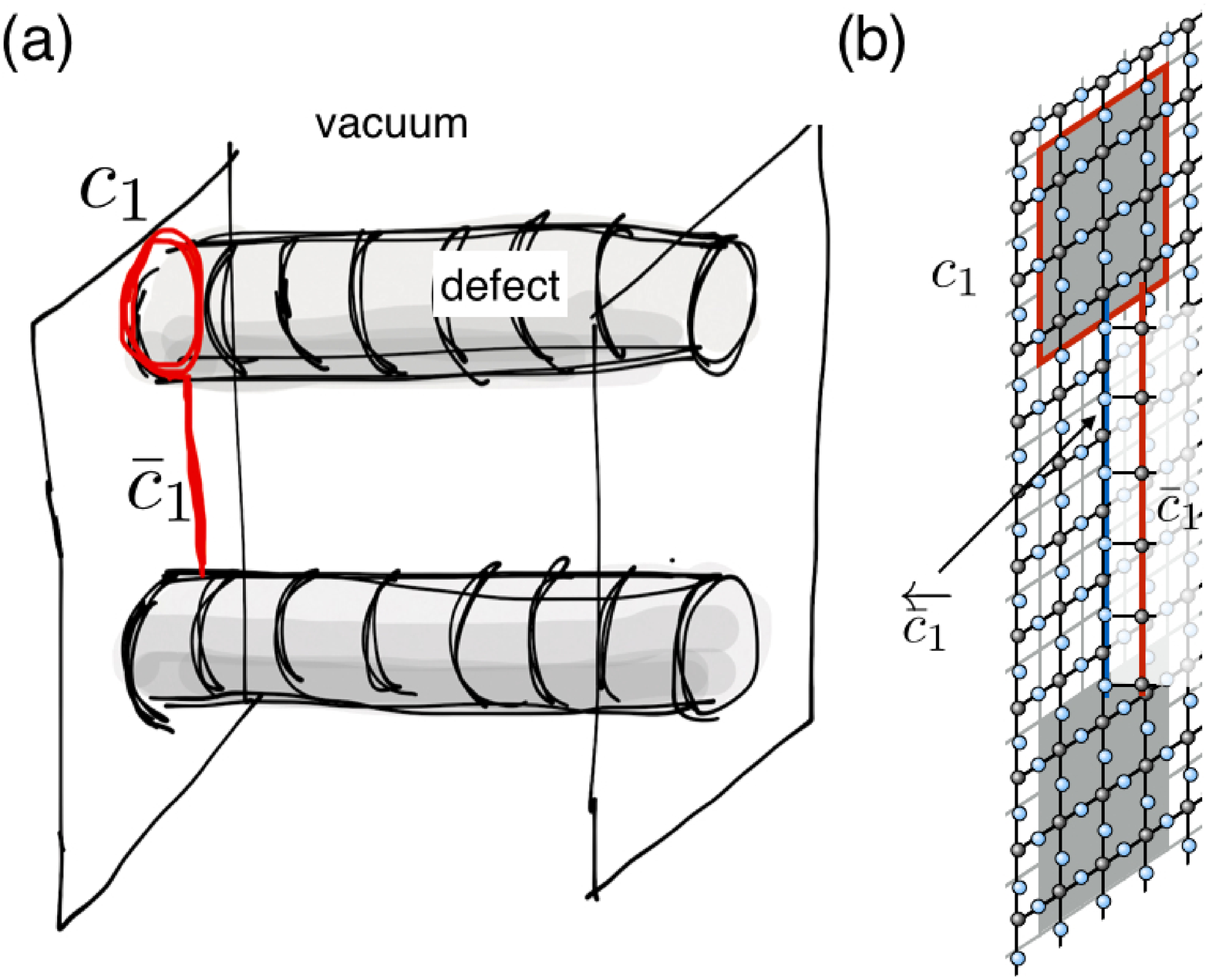}
\caption{(a) A defect pair logical qubit. 
(b) The logical operators $L_Z^{(t)}$ and $L_X^{(t)}$ at time step $t$.}
\label{figAp03}
\end{figure}
\section{Elementary operations in topological MBQC}
\subsection*{Definition of a logical qubit}
The logical information is encoded by using a pair of two defects as shown in Fig.~\ref{figAp03} (a) and (b),
where the measurements are done from left to right.
The logical degree of information at time step $t$ is described a primal 1-chain $c_1$surrounding the defect and a dual 1-chain $\bar c_1$ connecting the two defects, as shown in Fig.~\ref{figAp03} (a).
After measuring qubits up to the $(t-1)$th even and odd layers, according to the measurement patterns presented before, the following two operators may become logical operators, which commute with the stabilizer group of the remaining cluster state and are independent of it:
\begin{eqnarray}
L^{(t)}_Z= Z(c_1), \;\;\; L^{(t)}_X= X(\overleftarrow {\bar c_1}) Z(\bar c_1),
\label{eq:cluster_logical}
\end{eqnarray}
where $\overleftarrow{\bar c_1}$ indicates the dual face qubits on the even layer at time step $t$ 
that are the left neighbor of $\bar c_1$ , as shown in Fig.~\ref{figAp03} (b).
These two operators anticommute with each other and represent a logical qubit.
(If the cluster state ends at the even layer at the time step $t$, then the two logical operators are equivalent to the logical operators of the surface code.
Because there are the time-like CZ gates for the Hadamard gates,
the logical $X$ operator in Eq. (\ref{eq:cluster_logical}) accompanied
by the $Z$ operators.)

\begin{figure}[t]
\centering
\includegraphics[width=80mm]{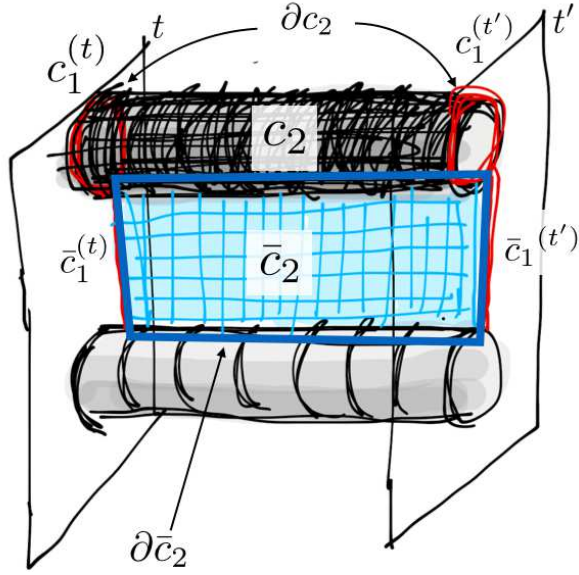}
\caption{A logical identity gate. 
The logical operators at time steps $t$ and $t'$ are related by the correlation surfaces $K(c_2)$ and $K(\bar c_2)$ via the measurements.
}
\label{figAp04}
\end{figure}
\subsection*{Identity gate}
Next, we will see how these logical operators evolve with the measurements.
We consider a correlation surface defined by a primal 2-chain $c_2$ and a dual 2-chain $\bar c_2$ as shown in Fig.~\ref{figAp04}.
A stabilizer operator $K(c_2)$ on the correlation surface $c_2$ surrounding the defect is obtained by multiplying the stabilizer generators $K_{f_m}$ on the primal 2-chain $c_2$:
\begin{eqnarray}
K(c_2 ) \equiv \prod _{m} K_{ f_m}^{z_m} = Z(\partial c_2) X(c_2).
\end{eqnarray}
Similarly, a stabilizer operator on the dual correlation surface $\bar c_2$ is defined by multiplying $K_{\bar f_m}$ on the dual 2-chain $\bar c_2$:
\begin{eqnarray}
K(\bar c_2 ) \equiv \prod _{m} K_{\bar f_m}^{z_m} = Z(\partial \bar c_2) X(\bar c_2).
\end{eqnarray}
Suppose measurements are done from the left to the right, except for those qubits on the final even layer.
Using the correlation surface, we obtain equivalence relations between the logical operators at time $t$ and $t'$:
\begin{eqnarray}
L_Z^{(t)} &\sim& Z(c_1^{(t)}) K(c_2 ) = Z(c_1^{(t')}) X(c_2).
\\
L_X^{(t)} &\sim& X(\overleftarrow {\bar c^{(t)}_1}) Z(\bar c^{(t)}_1) K(\bar c_2) = X(\overleftarrow {\bar c_1}) X(\bar c_2) Z(\bar c^{(t')}_1) 
\end{eqnarray}
Here, $ A\sim B$ means that $A$ and $B$ are equivalent up to a multiplication of the stabilizer operator of the cluster state, meaning that both $A$ and $B$ act the same on the cluster state.
When the qubits on $c_2$ have been measured in the $X$-basis, we obtain 
\begin{eqnarray}
L_Z^{(t)} &\sim& L_Z^{(t+1)},
\\
L_X^{(t)} &\sim& L_X^{(t+1)},
\end{eqnarray}
where we assumed that all measurement outcomes are $+1$ for simplicity.
This relation indicates that the logical information at time step $t$ is propagated to time step $t'$ without any operation, i.e., a logical identity operation.

\begin{figure}[t]
\centering
\includegraphics[width=80mm]{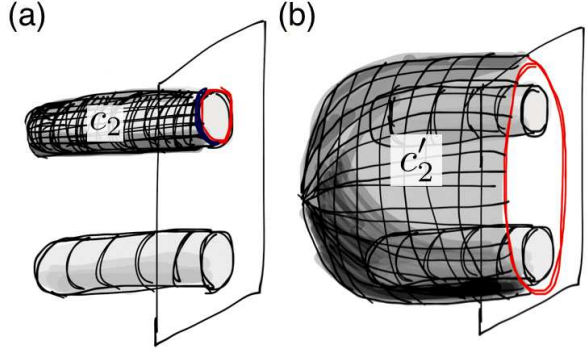}
\caption{(a) A logical $Z$-basis state preparation. 
(b) $K(c'_2)= X(c'_2) L_Z^{(t)} {L'_Z}^{(t)}$}
\label{figAp05}
\end{figure}
\subsection*{State preparation and measurement}
Next, we consider how the logical qubit is prepared from the vacuum.
To prepare the eigenstate of $L_Z^{(t)}$, we utilize the defect shown in Fig.~\ref{figAp05} (a).
By considering the correlation surface $c_2$, we obtain
\begin{eqnarray}
K(c_2) = X(c_2) L_Z^{(t)}.
\end{eqnarray}
Because $X(c_2)$ commutes with the measurements, the state at time step $t$ is stabilized by $L_Z^{(t)}$, and hence a logical $Z$-basis state is prepared.
Considering another surface $c'_2$, shown in Fig.~\ref{figAp05} (b), the state at time step $t$ is also stabilized by $Z(\partial c'_2)=L_Z^{(t)} {L'_Z}^{(t)}$.
Thus the pair of the defects is appropriately encoded into the code space.
Both $L_Z^{(t)}$ and ${L'_Z}^{(t)}$ act equivalently as logical $Z$ operators.

\begin{figure}[t]
\centering
\includegraphics[width=55mm]{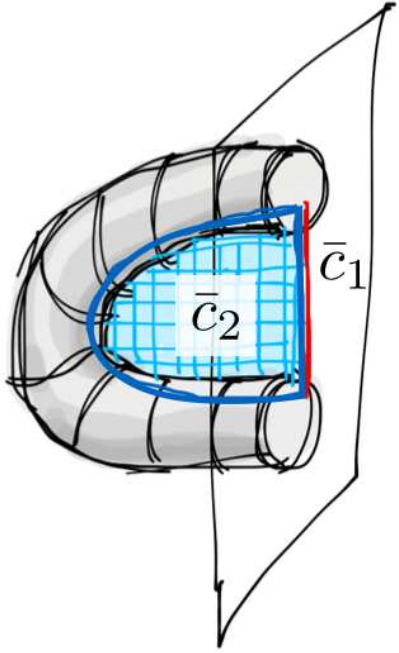}
\caption{A logical $X$-basis state preparation.}
\label{figAp06}
\end{figure}
Next, we consider the defect shown in Fig.~\ref{figAp06}.
Considering a correlation surface $\bar c_2$, we obtain
\begin{eqnarray}
K(\bar c_2) = Z(\partial \bar c_2) X(\bar c_2).
\end{eqnarray}
After the measurements, the state at time step $t$ is stabilized by $L_X^{(t)}= X(\overleftarrow{\bar c_1}) Z(c_1)$, where $c_1$ is a 1-chain on the $t$th even layer connecting the two defects.
Thus, a logical $X$-basis state is prepared.
Again, the logical state is stabilized by $Z(c_1)Z(c'_1)$, with $c_1$ and $c'_1$ being a cycle surrounding each defect.
Hence, we can choose either $L_Z^{(t)}=Z(c_1)$ or $L_Z^{(t')}=Z(c'_1)$ to serve as the logical operator.
The logical measurements of the defect pair qubits can be done with the same defects as the state preparations, but by reversing the time-like direction.

\begin{figure*}
\centering
\includegraphics[width=140mm]{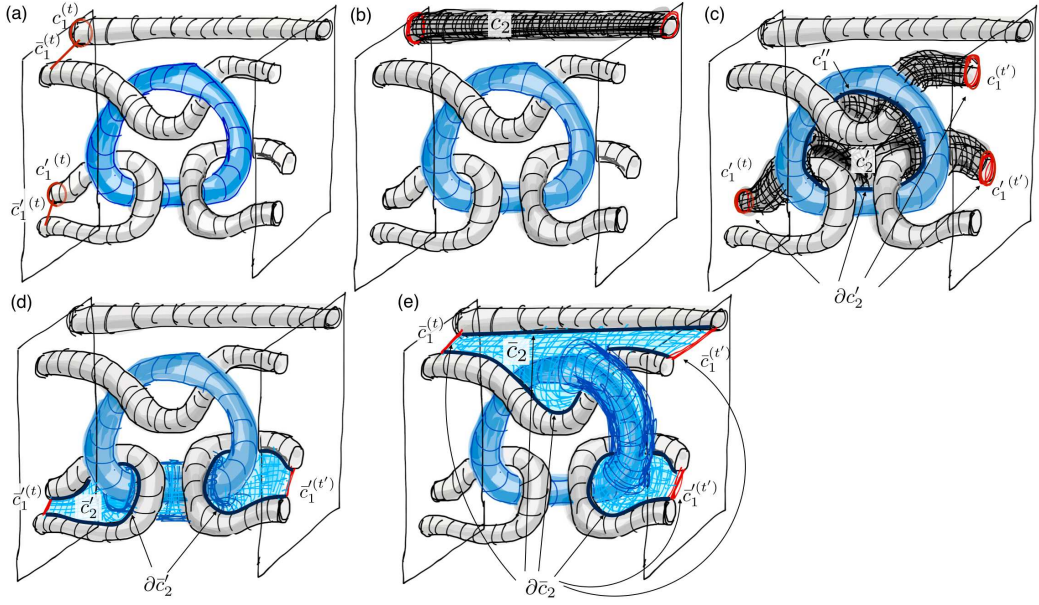}
\caption{(a) A diagram for defect braiding for a logical CNOT gate.
(b-e) Time evolutions of the logical operators and the corresponding correlation surfaces.}
\label{figAp09}
\end{figure*}
\subsection*{CNOT gate by braiding}
Let us consider primal defects braiding around a dual defect as shown in Fig.~\ref{figAp09} (a).
Similarly to the previous case, we calculate the time evolution of logical operators by the measurements.
The state at time step $t$ is described by $\{ L_Z^{(t)}, L_X^{(t)}\}$ and $\{ {L'_Z}^{(t)}, {L'_X}^{(t)}\}$ corresponding to $\{ c_1^{(t)}, \bar c_1^{(t)}\}$ and 
$\{ {c'_1}^{(t)}, {{\bar c}'_1}{}^{(t)} \}$, respectively.
We first consider a correlation surface $c_2$ with respect to $c_1^{(t)}$, as depicted in Fig.~\ref{figAp09} (b). 
Similarly to the identity gate, $L_Z^{(t)}$ is transformed into ${L'}_Z^{(t)}$.
An interesting thing happens when we consider the correlation surface $c'_2$ with respect to ${c'_1}^{(t)}$, as shown in Fig.~\ref{figAp09} (c).
The stabilizer operator on $c'_2$ is given by 
\begin{eqnarray}
K(c'_2) &=& Z(\partial c'_2) X( c'_2)
\nonumber \\
&=& Z({c'_1}^{(t)}) Z({c_1}^{(t')}) Z({c'_1}^{(t')}) Z(c''_1) X(c'_2) ,
\end{eqnarray}
where $c''_1$ is a cycle in the dual defect of a loop as shown in Fig.~\ref{figAp09} (c).
Then after the measurements, we obtain an equivalence relation,
\begin{eqnarray}
L_Z^{(t)} \sim {L}_Z^{(t')} {L'}_Z^{(t')}.
\end{eqnarray}
Note that inside the dual defect region, the dual face (primal edge) qubits are measured in the $Z$-basis, and hence we can obtain the eigenvalue of $Z(c''_1)$.
After a similar argument using a defect surface $\bar c_2$ and $\bar c'_2$ with respect to the dual 1-chain $\bar c_1$ and $\bar c'_1$, shown in Fig.~\ref{figAp09} (d) and (e), respectively, we obtain 
\begin{eqnarray}
L_X^{(t)} &\sim& {L}_X^{(t)},
\\
L_X^{(t')} &\sim& {L}_X^{(t')} {L'}_X^{(t')}.
\end{eqnarray}
These relations between the logical operators at time steps $t$ and $t'$ are equivalent to those for the CNOT gate.
Thus, the defect braiding in Fig.~\ref{figAp09} results in a logical CNOT gate.
Now we realize that the correlation surface 
introduced in Chapter~\ref{Chap:TQC} 
as a trajectory of the logical operator 
corresponds to the correlation surface defined by the 
stabilizer operator of the cluster state.

\subsection*{A singular qubit injection for magic state distillation}
\begin{figure}[t]
\centering
\includegraphics[width=70mm]{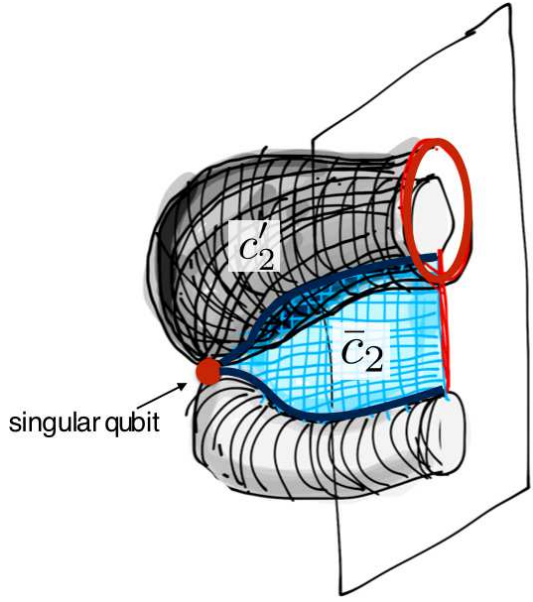}
\caption{A state injection on the singular qubit.}
\label{figAp10}
\end{figure}

So far, we have shown that the Clifford circuits, Pauli-basis preparation, measurements, and CNOT gate, can all be implemented in a topological way.
Unfortunately, these operations are not enough to generate universal quantum computation.
To implement universal quantum computation, we inject $Y$- and $(X+Y)/\sqrt{2}$-basis states by measuring singular qubits in the $Y$- and $(X+Y)/\sqrt{2}$-bases, respectively, as shown in Fig.~\ref{figAp10}.
Let us see how this measurement works.
Similarly to the previous case, we have two correlation surfaces $c_2$ and $\bar c_2$:
\begin{eqnarray}
K(c_2) &=& Z(\partial c_2) X(c_2) = Z(\partial c_2) X(c_2 \backslash s) X_s,
\\
K(\bar c_2) &=& Z(\partial \bar c_2) X(\bar c_2 ) =Z_s Z(\partial \bar c_2\backslash s) X(\bar c_2 ),
\end{eqnarray}
where $A_s$ is a Pauli operator on the singular qubit and $[\cdot]\backslash s$ indicates a chain with a removal of an element corresponding to the singular qubit.
Suppose the singular qubit is measured in the $Y$-basis.
After the measurements, the state at time step $t$ is stabilized by
\begin{eqnarray}
K(c_2) K(\bar c_2) \simeq L_X^{(t)}L_Z^{(t)} \equiv L_Y^{(t)}.
\end{eqnarray}
Thus, a logical $Y$-basis state is prepared.
When the singular qubit is measured in the $(X+Y)/\sqrt{2}$-basis, the state at time step $t$ is stabilized by
\begin{eqnarray}
[ K(\bar c_2) + K(c_2) K(\bar c_2)]/\sqrt{2} \simeq (L_X^{(t)}+ L_Y^{(t)})/\sqrt{2},
\end{eqnarray}
which means that a logical $(X+Y)/\sqrt{2}$-basis state is prepared.

These states are utilized to implement $S$, $T$, $HSH$, and $HTH$ gates using gate teleportation with the CNOT gate.
These gates form a universal set of gates.

\section{Topological quantum error correction in 3D}
Next, we will see how topological quantum error correction is
done in 3D.
Indeed, in the 3D case, 
the argument for the noisy syndrome measurements 
made in Chapter~\ref{Chap:TQC}
becomes more simple as follows.
All measurements in the vacuum region are done in the $X$-basis.
We consider a stabilizer operator on a unit primal cube $q$,
\begin{eqnarray}
K(\partial q_n) = \prod _{f_m \in \partial q_n } X_{f_m},
\end{eqnarray}
where there is no $Z$ operator due to $\partial \circ \partial q_n =0$.
This implies that, if there is no error, the parity of each six $X$-basis measurement outcomes on the primal cube is always even.
The errors are described by using a dual 1-chain $E=Z(\bar c_1)$.
At a unit cube $q_n$ belonging to $\partial \bar c_1$, 
we have 
$|q_n \cap \partial \bar c_1 |={\rm odd}$.
(Recall that the primal $3$-chain and the dual $0$-chain are identified.)
From a set of odd parity cubes $\partial \bar c_1$, we estimate the actual location of errors $E' = Z(\bar c'_1)$ such that $\partial \bar c_1 = \partial \bar c'_1$.
If the total of the actual and estimated error chains $\bar c_1 + \bar c'_1$ results in a trivial cycle, meaning that there is no defect inside the cycle, it can be contracted and removed by a continuous deformation.
If the total of the actual and estimated error chains $\bar c_1 + \bar c'_1$ results in a nontrivial cycle, meaning a cycle winding around a defect, $EE'=Z(\bar c_1 + \bar c'_1)$ may result in a logical operator. 
In such a case, the topological error correction has failed.
This property is completely the same as the topological quantum error correction
under faulty syndrome measurements 
argued in Sec.~\ref{Sec:TopoFaultTolerance}.
If the error probability is smaller than a constant value (the threshold), the failure probability of the topological quantum error correction decreases exponentially in the characteristic size and distance of the defects.

Inside the defect region, the face qubits are measured in the $Z$-basis.
Especially, the $Z$-basis measurement outcomes near the defect boundary are employed to evaluate the correlation surface.
Note that these $Z$-basis measurements and the removal of the corresponding bonds of the cluster state can instead be done by generating a cluster without connecting the corresponding bonds in advance.
In such a case, the errors on the $Z$-basis measurements do not appear.
We can obtain an additional parity $K_{f_m}$ and $K_{\bar f_{m'}}$ at the primal and dual faces on the boundary of the defects, respectively.
If the errors on the face qubits ${f_m}$ and $\bar f_{m'}$ are suppressed, the errors on the boundary are reduced into errors on a toric code on a 2D surface, $\partial D$ or 
$\partial \bar D$.
Again, if the error probability is sufficiently smaller than a constant threshold value, we can correct it faithfully.

The $X$- and $Z$-basis state preparations and measurements, and the CNOT gate obtained by braiding, are topologically protected because we can execute these topological operations by keeping the defect size and distance larger than an arbitrarily large constant length.
Unfortunately, through the state injection, we shrink the defect size into an elementary unit cell, where the defect size and distance become very small.
Thus, the topological protection is broken down around the singular qubit (see Fig.~\ref{figAp10} (a)). 
There will also be lower weight errors, which effectively increase the logical error probability on the injected logical states.
However, noisy injected states can be purified by using the topologically protected Clifford gates, the so-called magic state distillation.
The $Y$- and $(X+Y)/\sqrt{2}$-basis states are distilled by using the 7-qubit Steane and 15-qubit Reed-Muller codes, respectively
as explained.
The distilled states are utilized to implement non-Clifford gates via gate teleportation, as seen before.
In this way, universal quantum computation is executed with arbitrary accuracy.

\section{Applications for MBQC on thermal states}
Topologically protected MBQC in 3D is useful to study the quantum computational capacity of quantum many-body states at finite temperature.
Consider the stabilizer Hamiltonian of the 3D cluster state for topological MBQC \cite{MBQC_PRA,RaussendorfBravyi,BarrettThermal}:
\begin{eqnarray}
H_{\rm fc} = -J \left[ \sum _{f} K(f) + \sum_{\bar f}  K(\bar f) \right].
\end{eqnarray}
The thermal state at temperature $T=1/(\beta J)$ is given by
\begin{eqnarray}
\rho _{\rm fc} &=& e^{ - \beta H_{\rm fc} } / {\rm Tr}[e^{ - \beta H _{\rm fc} }].
\end{eqnarray}
Using a unitary operator $U_{CZ}= \prod _{(f_m,\bar f_l ) } \Lambda_{f_m,\bar f_l}(Z)$, consisting of $CZ$ gates on all nearest-neighbor two qubits, the thermal state can be mapped into the thermal state of an interaction-free spin model:
\begin{eqnarray}
U_{CZ} \rho _{\rm fc} U_{CZ}^{\dag}
=
e^{ - \beta H_{\rm f}   } / {\rm Tr}[e^{ - \beta H _{\rm f} }] ,
\end{eqnarray}
where 
\begin{eqnarray}
H_{\rm f} \equiv  -J \sum _{i} X_i = U_{CZ} H_{\rm fc} U_{CZ}^{\dagger}.
\end{eqnarray}
Because $H_{\rm f}$ is an interaction-free Hamiltonian, the stabilizer Hamiltonian, which we will call the {\it free cluster Hamiltonian}, does not undergo any thermodynamic phase transition.

The thermal state of the free cluster Hamiltonian is given as a product state of the single-spin density matrix:
\begin{eqnarray}
\rho _{\rm f} &=& e^{- \beta H_{\rm f}}/{\rm Tr}[e^{- \beta H_{\rm f}}] = \prod _{i} e^{\beta J X_i}/{\rm Tr}[e^{\beta J X_i}]
\\
&=& \prod _{i} \mathcal{E}_i(p_{\beta J}) (|+\rangle \langle +|)^{\otimes n},
\end{eqnarray}
where 
\begin{eqnarray}
\mathcal{E}_i(p) = (1-p) \rho + p Z_i \rho Z_i,
\end{eqnarray}
and $p_{\beta J} = e^{- 2 \beta J }/(1+e^{- 2 \beta J })$.
Because $\mathcal{E}_i$ and $U_{CZ}$ commute, the thermal state of $H_{\rm fc}$ is rewritten as
\begin{eqnarray}
\rho _{\rm fc} = U_{CZ} \rho _{\rm f} U_{CZ} ^{\dag} =\left[ \prod _{i} \mathcal{E}_i (p_{\beta J}) \right] U_{CZ} (|+\rangle \langle +|)^{\otimes n} U_{CZ}^{\dag}
=\left[ \prod _{i} \mathcal{E}_i (p_{\beta J}) \right]  |\Psi _{3D} \rangle \langle \Psi _{3D} |,
  \nonumber
\\
\end{eqnarray}
where $|\Psi _{3D}\rangle$ is the ground state of $H_{\rm fc}$, i.e., the 3D cluster state.
This means that the thermal state is given as an ideal 3D cluster state, followed by an independent dephasing for each qubit with probability 
$p_{\beta J}=e^{- 2 \beta J }/(1+e^{- 2 \beta J })$.
From the argument made in the previous section, if $p \leq 2.9-3.3\%$ and hence $T = 1/(\beta J) \leq 0.57-0.59$, we can then perform universal quantum computation reliably on the thermal state at a finite temperature, where the errors originating from the thermal excitations are corrected by the topological quantum error correction.
On the other hand, in the high temperature limit $T=1/(\beta J) \rightarrow \infty$, the thermal state is given by a completely mixed state, and hence MBQC on it can be simulated classically.

A projected-entangled-pair state (PEPS)\index{projected-entangled-pair state, PEPS} picture~\cite{PEPS,RaussendorfBravyi,BarrettThermal} allows us to obtain a lower bound for the possibility of a classical simulation.
In the PEPS picture, the 3D cluster state is described as follows.
A maximally entangled pair $|\psi _{\rm MES}\rangle \equiv (|0\rangle |+\rangle + |1\rangle |-\rangle)/\sqrt{2}$ is shared on each bond.
On each site consisting of halves of the entangled pair, a projection 
\begin{eqnarray}
|0\rangle \langle 00..0| + |1\rangle \langle 11..1|
\end{eqnarray}
is performed with an appropriate normalization.
The resultant state is the 3D cluster state.
Because the projection and the $Z$ error $\mathcal{E}_i$ commute, the effect of the thermal excitations on the shared entangled state can be determined beforehand:
\begin{eqnarray}
\rho_{\rm bond} =\mathcal{E}_a(p)\mathcal{E}_b(p) |\psi _{\rm MES}\rangle \langle \psi _{\rm MES}|
\end{eqnarray}
If $p \geq (2-\sqrt{2})/2$, the decohered entangled pair $\rho _{\rm bond}$ becomes a separable state.
If two bonds per site are made separable, the 3D cluster state becomes a separable state.
A sampling on such a resource state can be simulated efficiently classically~\cite{BarrettThermal,FujiiTamate}.
In this case, the $Z$ error probability per site has to be $p_{\beta J} \geq \sqrt{2}-1$, i.e., $T= 1/(\beta )=5.77J$.
The true critical temperature $T_c$ between the classically simulatable and universal quantum computational phases is located in the range $0.59J<T_c < 5.77J$.
Note that this model exhibits a transition of the computational capability, while there is no thermodynamic phase transition in the physical system~\cite{RaussendorfBravyi,BarrettThermal}.
\begin{figure}[t]
\centering
\includegraphics[width=100mm]{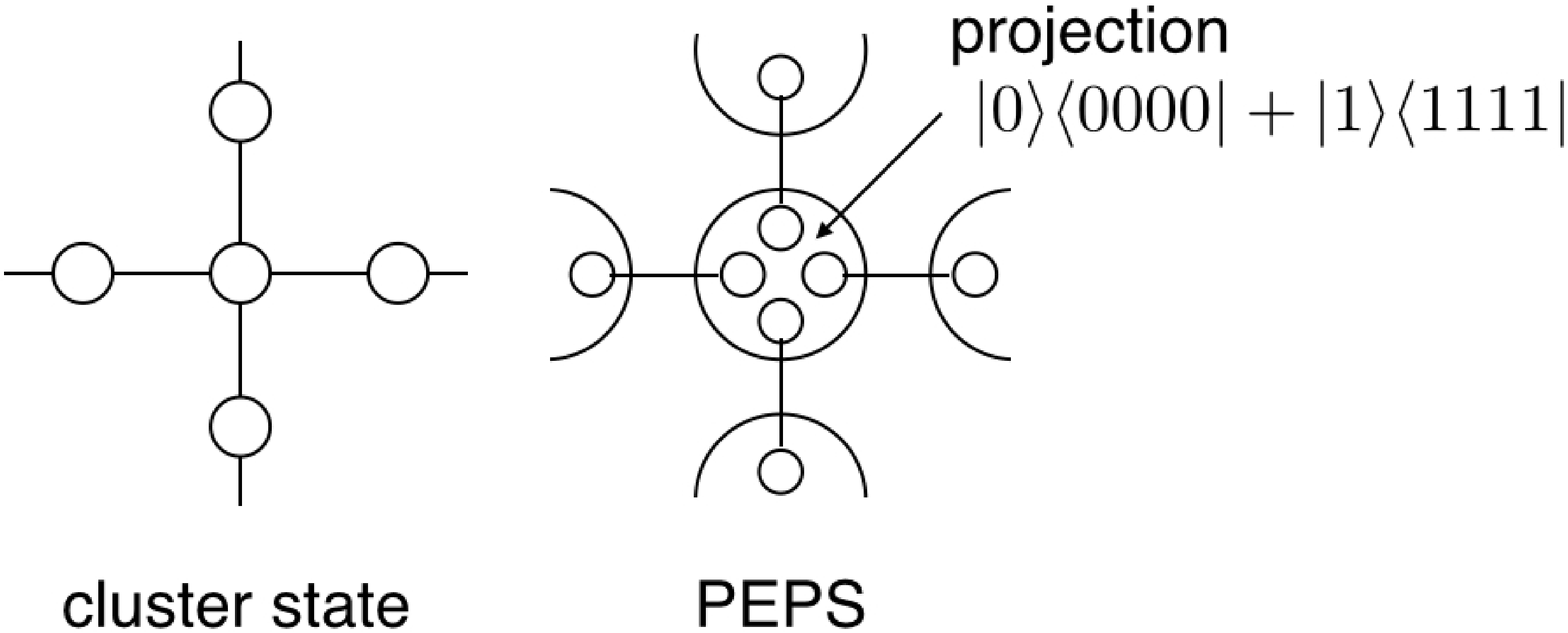}
\caption{A PEPS picture of the cluster state.}
\label{fig92}
\end{figure}

In classical information processing, a thermodynamic phase transition, or, more precisely, an ordered phase below a critical temperature is utilized for robust information storage in magnetic storage devises.
While there is no such long range order in the previous model, it is natural to ask whether or not a long range ordered phase is useful to enhance the measurement-based quantum computation on many-body thermal states for quantum information processing.
To address this issue, Fujii {\it et al.} proposed an interacting cluster Hamiltonian \cite{FujiiNakata},
\begin{eqnarray*}
H_{\rm ic} = -J \sum _{\langle f,\bar f \rangle } K(f)K(\bar f).
\end{eqnarray*}
Because interactions are introduced between the cluster stabilizers, this model is mapped by $U_{CZ}$ into an Ising model on a 3D lattice:
\begin{eqnarray*}
H_{\rm Ising} = U_{CZ}H_{\rm ic} U_{CZ}^{\dag} 
= -J \sum _{\langle f,\bar f \rangle } X_f X_{\bar f}.
\end{eqnarray*}
Thus, it undergoes a thermodynamic phase transition at a finite temperature.
The degenerate ground states
\begin{eqnarray}
U_{CZ}|+\rangle^{\otimes n} \textrm{ and } U_{CZ}|-\rangle^{\otimes n}
\end{eqnarray} 
are also the 3D cluster states, up to the simultaneous spin flipping due to the global symmetry.
Because the eigenvalues of the cluster stabilizer have a long range order (they are likely to be aligned in the same direction) in a ferromagnetic ordered phase, the topologically protected MBQC on the symmetry-breaking thermal state has a special robustness against the thermal excitations.
In Ref.~\cite{FujiiNakata}, topological quantum error correction of this model is mapped to a correlated random plaquette $\mathbf{Z}_2$-gauge model in 3D, where the disorder in the signs of the plaquettes has an Ising-type correlation.
By using this property and the gauge transformation on the Nishimori line \cite{NishimoriBook}, Fujii {\it et al.} showed that the critical temperature of this model, and hence the threshold temperature for topological protection, is equal to the critical temperature of the 3D Ising model, which is the unitary equivalent model of the interacting cluster Hamiltonian.
This means that the critical temperatures for the topological protection and the thermodynamic phase transition of the underlying physical system coincides exactly.
Due to this fact, we can improve the threshold temperature for topological protection by one order of magnitude.

\begin{figure}[t]
\centering
\includegraphics[width=130mm]{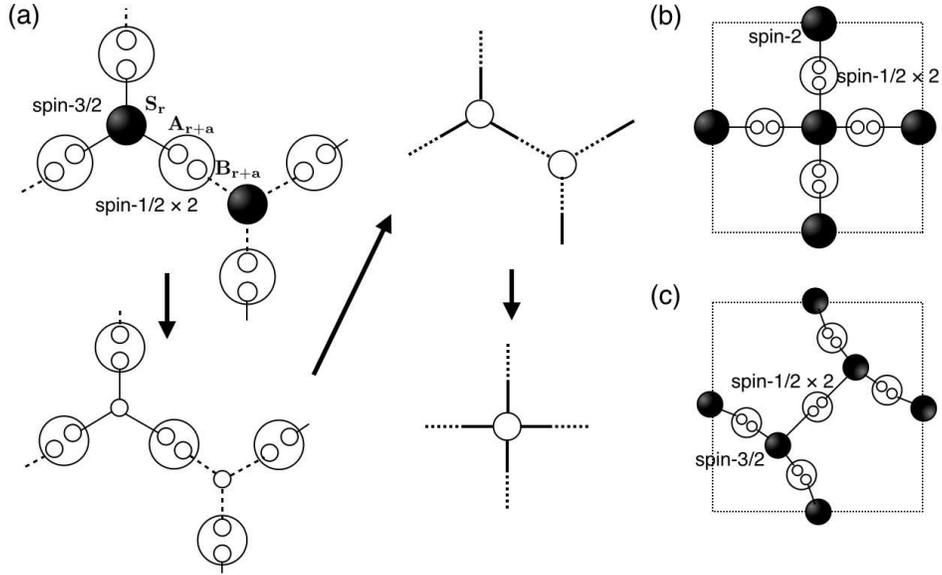}
\caption{(a) A system consisting of spin-3/2 particles and composite particles of two spin-1/2.
After the filtering operation on the ground state, we obtain cluster states.
(b) A system consisting of spin-2 particles and composite particles of two spin-1/2 for the 3D cluster state.
(c) A system consisting of spin-3/2 particles and composite particles of two spin-1/2 for the 3D cluster state.
}
\label{fig93}
\end{figure}

While the above Hamiltonian employs multi-body interactions, the 3D cluster state can be generated from the thermal states of a nearest-neighbor two-body Hamiltonian for spin-3/2 and composite spin-1/2 particles via local filtering operations \cite{LiKwek,FujiiMoriTher}.
Let us consider a system consisting of a spin-3/2 particle located at site $\mathbf{r}$ and a composite particle of two spin-1/2 particles located at the nearest-neighbor site $\mathbf{r}+\mathbf{i}$, with $\mathbf{i}=\mathbf{1},\mathbf{2},\mathbf{3}$, as shown in Fig~\ref{fig93} (a).
The Hamiltonian is given by
\begin{eqnarray*}
 H&=& \Delta  \sum _{\mathbf{r}} \vec{S} _{\mathbf{r}} \cdot (\vec{I}_{ \mathbf{r} + \mathbf{1}}+\vec{I}_{\mathbf{r} +\mathbf{2}}+ \vec{I}_{\mathbf{r} +\mathbf{3}} )
\end{eqnarray*}
where $\vec{S}_{\mathbf{r}} \equiv (S_{\mathbf{r}}^x,S_{\mathbf{r}}^y,S_{\mathbf{r}}^z)$ is the spin-3/2 operator of the center particle at the position $\mathbf{r}$ and $\vec{I}_{\mathbf{r} + \mathbf{a}} = \vec A _{\mathbf{r} + \mathbf{a}} $ or $\vec B_{\mathbf{r}+ \mathbf{a}}$ depending on the interaction types (line or dash). 
Here, $\vec A _{\mathbf{r} + \mathbf{a}}\equiv (A_{\mathbf{r} +\mathbf{a}}^x,A_{\mathbf{r} +\mathbf{a}}^y,A_{\mathbf{r}+\mathbf{a}}^z)$ and 
$\vec B_{\mathbf{r}+ \mathbf{a}}\equiv (B_{\mathbf{r} +\mathbf{a}}^x,B_{\mathbf{r} +\mathbf{a}}^y,B_{\mathbf{r}+\mathbf{a}}^z)$ are two independent spin-1/2 operators on the composite particle at the position $\mathbf{r} + \mathbf{a}$ ($\mathbf{a} = \mathbf{1} , \mathbf{2}, \mathbf{3}$).
The above Hamiltonian $H$ can be reformulated as
\begin{eqnarray*}
H
&=&  \sum _{\mathbf{r}} H_{\mathbf{r}} =\Delta /2 \sum _{\mathbf{r}}
( \vec{T}^2  _{\mathbf{r}} - \vec{S}^2 _{\mathbf{r}}  -  \vec {I}_{\mathbf{r}} ^2)
 \end{eqnarray*}
where $\vec{I}_{\mathbf{r}} \equiv  \vec{I} _{\mathbf{r} + \mathbf{1}}+\vec{I} _{\mathbf{r} + \mathbf{2}}+\vec{I} _{\mathbf{r} + \mathbf{3}}$ and 
$\vec T _{\mathbf{r}} \equiv \vec S_{\mathbf{r}} + \vec I _{\mathbf{r}}$.
The ground state $|G \rangle = \bigotimes _{\mathbf{r}} |g_{\mathbf{r}} \rangle$ is given by $T_{\mathbf{r}} =0$, $S_{\mathbf{r}} =3/2$, and $I_{\mathbf{r}} =3/2$, where $L_{\mathbf{r}}(L_{\mathbf{r}}+1)$ ($L=T,S,I$) is the eigenvalue of the operator $\vec L_{\mathbf{r}}^2$.
Each center particle in the ground state $|G\rangle$ is filtered by using the POVM measurement: 
\begin{eqnarray}
\{ F^{\alpha}= ({S_{\mathbf{r}} ^\alpha} ^2 - 1/4)/\sqrt{6} \} \;\;\; (\alpha =x,y,z).
\end{eqnarray}
If the measurement outcome is $\alpha =z$, we obtain a four-qubit GHZ (Greenberger-Horne-Zeilinger) state
~\cite{GHZ} as the post-POVM measurement state: 
\begin{eqnarray*}
|{\rm GHZ} ^4_{\mathbf{r}} \rangle  \equiv \frac{1}{\sqrt{2}} (|\tilde 0 +++\rangle + |\tilde 1 ---\rangle) ,
\end{eqnarray*}
where $|\tilde 1 \rangle  $ and $|\tilde 0 \rangle $ are eigenstates of $S^z$ with eigenvalues $+3/2$ and $-3/2$, respectively, and $|\pm\rangle$ are the eigenstates of $A^{z}$ or $B^{z}$ with eigenvalues $\pm 1$, respectively.
Even if we obtain other outcomes, we can transform the post-POVM measurement state to $|{\rm GHZ} ^4_{\mathbf{r}}\rangle$ by local operations.
The four-qubit GHZ state is subsequently used to construct the 2D honeycomb cluster state, which is a universal resource for MBQC, by measuring the operators $A^z \otimes B^x$ and $A^x \otimes B^z$ on the bond particle
as shown in Fig.~\ref{fig93} (a).

In the case of finite temperature, we have the thermal state $\bigotimes _{\mathbf{r}} \rho _{\mathbf{r}}$ with 
$\rho _{\mathbf{r}} \equiv e^{ - \beta  H_{\mathbf{r}} } /\mathcal{Z} $ instead of the ground state, where $\mathcal{Z}$ indicates the partition function and $\beta = T ^{-1}$ for a temperature $T$.  
Then, the GHZ state becomes a noisy, say thermal, GHZ state, 
$\sigma _{\mathbf{r}} \equiv F^{\alpha } \rho _{\mathbf{r}} {  F^{\alpha}} ^{\dag}/{\rm Tr}[ F^{\alpha } \rho _{\mathbf{r}}{  F^{\alpha}} ^{\dag}]$.
In the low temperature case, the thermal GHZ state 
is calculated, in the leading order, to be
 $\mathcal{E}_4 (|{\rm GHZ} ^4_{\mathbf{r}} \rangle  \langle {\rm GHZ} ^4_{\mathbf{r}}| )$ with 
\begin{eqnarray}
\mathcal{E}_4 &=& (1- q_1 - 3q_2 - 3q_3 ) [I] + q_1 [Z_{\mathbf{r}}  ]
\nonumber \\&&
+q_2 \sum _{\mathbf{a}=\mathbf{1},\mathbf{2},\mathbf{3}} [Z_{\mathbf{r} + \mathbf{a}}] 
+q_3 \sum _{\mathbf{a}=\mathbf{1},\mathbf{2},\mathbf{3}} [Z_{\mathbf{r} }Z_{\mathbf{r} + \mathbf{a}}] ,
\label{error}
\end{eqnarray}
where $q_1$, $q_2$, and $q_3$ are the error probabilities as functions of the temperature $T$, the Pauli $Z$ operator $Z_{\mathbf{b}}$ on the qubit at the position $\mathbf{b}$, and $[C]\rho \equiv C \rho C^{\dag}$, respectively.
The probability of other errors such as $Z_{\mathbf{r}} Z_{\mathbf{r} +\mathbf{a}} Z_{\mathbf{r} +\mathbf{a} '}$ is several orders of magnitude smaller than $q_{1,2,3}$.

To obtain the 3D cluster state for topological MBQC, as done in Ref.~\cite{LiKwek}, the five-qubit GHZ state $|{\rm GHZ}^5_{\mathbf{r}}\rangle$ is generated in a similar way by using spin-2 particles and composite particles of spin-1/2, as shown in Fig.~\ref{fig93} (b).
Instead of the spin-2 particles, spin-3/2 particles were employed in Ref.~\cite{FujiiMoriTher} to obtain the 3D cluster state shown in Fig.~\ref{fig93} (c).
After the filtering operation and local operations, the two four-qubit GHZ states are connected to obtain the five-qubit GHZ state for building the 3D cluster state.

By using the threshold for topologically protected MBQC, we can calculate the threshold temperatures $T=0.21\Delta$ and $T=0.18\Delta$ for the cases of spin-2 and spin-3/2 center particles, respectively~\cite{LiKwek,FujiiMoriTher}.
Accordingly, we can perform 
fault-tolerant universal measurement-based quantum computation
even on the thermal states of local two-body Hamiltonians 
at finite temperature.

\appendix
\chapter{Fault-tolerant quantum computation}

\section{Fault-tolerant syndrome measurements}
\label{Ap:FT_syndrome_meas}
In the case of the stabilizer code, the syndrome measurements are done by simply measuring the stabilizer operators.
Several QEC gadgets have been proposed to implement the stabilizer measurement fault-tolerantly \cite{DiViShor,KnillNature,Steane97}.

{\it DiVincenzo-Shor's gadget---}
The first QEC gadget was proposed by David DiVincenzo and Peter Shor, who used cat states as ancillae for the syndrome measurement \cite{DiViShor}.
It is based on an indirect measurement of the observable $A$ with the eigenvalues $\pm 1$:
\begin{equation}
\includegraphics[width=70mm]{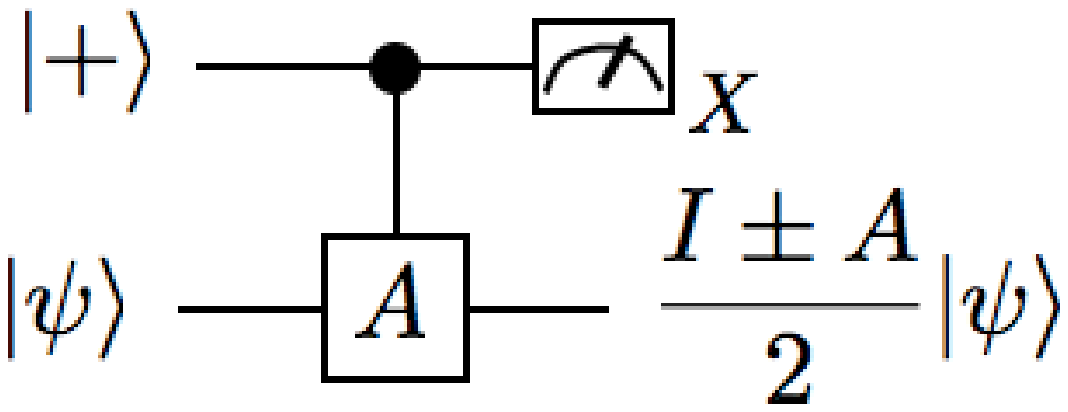}
\label{eq:proj_meas}
\end{equation}
For example, the stabilizer $S_{1}$ of the seven-qubit code can be measured as the transversal $X$ measurements of the corresponding physical qubits in the code block:
\begin{equation}
\includegraphics[width=70mm]{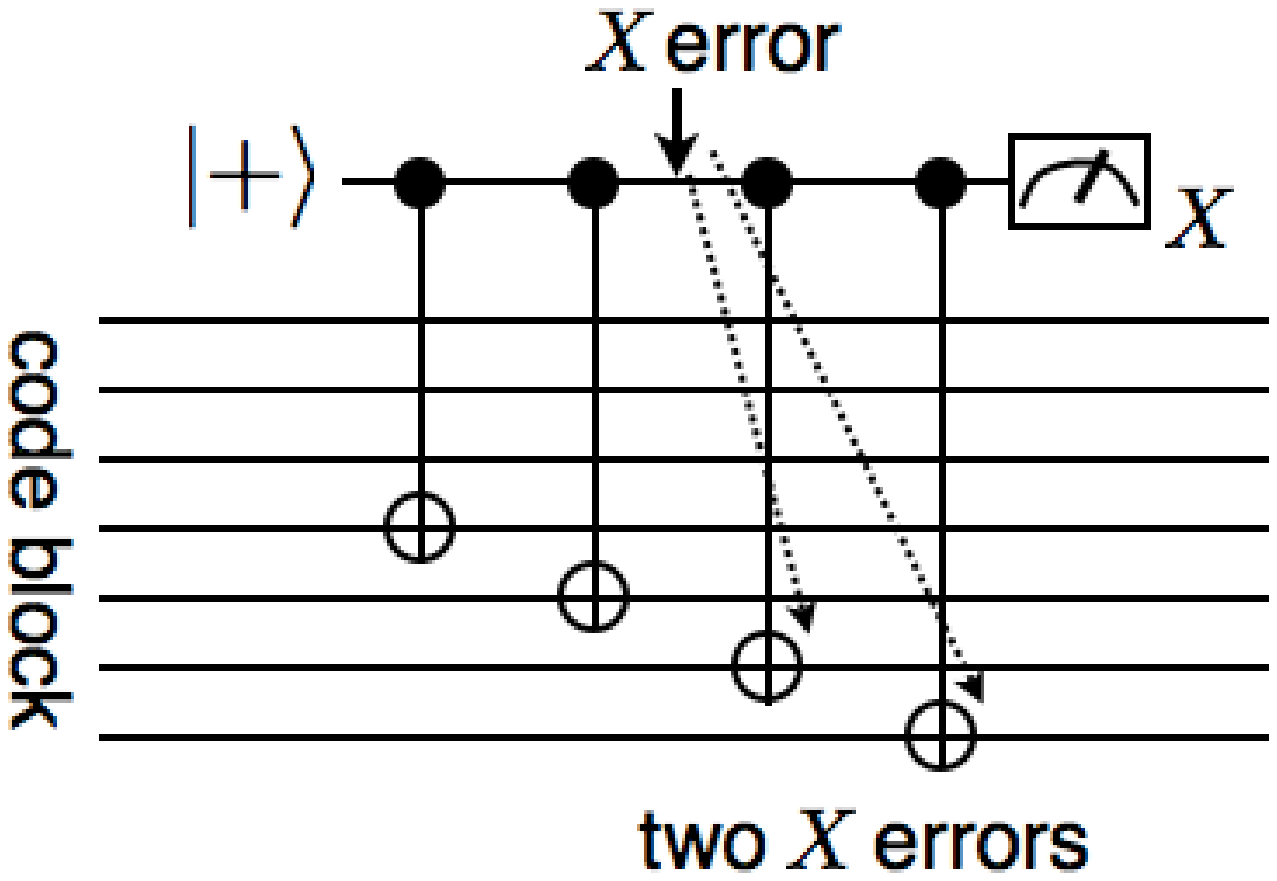}
\end{equation}
Unfortunately, this measurement is not fault tolerant, because the errors in the CNOT gates [$A=X$ in Eq.\ (\ref{eq:proj_meas})] are spread by the following CNOT gates, as shown in the above circuit.
To make it fault-tolerant, a cat state $|{\rm cat} \rangle = ( | 000 \cdots 0\rangle + | 111 \cdots 1\rangle)/\sqrt{2}$ is used as an ancilla for the measurement as follows:
\begin{equation}
\includegraphics[width=90mm]{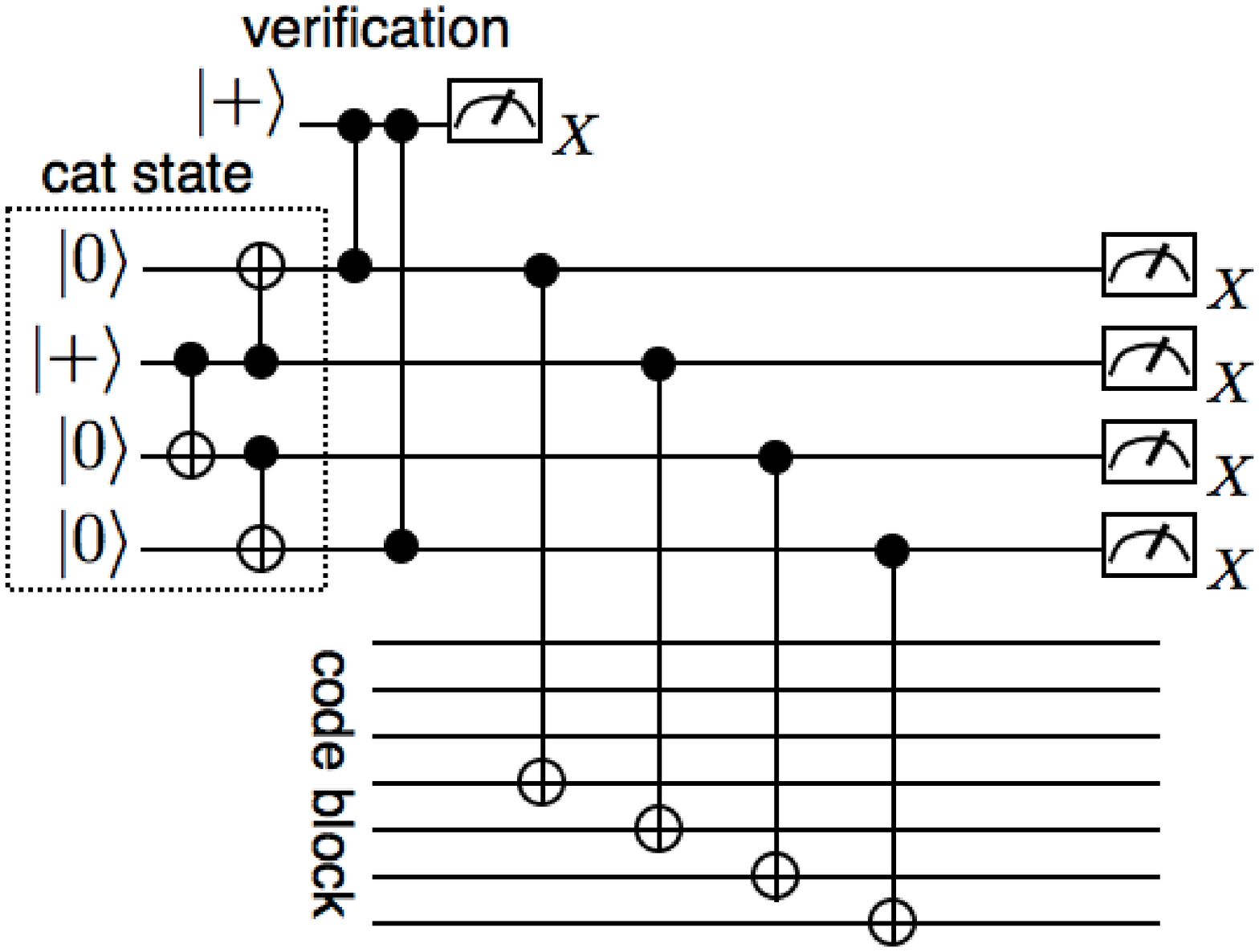}
\end{equation}
where the cat state is verified before connecting with the code state.
Because the qubits in the code block interact with different ancilla qubits, this measurement does not spread the errors in the CNOT gates.
Similarly, other stabilizers $S_{2}, \cdots, S_{6}$ are measured fault-tolerantly to obtain the error syndrome.
Instead of the verification, one can perform a suitable recovery operation by postprocessing the ancilla state after its interaction with the code state \cite{DiVincenzo07}:
\begin{equation}
\includegraphics[width=90mm]{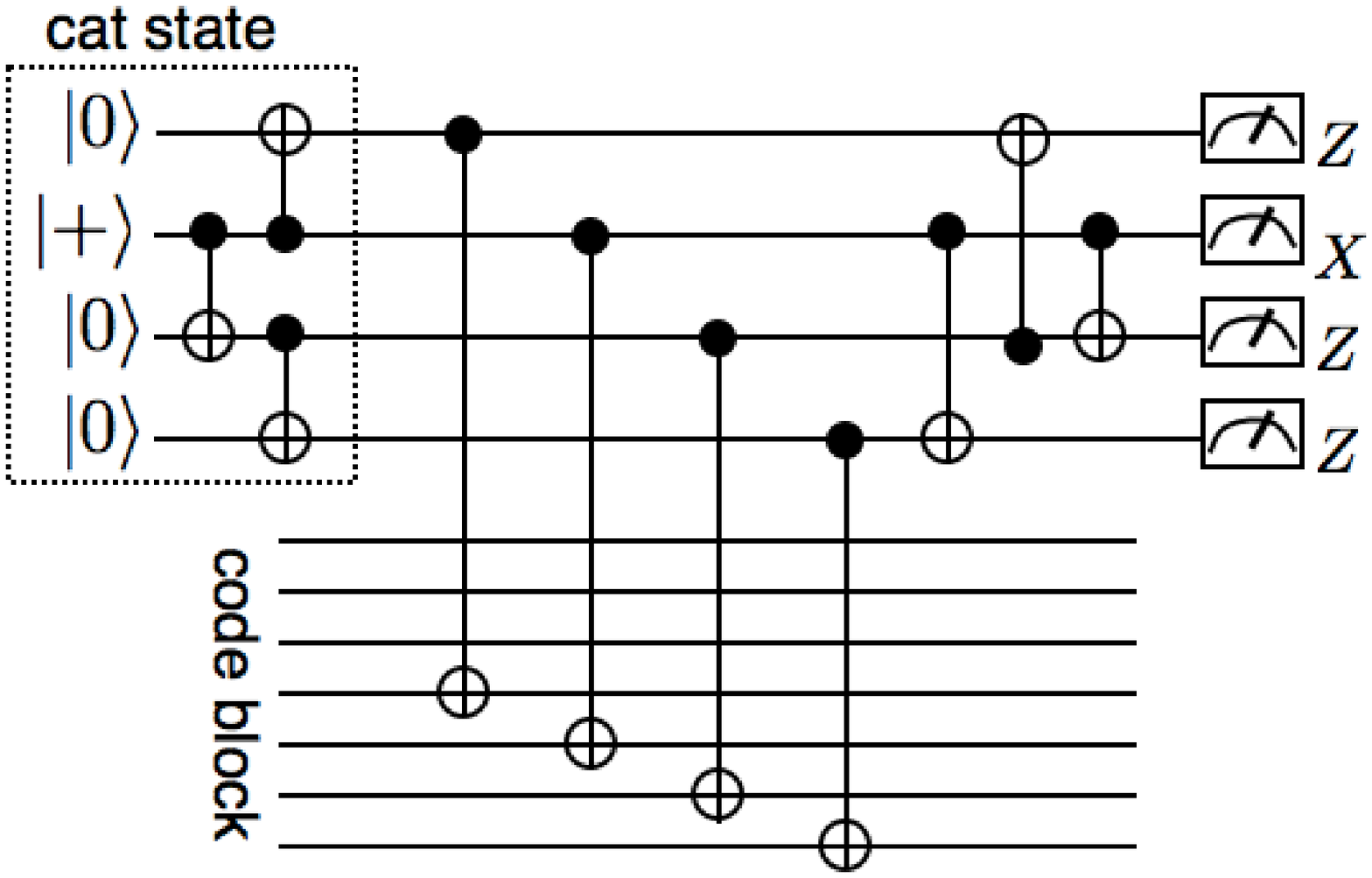}
\end{equation}
The DiVincenzo-Shor's QEC gadget and its improved version both require a lot of physical gate operations, which results in deterioration of the performance.

{\it Steane's gadget---}
Subsequently, a relatively simple QEC gadget was proposed by Andrew Steane \cite{Steane97}, where encoded ancilla states are used to extract the syndrome with transversal operations.
In particular, for the case of the CSS code, the logical code states can be used as ancilla states.
The following circuit executes the $Z$ and $X$ error syndrome extractions by using the ancilla $ | 0_L \rangle $ states,
\begin{equation}
\includegraphics[width=100mm]{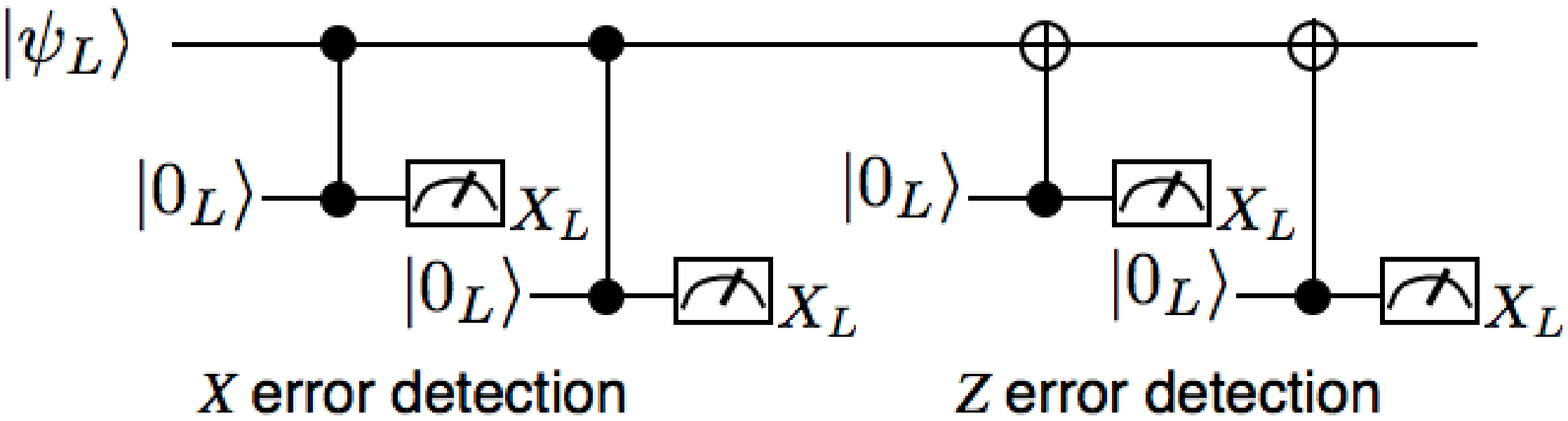}
\end{equation}
Because the ancilla states are the logical code states, one can obtain the error syndrome by simply measuring the ancilla states.
The syndrome extraction is repeated a couple of times to extract reliable error information.
An optimized way to extract the syndrome information was proposed in Ref.~\cite{Plenio97}, where the subsequent syndrome extraction is performed conditionally according to the preceding syndrome information.
For these schemes to work fault-tolerantly, the encoded ancilla $|0_{L}\rangle$ states have to be prepared with high fidelity.
This is achieved by using either verification or entanglement purification \cite{KnillNature,Steane97,DAB03,ADB05}.

{\it Knill's gadget---}
Another interesting QEC gadget was proposed by Emanuel Knill \cite{KnillNature}.
It is based on quantum teleportation as illustrated in the following circuit:
\begin{equation}
\includegraphics[width=100mm]{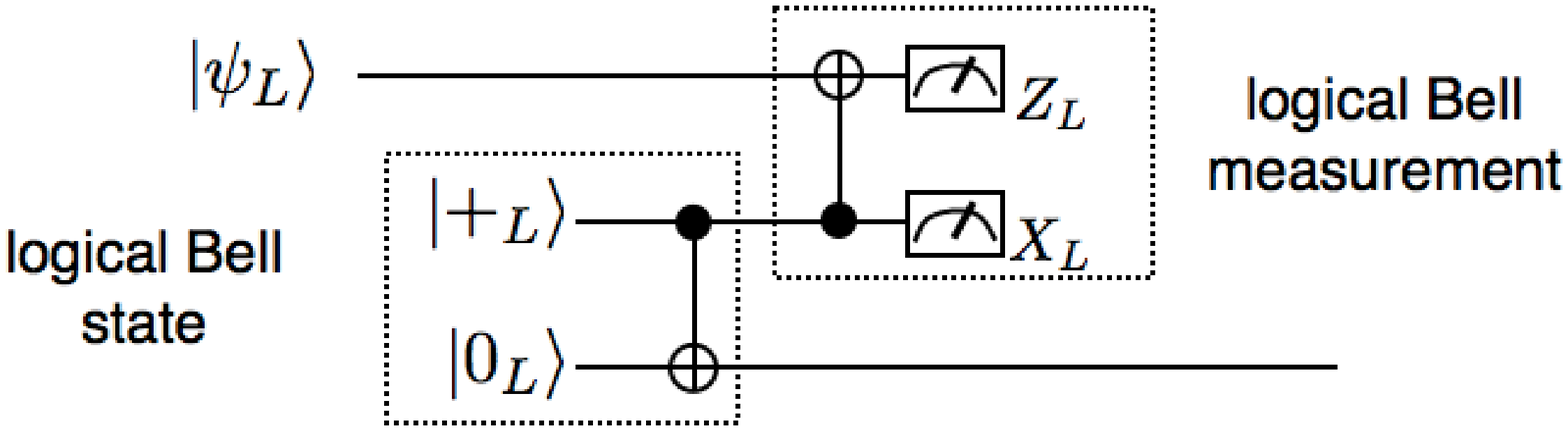}
\end{equation}
Here, the encoded data qubit $|\psi _{L}\rangle$ is teleported to the fresh encoded qubit of the ancilla Bell state.
Thus, the encoded ancilla Bell state has to be prepared with high fidelity by using verification or entanglement purification, similarly to the Steane's gadget.
The logical Bell measurement completes the teleportation, namely {\it error-correcting teleportation}.
There is no need to identify the error syndrome, but 
it is sufficient to find the logical measurement outcomes of the logical Bell measurement.
Thus, it is not necessary to repeat the syndrome extraction in this QEC gadget.
The outcome of the Bell measurement is properly propagated to the subsequent computation as the Pauli frame \cite{KnillNature,Dawson06a,Dawson06b}.

\section{Fault-tolerant gate operations}
Fault-tolerant computation is now executed by the logical gate operations that are followed by the QEC gadgets.
This is illustrated for the fault-tolerant CNOT gate as follows:
\begin{equation}
\includegraphics[width=70mm]{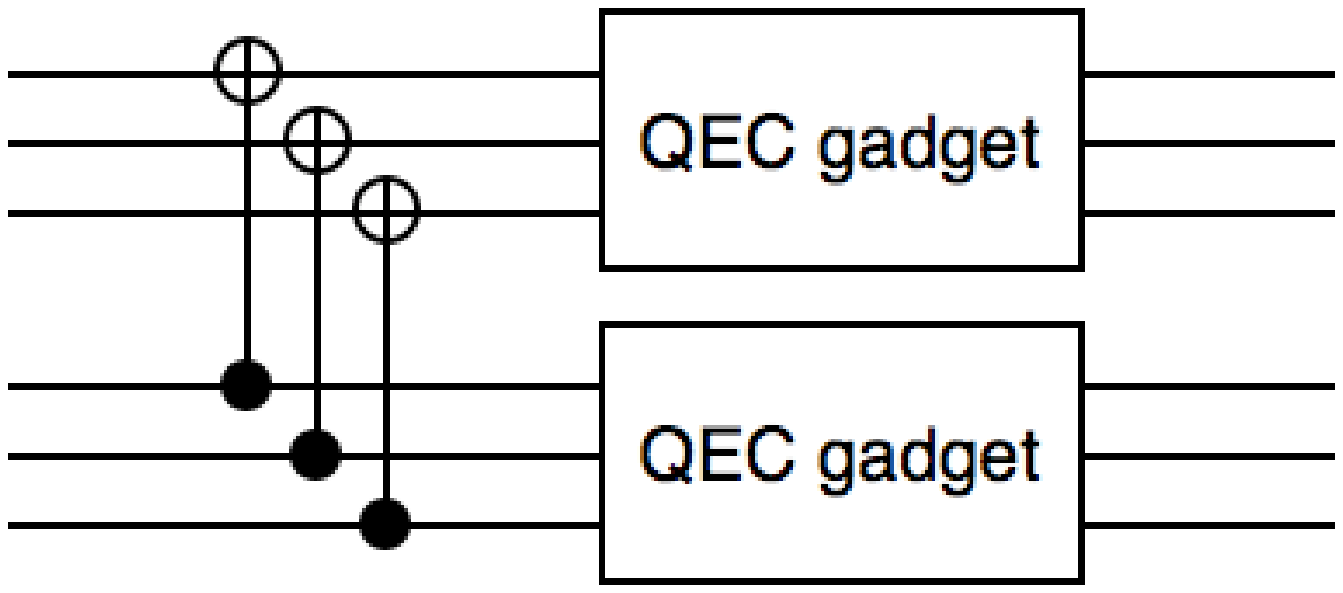}
\label{fig:logicalgate}
\end{equation}
where the code block is depicted as though it is a three-qubit code.
A QEC gadget is attached to each logical output of the transversal CNOT gate.
Because a single error never will be propagated as multiple errors in a fault-tolerant gate, a logical error is caused by two (or more) simultaneous physical errors. 
Denoting the number of such faulty pairs of error locations as $C$, the logical error probability is given by $Cp^2$, with $p$ being the physical error probability.
If the physical error probability is sufficiently small, $p < 1/C$, one can improve the accuracy of the gate and achieve a fault-tolerant computation.

\section{Concatenated quantum computation}
For a reliable computation of a large size, the logical error probability should be reduced arbitrarily.
This is done by a concatenated fault-tolerant computation \cite{Knill98a,Knill98b,Aharonov97,Aharonov08}.
In the concatenated computation, each physical gate operation is repeatedly replaced by a logical gate operation followed by QEC gadgets.

Suppose $A$ is a quantum computation, which consists of some physical gate operations.
Then the first level concatenated computation is defined by $\mathcal{C} (A)$, where the operation $\mathcal{C}$ indicates replacing each physical gate with a logical one, followed by QEC gadgets. 
For example, $\mathcal{C}(\textrm{CNOT})$ is described in the diagram (\ref{fig:logicalgate}).
The second level concatenated CNOT gate $\mathcal{C} \circ \mathcal{C} (\textrm{CNOT})$ is also described as follows:
\begin{equation}
\includegraphics[width=130mm]{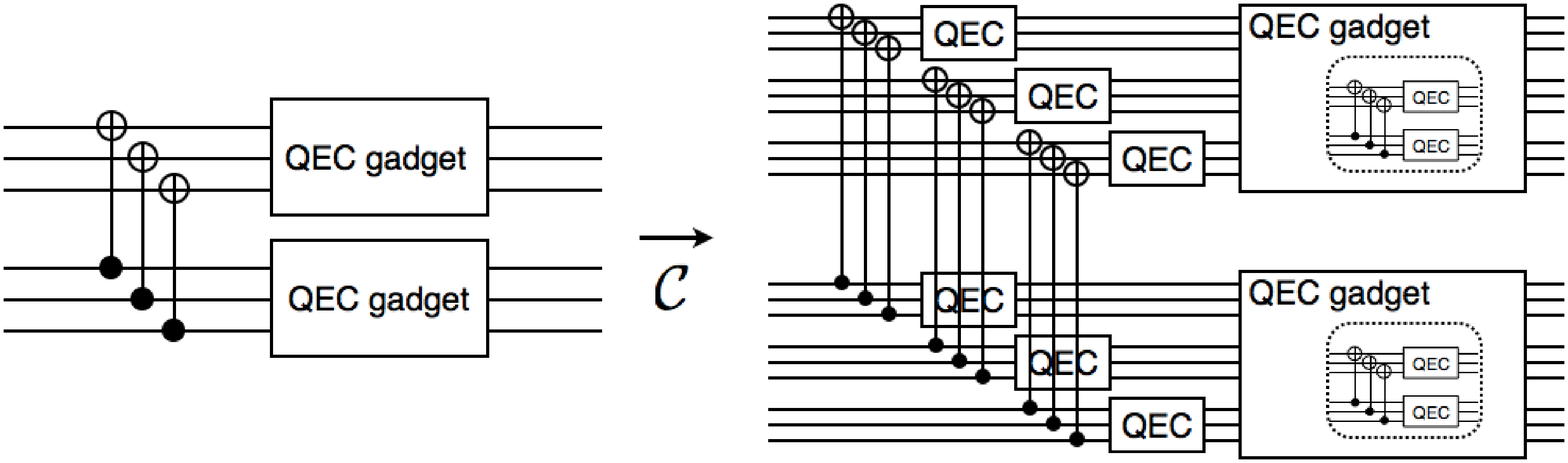}
\label{fig:concatenated gate}
\end{equation}
where each physical gate in the QEC gadgets is replaced by the logical one, followed by the QEC gadgets.
By repeating this procedure, the $l$th level concatenated computation of $A$ is given by $\mathcal{C}^{l}(A)$.
Specifically, for a physical gate operation $G$ (e.g., Hadamard, CNOT gate, etc.), $\mathcal{C}^{l}(G)$ is called the level-$l$ $G$ gate.
The $l$th level concatenated code state is called the level-$l$ qubit, denoted by$|0^{(l)}\rangle$, $|+^{(l)}\rangle$.

As mentioned previously, the logical error probability $p_{g}^{(1)}$ of the level-1 gate is given by 
\begin{equation}
p^{(1)} = C (p^{(0)})^{2},
\end{equation}
where $p^{(0)}=p$ and $C$ denotes the number of all faulty pairs of error locations.
The constant $C$ differs between the logical gates.
It is, however, sufficient to choose the maximum value.
Due to the self-similarity of the concatenation, the logical error probability of the level-2 gate is given in terms of $p^{(1)}$ by 
\begin{equation}
p^{(2)} = C (p^{(1)})^{2}.
\end{equation}
Similarly, the logical error probability $p^{(l)}$ of the level-$l$ gate is obtained recursively as
\begin{eqnarray}
p^{(l)} &=& C (p^{(l-1)})^{2}
\nonumber \\
&=& (Cp^{(0)})^{2^{l}}/C.
\end{eqnarray}
We conclude that if $p^{(0)}<p_{\textrm{th}} \equiv 1/C$, the logical error probability can be reduced super-exponentially with the concatenation level $l$.
This is the so-called {\it threshold condition}.

On the other hand, the resources usage, $R^{(l)}$, consumed for the level-$l$ gate is estimated roughly as
\begin{equation}
R^{(l)}=N^{l},
\end{equation}
where $N$ indicates the total number of physical gates in the level-1 gate.
Suppose that the size of the computation is $10^{n-1}=M$.
Then an accuracy of $p^{(l)} < 10^{-n}$ is required for each logical gate at the highest level. 
The total resources to perform a reliable quantum computation of size $M$ amount to 
\begin{eqnarray}
R_{\textrm{tot}}&=&N^{\bar{l}} M = \left( \frac{n}{\log _{10} (Cp^{(0)})^{-1}} \right)^{\log _{2} N}M = \textrm{poly}( \log (M))M,
\end{eqnarray}
where $\bar{l} \simeq \log _{2} \left[ n/\log _{10}(Cp^{(0)}) \right] $ is the number of levels necessary to achieve the required accuracy.
This result clearly shows that if the physical error probability $p^{(0)}$ is smaller than $p_{\textrm{th}}$, one can execute quantum computation to an arbitrary accuracy with only polylogarithmic overhead.
This is the celebrated {\it threshold theorem} and the critical value $p_{th}$ is called the {\it noise threshold} \cite{NielsenChuang,Kitaev97,Knill98a,Knill98b,Aharonov97}. 
The noise thresholds have been calculated to be about $10^{-4}-10^{-2}$ for several fault-tolerant schemes under varying degrees of assumption and rigor
\cite{Knill98a,Knill98b,Aharonov97,Aharonov08,Steane03,KnillNature,Aliferis06,Aliferis07,Aliferis08,Aliferis09,Steane99,Cross09}.

\chapter{Decoding stabilizer codes}
\label{Ap:Decode}
Consider an $n$-qubit stabilizer code, whose stabilizer group is given by $\mathcal{G}=\{ G_i \} $.
The group consisting of the logical operators is denoted by $\mathcal{L}=\{ L_i \}$.
Suppose a Pauli product $E \in \{ I,X,Y,Z\}^{\otimes n}$ acts as an error on the stabilizer code state.
The error syndrome, a set of eigenvalues of the stabilizer generators, of an error $E$ is denoted by $S=\mathcal{S}(E)$. 	
For each error syndrome $S$, we define a pure error operator $R(S) \in \{ I,X,Y,Z\}^{\otimes n}$ such that $S= \mathcal{S}[R(S)]$.
The pure error operator $R(S)$ is chosen arbitrarily as long as $S= \mathcal{S}[R(S)]$.
The error $E$ is decomposed uniquely into logical, stabilizer, and pure error operators:
\begin{eqnarray}
E=L_i G_j R[\mathcal{S}(E)].
\end{eqnarray}
We define a decoding map $\mathcal{D}$, which computes the logical operator $L_i$ from the error $E$, i.e., $L_i=\mathcal{D}(E)$.

The decoding problem consists of finding an optimal logical operator $L_i$ that maximizes the posterior probability $P(L|S)$:
\begin{eqnarray}
{\rm arg} \max _{L_i \in \mathcal{L}} P(L_i|S).
\end{eqnarray}
The posterior probability is given by
\begin{eqnarray}
P(L_i| S) &=& \frac{1}{\mathcal{N}}\sum _{E} P(E) \delta [L_i=\mathcal{D}(E)] \delta [S=\mathcal{S}(E)]
\\
&=&  \frac{1}{\mathcal{N}} \sum _{G_j \in \mathcal{G}} P[L_i G_j R(S)].
\end{eqnarray}
where $1/\mathcal{N}$ is a normalization factor, $\delta (\cdots)$ is an indicator function, and $P(E)$ is the probability of the error $E$.
In general, computing $P(L|S)$ is hard because the summation over all stabilizer operators, $\sum _{G_j}$, takes an exponential time.
However, if the code has a good structure, such as a concatenated code, we can manage it efficiently.

A concatenated code is defined recursively using the logical qubits at the lower level as physical qubits at the higher level.
Using the logical Pauli operators $L_{i} ^{(k)} \in \mathcal{L}^{(k)}$ at the $k$th level ($k=1,2,..$), we define a stabilizer group $\mathcal{G}^{(k+1)}=\{G_{j}^{(k+1)}\}$ and logical operators $\mathcal{L}^{(k+1)}=\{ L_{i}^{(k+1)}\}$ at the $(k+1)$th level.
The whole stabilizer group of the concatenated code is given by the union of all stabilizer groups $ \cup _{k=1}^{l} \mathcal{G}^{(k)}$, where we consider the $l$th concatenated code.
A logical operator of the $l$th concatenated code is given by a logical operator at the highest level, $L_{i}^{(l)} \in \mathcal{L}^{(l)}$.
At each level, we define a level-$k$ error $E^{(k)}$.
In the case of $k=0$, the level-$0$ error $E^{(0)}\in \{ I, X, Y, Z\}^{\otimes n}$ is a physical error.
At a higher level, the level-$k$ error $E^{(k)}$ is a level-$k$ logical operator, which will be defined later.
At each level, we define an error syndrome $S^{(k)} = \mathcal{S}^{(k)} (E^{(k-1)})$, i.e., a set of eigenvalues of the level-$k$ stabilizer operators $\{G_j^{(k)}\}$ under the level-$(k-1)$ error $E^{(k-1)}$.
The level-$k$ pure error operator $R^{(k)}(S^{(k)})$ is defined arbitrarily such that $\mathcal{S}^{(k)} [R^{(k)}(S^{(k)})] = S^{(k)}$.

The level-$k$ error $E^{(k)}$ is now defined recursively as follows.
Any physical error $E^{(0)}$ can be decomposed into the stabilizer, logical, and pure error operators of level-1:
\begin{eqnarray}
E^{(0)} = G_{j}^{(1)} L_{i}^{(1)} R^{(1)}(S^{(1)}).
\end{eqnarray}
The obtained level-1 logical operator $L_{i}^{(1)} = E^{(0)}G_{j}^{(1)}R^{(1)}(S^{(1)})$ is further regarded as the level-1 error $E^{(1)} \equiv L_i^{(1)}$.
Recursively, the level-$k$ error $E^{(k)}$ is decomposed into the stabilizer, logical, and pure error operators of level-$(k+1)$:
\begin{eqnarray}
E^{(k)}  = G_{j}^{(k+1)} L_{i} ^{(k+1)} R^{(k+1)}(S^{(k+1)}).
\label{eq:level-k-error}
\end{eqnarray}
Then, we define the level-$(k+1)$ error $E^{(k+1)} = L_i^{(k+1)}=E^{(k)}G_{j}^{(k+1)} R^{(k+1)}(S^{(k+1)})$.
Let $\mathcal{D}^{(k)}$ be such a level-$k$ decoding map $L_{i}^{(k)} = \mathcal{D}^{(k)} (E^{(k-1)})$, which computes the level-$k$ logical operator from the level-$(k-1)$ error $E^{(k-1)}$.
At the highest level, we have a decomposition
\begin{eqnarray}
E^{(0)}=L_{i}^{(l)} \prod _{k=1}^{l} G_{j_k}^{(k)} R^{(k)}[\mathcal{S}^{(k)}(E^{(k-1)})],
\label{eq:error_decomposition}
\end{eqnarray} 
where $E^{(k)}$ is defined by Eq.\ (\ref{eq:level-k-error}).

The decoding is performed by maximizing the posterior probability of the logical operator $L_i^{(l)}$, conditioned on the union of the error syndrome 
$\bar S^{(l)} = \cup_{k=1}^{l} S^{(k)}$ up to the $l$th concatenation level,
\begin{eqnarray}
P(L_i^{(l)}|\bar S^{(l)}) =\frac{1}{\mathcal{N}} \sum _{G_{j_l}^{(l)},G_{j_{l-1}}^{(l-1)},...,G_{j_{1}}^{(1)}} P(E^{(0)}) ,
\label{eq:posterior_stab} 
\end{eqnarray}
where $E^{(0)}$ is given by Eq.\ (\ref{eq:error_decomposition}).
Using the hierarchal structure, Eq.\ (\ref{eq:posterior_stab}), can be rewritten as
\begin{eqnarray}
&&P(L_i^{(l)}|\bar S^{(l)})
\\
&=&\sum _{L_{i_{l-1}}^{(l-1)} \in \mathcal{L}^{(l-1)}}  P(L_i^{(l)}|\bar S^{(l)},L_{i_{l-1}}^{(l-1)}) P(L_{i_{l-1}}^{(l-1)}|\bar S^{(l)})
\nonumber \\
&=&
\sum _{L_{i_{l-1}}^{(l-1)}\in \mathcal{L}^{(l-1)}} \delta \left[  L_{i_{l}}^{(l)} = \mathcal{D}^{(l)}(L_{i_{l-1}}^{(l-1)}) \right] 
\frac{P(L_{i_{l-1}}^{(l-1)},\bar S^{(l)})}{P(\bar S^{(l)})}
\nonumber \\
&=&
\sum _{L_{i_{l-1}}^{(l-1)}\in \mathcal{L}^{(l-1)}}  \delta [  L_{i_{l}}^{(l)} = \mathcal{D}^{(l)}(L_{i_{l-1}}^{(l-1)}) ] 
\delta [  S^{(l)} = \mathcal{S}^{(l)}(L_{i_{l-1}}^{(l-1)}) ] 
\frac{P(L_{i_{l-1}}^{(l-1)},\bar S^{(l-1)})}{P(\bar S^{(l)})}
\nonumber \\
&=&
\sum _{L_{i_{l-1}}^{(l-1)}\in \mathcal{L}^{(l-1)}}  \frac{\delta [  L_{i_{l}}^{(l)} = \mathcal{D}^{(l)}(L_{i_{l-1}}^{(l-1)}) ] \delta [S^{(l)}= \mathcal{S}^{(l)}(L_{i_{l-1}}^{(l-1)}) ]} {P(\bar S^{(l)}|\bar S^{(l-1)})} P(L_{i_{l-1}}^{(l-1)}| \bar S^{(l-1)})
 \nonumber \\
\end{eqnarray}
If the physical errors occur independently for each qubit, we can factorize the posterior probability of the $(l-1)$th level into posterior probabilities for level-$(l-1)$ code blocks
\begin{eqnarray}
 P(L_{i_{l-1}}^{(l-1)}| \bar S^{(l-1)}) = \prod _{j} P(L_{i_{l-1}}^{(l-1,j)}| \bar S^{(l-1,j)}),
\end{eqnarray} 
where $L_{i_{l-1}}^{(l-1,j)}$ is a level-$(l-1)$ logical operator acting on the $j$th code block, and $\bar S^{(l-1,j)}$ is a level-$(l-1)$ syndrome with respect to the stabilizer operator on the $j$th code block.
By repeating this procedure, we can rewrite the posterior probability Eq.\ (\ref{eq:posterior_stab}) as a summation over logical operators defined at each concatenation level $L_{i_{k}}^{(k,j)} \in \mathcal{L}^{(k,j)}$, where we have defined the group of the level-$k$ logical operators on the $j$th code block, $\mathcal{L}^{(k,j)}$.

The summation can be reformulated as a marginalization problem on a factor graph, which is a bipartite graph consisting of two types of nodes, circles, and boxes~\cite{Poulin06}.
The variable $x_c$ and a function $f_b(\partial b)$ are assigned on each circle $c$ and box $b$.
Here, $\partial b$ is a set of circles neighboring $b$.
Specifically, in the present case, the factor graph is a tree graph.
The marginal distribution on a tree factor graph can be computed efficiently by using the brief-propagation method as follows.
From circles to boxes, we pass a message,
\begin{eqnarray} 
\mu _{c\rightarrow b}(x_c)= \prod _{b' \in  \delta c \backslash b} \nu _{b'\rightarrow c}(x_c).
\end{eqnarray}
Then from boxes to circles, we pass another message
\begin{eqnarray} 
\nu _{b\rightarrow c}(x_c)= \sum _{\delta b \backslash x_c}\prod _{b} f_{b}(\delta b) \prod _{c'\in \delta b \backslash c} \mu _{c' \rightarrow b}(x_c').
\end{eqnarray}
By repeating these procedures alternatively, we can obtain $P(L_i^{(l)}|\bar S^{(l)})$ as a message $\nu_{b\rightarrow c}(x_c)$, where the $\mu _{c\rightarrow b}(x_c)$ and 
$\nu _{b\rightarrow c}(x_c)$ are updated from the bottom leaf nodes to the top root node, while using marginal distributions.
By replacing $\mu _{c\rightarrow b}(x_c)$ and $f_b(\delta b)$ with $P(L_{i_{k-1}}^{(k-1,j)}| \bar S^{(k-1,j)})$ and 
$\delta [  L_{i_{k}}^{(k)} = \mathcal{D}^{(k)}(L_{i_{k-1}}^{(k-1)}) ] \delta [S^{(k)}= \mathcal{S}^{(k)}(L_{i_{k-1}}^{(k-1)}) ]$, we obtain the posterior probability $P(L_i^{(l)}|\bar S^{(l)})$ as $\nu _{b\rightarrow c}(x_c)$ at the top root node.
In this way, decoding by maximizing the posterior probability can be executed efficiently for the concatenated stabilizer codes~\cite{Poulin06}.
\printindex

\bibliographystyle{alpha}
\bibliography{book}


\end{document}